\documentclass{jfm}
\usepackage{graphicx,color}
\usepackage[shortlabels]{enumitem}
\usepackage{epstopdf, epsfig,comment}
\usepackage{mathtools}
\usepackage{bm}
\usepackage{amsmath, amssymb, xfrac, enumitem}
\usepackage[dvipsnames]{xcolor}
\usepackage[normalem]{ulem}
\usepackage{lscape}
\usepackage{appendix}
\usepackage{tikz}

\newcommand{\rhom}{\overline{\rho}}
\newcommand{\At}{$A$ }

\newcommand{\ret}{$Re_t$ }
\newcommand{\rez}{$Re_0$ }

\def\lp{\langle}
\def\rp{\rangle}

\shorttitle{Atwood and Reynolds number effects on HVDT}
\shortauthor{Denis Aslangil, Daniel Livescu and Arindam Banerjee}

\title{Atwood and Reynolds numbers effects on the evolution of buoyancy-driven homogeneous variable-density turbulence}

\author{Denis Aslangil \aff{1,2} \thanks{denis.aslangil@gmail.com},
  Daniel Livescu \aff{2} \thanks{livescu@lanl.gov}
 \and Arindam Banerjee \aff{1} \thanks{arb612@lehigh.edu}
}
\affiliation{\aff{1}Department of Mechanical Engineering and Mechanics, Lehigh University,
Bethlehem, PA 18015, USA
\aff{2}Los Alamos National Laboratory, Los Alamos, NM 87545, USA}
\begin{document}

\maketitle

\begin{abstract}
The evolution of buoyancy-driven homogeneous variable-density turbulence (HVDT) at Atwood numbers up to 0.75 and large Reynolds numbers is studied by using high-resolution Direct Numerical Simulations. 
To help understand the highly non-equilibrium nature of buoyancy-driven HVDT, the flow evolution is divided into four different regimes based on the behavior of turbulent kinetic energy derivatives. The results show that each regime has a unique type of dependency on both Atwood and Reynolds numbers. It is found that the local statistics of the flow based on the flow composition are more sensitive to Atwood and Reynolds numbers compared to those based on the entire flow. It is also observed that at higher Atwood numbers, different flow features reach their asymptotic Reynolds number behavior at different times. 
The energy spectrum defined based on the Favre fluctuations momentum has less large scale contamination from viscous effects for variable density flows with constant properties, compared to other forms used previously. The evolution of the energy spectrum highlights distinct dynamical features of the four flow regimes. Thus, the slope of the energy spectrum at intermediate to large scales evolves from $-7/3$ to $-1$, as a function of the production to dissipation ratio. The classical Kolmogorov spectrum emerges at intermediate to high scales at the highest Reynolds numbers examined, after the turbulence starts to decay. Finally, the similarities and differences between buoyancy-driven HVDT and the more conventional stationary turbulence are discussed and new strategies and tools for analysis are proposed.

\end{abstract}

\section{\label{sec:Intro} Introduction}

The mixing of two or more miscible fluids with different densities (or molar masses) is of fundamental interest due to the occurrence in atmospheric and oceanic flows \citep{MOLCHANOV2004559,Adkins1769,Wunsch_2004_doi:10.1146/annurev.fluid.36.050802.122121}, supernova formations \citep{Gull_1975_doi:10.1093/mnras/171.2.263,Colgate_1966_ApJ...143..626C,Nouri_etal_PAS19}, combustion applications in ramjet engines \citep{GIVI1989,clemens_mungal_1995,Sellers_Chandra_1997_doi:10.1108/02644409710157596} and high energy density processes like inertial confinement fusion \citep{Lindl_1995_doi:10.1063/1.871025,lindl1998inertial,Nakai_1996_0034-4885-59-9-002,Nakai_2004_0034-4885-67-3-R04}. These flows are traditionally referred to as variable density (VD) flows in the scientific literature. Unlike incompressible single-fluid flows, the velocity field in VD flows is tightly coupled to the density field and is not divergence free, even in the incompressible limit. The VD mixing process is directly related to flow dynamics and plays a vital role in the flow evolution, as compositional changes lead to significant  effects on both the mixing behavior and the resultant turbulence structure \citep{livescu2007,chung_pullin_2010,gat_2017,rao_caulfield_gibbon_2017, aslangil_book_ch, aslangilPhysicaD, Livescu2020}. Here, we investigate buoyancy-driven homogeneous VD turbulence (henceforth referred to as HVDT) by high-resolution Direct Numerical Simulations (DNS) in triply-periodic domain sizes up to $2048^3$.

Introduced by \citet{batchelor1992} to investigate buoyancy-driven turbulence under Boussinesq approximation, homogeneous buoyancy driven turbulence is a canonical fluid flow problem; the presence of triply-periodic boundaries eliminates the inhomogeneities that may arise in the flow due to the mixing layers edge and/or wall effects \citep{batchelor1992,SandovalPhd,livescu2007}. Initial large pure fluid regions with different densities start to move in opposite directions when acceleration is applied to the domain. The main turbulent kinetic energy production mechanism is through the product of the mean pressure gradient and the mass flux \citep{livescu2007}. At the same time, vorticity is produced through the baroclinic mechanism due to misalignment of pressure and density gradients. The non-dimensional number representing the density contrast between the two fluids is the Atwood number, $A$:
\begin{equation} \label{Eq:At}
A = \frac{\rho_2-\rho_1}{\rho_2+\rho_1}\qquad \Rightarrow \qquad \frac{\rho_2}{\rho_1}=\frac{1+A}{1-A}
\end{equation}
where, $\rho_1$ and $\rho_2$ are the densities of the light and heavy fluids, respectively. This study covers a broad range of non-dimensional density ratios from $1.105:1$ to $7:1$ that corresponds to Atwood number values from $0.05$ to $0.75$, and spans both the traditional Boussinesq case ($A=0.05$) and the strongly non-Boussinesq case ($A=0.5$ and $0.75$).

HVDT is of interest to the turbulence community also because it mimics the core of the mixing layer formed by the acceleration-driven Rayleigh-Taylor instability (RT/RTI) \citep{LordRayleigh,LordTaylor} and the shock-driven Richtmyer-Meshkov instability (RM/RMI) \citep{Richtmey_doi:10.1002/cpa.3160130207,Meshkov1969}. In addition, it has some similarities with VD jet flows \citep{charonko_prestridge_2017} and VD shear (mixing) layers \citep{almagro_garcia-villalba_flores_2017,schwarzkopf2016tls,baltzer_livescu_2018}.
In RTI, VD mixing occurs between light and heavy fluids in the presence of an acceleration field, where the heavy fluid sits on top of the light fluid \citep{ristorcelli_clark_2004,BANERJEE20093906}. The generality of the HVDT findings was later confirmed by comparisons with the classical RTI studies by \citet{Livescu_Cabot_Cook,Livescu_HVDT_New_Ph}. Higher Atwood number simulations by \citet{Livescu_variable_accel_2011}, $A$ up to $0.75$, and RTI gas-channel experiments by \citet{banerjee_kraft_andrews_2010}, $A$ up to $0.6$, and \citet{akula_ranjan_2016}, $A$ up to $0.73$, also confirm those findings. An important aspect of these flows is the heavy-light fluid mixing asymmetry \citep{livescu2008,livescu2009mav,livescu2013nst}, manifested, for example, in different penetration levels of light and heavy fluids, indicating different turbulent features of the flow for the bubble side (mostly composed of light fluid) and spike side (mostly composed of heavier fluid). In some important applications like inertial confinement fusion, blast waves, and astrophysical flows such as Type Ia supernovae, RTI is driven by time varying accelerations. Such effects have been studied in the context of RTI by \citet{Dimontte_Schneider,Livescu_variable_accel_2011,livescu_variable_accel_2012,Ramaprabhu_ADA,Denis_PRE}. HVDT has certain connections with RTI under variable-acceleration as the HVDT evolution has unsteady turbulent kinetic energy ($E_{TKE}$) behavior with $dE_{TKE}/dt$ and/or $d^2E_{TKE}/dt^2$ phases which are reminiscent of the RTI problem with $ d g_i/dt  \neq 0$. VD RMI is also another application of interest; the flow is driven by one or more-shocks that pass through the interface between light and heavy fluids \citep{Martin_Broillette_RMI_rev_2002,O_S_RMI,SCHILLING2010595,ZHOU2017_1,ZHOU2017_2}. In RMI, shock generated turbulence decays, while the fluids molecularly mix \citep{Nishihara_2010_RMI,Bailie2012_RMI}. This behavior has similarities with the late-time decay of HVDT where the buoyancy forces weaken and buoyancy-generated VD turbulence decays. The late time decay stage of buoyancy-driven HVDT, where the VD effects are minimal, also has some similarities with buoyancy-driven Rayleigh-Be\'nard convection (RB), which occurs between two plates such that the bottom plate is heated \citep{RBI_book_doi:10.1142/3097, RBI_RevModPhys.81.503}. 

Several studies have used the homogeneous configuration to investigate VD effects on buoyancy-driven turbulence \citep{SandovalPhd,Sandoval1997,livescu2007,livescu2008}. \citet{SandovalPhd} and \citet{Sandoval1997} studied VD dynamics of the buoyancy-driven turbulence at fairly low-Reynolds numbers ( domain size $\approx 128^3$). \citet{livescu2007,livescu2008} used the homogeneous configuration to investigate VD effects on buoyancy-driven turbulence and turbulent mixing with density-ratios up to $3:1$ and higher Reynolds numbers. \citet{chung_pullin_2010} studied VD mixing in a stationary HVDT configuration, where the flow is continuously fed pure heavy fluid from the top and pure light fluid from the bottom of the domain leading to a stationary state. More recently, \citet{gat_2017} numerically studied VD turbulence in vertical fluid columns with different densities that are moving in opposite directions due to the presence of an acceleration field. In their study, the buoyancy forces generate a VD shear layer, while the two fluids are molecularly mixing.

\subsection {\label{sec:highlights} Highlights of this study}

In this paper, buoyancy-driven HVDT is investigated by high-resolution DNS, up to domain sizes of $2048^3$, and covers a broad range of density ratios from $1.105$$:$$1$ (corresponding to Atwood number value of $0.05$, which is close to the traditional Boussinesq case) to $7$$:$$1$ (corresponding to Atwood number value of $0.75$). The simulations significantly extend the range of parameters from \citet{livescu2007,livescu2008} in terms of both the Reynolds and Atwood numbers. This HVDT flow provides unique challenges for turbulence modeling compared to,  for  example, the more complex RTI because it does not reach a self-similar stage. Due to the unsteady nature of the mean variables, since turbulence  undergoes a rapid initial increase in the kinetic energy followed by a buoyancy mediated decay, the flow encompasses the generation, growth, and decay of buoyancy-driven turbulence. Such a behavior allows us to compare this canonical flow to a wide range of VD flow applications such as RTI under constant acceleration, RTI under time varying acceleration, RMI,  VD  jets, and VD shear  layers. 

To streamline the connection between the idealized HVDT flow and various applications, as well as isolate specific physical behaviors, we have identified four different regimes of the flow based on time derivatives of turbulence kinetic energy ($E_{TKE}$). These regimes are named according to the growth/decay of $E_{TKE}$ in the flow as:

\begin{enumerate}[I] \addtolength{\itemindent}{1cm}
\item ~~~~Explosive growth, when $dE_{TKE}/dt>0$, $d^2E_{TKE}/dt^2>0$,
\item ~~~~Saturated growth, when $dE_{TKE}/dt>0$, $d^2E_{TKE}/dt^2<0$, 
\item ~~~~Fast decay, when $dE_{TKE}/dt<0$, $d^2E_{TKE}/dt^2<0$, and 
\item ~~~~Gradual decay, when $dE_{TKE}/dt<0$, $d^2E_{TKE}/dt^2>0$. 
\end{enumerate}

Both high-Reynolds number and large density-ratio effects on buoyancy-driven HVDT, that add additional non-linearities to the problem by causing significant asymmetries within the flow, are investigated. Unlike previous studies that mostly examined the global behavior, we compare different regions in the flow by using conditional expectations to evaluate the asymmetric behavior of density and velocity fields. In addition,  the transport equation for the density-velocity joint mass density function (jMDF) is analyzed. Such information can also be used in the context of modeling using PDF methods \citep{POPE1985119}. Furthermore, the energy conversion rates are compared for different \At and turbulent Reynolds numbers during the growth regimes. Our largest resolution simulations with domain sizes of $1024^3$ and $2048^3$ show that the results become converged with increasing the Reynolds number. This allows us to comment on the notion of the mixing transition for VD turbulence. The last regime, gradual decay, exhibits long time buoyancy-assisted turbulence decay, which is different than the traditional non-buoyant turbulence decay. The results show that the buoyancy-assisted $E_{TKE}$ decay occurs at non-decaying turbulent Reynolds number. Finally, we suggest the Favre fluctuation momentum as a proper way of investigating the evolution of the energy spectra in VDT and the turbulence spectral evolution during the four flow regimes is  characterized. 

The rest of the paper is organized as follows. In Section 2, we first present the governing equations, computational approach, and simulation cases. Next, we define some useful mathematical tools to analyze our flow. Section 3 presents the global evolution of the buoyancy-driven HVDT. Connections to applications such as RTI and RMI are also introduced in Section 3. The flow asymmetries during the four different regimes (I-IV) are discussed in Section 4.
Section 5 examines the spectral evolution of HVDT. Finally, section 6 summarizes our findings, and discusses the main conclusions of the paper.

\section{Problem formulation, simulations cases, and tools to analyze the flow} \label{Sec:useful}

Throughout the paper, the superscript $^*$ is used for instantaneous values; capital Roman letters or angle brackets are used for mean values, and lower-case Roman letters or primes are used for Reynolds fluctuations. The velocity decomposition in index notation is $u^*_i=U_i+u_i$, while the density decomposition is $\rho^*=\rhom +\rho$. Moreover, Favre (density weighted) averaged values are denoted using tilde $\tilde{}$ and Favre fluctuations are denoted using double primes $^{''}$, such that $u^*_i=\tilde{U}_i+u^{''}_i$, with $\tilde{U}_i=\langle \rho^* u^*_i\rangle /\rhom$.

\subsection {\label{sec:governing} Governing equations}

The governing equations describing the mixing of two fluids with different densities can be derived from the fully compressible Navier-Stokes equations with two miscible fluids (species) with full diffusion and heat flux operators, as the limit of infinite speed of sound \citep{livescu2013nst}. In non-dimensional form, they can be written as \citep{cook_dimotakis_2001,livescu2007}:

\begin{equation} \label{Eq:continuity}
\rho^*_{,t}+(\rho^*u^*_j)_{,j}=0,
\end{equation}

\begin{equation} \label{Eq:moment}
(\rho^*u^*_i)_{,t}+(\rho^*u^*_iu^*_j)_{,j}=-p^*_{,i}+\tau^*_{ij,j}+\frac{1}{Fr^2}\rho^*g_i,
\end{equation}
where, $\rho^*$ is the density, $u_i^*$ is the velocity in direction $i$, $p^*$ is the pressure, $g_i$ is the gravity (acceleration) in direction $i$, and the stress tensor is assumed Newtonian, $\tau^*_{ij}=(\rho^*/Re_0)(u^*_{i,j}+u^*_{j,i}-(2/3)u^*_{k,k}\delta_{ij}$). 
In incompressible VD turbulence, the divergence of velocity is not zero due to the change in specific volume during mixing and can be written as:
\begin{equation} \label{Eq:divergence}
u^*_{j,j}=-\frac{1}{Re_0Sc}ln\rho^*_{,jj}.
\end{equation}
This relation can also be derived either from the mass fraction transport equations or energy transport equation as the infinite speed of sound limit \citep{livescu2013nst}. For the binary case, mass conservation for two fluids with constant densities, $\rho_1$ and $\rho_2$, requires:

\begin{equation}
\label{Eq:massfraction}
    \frac{1}{\rho^*}=\frac{Y_1^*}{\rho_1}+\frac{Y_2^*}{\rho_2},
\end{equation}
where $Y_1$ and $Y_2$ are the mass fractions of the two fluids. This relation also represents the infinite speed of sound limit of the ideal gas equation of state for the mixture \citep{livescu2013nst}. Since $Y_1^*+Y_2^*=1$, relation (\ref{Eq:massfraction}) becomes a diagnostic equation for mass fractions.

The non-dimensional parameters in equations (\ref{Eq:continuity})-(\ref{Eq:divergence}) are the computational Reynolds number, $Re_0$, Schmidt number, $Sc$, and Froude number, $Fr$, defined as:
\begin{equation} \label{Eq:Reynolds}
Re_0=\rho_0L_0U_0/\mu_0,
\end{equation}
\begin{equation} \label{Eq:Schmidt}
Sc=\mu_0/\rho_0D_0,
\end{equation}
\begin{equation} \label{Eq:Froude}
Fr^2=U^2_0/gL_0,
\end{equation}
where $\rho_0=(\rho_1+\rho_2)/2$ is the reference density, and for the initial conditions in this paper it is equal to the mean density (calculated as the volumetric average $\rhom=\frac{1}{\cal{V}}\int_{\cal{V}} \rho^* d{\cal{V}}$ due to the homogeneity of the flow), $g$ is the magnitude of acceleration field, $\mu_0$ is the reference dynamic viscosity, $D_0$ is the diffusion coefficient and $L_0$ and $U_0$ are the reference length and velocity scales.  For the cases investigated in this paper, $D_0$ is constant, while the instantaneous dynamic viscosity, $\mu^*=\mu_0 \rho^*/\rho_0=\nu_0 \rho^*$, where $\nu_0$ is reference kinematic viscosity and is constant (see Table \ref{Table:cases}). This ensures that the instantaneous Schmidt number is uniform and constant during the flow evolution. For all cases considered, $Sc=1$. The mixture rule implied by the instantaneous dynamic viscosity relation is 
$1/\mu^*=Y_1^*/\mu_1+Y_2^*/\mu_2$, where $\mu_1$ and $\mu_2$ are the dynamic viscosity of the two fluids. Similarly, for all cases considered in the paper, $Fr=1$. As shown below, it is useful to further scale the results by the velocity scale $U_r=\sqrt{A/Fr^2}$ and time scale $t_r=\sqrt{Fr^2/A}$, as some of the quantities discussed collapse with this scaling. The corresponding dimensional quantities are $U_r^\dagger=U_r U_0=\sqrt{A g L_0}$ and $t_r^\dagger=t_r L_0/U_0=\sqrt{L_0/(A g)}$. In the triply periodic case, there is no intrinsic length scale; however, to account for the initial fluid distribution and facilitate the comparison with other flows, such as RTI and RMI, $L_0$ can be taken as the initial density integral scale. Using $L_0$, $U_r^\dagger$, and $t_r^\dagger$ for non-dimensionalization changes the non-dimensional parameters in the governing equations to $Fr^{\dagger\ 2}=A$ and $Re_0^\dagger=Re_0\sqrt{A/Fr^2}=\rho_0L_0\sqrt{AgL_0}/\mu_0$.

In HVDT, due to the periodic boundary conditions, the pressure can only be determined up to a constant gradient. Thus, the mean pressure gradient needs to be specified. Similar to \citet{livescu2007, livescu2008}, the mean pressure gradient is chosen to obtain a maximally unstable flow (understood as the time derivative of the mass flux attaining its maximum absolute value):
\begin{equation}
    \label{Eq:mean_Pgrad}
    P_{,i}=\frac{1}{V}\Big(\frac{1}{Fr^2}g_i-\langle vp_{,i} \rangle + \langle u_iu_{j,j}\rangle + \langle v\tau_{ij,j}\rangle \Big),
\end{equation}
where $V$ is the mean specific volume ($v^*=1/\rho^*=V+v$). This also leads to $U_i=0$; hence, in this study $u^*_i=u_i$. This choice of  $P_{,i}$ is consistent with previous studies of HVDT by \citet{Sandoval1997}, where similar arguments are used but under Boussinesq approximation. Detailed derivations to this effect can be found in section 2.2 of \citet{livescu2007}. As noted in the 2007 paper, the choice of the maximally unstable flow sets the mean velocity to the value of zero; this is similar to observations by \citet{Cabot_Cook} and \citet{Livescu_Cabot_Cook} in the core region of the Rayleigh-Taylor instability where the $P_{,i}$ is set by the non-periodic boundary conditions. Thus, this flow has significant similarities to the Rayleigh-Taylor instability, as explained in \citet{Livescu_Cabot_Cook}. The choice $P_{,i}=0$ has been made by \citet{gat_2017} in their acceleration driven mixing layer study. In that study, which had non-zero mean velocity, the vertical growth of the mixing layers is minimized, while the horizontal growth is maximized, which has analogies with vertical convection or differential heated cavities (see also \citet{Livescu2020}).

\subsection{Computational approach and simulation cases} \label{Sec:sim cases}

Equations (\ref{Eq:continuity}) and (\ref{Eq:moment}), together with the divergence condition (\ref{Eq:divergence}), are solved in a triply periodic domain $[(2\pi)^3]$ using the CFDNS code, as described in \citet{livescu2007}. The spatial derivatives are evaluated using Fourier transforms and the time advancement is performed with the variable time step third order Adams-Bashforth-Moulton scheme, coupled with a fractional time method. To minimize the aliasing errors, the advection terms are written in the skew-symmetric form. 

\begin{table}
  \begin{center}
\def~{\hphantom{0}}
\setlength{\tabcolsep}{18pt}
  \begin{tabular}{cccccc}
      Cases   		& $A$ 		& $Re_0$      & $Re_{b0}$    & $Re_{\lambda,max}$      & Resolution\\      [3pt]
      
      A1Re5  		& 0.05		&	20000	&	7014  & $490$	& $2048^3$\\
      ~A1Re4*       & 0.05		&	10000		&	3507  & $298$	& $1024^3$\\
      A1Re3  		& 0.05		&	~4000	&	1403  & $139$	& ~$512^3$\\
      A1Re1  		& 0.05		&	~1563		&	~548  & ~$69$	& ~$256^3$\\
      A2Re3 		& 0.25		&	~4000	&	3137  & $261$	& $1024^3$\\
      A2Re1  		& 0.25		&	~1563		&	1225  & $120$	& ~$512^3$\\
      A3Re2  		& 0.5~		&	~3125		&	3466  & $236$	& $1024^3$\\
      A3Re1  		& 0.5~		&	~1563		&	1733  & $144$	& $1024^3$\\
      A3Re0  		& 0.5~		&	~~556		&	~616  & ~$62$	& ~$512^3$\\
      A4Re2        & 0.75      &	~3125     &	4245  & $191$	& $2048^3$\\
      A4Re1  		& 0.75		&	~1563	&	2123  & $122$		& $1024^3$\\
      A4Re0  		& 0.75		&	~~556		&	~755  & ~$58$	& ~$512^3$\\
  \end{tabular}
  \caption{Parameters for the DNS cases. *Similar results for this case, with a slightly different initialization, are available through the Johns Hopkins Turbulence Database \citep{JHTD}. }
  \label{Table:cases}
  \end{center}
\end{table}

Table \ref{Table:cases} lists the various cases that were chosen to investigate the influence of the Atwood and Reynolds numbers on HVDT. In the nomenclature chosen for the case names, the index following the first letter, $A$, varies from 1 to 4, denoting the Atwood numbers, $0.05$, $0.25$, $0.5$, and $0.75$, respectively. In addition, case names include the \rez index from 0 to 5 corresponding to six different values of \rez in increasing order,  $556$, $1563$, $3125$, $4000$, $10000$, and $20000$, respectively. Also tabulated is a static buoyancy Reynolds number $Re_{b0}$ defined as \citep{batchelor1992,livescu2007}: $Re_{b0}=Re_0\sqrt{\mathcal{L}_\rho^3 A/Fr^2}$ where $\mathcal{L}_\rho$ is the initial density integral length-scale. The Taylor Reynolds number is calculated from the turbulence Reynolds number, $Re_t$, using the isotropic turbulence formula $Re_{\lambda}=\sqrt{20/3Re_t}$, where
\begin{equation}
    Re_t=Re_0\langle\rho^*u^{''}_iu^{''}_i\rangle^2/(\rhom \epsilon),
\end{equation}
and $\epsilon=-\langle u_{i,j}\tau_{ij}\rangle$ is the dissipation in the turbulent kinetic energy equation.

The density field in all simulations is initialized as a Gaussian random field with top-hat energy spectrum between wave numbers 3 to 5. After transforming into the real space, the negative values are assigned as 1, and the positive values are assigned as $(1+A)/(1-A)$; the pure fluid densities thus yield the desired Atwood number. To ensure that the mixing layer between the pure fluid regions is captured on the grid, a Gaussian filter is used to smoothen the density field, thereby preserving the bounds. The width of the Gaussian filter is $1.1 \Delta x$ and is applied once for $256^3$ and $512^3$, four times for $1024^3$ (resulting in a total width of $\approx 2.2 \Delta x$), and sixteen times for $2048^3$ resolutions (resulting in a total width of $\approx 4.4 \Delta x$); the density integral length-scale and mixing-state parameter (defined below) are similar for all cases.

After the initialization procedure, the non-dimensional initial density integral length-scale, which is calculated from the resultant density spectra by:
\begin{equation}\label{Eq:int_Lr}
\mathcal{L}_{\rho}=2\pi\int_0^{\infty} \frac{E_{\rho ^{'}}(\mathcal{K})}{\mathcal{K}}d\mathcal{K}\Big/\int_0^{\infty}E_{\rho ^{'}}(\mathcal{K})d\mathcal{K},
\end{equation}
is $1.3-1.4$ for all cases. Initial mixing-state is represented by a commonly used mixing-state parameter -$\theta$- \citep{Youngs91,linden_redondo_youngs_1994} defined as:
\begin{equation}\label{Eq:theta}
    \theta=1-\frac{\langle\rho^{2}\rangle}{(\rhom-\rho_1)(\rho_2-\rhom)},
\end{equation}
where $\rhom$ is the mean density. The initial value of $\theta$ is $\theta_0=0.068-0.07$ for cases considered in Table 1 except for the case A1Re1 for which $\theta_0=0.14$. 

The non-dimensional acceleration field is $g_i=(-1,0,0)$ and is gradually applied to the flow between $t/t_r=0$ to $0.1$.  A $5^{\text{th}}$ order polynomial equation is used to allow the flow to have a smooth transition from rest to the accelerated phase. This is especially important at high Atwood numbers, where exceedingly small time steps would otherwise be required at initial times for accuracy. All simulations are well resolved, with $\eta k_{max} > 2$ at all times during the flow evolution, except the A3Re2 case, for which $\eta k_{max} > 1.7$. Here, $\eta=\left(1/[Re_0^3 (\epsilon/\rho_0)]\right)^{1/4}$ is the Kolmogorov micro-scale, and $k_{max}= \pi N /\mathcal{L} = N/2$ is the largest resolved wave number. All data presented from the $256^3$ resolution cases represent averages over 10 realizations, while data from $512^3$ resolution cases represent averages over 3 realizations. The initial conditions for different realizations were generated using different random number seeds. Because of the computational cost, cases with higher resolutions ($1024^3$ and $2048^3$) represent only one realization.

\subsection{Energy budgets}\label{Sec:Energy_bud}

To help describe the turbulence evolution, we define below the scalar energy (half density variance), the total and turbulent kinetic energies, as well as the potential energy, and discuss their transport equations. 

\subsubsection{Scalar energy ($E_{\rho}$)}

Here, the scalar energy refers to the density variance and is defined as:
\begin{equation}
   E_\rho=\frac{1}{2}\langle\rho^2\rangle.
\label{Eq:scalar_en}
\end{equation}
In HVDT, during the flow evolution, $\rhom$ remains constant due to homogeneity, so that $\rho^*_{,t}=\rho_{,t}$ and the rate of change of scalar energy per unit volume can be calculated by multiplying Eq. \ref{Eq:continuity} by $\rho$. Using homogeneity [$\lp{()_{,j}}\rp=0$], the transport equation turns into:

\begin{equation}\label{Eq:Scalar_transport}
\begin{split}
\frac{1}{2}\lp{\rho^2_{,t}}\rp= &-\lp{(\rho^*-\rhom)(\rho^*u^*_j)_{,j}}\rp=-\lp{[(\rho^*-\rhom)\rho^*u^*_j]_{,j}}\rp+\frac{1}{2}\lp{\rho^{*2}_{,j}u^*_j}\rp\\
& = \frac{1}{2}\lp{(\rho^{*2}u^*_{j})_{,j}}\rp-\frac{1}{2}\lp{\rho^{*2}u^*_{j,j}}\rp\\
& = \frac{1}{2Re_0Sc_0}\lp{\rho^{*2}(ln\rho^*)_{,jj}}\rp\\
&=-\frac{1}{Re_0Sc_0}\langle\rho_{,j}\rho_{,j}\rangle.
\end{split}
\end{equation}
The resultant dissipation rate, $\chi=\frac{1}{Re_0Sc_0}\langle\rho_{,j}\rho_{,j}\rangle$, is similar to the dissipation rate for a passive scalar \citep{livescu_jaberi_madnia_2000,DonDaniel_ReactionForcing}. 

\subsubsection{Total kinetic energy ($E_{KE}$)} \label{Sec:KE}

The total kinetic energy is defined as:

\begin{equation}
    E_{KE}=\frac{1}{2}\langle \rho^*u^*_iu^*_i\rangle .
\end{equation}
The rate of change of kinetic energy per unit volume is obtained by multiplying the momentum eq. (\ref{Eq:moment}) by $u^*_i(=u_i)$ and averaging the resulting equation:
\begin{equation}
    E_{KE,t}=\frac{g_i}{Fr^2}\langle \rho^*u_i\rangle +\langle pu_{j,j}\rangle -\langle u_{i,j}\tau^{*}_{ij}\rangle .
    \label{Eq:KE_evolv}
\end{equation}
The first term on the right-hand side is the buoyancy-production term and is proportional to the mass flux defined by:
\begin{equation} \label{Eq:mass_flux}
a_i=\frac{\langle \rho u_i\rangle }{\overline{\rho}}=\tilde{U}_i-U_i=-\langle u''_i\rangle .
\end{equation} 
The second term (pressure-dilatation) is the trace of the pressure strain tensor and is the gain or loss of kinetic energy due to work done by the change of specific volume during mixing. For the parameters considered here, this term is negligible compared to the other terms. However, since the flow is anisotropic, the components of the pressure strain tensor (e.g., $\langle pu_{1,1}\rangle$) are not negligible and they represent the primary mechanism of redistributing kinetic energy in different directions. The last term on the right-hand side is the total kinetic energy dissipation rate, $\epsilon_{tot}$. 
\subsubsection{Favre averaged turbulent kinetic energy ($E_{TKE}$)} \label{Sec:TKE}
The Favre averaged turbulent kinetic energy $(E_{TKE})$ is defined as:
\begin{equation} \label{Eq:tke}
    E_{TKE}=\frac{1}{2}\langle \rho^*u''_iu''_i\rangle =\frac{1}{2}\Big(\langle \rho^*u^*_iu^*_i\rangle  - \rhom a_ia_i\Big).
\end{equation}
The terms on the right-hand side of equation (\ref{Eq:tke}) are the total kinetic energy ($E_{KE}$) and mean kinetic energy ($E_{MKE}$), respectively. 

The transport equation for $E_{TKE}$ can be written as \citep{livescu2007}:

\begin{equation}
    E_{TKE,t}=a_iP_{,i}+\langle p u^{''}_{j,j}\rangle -\langle u^{''}_{i,j}\tau^*_{ij}\rangle .
    \label{Eq:TKE_evolv}
\end{equation}
The first term on the right hand side is the TKE production and the last term on the right hand side is the turbulent kinetic energy dissipation rate ($\epsilon$). Due to homogeneity, $\epsilon_{tot}=\epsilon=\langle u^*_{i,j}\tau^*_{ij}\rangle$. Similarly, $\langle p u^{''}_{j,j}\rangle=\langle p u_{j,j}\rangle$.

\subsubsection{Potential energy ($E_{PE}$)} \label{Sec:PE}

The total available potential energy ($E_{PE}$) for the triply periodic volume ($V$) is calculated as:
\begin{equation}
    E_{PE}(t)=-\frac{g_i}{\mathcal{V} Fr^2}\int_{\mathcal{V}}(\rho^*-\rhom)x_id\mathcal{V}-\int_{t=0}^{t\to\infty}\mathcal{F}_{E^*_p}dt.
    \label{Eq:potential}
\end{equation}
where $x_i$ is the relative height that varies from $0$ to $2\pi$ in the direction opposite to the acceleration vector $\mathbf{g}$, and $\mathcal{F}_{E^*_p}=-\frac{g_i}{\mathcal{V} Fr^2}\int_Su_j(\rho^*-\rhom)x_idS_j$ is the flux of the potential energy through the boundaries. The change of the available potential energy can be derived as \citep{livescu2007}:
\begin{equation}
\begin{split}
    E_{PE,t}=&-\frac{\partial}{\partial t}\Big(\frac{g_i}{\mathcal{V} Fr^2}\int_{\mathcal{V}}(\rho^*-\rhom)x_id\mathcal{V}\Big)-\mathcal{F}_{E^*_p}\\
    &=-\frac{g_i}{Fr^2}\langle \rho^*u_i\rangle +\frac{g_i\rhom}{\mathcal{V} Fr^2}\int_{S} u_jx_idS_j
\end{split}
    \label{Eq:pet}
\end{equation}
where the first term is identical to the buoyancy-production term in eq. (\ref{Eq:TKE_evolv}), but with opposite sign, and the second term is a surface integral that goes to zero after averaging the results over different realizations.

\subsection{Energy conversion rates in buoyancy-driven HVDT}\label{Sec:mixing_efficiency}

In RTI under constant acceleration, the energy conversion ratio has been related to the growth rate of the RTI mixing layer width in the alpha-group study \citep{Alpha_groupRTI2004}. They reported an energy conversion rate from potential to kinetic energy ($E_{KE}/\delta E_{PE}$) of  $0.46\pm0.04$ when $\%80$ of the flow is molecularly mixed ($\theta \approx 0.8$) within the mixing layer during the self-similar regime of RTI. However, this asymptotic behavior remains an open question as \citet{Cabot_Cook} found that after a relatively flat stage, $E_{KE}/\delta E_{PE}$ starts to increase again using a $3072^3$ DNS with \At$=0.5$; this change in behavior corresponds to the end of the alpha-group simulations where the  layer reaches the domain boundaries. In HVDT, since there is no self-similar stage during the growth regimes, we report the energy conversion rate separately for the two growth regimes (explosive and saturated growths) to explore \At and \rez numbers effects on this ratio. Similar to RTI, we define $\beta_{KE}$ as the ratio of the change in $E_{KE}$ to the change in available $E_{PE}$ as:

\begin{equation}\label{Eq:mixing_eff}
    \beta_{KE}=\frac{\Delta E_{KE}}{\Delta E_{PE}}=(E_{KE_{t2}}-E_{KE_{t1}})/\int_{t1}^{t2} E_{PE,t}(t) dt,
\end{equation}
where $\Delta E_{KE}$ is the change in $E_{KE}$ and $\Delta E_{PE}$ is the change in available $E_{PE}$ between any two time instants chosen for analysis. We will also discuss this ratio for the turbulent kinetic energy, ($\beta_{TKE}$) and mean kinetic energy ($\beta_{MKE}$). The $\beta$ values are calculated by setting $t_1$ and $t_2$ in eq. (\ref{Eq:mixing_eff}) to the normalized start and end times for the corresponding regime. For example, to calculate $\beta_{TKE}$ for the explosive growth regime, the start time is set as $t_1=t/t_r=0.1$; while the end time $t_2$ is equal to the normalized time when the condition $dE_{TKE}/dt>0$ and $d^2E_{TKE}/dt^2=0$ is reached.

\subsection{Transport equation for velocity-density joint mass density function (jMDF)} \label{Sec:joint_PDF}

Turbulent mixing as occurs in HVDT is a dynamic process where both local velocities and the fluid composition are coupled and integral to the mixing process. Here, in order to investigate the coupled effects of the composition (density) and velocity fluctuations on VD mixing, the transport equation of the density-weighted  velocity-density joint-PDF (or velocity-density joint mass density function, jMDF) is derived. As noted by \citet{POPE1985119} and \citet{HAWORTH2010168}, for VD turbulence, mass density function (MDF) is more useful compared to the PDF because the resulting equations are simpler. In addition, velocity-density joint MDF (jMDF) is an effective way of exploring the coupled effects of composition  and velocity fluctuations which provides comprehensive information about the buoyancy-driven VDT evolution. The jMDF ($\mathcal{F}$) is defined as:

\begin{equation}
    \label{Eq:MDF_def}
    \mathcal{F}_{u_i\rho^*}({V_i,R;x_i,t})=Rf_{u_i\rho^*}({V_i,R;x_i,t})
\end{equation}
where $f_{u_i\rho^*}({V_i,R;x_i,t})$ is the joint probability density function (jPDF) of the velocity and density fields. Following \citet{POPE1985119}, the jPDF can be calculated using the fine-grained PDF, $f^*$, as: $f=<f^*>=<\delta(u^*_i-V_i)\delta(\rho^*-R)>$, where $\delta$ is the delta function and $V_i$ and $R$ are the independent sample space variables. The details of the derivation can be found in Appendix A; the final version can be written as:
\begin{equation}
\begin{split}
\label{Eq:joint}
      \frac{\partial \mathcal{F}}{\partial t}=&-\frac{\partial}{\partial V_i}\bigg[\mathcal{F} \bigg\langle-\frac{p_{,i}}{\rho^*}\bigg\vert_{V_i, R}-\frac{P_{,i}}{\rho^*}\bigg\vert_{V_i, R}+\frac{\tau^*_{ij,j}}{\rho^*}\bigg\vert_{V_i, R}+\frac{1}{Fr^2}g_i \bigg\vert_{V_i, R}\bigg\rangle\bigg]\\
    &-\frac{\partial}{\partial R} \bigg[\mathcal{F} \bigg\langle -\rho^*u^*_{j,j} \bigg\vert_{V_{i},R}\bigg\rangle\bigg].
\end{split}
\end{equation}
where $<Q |_{V_i,R}>$ is the conditional expectation of the function $Q(\rho^*,{u^*_i};{x_i}, t)$ for specific velocity ($u^*_i=V_i$) and density ($\rho^*=R$) values. For the homogeneous case, the explicit spatial dependence drops from the averages. Thus, in homogeneous VDT, equation (\ref{Eq:joint}) is four dimensional. The first term on the right-hand side represents the transport of $\mathcal{F}$ in the velocity sample space, while the second term represents the transport of $\mathcal{F}$ in the density sample space \citep{POPE1985119}. The presence of the two terms indicates that any asymmetry of the density PDF is carried over to the velocity PDF (or vice versa). Both $g_i$ and $P_{,i}$ are constant everywhere within the domain. However, since the pressure term is multiplied by the specific volume ($1/\rho ^*$), it leads to significant asymmetric behavior at high $A$. 

\subsection{Conditional expectations} \label{Sec:cond_def}

Due to the large density variations among different fluid regions, differential inertial forces affect mixing \citep{Livescu_HVDT_New_Ph}, as well as flow dynamics. \citet{banerjee_kraft_andrews_2010} utilized conditional expectations by using the density field as a fluid marker to study the dynamics of bubbles and spikes for their large \At number RTI experiments. They found significant differences between the statistics calculated at the bubble side (defined as $\rho < 0$) compared to the spike side (defined as $\rho > 0$). 

To further explore the structure of the HVDT flow, the conditional expectations of several quantities are discussed for each of the four flow regimes. These are related to the large (i.e. turbulent kinetic energy) as well as small (i.e. dissipation and enstrophy) scales. In particular, conditional dissipation highlighted in equation (\ref{Eq:joint}), is an important quantity in combustion models \citep{Klimenko2003}. For binary VD flows in the incompressible limit, the mass fractions do not appear explicitly in the governing equations and can be completely determined from density.  Therefore, the conditional expectations of $E_{TKE}$, dissipation of $E_{TKE}$, and enstrophy ($\omega^2$) are evaluated with respect to the local density (i.e., $\langle \rho^*u^{''}_iu^{''}_i \Big\vert_{\rho^*= R} \rangle$).

\section{Flow evolution}\label{Sec:Flow_evolve}

\begin{figure}
\hspace{1.5cm}(\emph{a}) \hspace{4cm} (\emph{b})  \hspace{4cm} (\emph{c}) \\
 \centerline{
 \includegraphics[width=4.8cm]{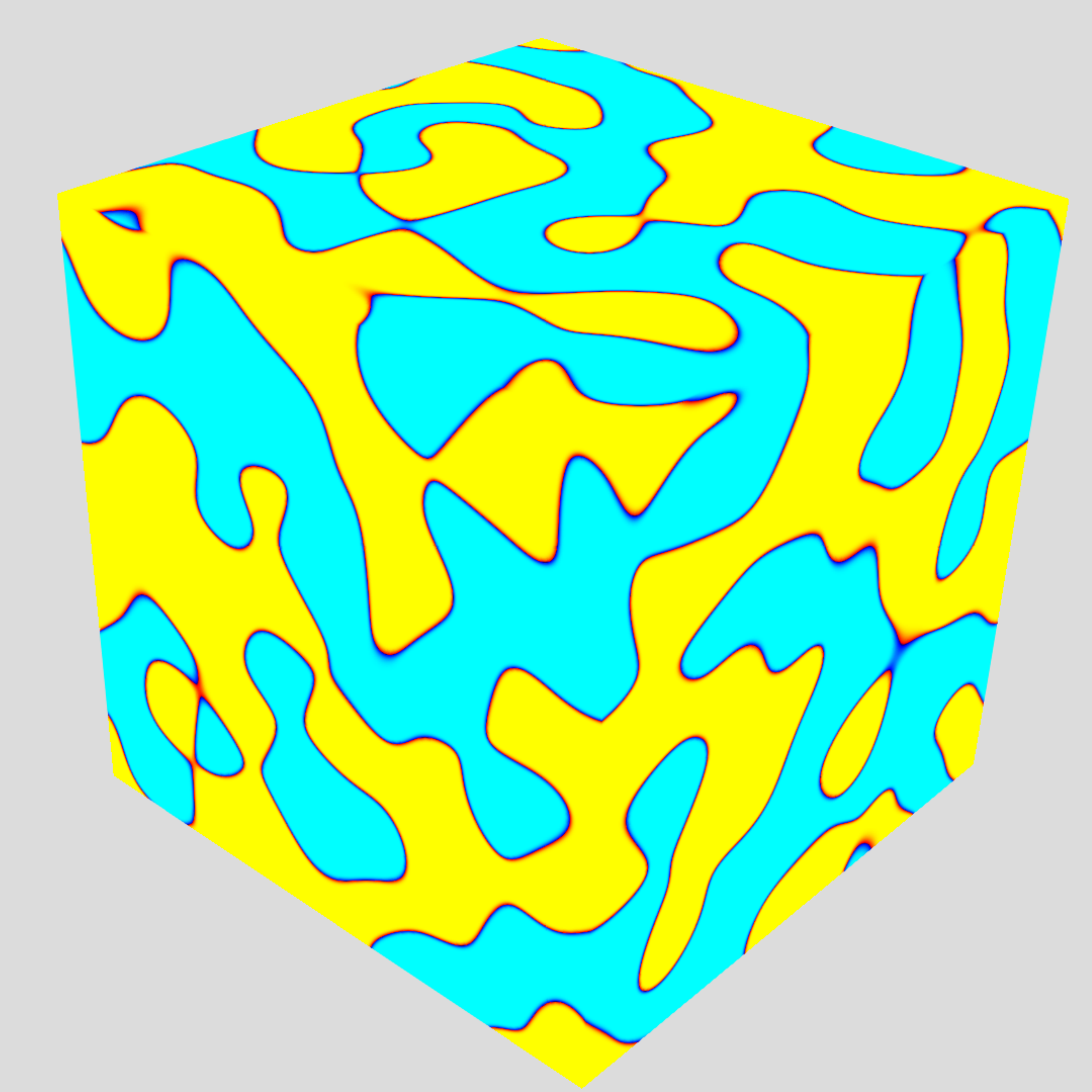}\includegraphics[width=4.8cm]{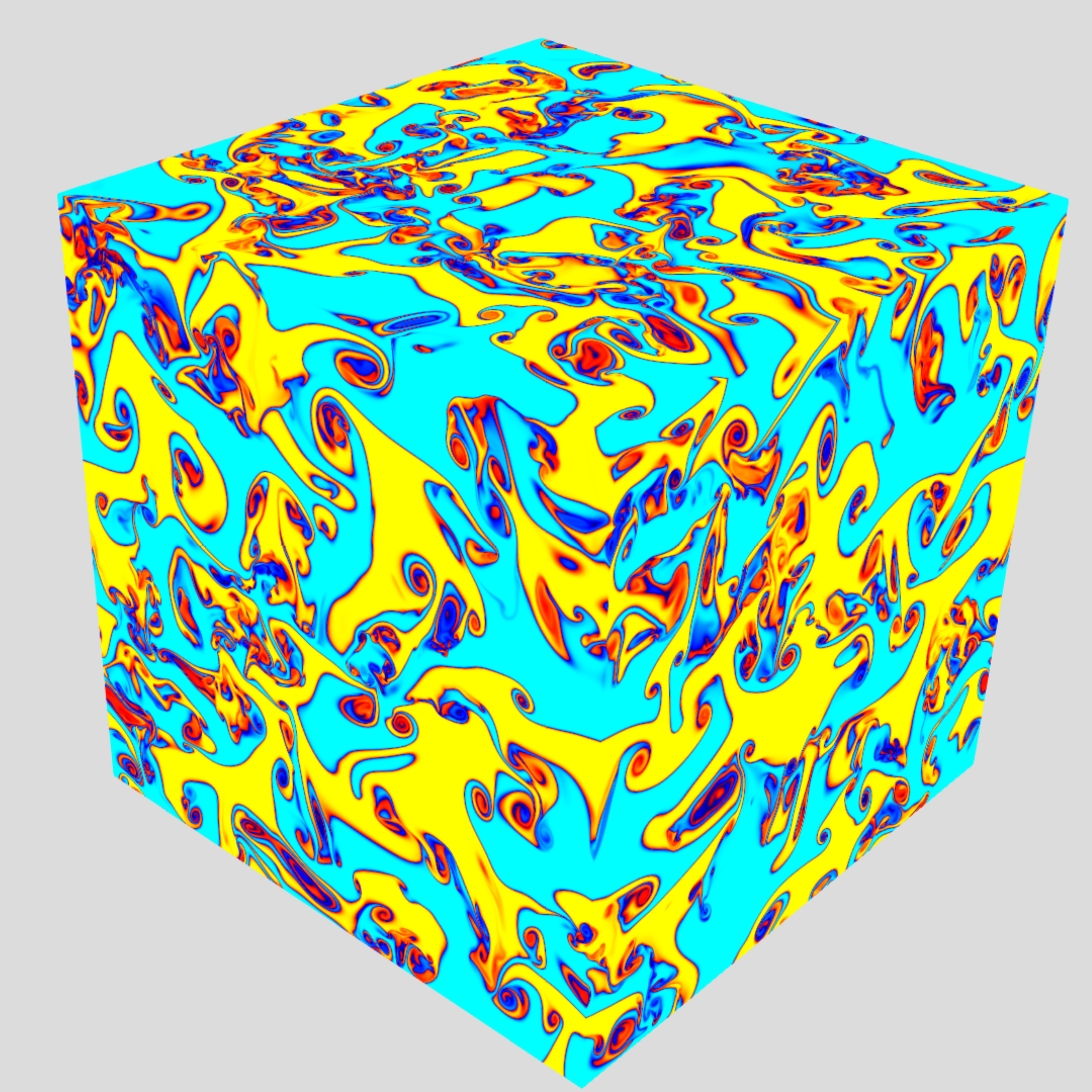}\includegraphics[width=4.8cm]{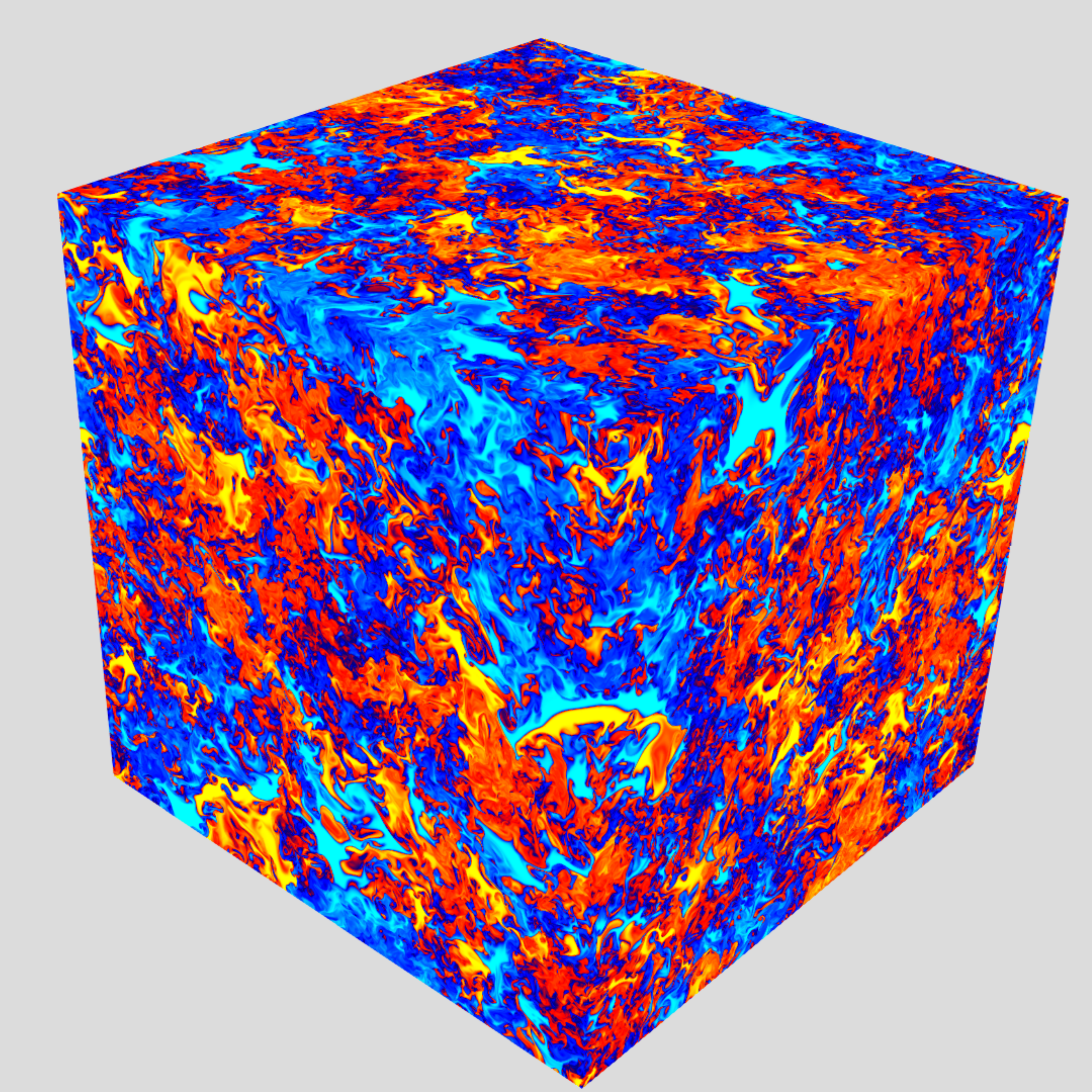}}
 \caption{3D visualization of the density field for A1Re0 case. (a) Initially segregated patches of heavy and light fluids at $t/t_r=0$; (b) at intermediate times mixing is induced by differential buoyancy forces ($t/t_r\approx2$); (c) at late time ($t/t_r\approx3$) buoyancy-forces decrease and this leads to decay of $E_{TKE}$.} 
\label{Fig_3D_intro}
\end{figure}

In HVDT, two fluids with different densities are initially segregated into random regions in a triply-periodic domain and are subjected to an acceleration $g_1$ (Figure \ref{Fig_3D_intro}a). At early times, turbulence is generated as the two fluids start moving in opposite directions due to differential buoyancy forces (figure \ref{Fig_3D_intro}b). 
    Meanwhile, mixing is initiated by molecular diffusion and enhanced by stirring induced by buoyancy-generated motions. As the fluids become molecularly mixed, the buoyancy forces decrease and, at some point, $E_{TKE}$ dissipation overcomes $E_{TKE}$ production.  This leads to a decay of $E_{TKE}$ (see figure \ref{Fig_3D_intro}c).

\begin{figure}
\hspace{1.5cm}(\emph{a}) \hspace{6cm} (\emph{b}) \\
 \centerline{
 \includegraphics[height=4.2cm]{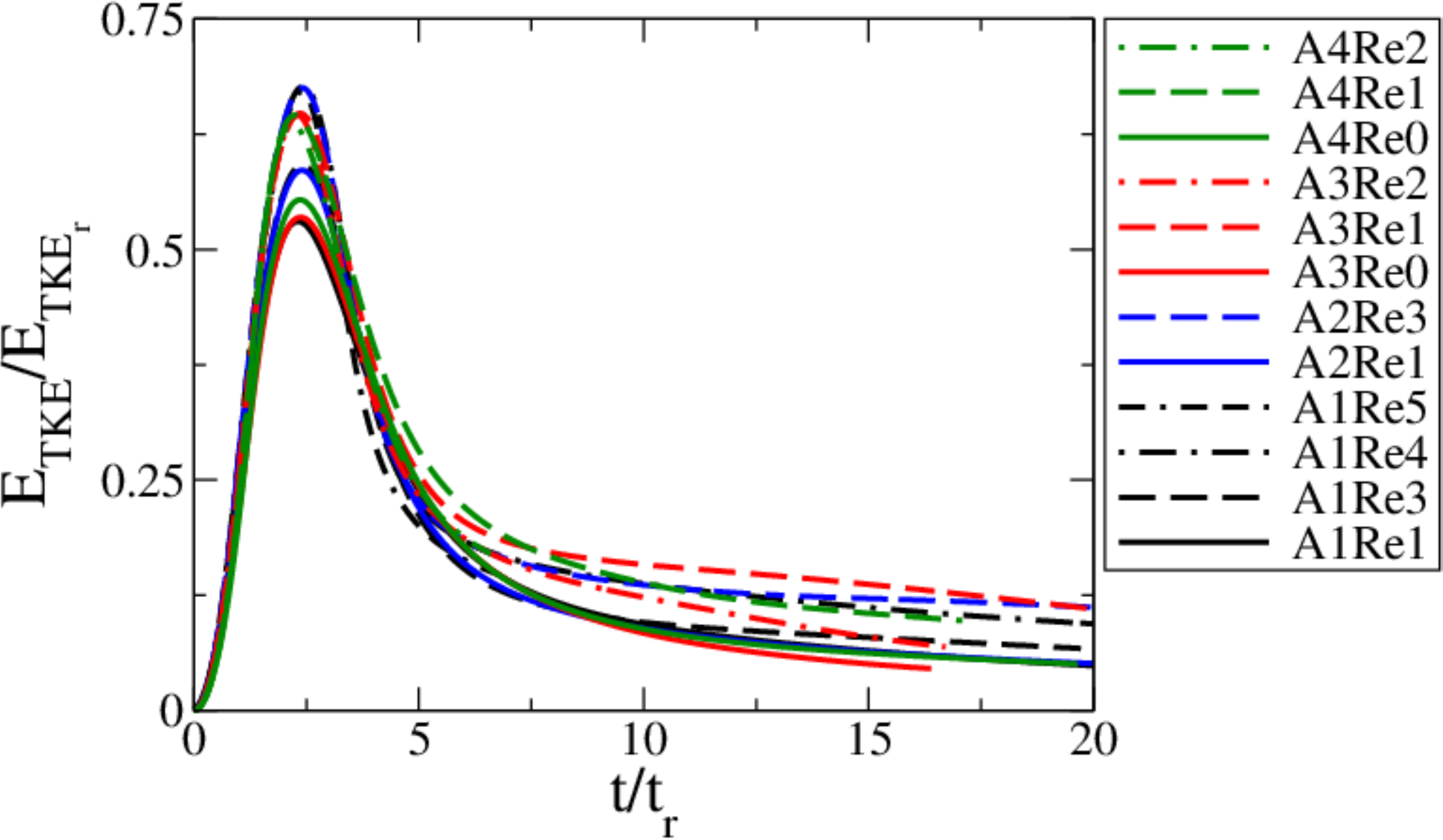}
 \includegraphics[height=4.2cm]{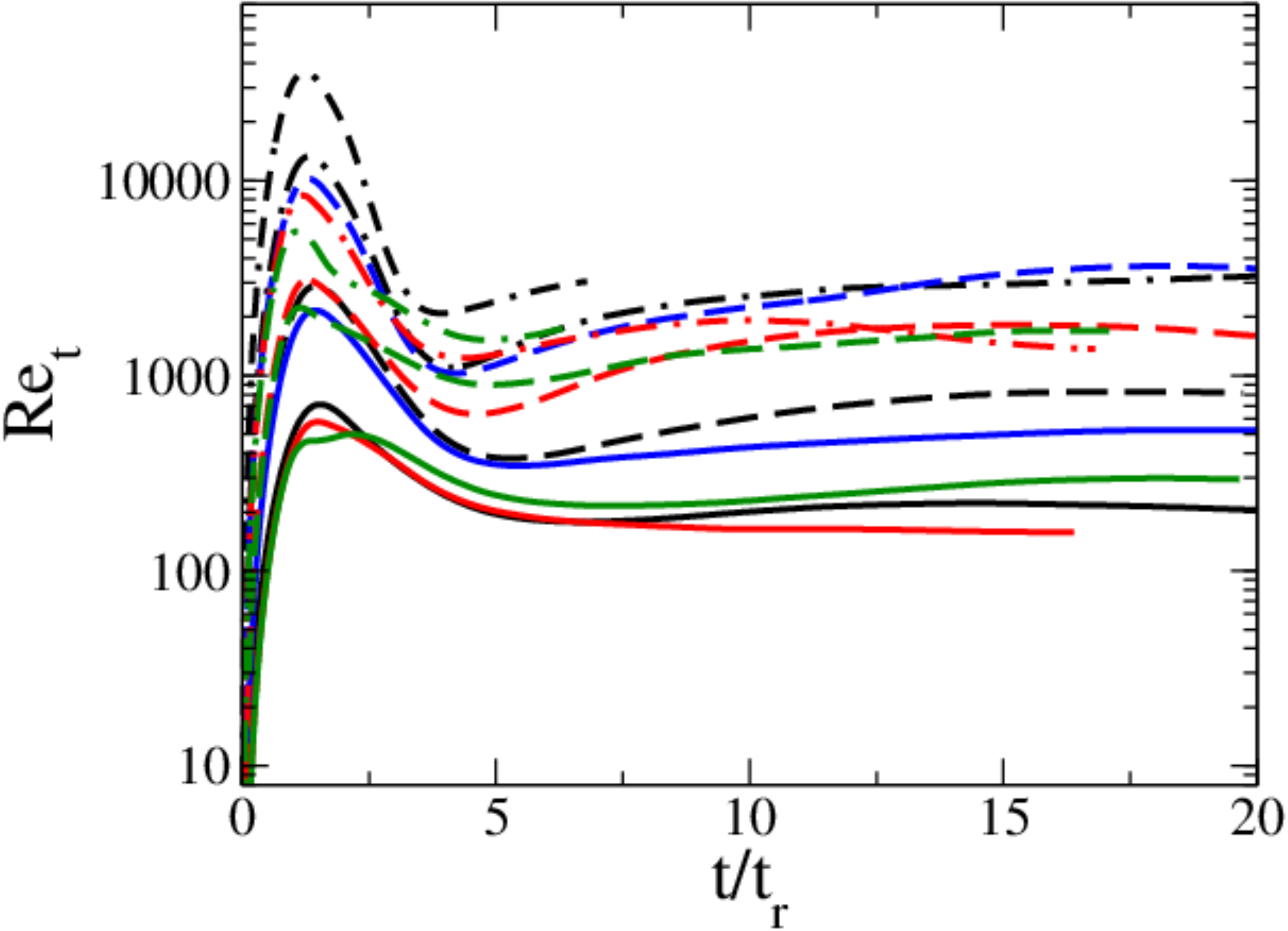}}
\caption{Evolution of (a) the normalized Favre turbulent kinetic energy ($E_{TKE}/E_{TKE_r}$), and, (b) turbulent Reynolds number ($Re_t$).} 
\label{Fig:TKE-Ret}
\end{figure}

The time evolution of the $E_{TKE}/E_{TKE_r}$ ($\langle \rho^*u^{''}_iu^{''}_i\rangle /\rhom U^2_r$), and turbulent Reynolds number ($Re_t$) are shown in Figure \ref{Fig:TKE-Ret}. The values are plotted as a function of time ($t/t_r$). As seen in figure \ref{Fig:TKE-Ret}a, the maximum of $E_{TKE}/E_{TKE_r}$ asymptotes to a finite value as $Re_{b0}$ increases; this observation is consistent with the prediction by \citet{batchelor1992} at very large $Re_{b0}$ values for the Boussinesq limit. In addition, \citet{batchelor1992} also predicted that the $E_{TKE}$ maxima occur at the same normalized time instant, if $Re_{b0}>256$. For our VD cases at $Re_{b0}>512$, all $E_{TKE}$ maxima occur at approximately $t/t_r\approx 2.3$, as seen in Fig. \ref{Fig:TKE-Ret}a (this observation was also hinted at by \citet{livescu2007} using low $Re$ data);  this further justifies the scaling used here.
        
Figure \ref{Fig:TKE-Ret}b presents the evolution of $Re_t$; it is observed that an increase in $Re_0$ leads to significant increase in $Re_t$. For the simulations reported in this paper, the largest $Re_t$ value is $\sim 36000$, corresponding to a Taylor Reynolds number ($Re_{\lambda}$) of $490$, for the lowest $A(=0.05)$ number case with $2048^3$ resolution (A1Re5 case). In addition, for the largest $A$ (=0.75) number case with the $2048^3$ resolution (A4Re2 case), $Re_t$ reaches a maximum value of $5500$, with $Re_{\lambda}=191$. The variation of $Re_t$ with $A$, at the same $Re_0$, is non-monotonic. Thus, the largest $Re_t$ values are obtained for $A=0.5$. This behavior is due to the opposing influences of buoyancy induced stirring (which increases at early times with $A$) and molecular mixing (which is enhanced by stirring), and is explained in more detail below. At late times, \ret is non-decreasing, as suggested by \citet{batchelor1992}, even though the effective Atwood number asymptotes to zero. Thus, buoyancy mediated turbulence decay is fundamentally different than regular turbulence decay.

\begin{figure}
\centerline{\includegraphics[height=4.6cm]{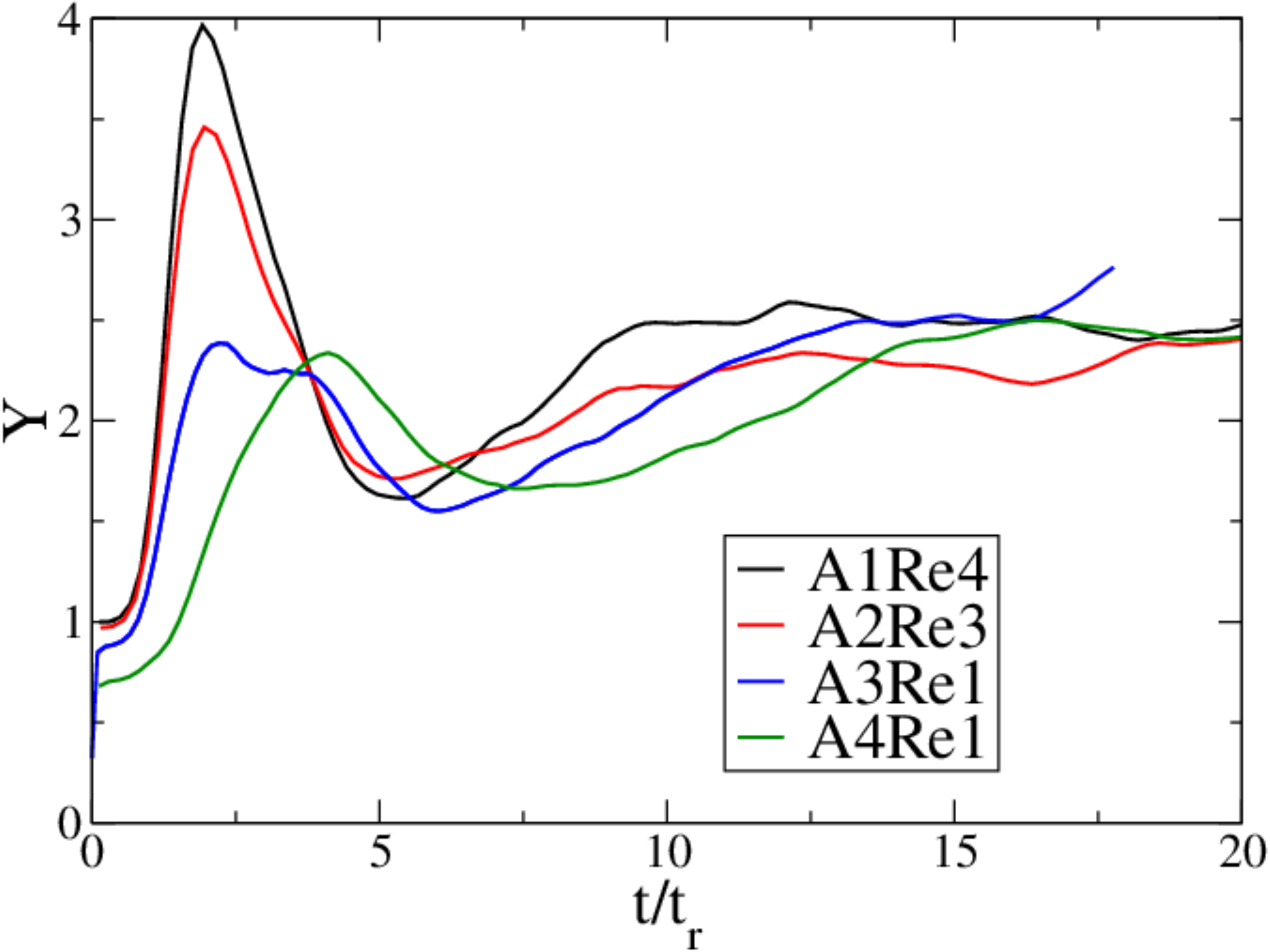}}
\caption{Evolution of the ratio of the time scales for turbulent kinetic energy and density variance equations ($\Upsilon$) for different $A$ numbers.}
\label{Fig:eddy}
\end{figure}

Figure \ref{Fig:eddy} presents the ratio of the time scales for turbulent kinetic energy and density variance equations, which is defined as:
    \begin{equation}
    \Upsilon=\frac{E_{TKE}}{\epsilon}\Bigg/\frac{E_{\rho}}{\chi}=\frac{E_{TKE}\chi}{E_\rho\epsilon}.
        \label{Eq:eddy}
\end{equation}
In low order turbulent mixing models, $\Upsilon$ is assumed to be a constant and the scalar dissipation ($\chi$) is estimated using the turbulent kinetic energy dissipation ($\epsilon$) \citep{livescu_jaberi_madnia_2000,Kolla_scalar,DonDaniel_ReactionForcing}. For passive scalar mixing with mean scalar gradient forcing $\Upsilon\approx2$  \citep{Overholt_Pope_passive_scalar_mean_gradient,Kolla_scalar}; other forcing mechanisms or flow conditions can lead to different values \citep{DonDaniel_ReactionForcing}. For reacting flows, much larger values can be obtained \citep{livescu_jaberi_madnia_2000}. In HVDT, $\Upsilon$ is a dynamic parameter during the earlier evolution of the flow, and the dissipation of the density field and $E_{TKE}$ must be captured separately until the gradual decay regime where buoyancy-forces weaken. 

As observed in figures \ref{Fig_3D_intro} and \ref{Fig:TKE-Ret}, HVDT includes the birth, growth, and decay of buoyancy-driven turbulence; the turbulence generation and decay stages contain unique physics that can be related to a wide range of flows discussed previously. Based on the flow behavior, we have sub-divided the flow evolution into four distinct regimes: (I) explosive growth, (II) saturated growth, (III) fast decay, and (IV) gradual decay.
This classification allows us to study the connection between this idealized flow and RTI, RMI as well as other flows that contain VD dynamics. 

\begin{figure}
(\emph{a}) \hspace{6.5cm}  (\emph{b}) \\
    \includegraphics[width=6.5cm]{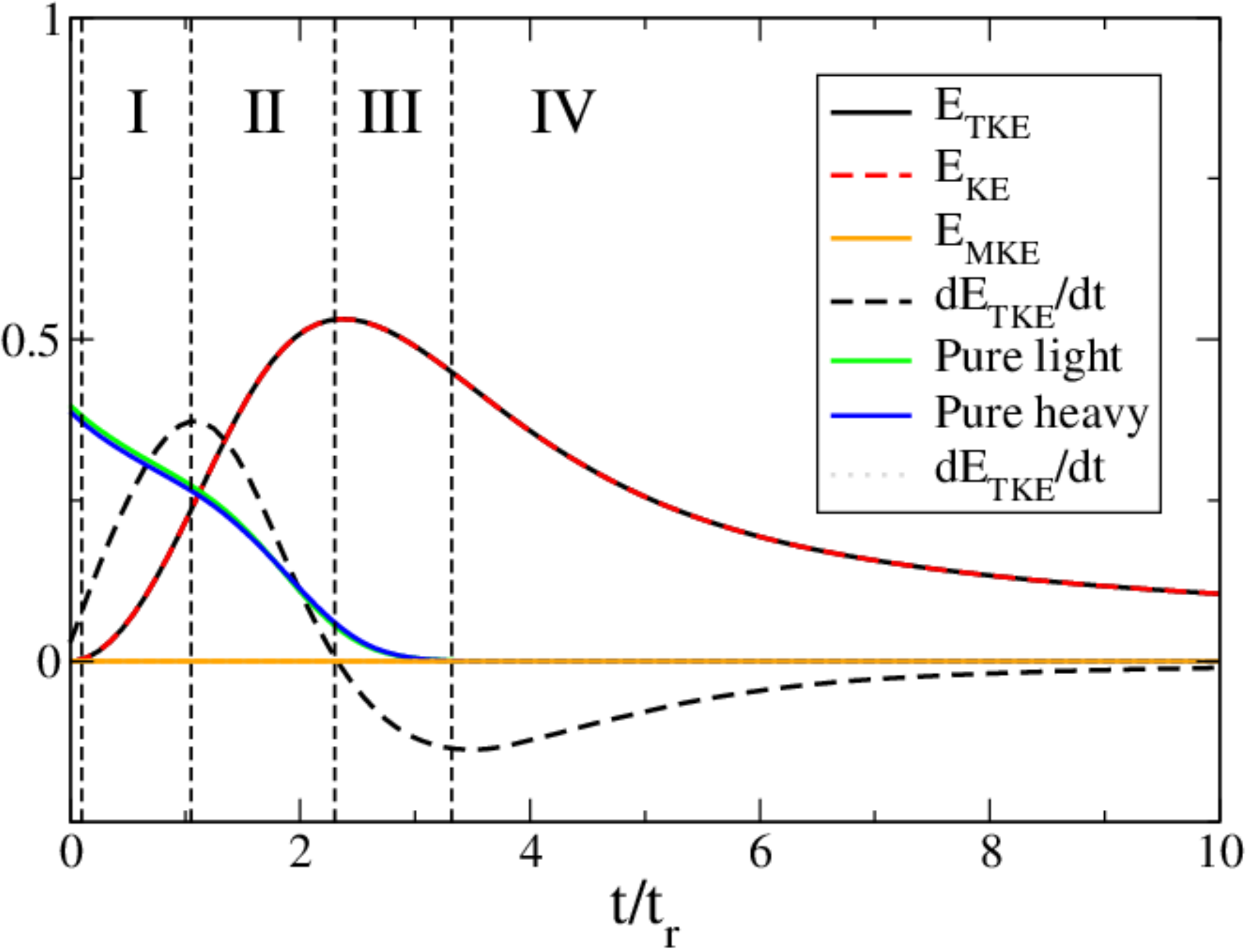}
    \includegraphics[width=6.5cm]{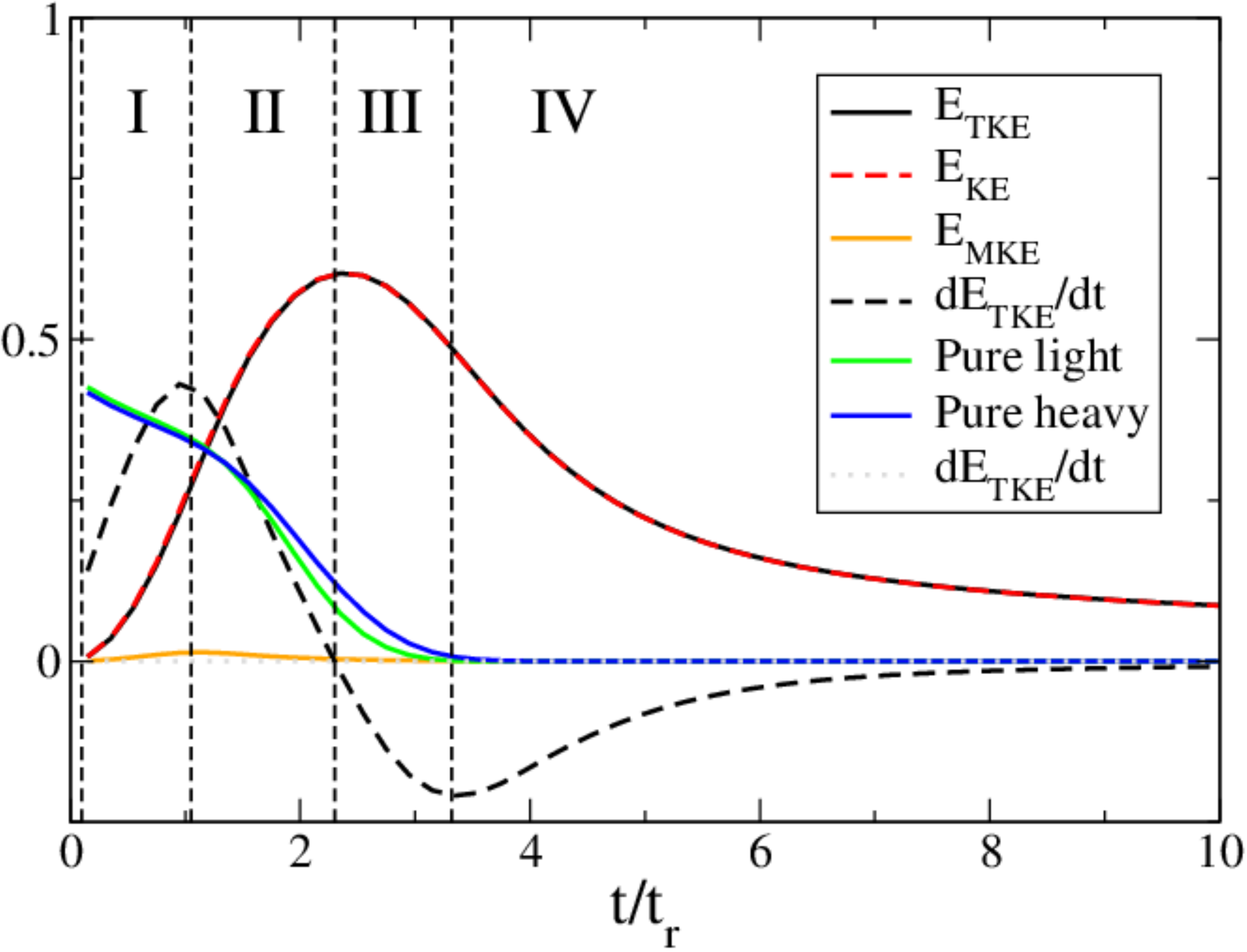}\\
(\emph{c}) \hspace{6.5cm} (\emph{d}) \\
    \includegraphics[width=6.5cm]{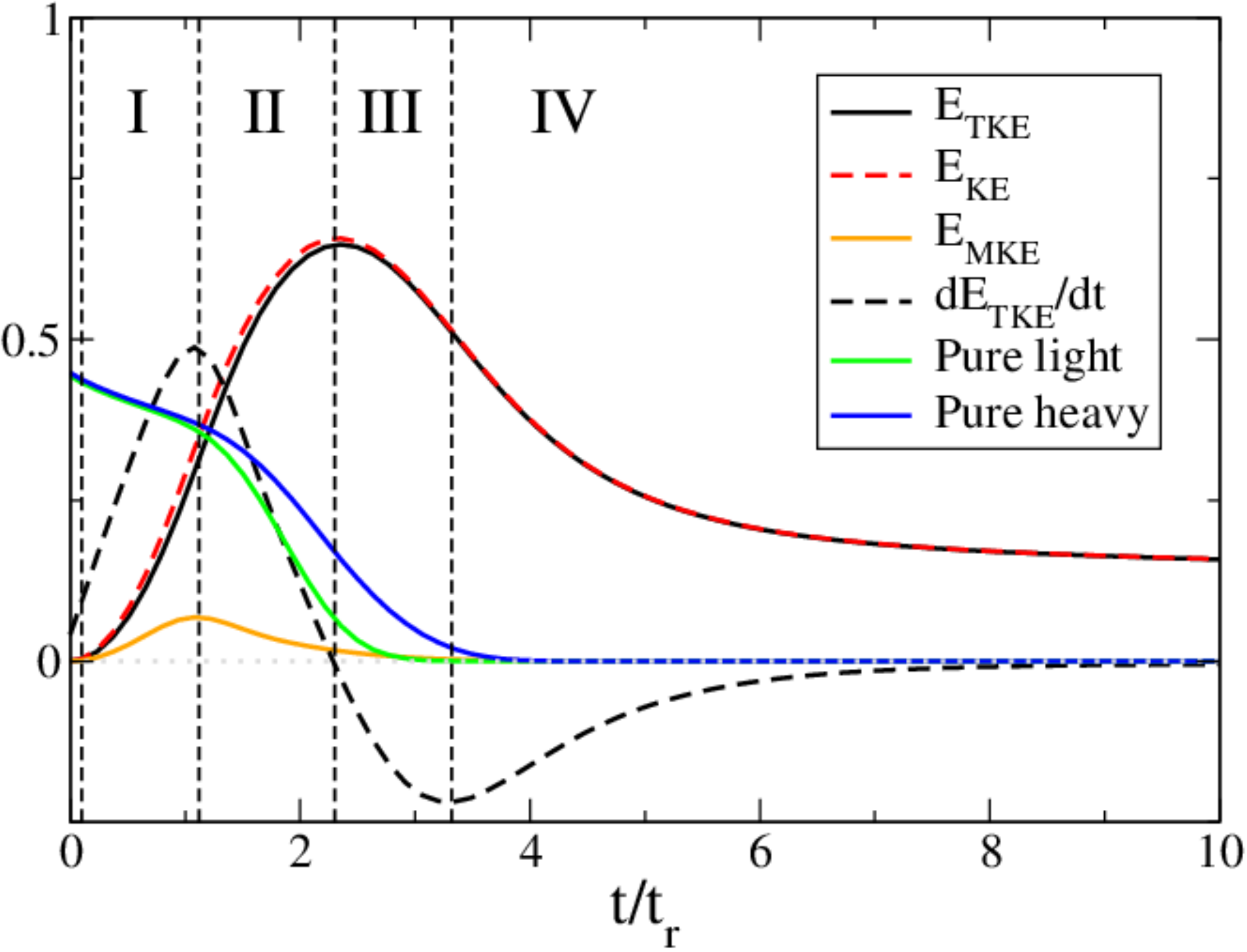}
    \includegraphics[width=6.5cm]{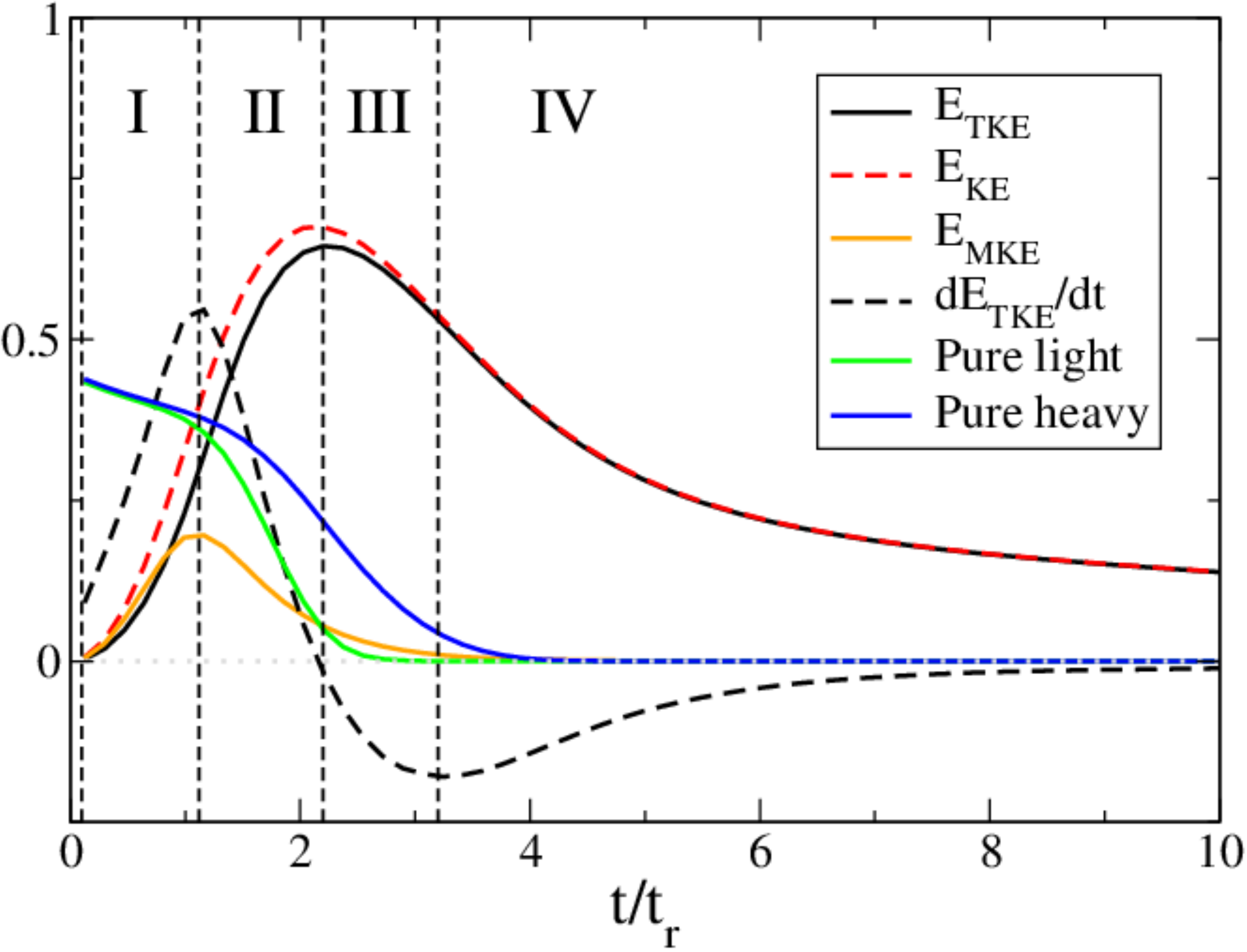}
\caption{Evolution of the normalized $E_{TKE}$, its time derivative,  $E_{KE}$, $E_{MKE}$ and the volume fractions of the pure light and heavy fluids for the cases (a) A1Re1, (b) A2Re1, (c) A3Re1, and (d) A4Re1.}
\label{Fig:Regimes}
\end{figure}

    To illustrate these regimes, figure~\ref{Fig:Regimes} shows the time evolution of $E_{TKE}/E_{TKE_r}$, its time derivative, $E_{KE}/E_{KE_r}$ ($\langle \rho^*u^*_iu^*_i\rangle /\rhom U^2_r$), and $E_{MKE}/E_{MKE_r}$ ($a_ia_i/U^2_r$) for different \At numbers. $E_{MKE}$ values are higher and lead to a phase difference between the $E_{TKE}$ and $E_{KE}$ evolutions for larger \At numbers, where the inertial differences between the light and heavy fluid regions are important; such differences are negligible at low \At numbers.
The amounts of pure light and heavy fluids are also plotted in figure~\ref{Fig:Regimes}. The 5\% and 95\% density cut-offs are used to define the pure light and heavy fluids. The pure light fluid is considered as having a density below $\rho_{pl}$, where $\rho_{pl}=\rho_{1}+0.05(\rho_{2}-\rho_{1}$), and the pure heavy fluid is defined as having a density above $\rho_{ph}$, where $\rho_{ph}=\rho_{1}+0.95(\rho_{2}-\rho_{1}$).

    At $t/t_r=0$ molecular diffusion is important, since the flow starts from rest. However, as the fluids are accelerated, i.e. $g_1$ increases from $0$ to $1$ in $t/t_r=0.1$; turbulence diffusion starts dominating molecular diffusion. Turbulence birth is followed by intense turbulence generation, which is divided into two sub-regimes: (I) explosive growth, where $E_{TKE}$ growth accelerates as $d^2 E_{TKE}/dt^2>0$; and, (II) saturated growth, where the rate of increase of turbulence fluctuations starts decreasing ($d^2 E_{TKE}/dt^2<0$). Regime (I) ends at $t/t_r\approx1.1$, when $d E_{TKE}/dt$ reaches its maximum, while regime (II) ends at $t/t_r\approx 2.3$, when $d E_{TKE}/dt=0$. The amount of pure heavy fluids remains significant during these two regimes; however, at high Atwood numbers there is also a significant asymmetry between pure light and heavy fluid volumes, which starts to develop during saturated growth. Eventually, the cumulative effect of molecular diffusion becomes large enough that, once again, it dominates the turbulence production and turbulence starts decaying. Flow characteristics also change during turbulence decay so that this part of flow evolution is broken up into two sub-regimes: (III) fast decay, where $d^2 E_{TKE}/dt^2<0$; and, (IV) gradual decay, where $d^2 E_{TKE}/dt^2>0$ and the decay process becomes slow and lengthy. Regime (III) lasts until $t/t_r\approx 3.2$, when $d E_{TKE}/dt$ reaches its minimum. Regime (IV) is the only part of flow evolution that becomes self-similar. During fast decay, there is still a sizable amount of pure heavy fluid, while the pure light fluid vanishes from the flow. Eventually, the pure heavy fluid also vanishes, as the gradual decay regime starts.
    
    Atwood number has limited effect on the normalized time instants where the regimes start, even though the flow structure dramatically differs by increasing the density ratio between the two fluids. \rez also does not have any significant effect on these normalized time instants and is not shown here for brevity. All regimes have their own characteristics concerning molecular mixing, energy conversion rates, dependency on \At and \rez numbers. The flow physics are discussed in detail in the next section  for each of the four regimes.
    
While the initial conditions are random, the flow is not turbulent from the start. This raises an important question:  when does the turbulence become fully developed? This question is also related to the concept of mixing transition \citep{dimotakis_2000}. Above the mixing transition, the flow characteristics become independent of $Re_t$. Here, the Reynolds number increases during the early stages of the flow evolution and reaches larger values for cases with smaller viscosity (see Figure \ref{Fig:TKE-Ret}.b). For each Atwood number, we investigate separately the convergence of turbulence statistics as the Reynolds number increases. Due to highly dynamic nature of the flow and highly asymmetric evolution for the high $A$ number cases, we use direct comparisons between the cases with lower and larger $Re_t$ values (especially comparing $1024^3$ simulations versus $2048^3$ simulations) to determine whether any flow quantity becomes insensitive to increase in $Re_t$ in different flow regions. As a result, we show (in the next section) that different turbulent quantities reach their asymptotic behavior at different time instants, which also depends on the Atwood number. Moreover, for the higher \At number cases, this occurs at different times for different density regions.

\section{Flow regimes} \label{Sec:Flow_regimes}
\subsection{Explosive growth}\label{Sec:Explosive}
\begin{figure}
    \centering
    \includegraphics[width=9cm]{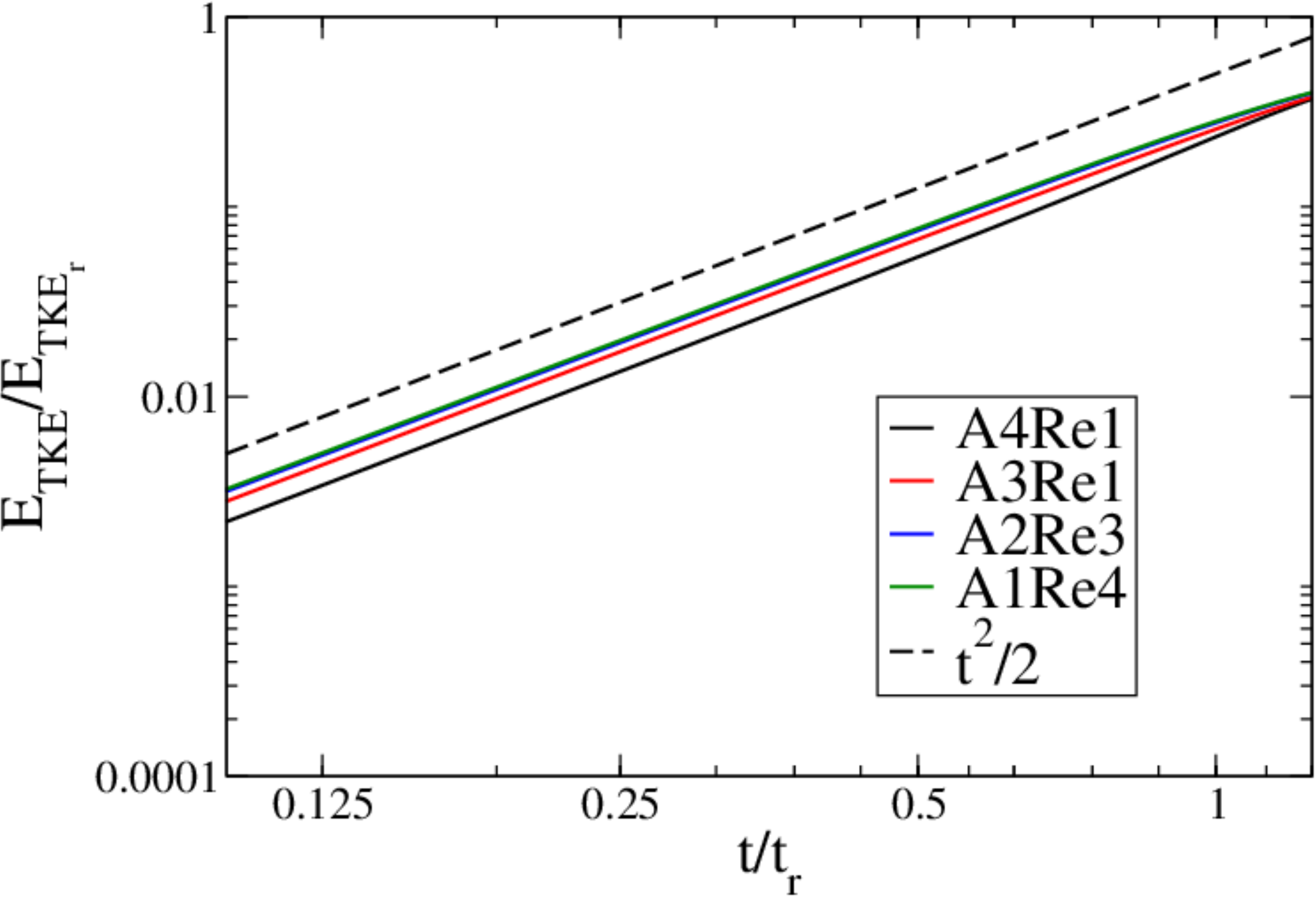}
    \caption{Growth of the $E_{TKE}$ during explosive growth.}
    \label{Fig:growth_rate}
\end{figure}

Explosive growth is initiated when the large structures in the domain start to move fast enough to generate turbulence. As the generation of $E_{TKE}$ is accelerated in this regime ($d^2E_{TKE}/dt^2>0$), there are certain similarities with the core region of the RTI mixing layer. In RTI, when the flow becomes self-similar, h (the mixing layer width) grows quadratically, with the leading order term of the form $A g t^2$. Consequently, the turbulent kinetic energy variation, which can be estimated as $\approx 1/2\Dot{h}^2$ also grows quadratically. Figure \ref{Fig:growth_rate} shows that the growth of the $E_{TKE}$ during explosive growth scales as $\approx t^2$, similar to RTI. 

During this regime, the amounts of pure fluids start decreasing slowly as mixing is initiated (Figure \ref{Fig:Regimes}).  This decay is slow, as molecular mixing occurs mostly at the interface of the pure fluids, where stirring first develops. No significant differences between the behavior of the pure -light and -heavy fluids are observed for the \At numbers investigated in this paper.

\begin{figure}
\hspace{3.3cm}(\emph{a}) \hspace{6.1cm} (\emph{b})

\vspace{0.5cm}
\hspace{.9cm}   Case:A1Re5 \hspace{1.32cm} Case:A4Re2 \hspace{1.45cm}  Case:A1Re5 \hspace{1.3cm} Case:A4Re2 \\
\rotatebox{90}{\hspace{0.9cm}$t/t_r=0$}\includegraphics[width=3.2cm]{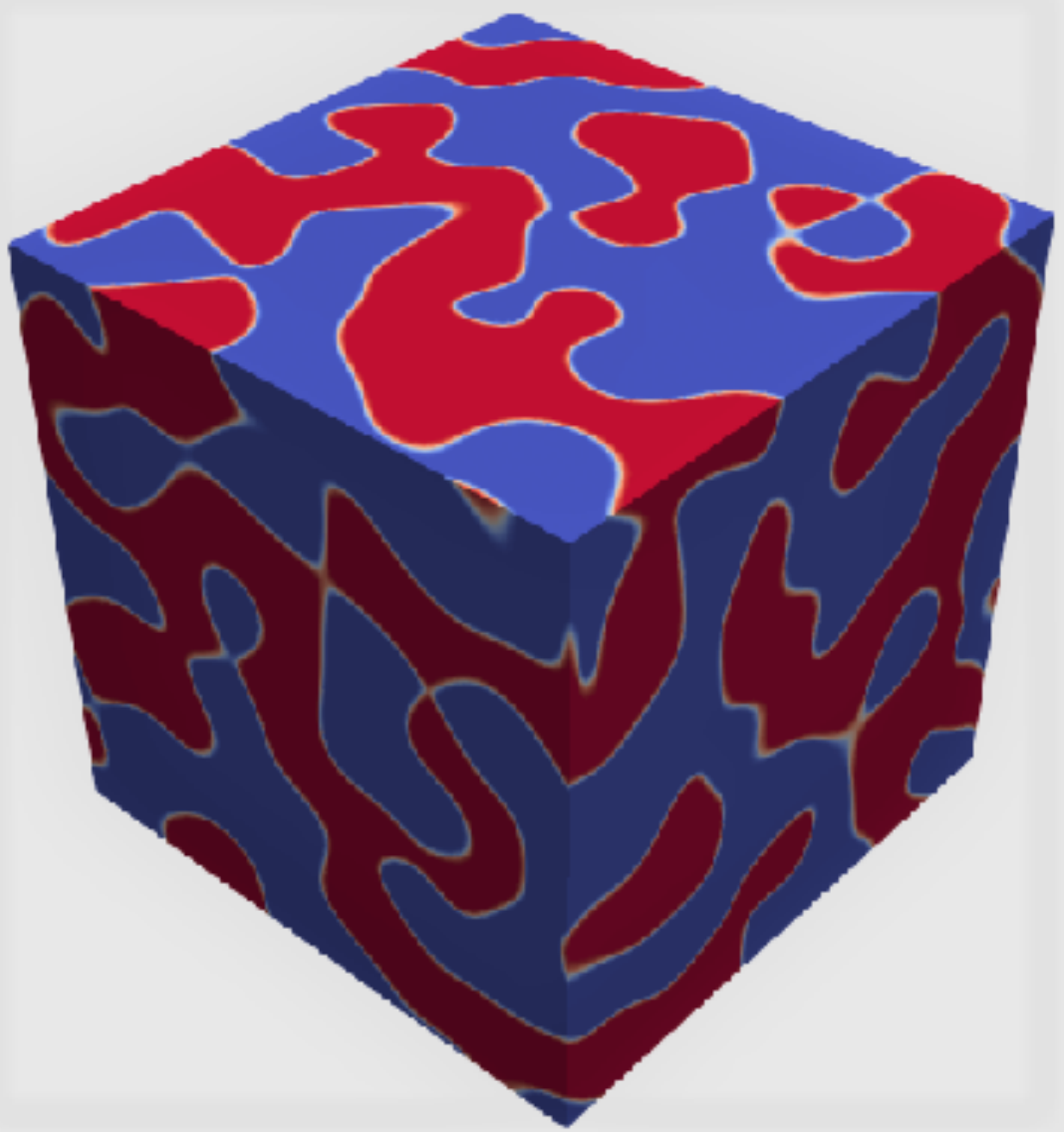}
\includegraphics[width=3.2cm]{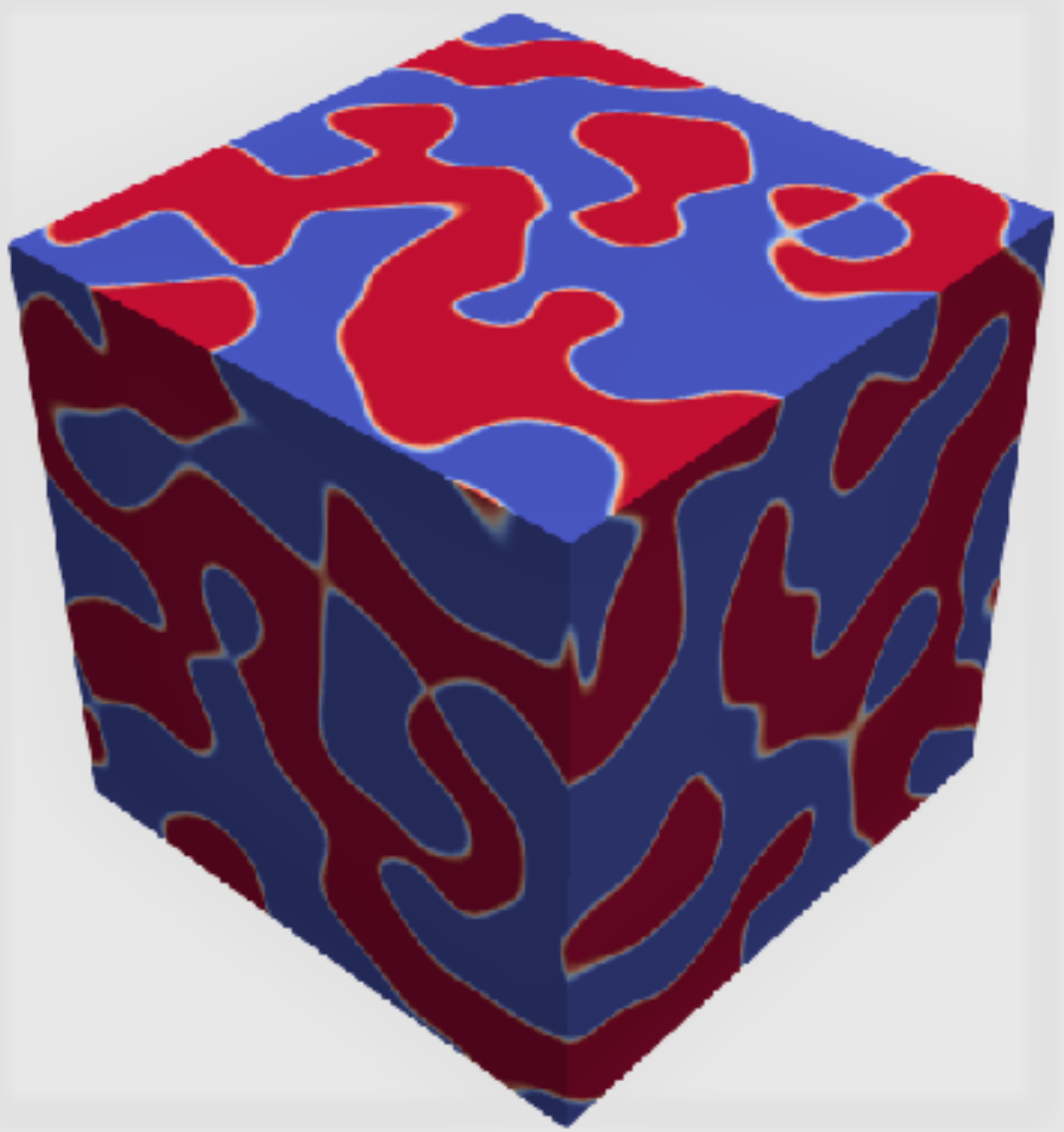}~~
\includegraphics[width=3.2cm]{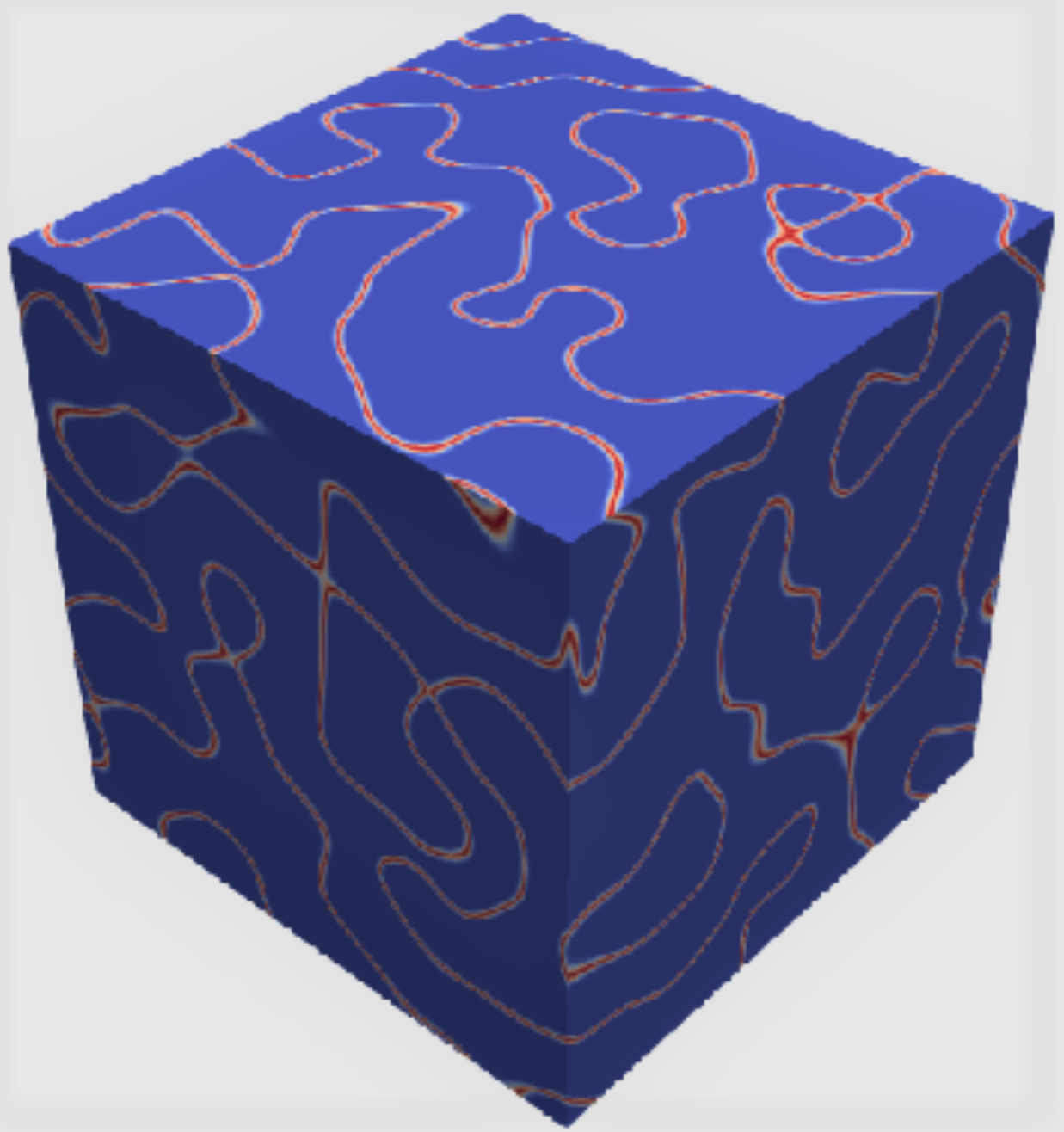}
\includegraphics[width=3.2cm]{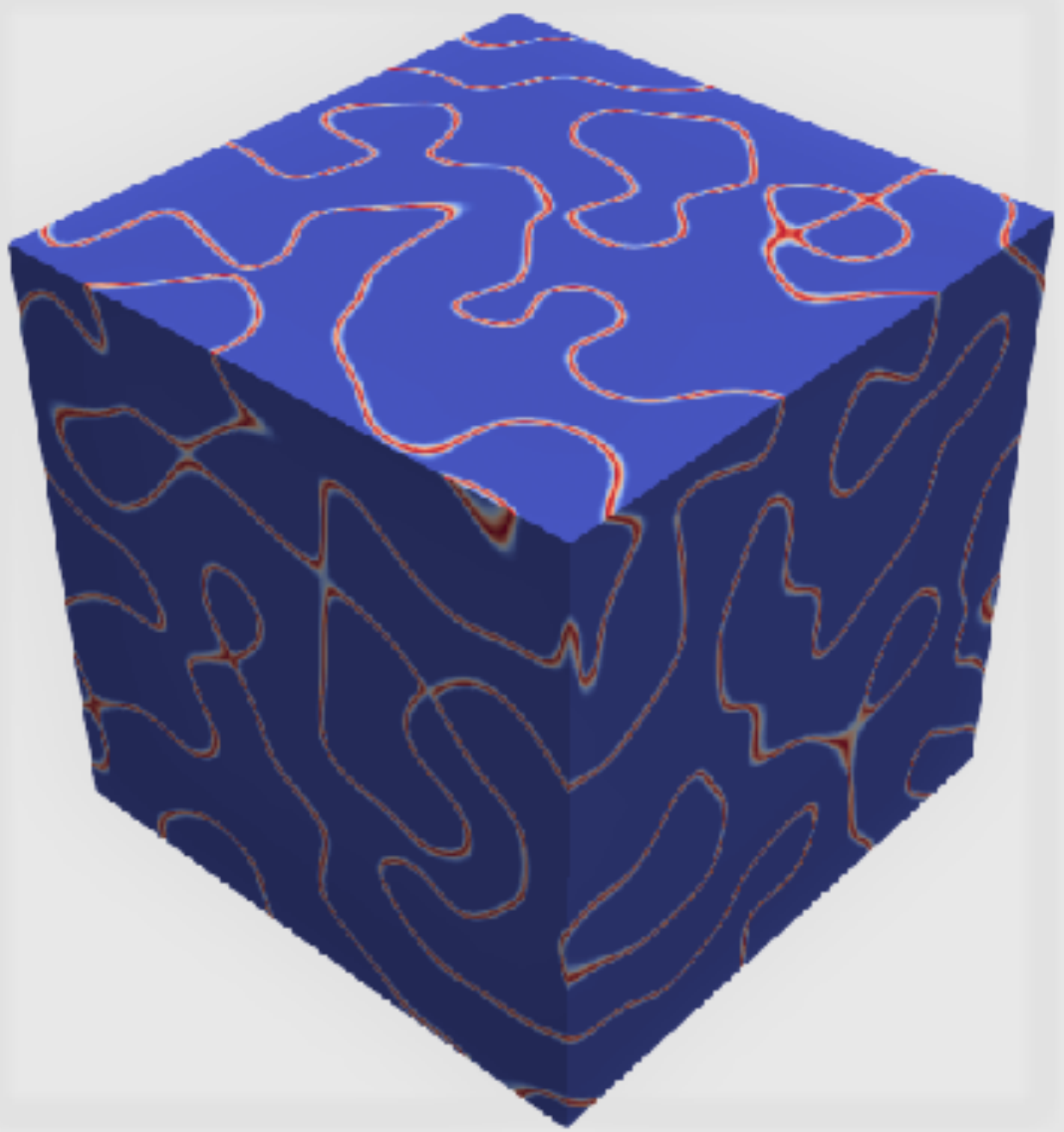}\\
\rotatebox{90}{\hspace{0.9cm}$t/t_r=0.65$}\includegraphics[width=3.2cm]{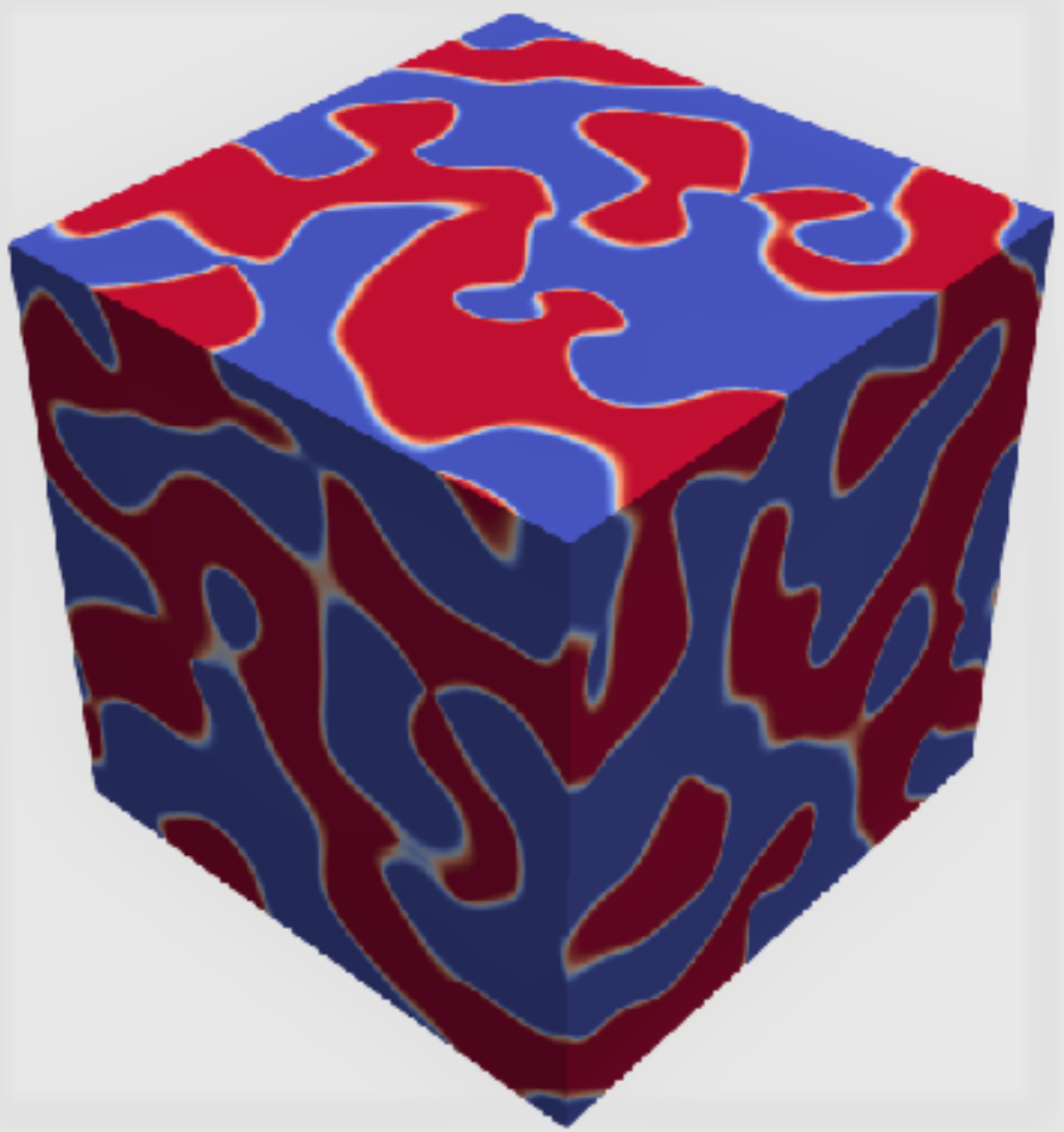}
\includegraphics[width=3.2cm]{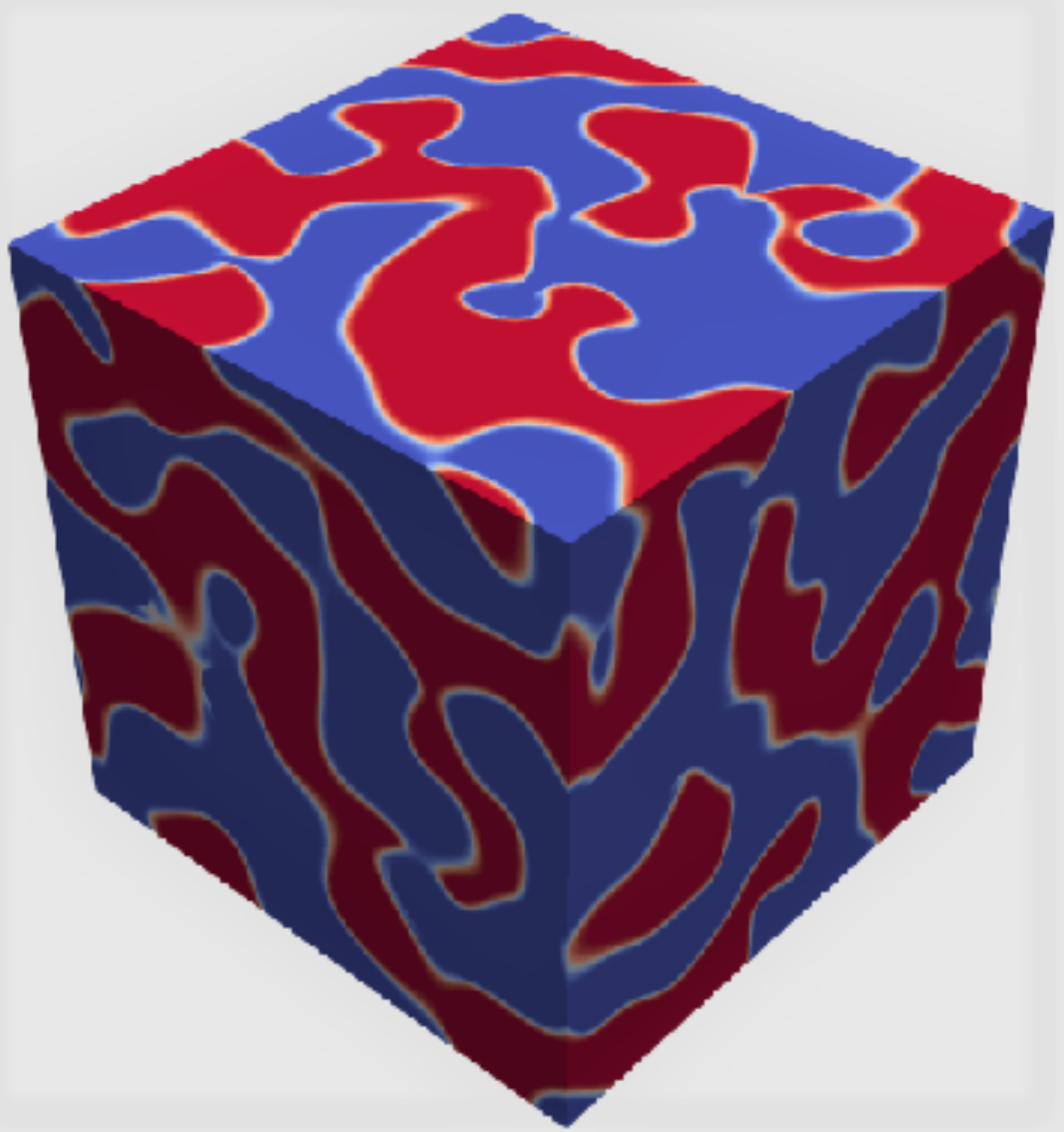}~~
\includegraphics[width=3.2cm]{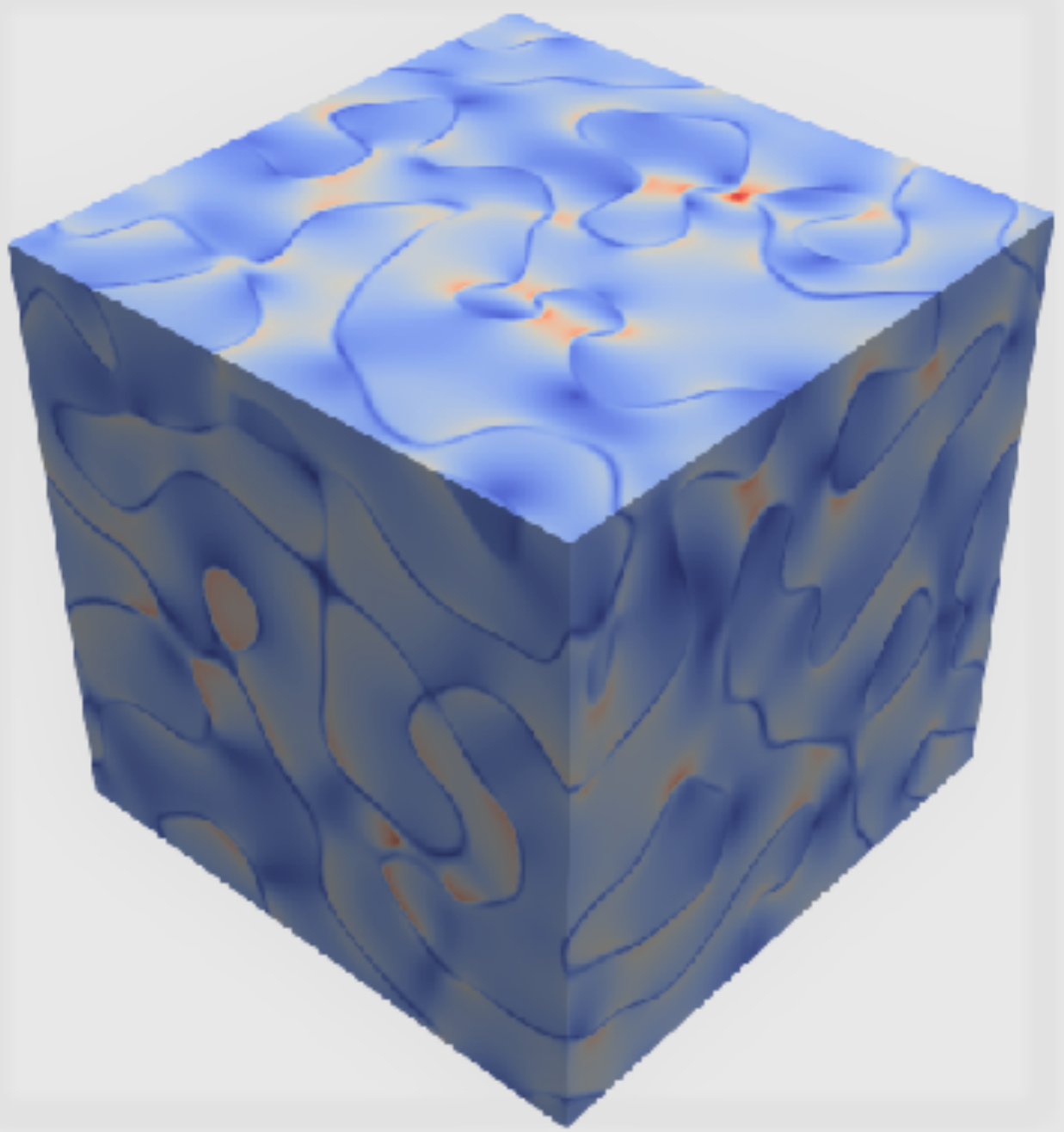}
\includegraphics[width=3.2cm]{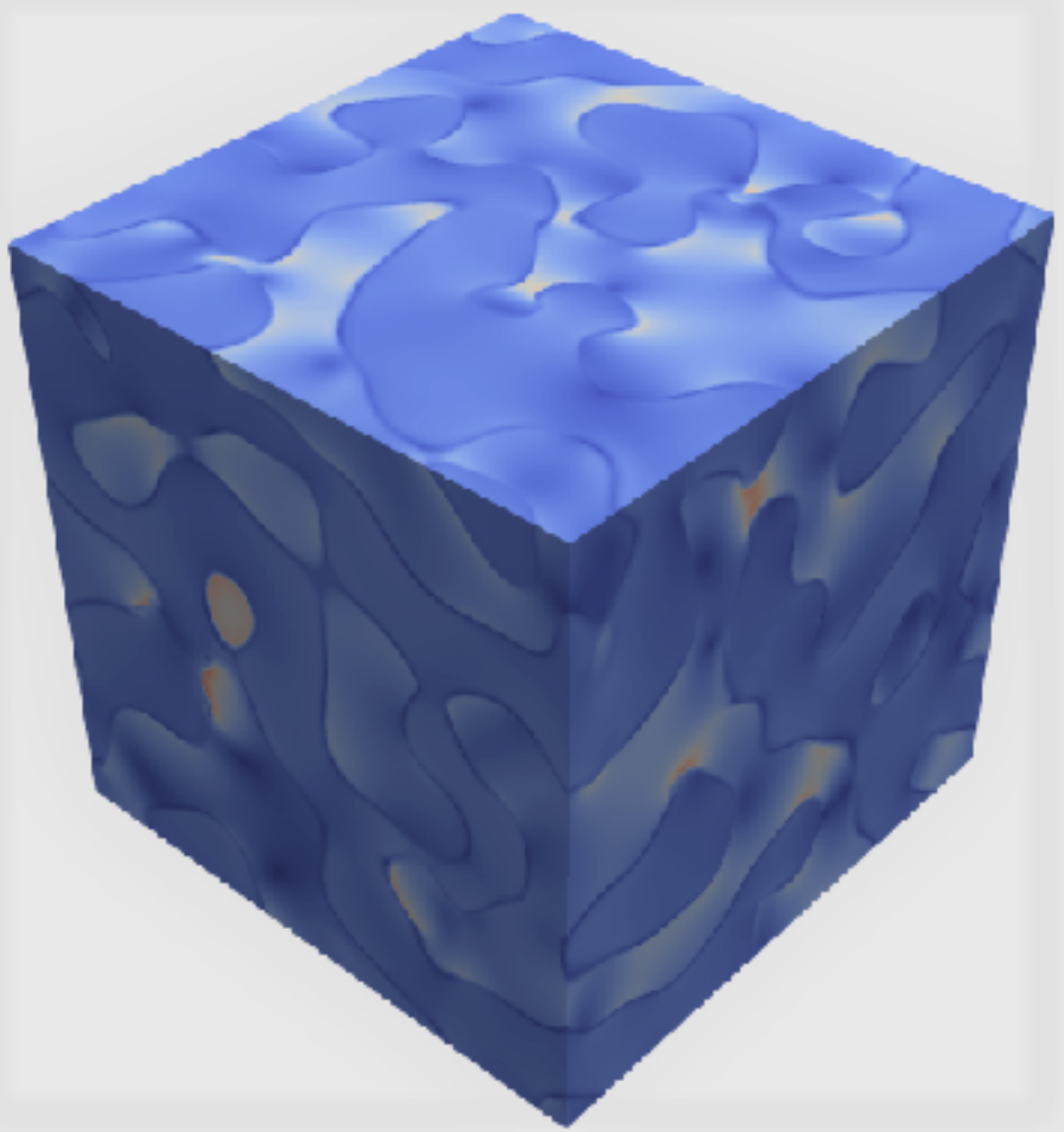}
 \caption{3D visualization of (a) the mole fraction where blue represent the pure light fluid ($\chi_h=0, \chi_l=1$) and red represents the pure heavy fluid ($\chi_h=1, \chi_l=0$), and (b) the velocity magnitude ($\sqrt{u^2_1+u^2_2+u^2_3}$) where blue represent the minimum velocity magnitude and red represents the maximum velocity magnitude for the cases $A=0.05$ (A1Re5) and $A=0.75$ (A4Re2). The top row shows 3D contours at $t/t_r=0$ while the bottom row displays contours at $t/t_r=0.65$. }
\label{Fig:3Devolve_T1}
\end{figure}

Figure \ref{Fig:3Devolve_T1} shows the 3D evolution of (a) the mole fraction, and, (b) the velocity magnitude for $A=0.05$ and $A=0.75$ (low and high density ratios) at different time instants. At $t/t_r=0$, the domain consists of pure light and heavy fluid patches separated from each other by thin layers; the mole fraction distribution is identical for both cases. At $t/t_r=0.65$, when the flow experiences a period of explosive growth, the behavior of the density field (or mole fraction) is still similar for both low and high \At numbers. 
    The large structures are observed to conserve their shapes during this regime and do not attend to molecular mixing; this is consistent with the slow decrease of the volume fractions of pure fluids in figure \ref{Fig:Regimes}.
However, at $t/t_r=0.65$, the velocity field is very different for low and high \At numbers. At first glance, the velocity field is more homogeneously distributed for the low \At number case; whereas for the high \At number case, the largest velocity magnitudes are more concentrated within regions that are occupied by the lighter than average fluid. 
    
\subsubsection{Energy conversion rates}\label{Sec:Mixing_eff1}

\begin{figure}
    \centerline{\includegraphics[width=8cm]{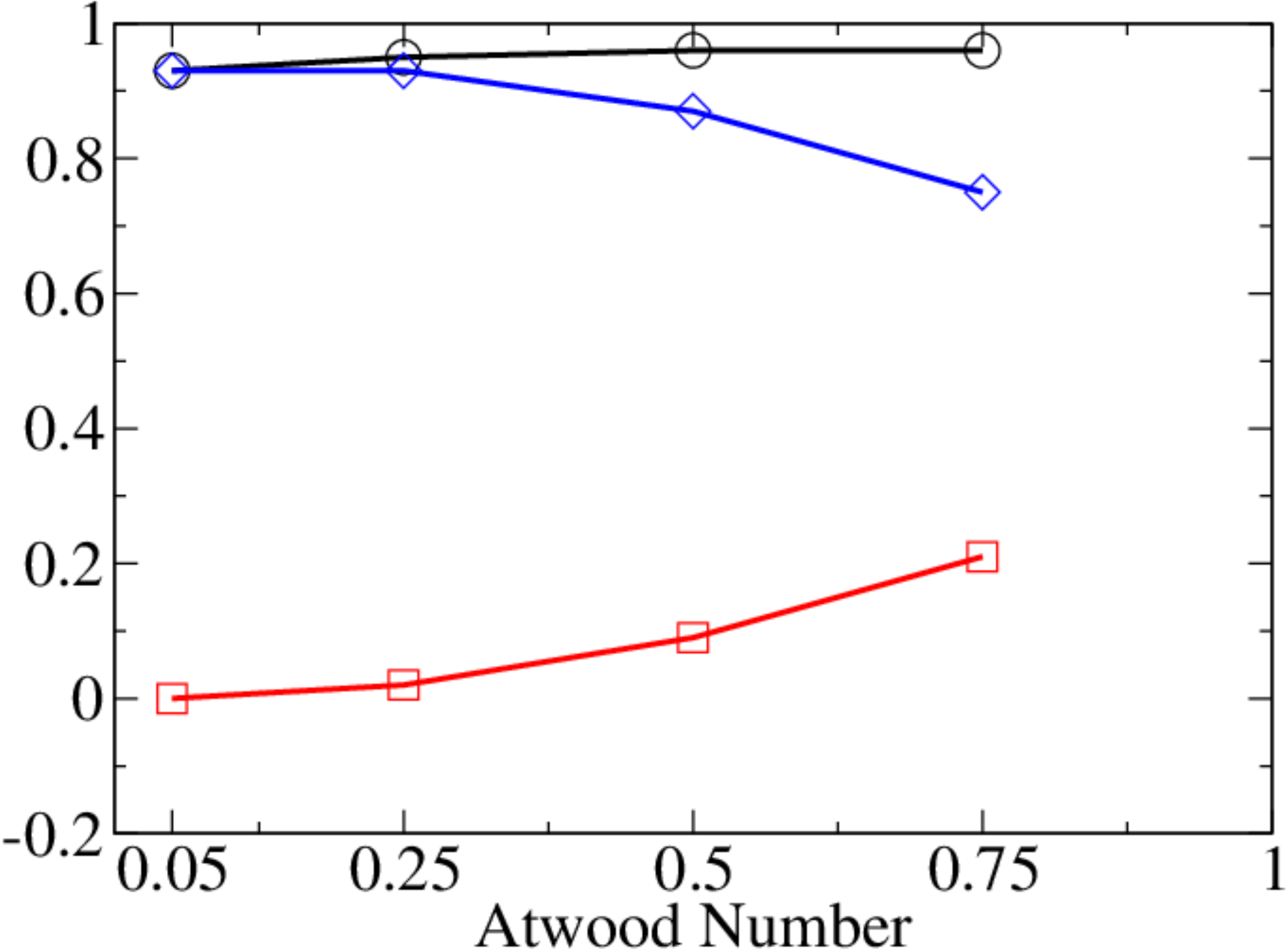}}
\caption{Atwood number effects on energy conversion rates ($\beta_{KE}$ - black line), ($\beta_{TKE}$ - blue line and $\beta_{MKE}$ - red line) during explosive growth.}
\label{Fig:Eff_At_1}
\end{figure}

\begin{figure}
(\emph{a}) \hspace{6cm}  (\emph{b})\\
    \includegraphics[width=6.5cm]{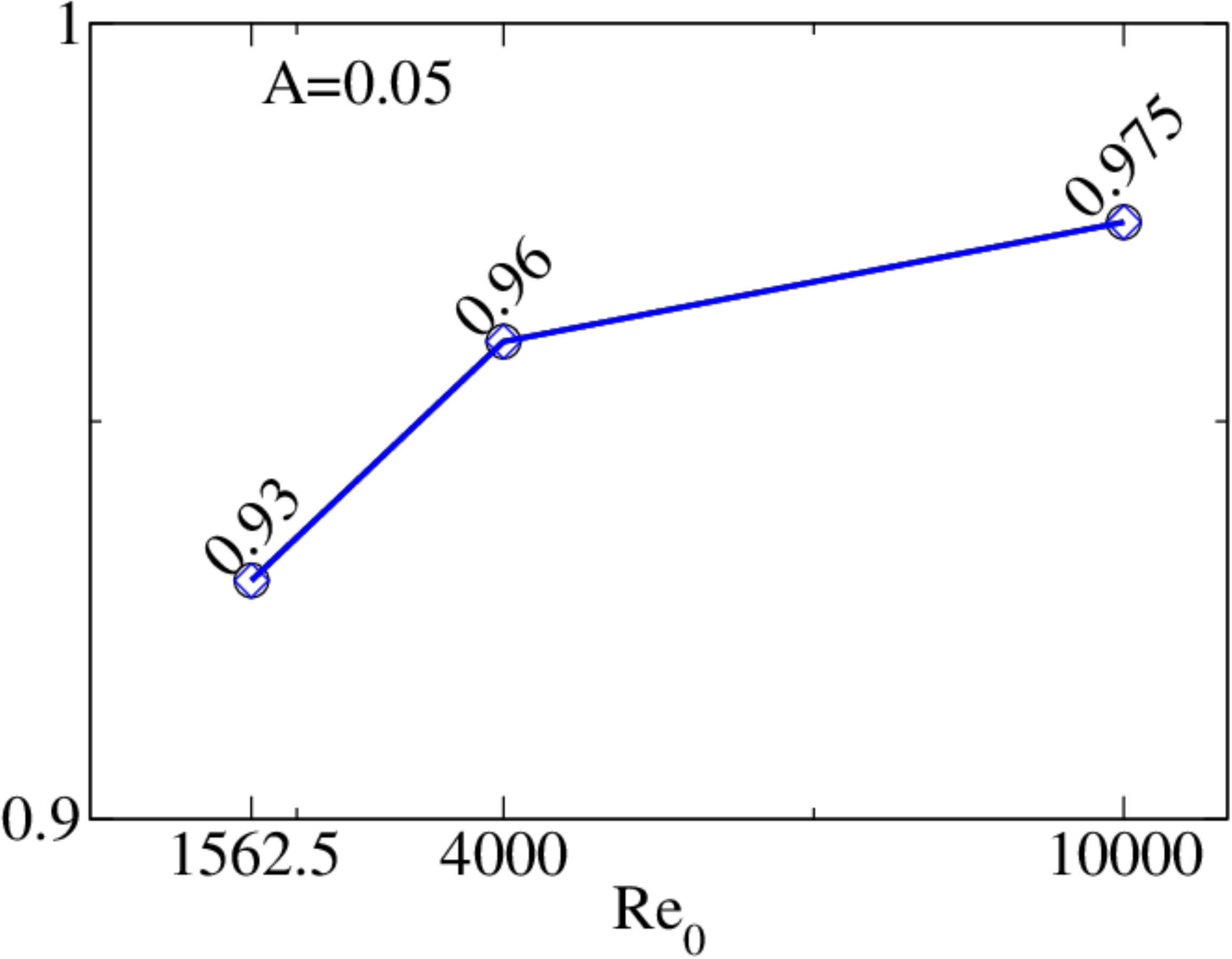}
    \includegraphics[width=6.5cm]{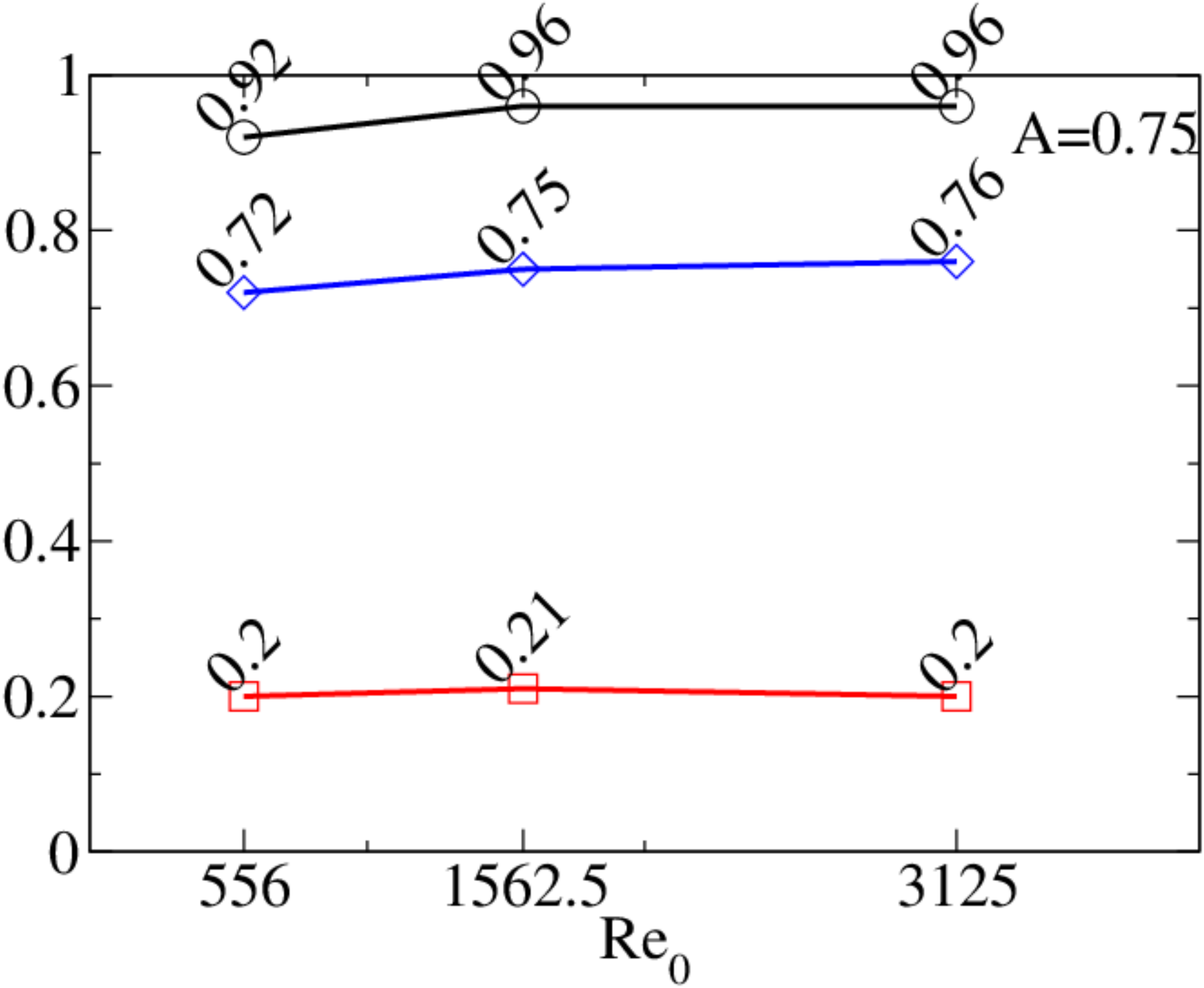}
\caption{Reynolds number effects on energy conversion rates ($\beta_{KE}$ - black line), ($\beta_{TKE}$ - blue line and $\beta_{MKE}$ - red line) during explosive growth regime for (a) \At $=0.05$ (blue and black lines are identical, while the red line is close to zero) and (b) \At $=0.75$.}
\label{Fig:Eff_Re_1}
\end{figure}

Using eqs. (\ref{Eq:mixing_eff}), the energy conversion rates are calculated for different \At and \rez numbers. Figures \ref{Fig:Eff_At_1} and \ref{Fig:Eff_Re_1} show the variation $\beta_{KE}$, $\beta_{TKE}$ and $\beta_{MKE}$ with \At and \rez numbers. During explosive growth, high levels of turbulence generation occurs as most of the potential energy lost gets converted to kinetic energy ($\beta_{KE}>\%90$). For cases with the same \rez, $\beta_{KE}$ is similar for all \At numbers; however, it increases slightly with an increase in the value of \rez. Meanwhile, $\beta_{TKE}$ is found to dramatically decrease upon increase of the \At number. 
    As also observed in Fig. \ref{Fig:Regimes}, for moderate ($0.5$) and high ($0.75$) \At numbers, there is a delay in $E_{TKE}$ growth compared to $E_{KE}$. This delay might be attributed to the larger inertial differences between heavy and light fluids regions. Initially, heavy fluid regions may not be stirred as efficiently as the light fluid regions due to their larger inertia and this causes a decrease in $\beta_{(TKE)}$ for larger \At numbers. In addition, the difference between $E_{KE}$ and $E_{TKE}$ is stored within $E_{MKE}$, which acts as a reservoir for $E_{TKE}$ for the subsequent regimes in the flow evolution. Furthermore, for the low $A$ number case, the energy conversion rates tend to asymptote to constant values, while for the high $A$ number case, they asymptote to constant values when $Re_0$ is increased (see Fig. \ref{Fig:Eff_Re_1}). This indicates that the flow undergoes the mixing transition in terms of energy conversion rates as the effects of the increase in the $R_0$ become negligible.

\vspace{0.5cm}
\subsubsection{Atwood number effects on PDF evolutions}\label{Sec:PDF_1}
\vspace{3mm}
\noindent\textit{Density PDF} 

\begin{figure}
   (\emph{a}) \hspace{6.5cm}  (\emph{b}) \\
    \centerline{\includegraphics[width=6.5cm]{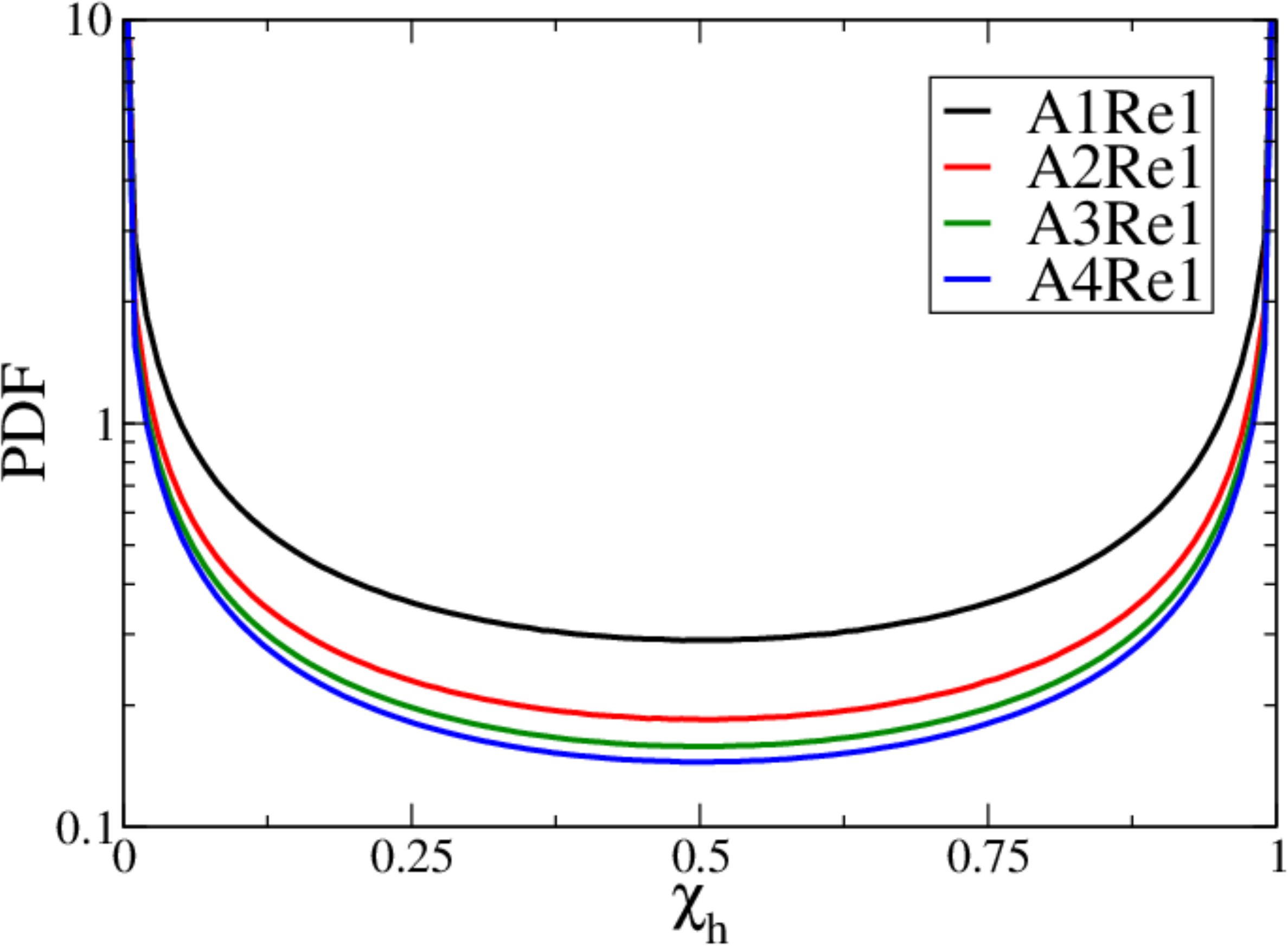}
    \includegraphics[width=6.5cm]{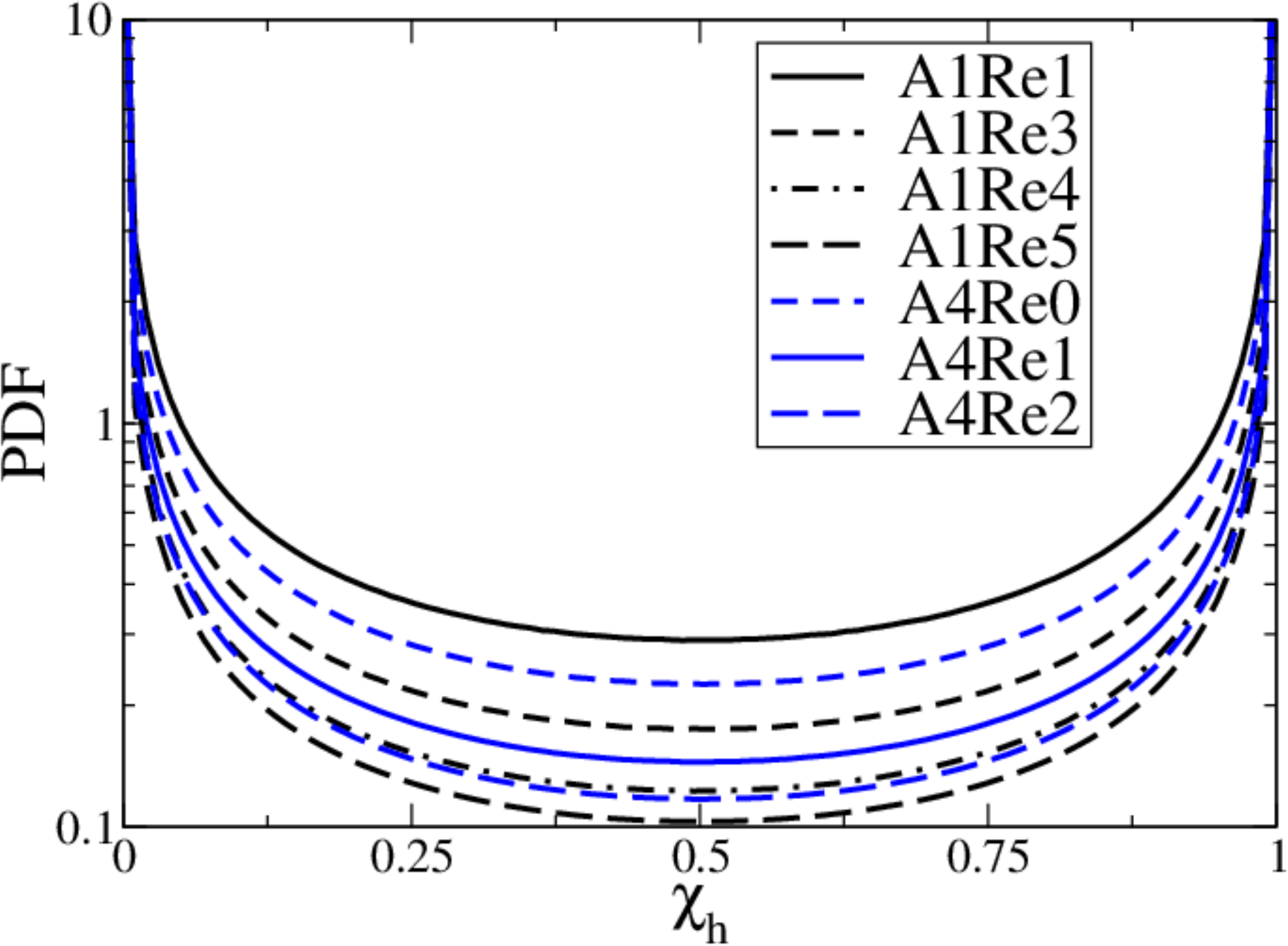}}
 \caption{PDF of density field for the different (a) \At and (b) \rez values at $t/t_r=0.65$.}
\label{Fig:dens_PDF_1}
\end{figure}

Figure \ref{Fig:dens_PDF_1} presents \At and \rez numbers effects on the density PDF during explosive growth ($t/t_r=0.65$). In all figures, the density field is represented by the heavy fluid's mole fraction ($\chi_h$); due to incompresibility, $\chi_h=(\rho^{*}-\rho_1)/(\rho_2-\rho_1)$. The PDF behaves similarly for low and high \At numbers. In addition, \rez number does not have any significant effect on these PDFs. The A1Re1 case is slightly more mixed compared to the other cases due to its higher initial mix-state as described in section \ref{Sec:sim cases}. Weak \At number dependency on density PDF is consistent with the observations reported for Figure \ref{Fig:3Devolve_T1}. The mixing rates are smaller during this regime and stirring is mostly localized; as a result \At number effects on density PDF are not prominent.

\vspace{3mm}
\noindent\textit{Velocity-density jMDF} 
 
\begin{figure}
\vspace{0.5cm}
\hspace{2.6cm}   Case:A1Re5 \hspace{3.8cm} Case:A4Re2  \\
\centerline{\includegraphics[width=7.0cm]{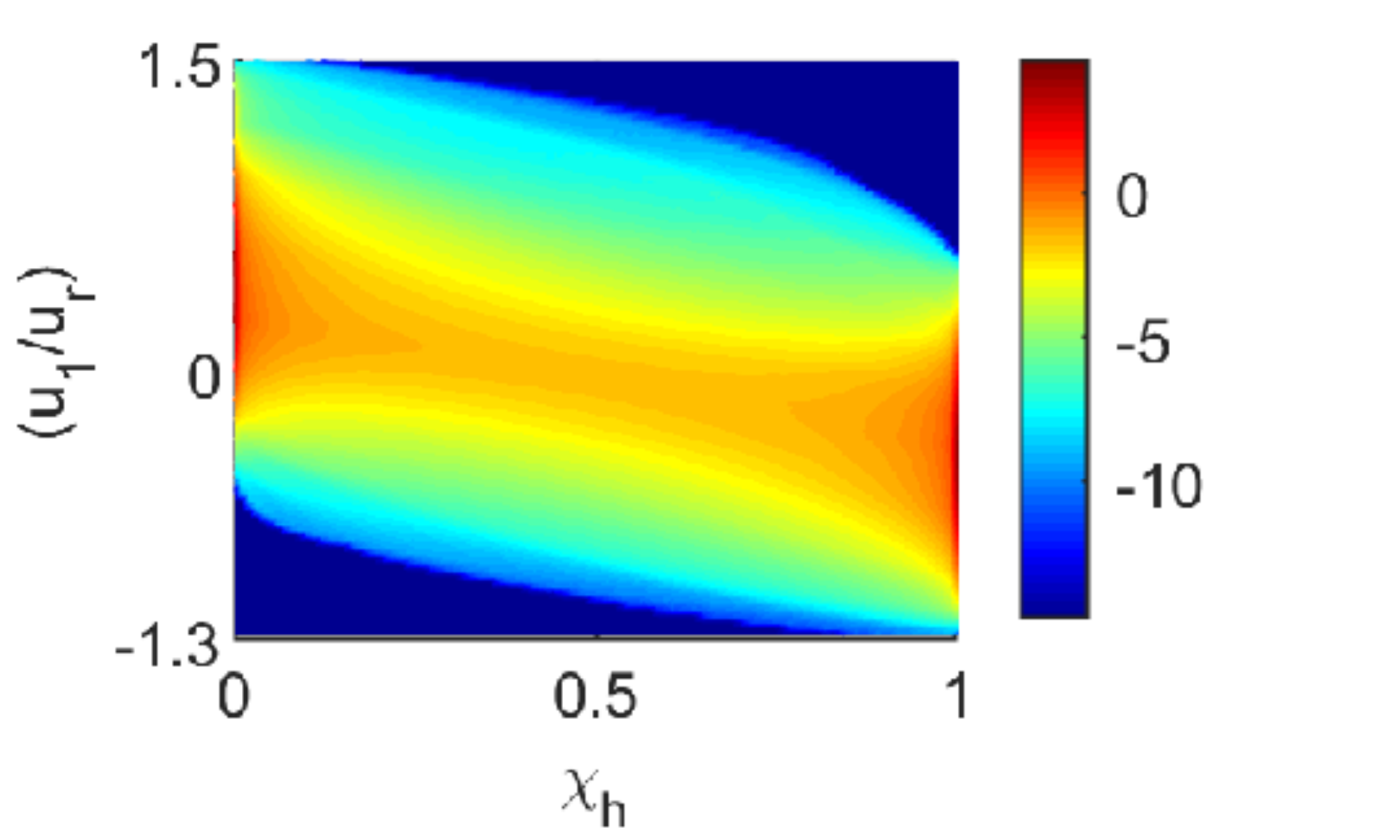} \includegraphics[width=7.0cm]{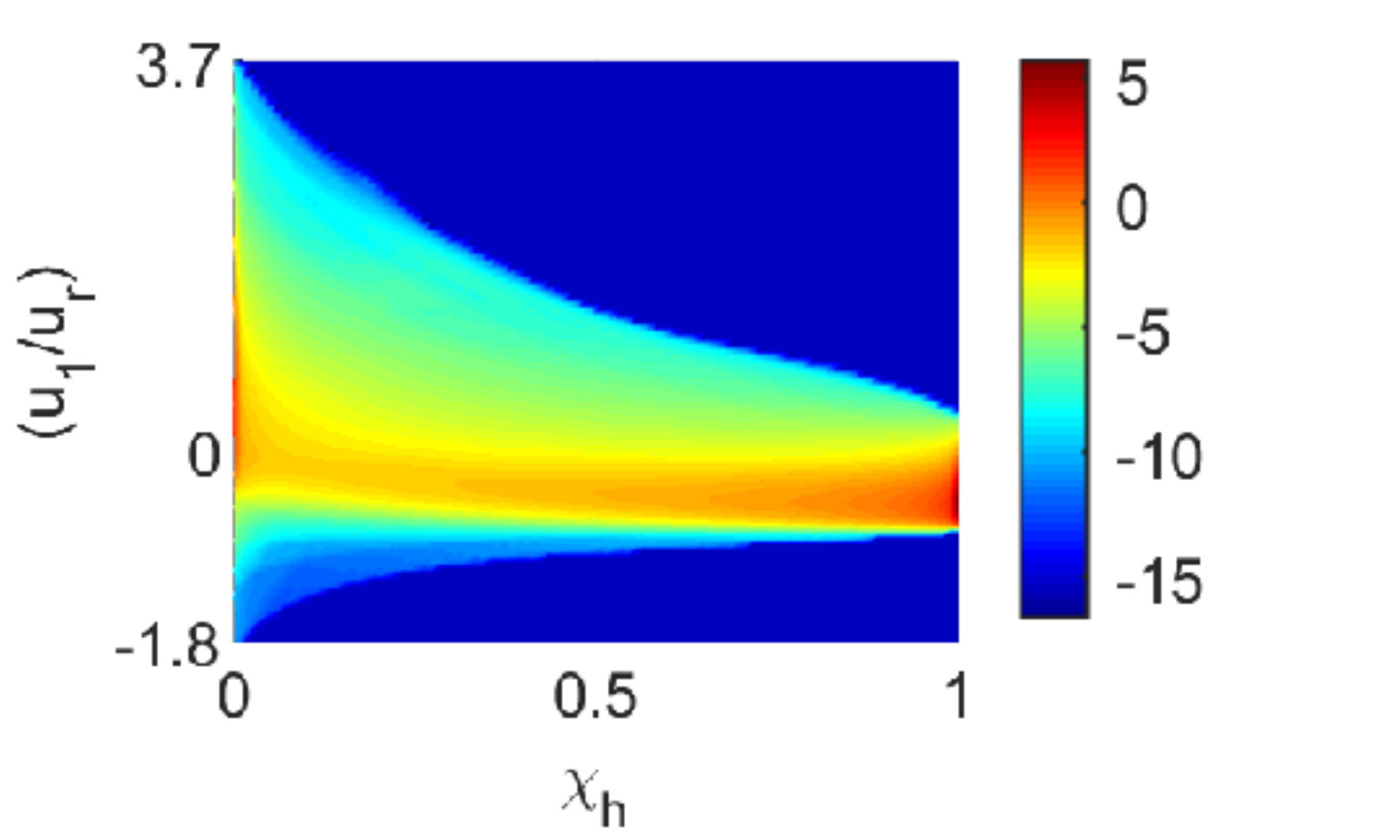}}
 \caption{Normalized jMDFs [$\log(\mathcal{F}/\rhom)$] for (a) $A=0.05$ (A1Re5) and (b) $A=0.75$ (A4Re2) cases displayed at $t/t_r=0.65$.}
\label{Fig:Joint_1}
\end{figure}

As observed in Figure \ref{Fig:3Devolve_T1}, the velocity field starts to behave differently for low and high \At numbers right after the acceleration ramp-up; this is noticeably earlier than the density field. The flow is highly anisotropic in the vertical (accelerated) direction \citep{livescu2008}. Figure \ref{Fig:Joint_1} shows the normalized jMDF [$\log(\mathcal{F}/\rhom)$] corresponding to the vertical component of the velocity field ($u_1/u_r$) for the low and high \At number cases (at $t/t_r=0.65$). 
    It is observed that for both low and high \At numbers, lighter fluid regions mostly move in the opposite direction of acceleration field, while heavier fluid regions mostly move in the acceleration field direction. During explosive growth, the jMDF has double-delta shape with two peak points, as a result of initial double-delta density PDF; the peak points are on the negative and positive velocity sides. These two peaks indicate that there are two different velocity sub-distributions within the flow. Similar to RTI, the negative side represents the velocity distribution within pure heavy fluid regions (which are mostly going down) and the positive side represents the velocity distribution within pure light fluid regions (which are mostly going up).
 This is also consistent with high \At number experiments \citep{banerjee_kraft_andrews_2010}, where it is reported that lighter fluid regions mostly move in the opposite direction of the acceleration field as they are associated with the rising bubbles, and heavier fluid regions mostly move in the same direction of the acceleration field as they are associated with the dropping spikes.
        Moreover, for low \At number cases, the velocity distribution is almost symmetric within the different regions of the flow (for the different levels of $\chi_h$) and density and velocity fields are moderately correlated. However, for high \At number case, the velocity distribution within the different regions of the flow becomes highly-asymmetric and the pure light fluid regions move much faster than the pure heavy fluid regions. 
Those faster motions for larger \At numbers can be attributed to the $P_{,i}/\rho^*$ term in Eq. (\ref{Eq:joint}), which is asymmetric due to the specific volume variation. In the light fluid, $P_{,i}/\rho^*$ becomes large and, since turbulence is not developed yet, the non-linear terms do not compensate for this variation.

\subsubsection{Conditional expectations}\label{Sec:Cond_1}
In this subsection, we present the conditional expectations of $E_{TKE}$, $E_{TKE}$, dissipation, and enstrophy to study \At and \rez effects on the asymmetry with respect to the density field. 

\vspace{3mm}
\noindent\textit{Conditional expectation of $E_{TKE}$} 

\begin{figure}
(\emph{a}) \hspace{6.5cm}  (\emph{b}) \\
    \centerline{\includegraphics[width=6.4cm]{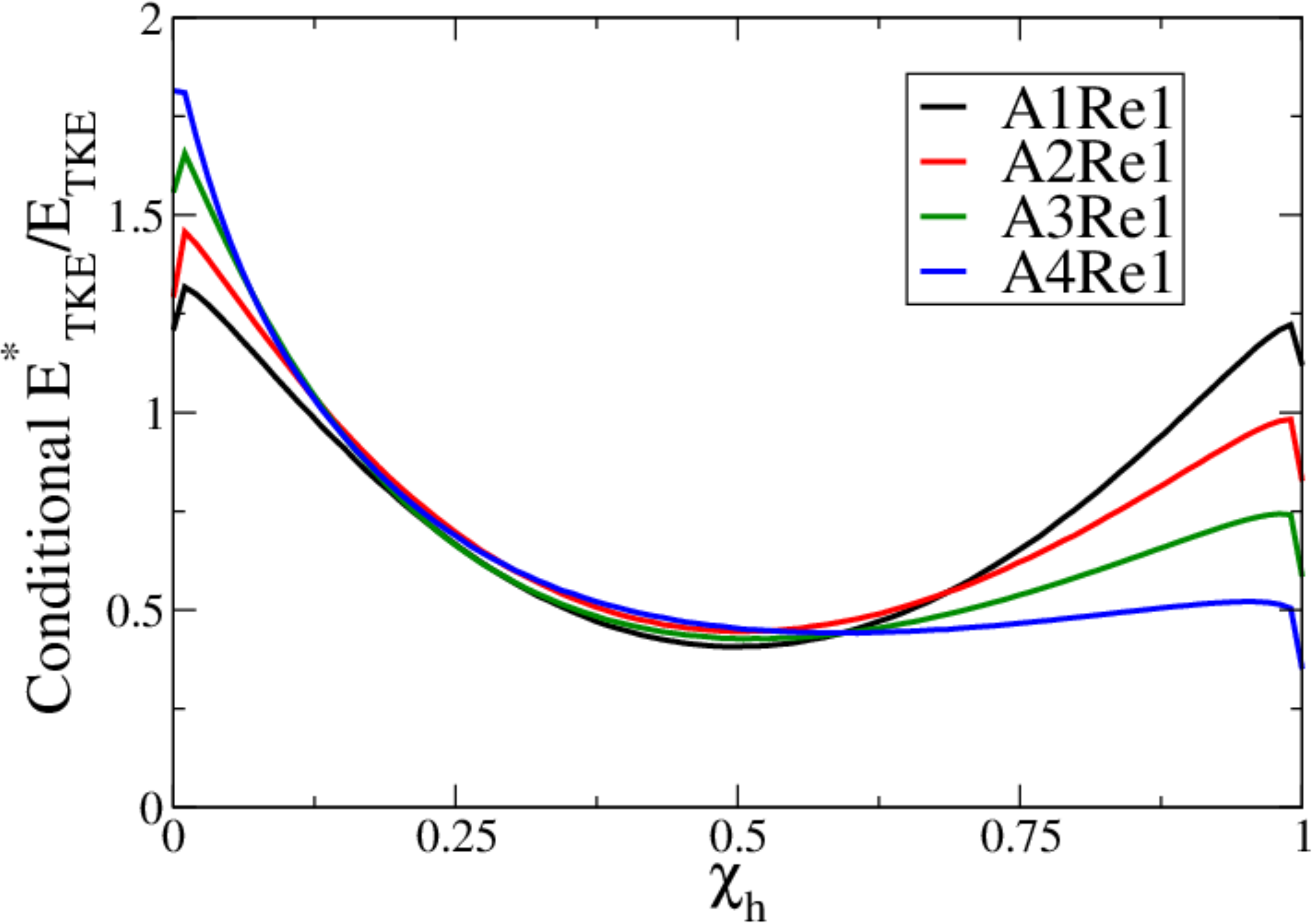}
    \includegraphics[width=6.4cm]{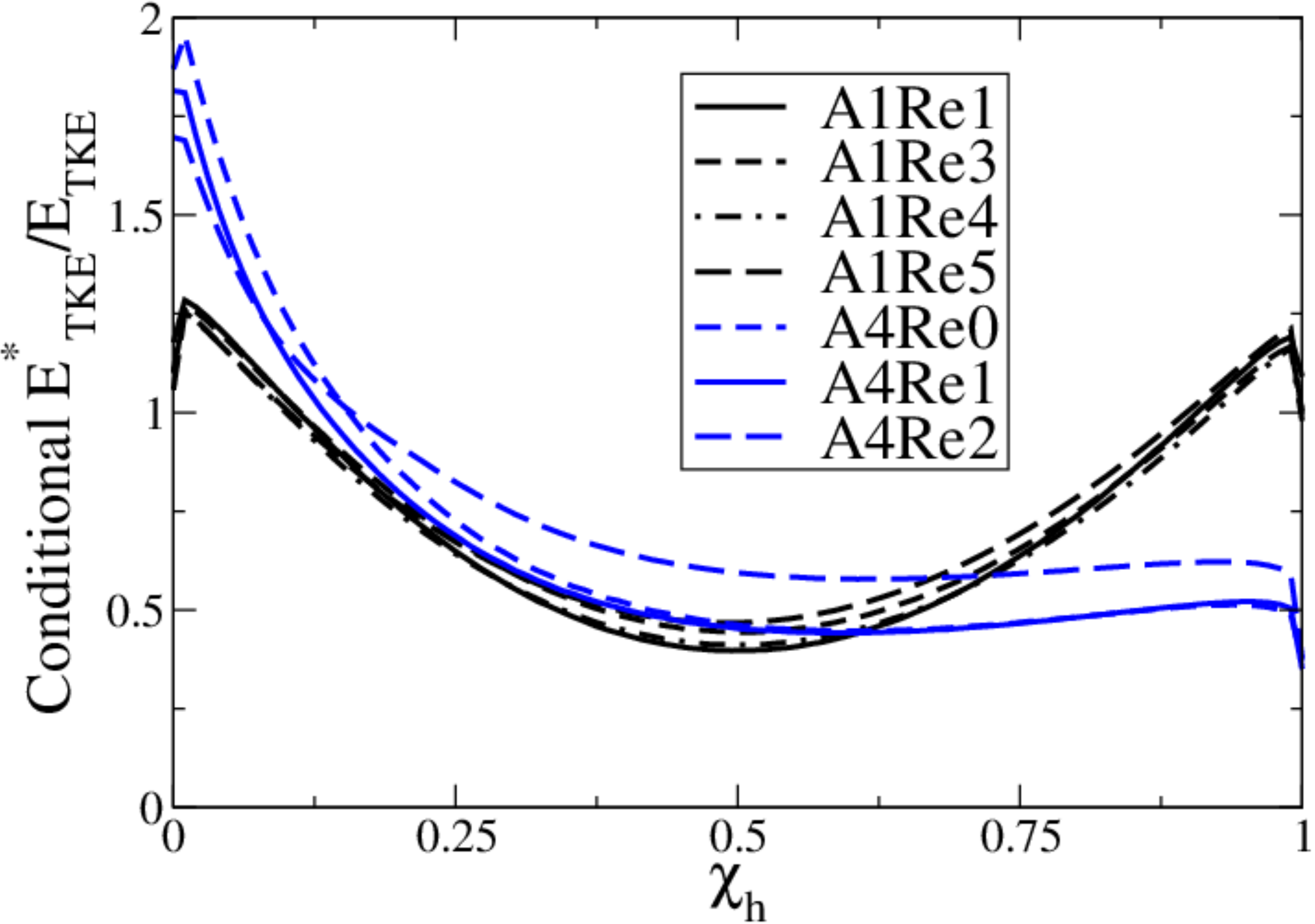}}
\caption{Atwood and Reynolds numbers effects on conditional expectation of $E_{TKE}$ during explosive growth regime ($t/t_r=0.65$).}
\label{Fig:cond_tke_1}
\end{figure}

The effects of variation of \At and \rez numbers on conditional expectations of $E_{TKE}$ are shown in Fig. \ref{Fig:cond_tke_1}. For higher \At numbers, light fluid regions become more stirred compared to heavy fluid regions. This is also illustrated in the asymmetric shape of the jMDF (see Fig. \ref{Fig:Joint_1}). Thus, $\langle\rho^* u_i^{''}u_i^{''}|_{\rho^*=R}\rangle$ within the lightest regions ($\sim \rho_1$) is larger than $E_{TKE}$ for the high $A=0.75$ case. \rez has a relatively small effect on the distribution of the $E_{TKE}$ within the different regions of the flow during explosive growth. In addition, the $E_{TKE}$ values are larger in the lighter fluid regions compared to the heavy fluid regions, irrespective of the Reynolds number. This effect is the largest for the $A=0.75$ cases. The larger $E_{TKE}$ values may also lead to earlier mixing transition in those regions. This is discussed in detail below.

\vspace{3mm}
\noindent\textit{Conditional expectation of $E_{TKE}$ dissipation}

\begin{figure}
(\emph{a}) \hspace{6.5cm}  (\emph{b}) \\
    \centerline{\includegraphics[width=6.4cm]{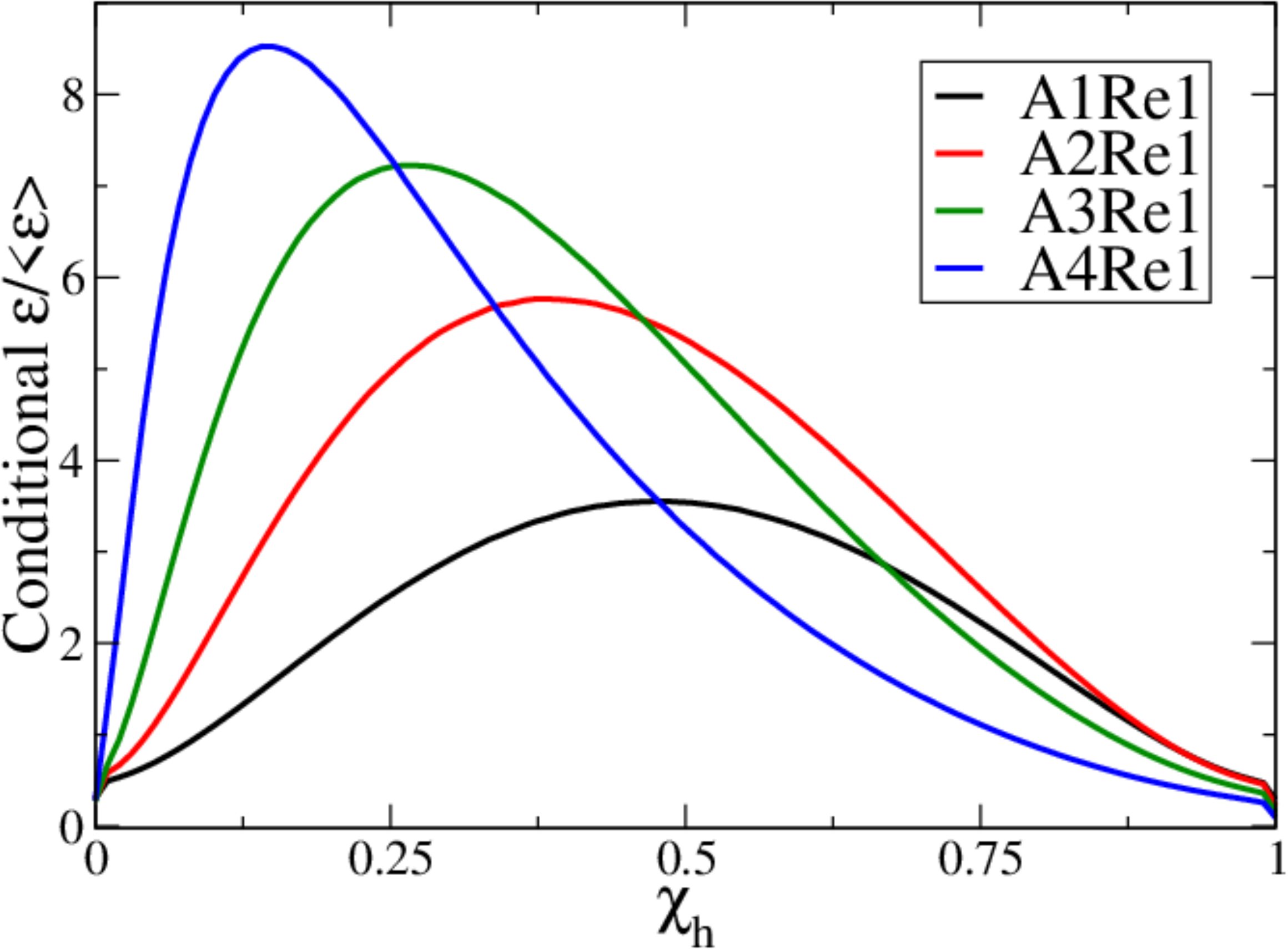}
    \includegraphics[width=6.4cm]{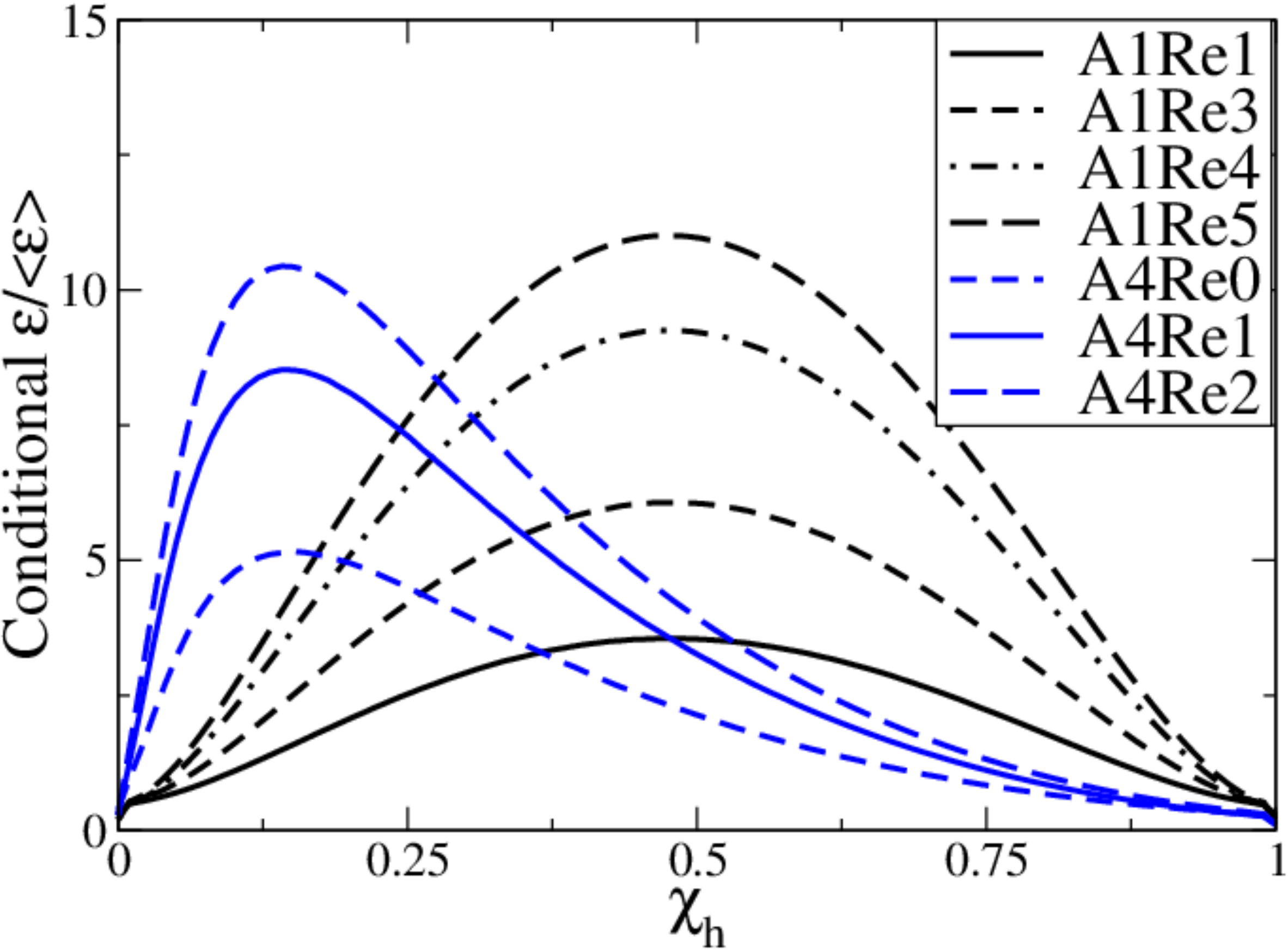}}
\caption{Atwood and Reynolds numbers effects on conditional expectation of $E_{TKE}$ dissipation during explosive growth regime ($t/t_r=0.65$).}
\label{Fig:cond_eps_1}
\end{figure}

The conditional expectation of the dissipation of $E_{TKE}$ is plotted in Fig. \ref{Fig:cond_eps_1}. For the lowest \At number case, dissipation takes its highest values around the mean density. However, for the larger \At number cases, the peak of dissipation moves to lighter fluid regions. This observation is consistent with Fig. \ref{Fig:Joint_1} where we observed the velocity distribution to be much wider in the lighter fluid region, thereby causing larger velocity gradients within the region.
    
    Increasing \At and / or \rez increases the local values more than statistics based on the whole volume. When local values become much larger than the average value, these are usually called 'extreme events' within the flow. For the same \rez values, the maximum local average (of dissipation) is larger compared to cases with lower \At numbers (see Figure \ref{Fig:cond_eps_1}a).
Moreover, for the A1Re5 case, $E_{TKE}$ dissipation within the fully mixed regions is more than ten times larger than its volume average; while for the A1Re1 case, $E_{TKE}$ dissipation within the fully mixed regions is only three times larger than its volume average (see Figure \ref{Fig:cond_eps_1}b)). These larger local values indicate that both increasing \At and / or \rez numbers may increase the probability of occurrence of 'extreme events' within the flow.

\vspace{3mm}
\noindent\textit{Conditional expectation of enstrophy} 

\begin{figure}
(\emph{a}) \hspace{6.5cm}  (\emph{b}) \\
    \centerline{\includegraphics[width=6.4cm]{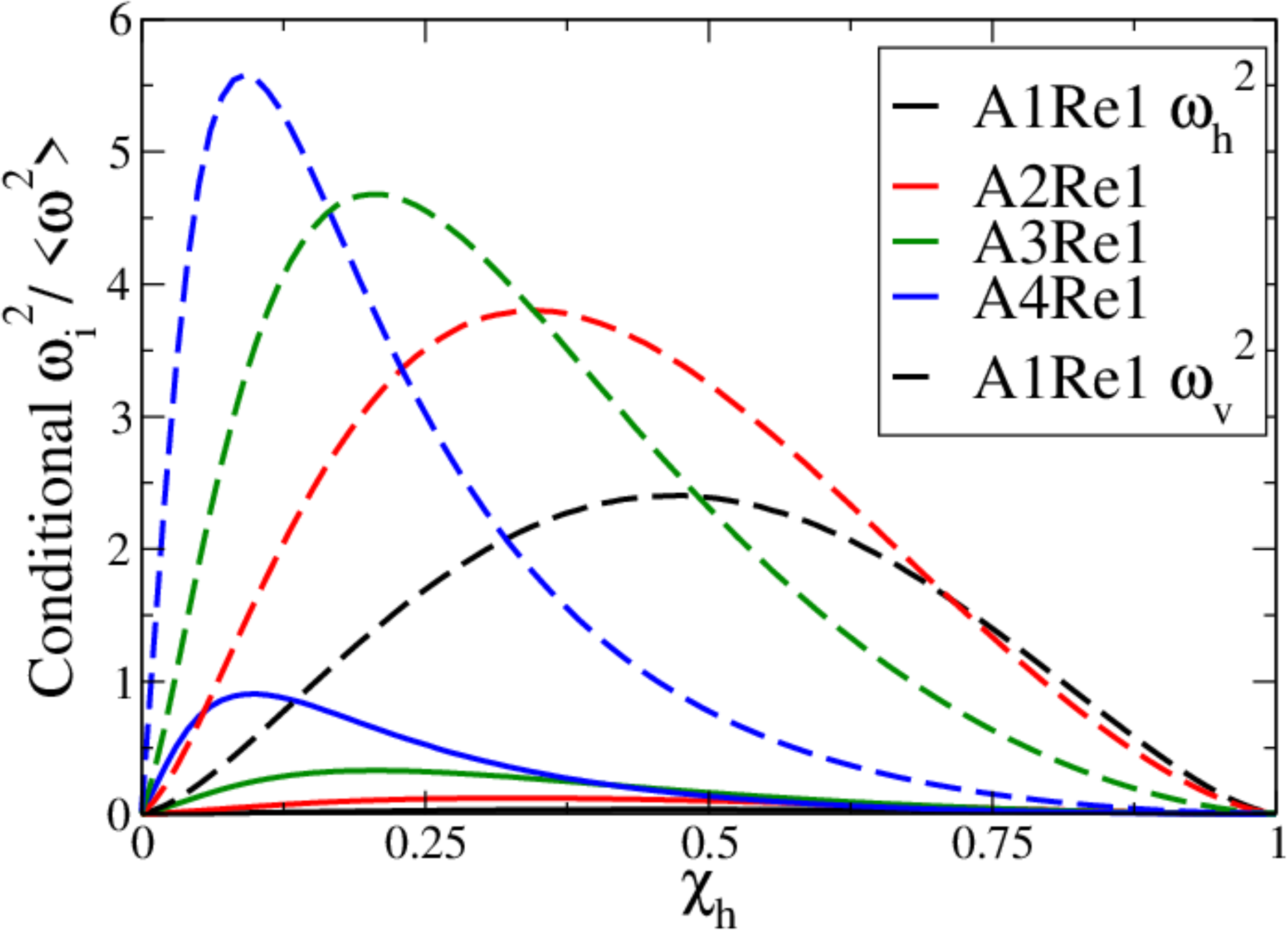}
    \includegraphics[width=6.4cm]{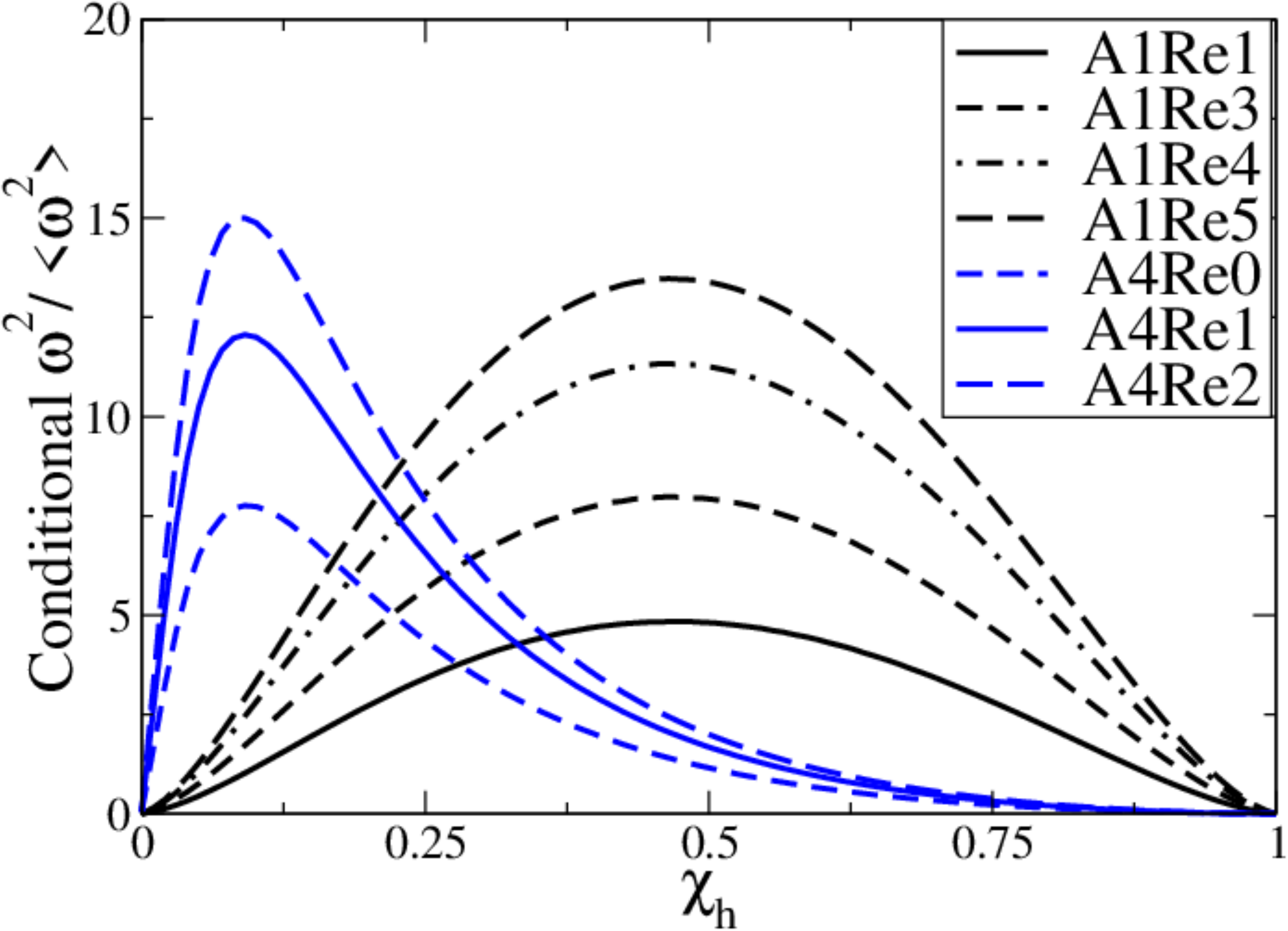}}    
\caption{Atwood and Reynolds numbers effects on conditional expectation of (a) vertical ($\omega^2_v=\omega^2_1$) and horizontal ($\omega^2_h=(\omega^2_2+\omega^2_3)/2$) enstrophy components and (b) total enstrophy during (I) explosive growth ($t/t_r=0.65$).}
\label{Fig:cond_ens_1}
\end{figure}

The conditional expectations of the components of the enstrophy ($\omega^2$) in vertical and horizontal directions and the total enstrophy are plotted in figs. \ref{Fig:cond_ens_1}a and \ref{Fig:cond_ens_1}b respectively. During this regime, the enstrophy behaves very similarly to dissipation, and becomes asymmetric at larger \At numbers. In addition, consistent with observations reported for dissipation, the local enstrophy values become larger than the volume average values upon increasing both \rez and \At numbers. Both dissipation and enstrophy are considered indicative of small-scale behavior \citep{pope2000turbulent}; thus, the asymmetric behavior of these quantities indicates asymmetry even at small scale motions at large \At numbers ($A \geq 0.5$). It is also observed (see Fig. \ref{Fig:cond_ens_1}a) that the vertical component of the enstrophy is much smaller than horizontal components; for example, for the A1Re1 and A4Re1 cases, $\omega^2_h$ are ten and five times larger than $\omega^2_v$, respectively. This implies that the flow is anisotropic even at small scales. This aspect is discussed in the spectral evolution section \ref{Sec:Spectral_evolve}.

\subsection{Saturated growth}\label{Sec:Saturated_growth}
\begin{figure}
\hspace{3.3cm}(\emph{a}) \hspace{6.1cm} (\emph{b})

\vspace{0.5cm}
\hspace{.9cm}   Case:A1Re5 \hspace{1.32cm} Case:A4Re2 \hspace{1.45cm}  Case:A1Re5 \hspace{1.3cm} Case:A4Re2 \\
\rotatebox{90}{\hspace{0.9cm}$t/t_r=1.75$}\includegraphics[width=3.2cm]{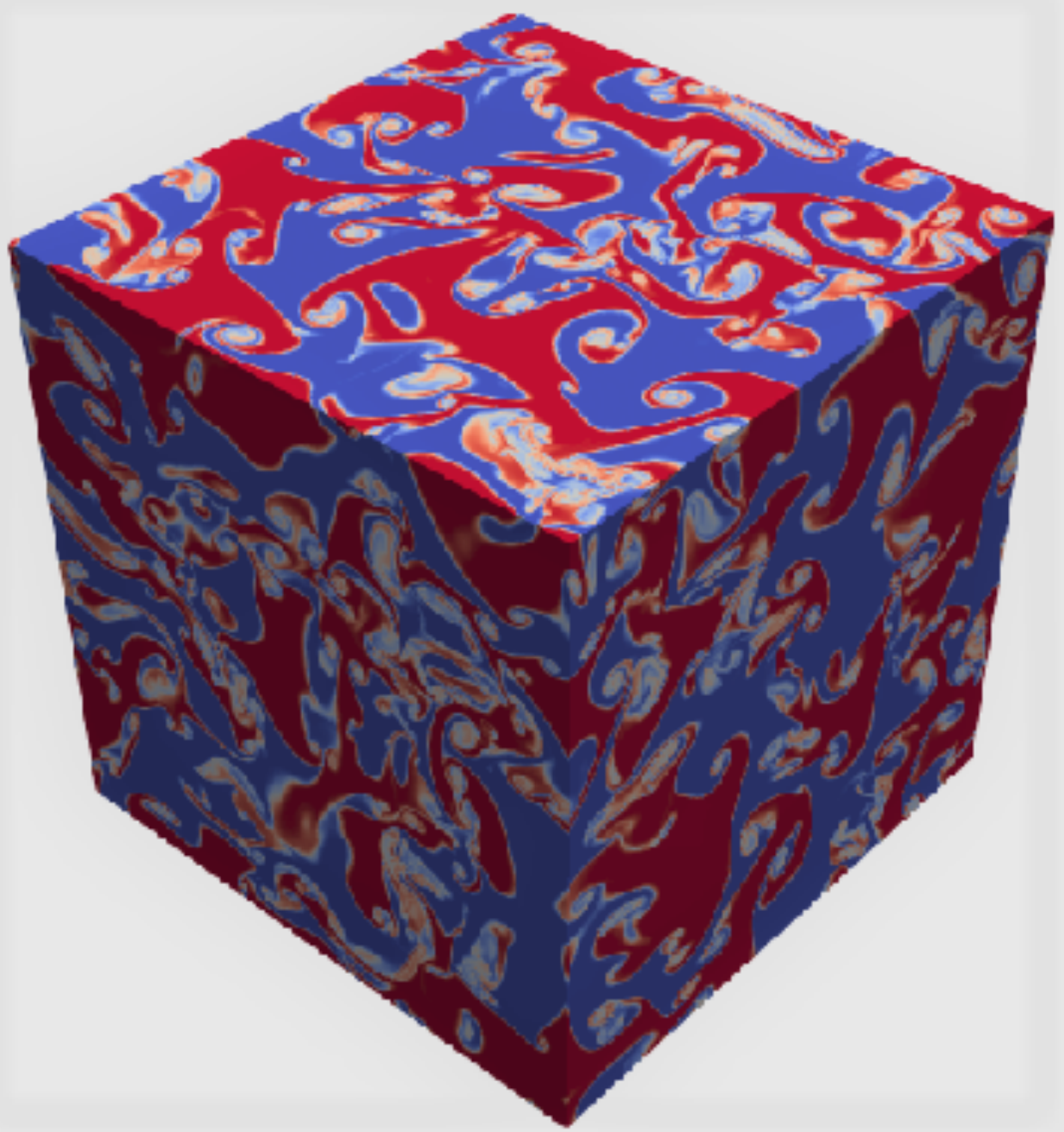}
\includegraphics[width=3.2cm]{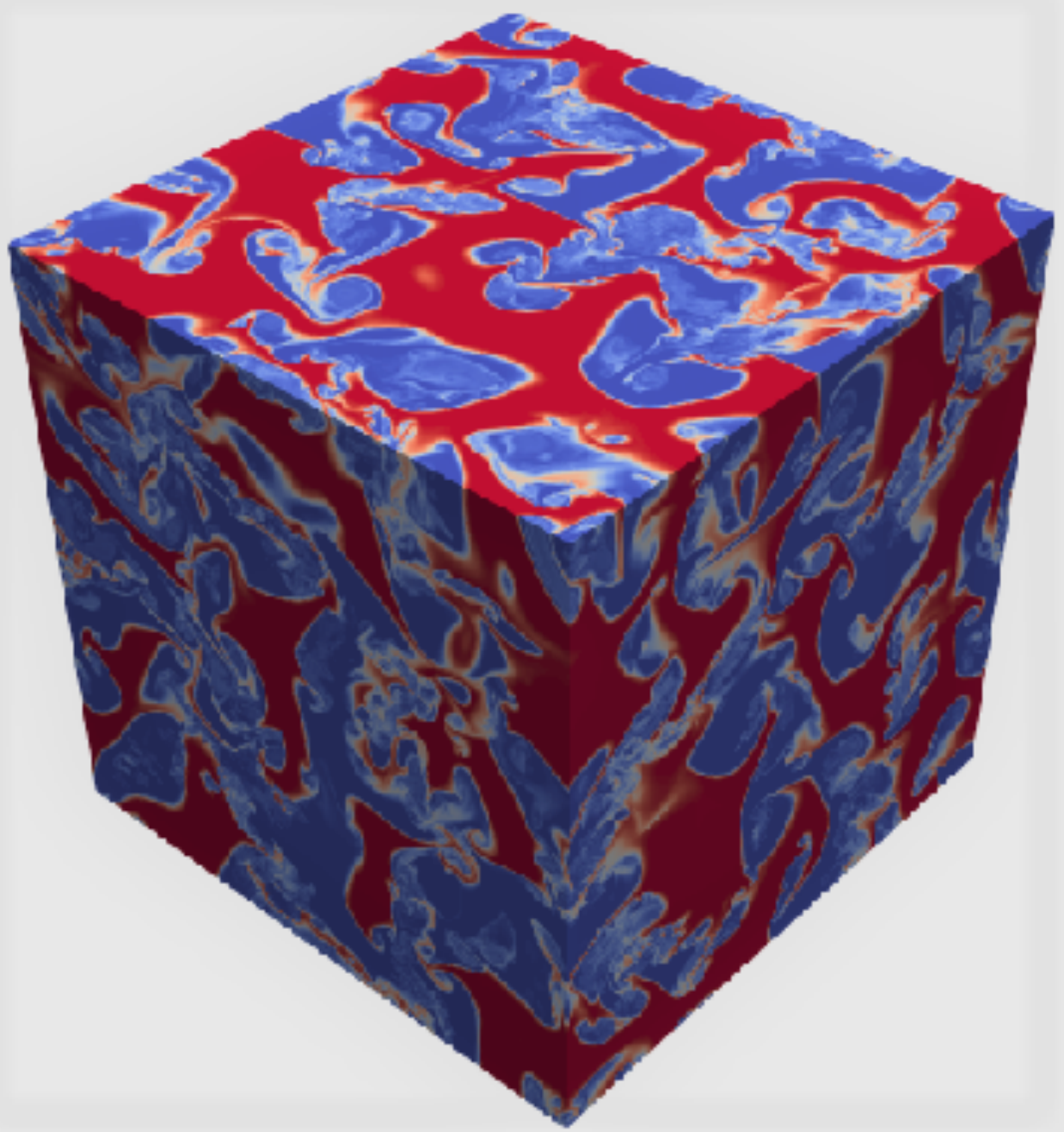}~~
\includegraphics[width=3.2cm]{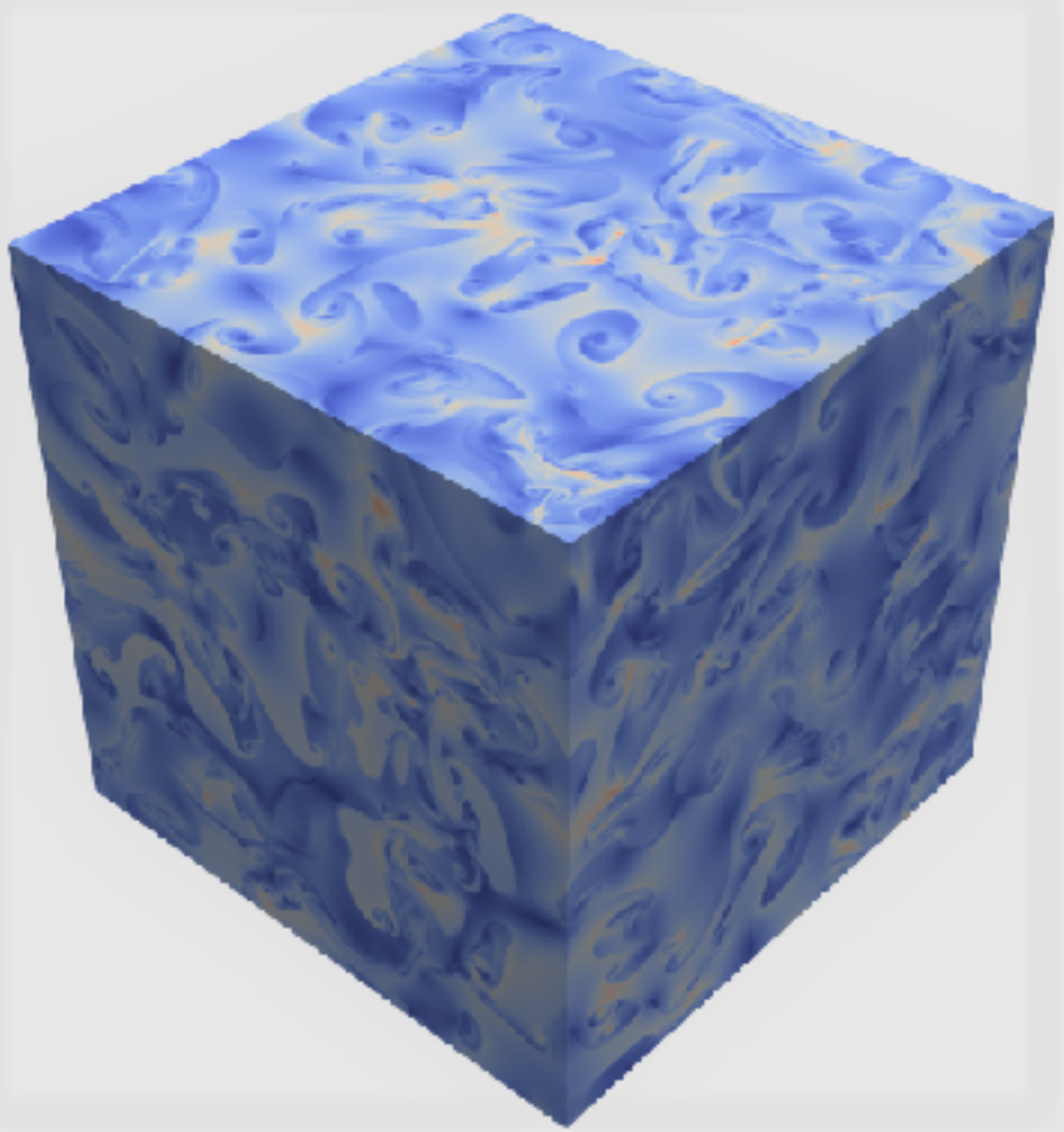}
\includegraphics[width=3.2cm]{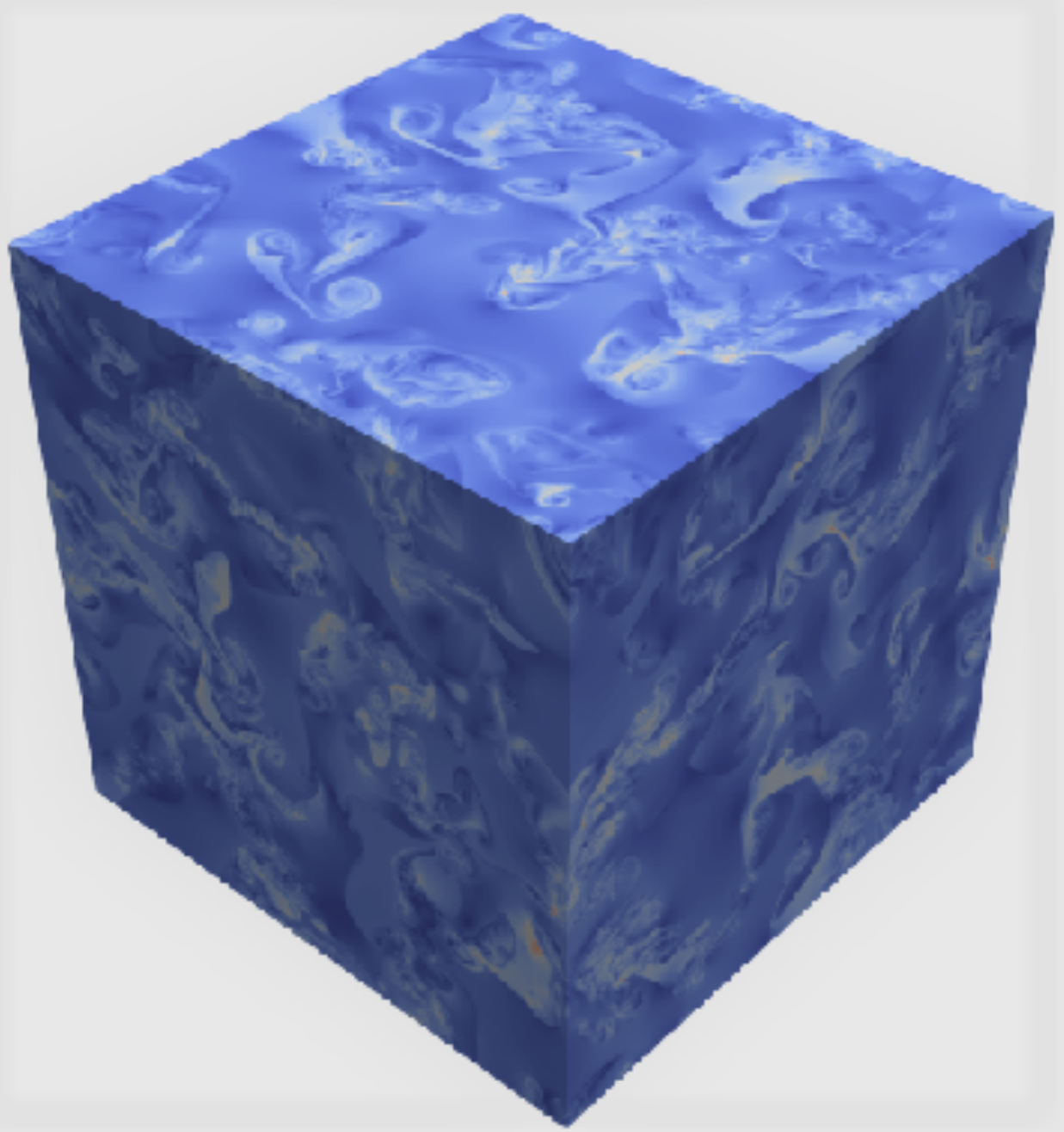}\\
\caption{3D visualization of (a) the mole fraction, and (b) the velocity magnitude ($\sqrt{u^2_1+u^2_2+u^2_3}$) for the cases $A=0.05$ (A1Re5) and $A=0.75$ (A4Re2) are displayed at $t/t_r=1.75$.}
\label{Fig:3Devolve_T2}
\end{figure}

The time instant when $E_{TKE}$ growth starts to slow down (such that $d^2(E_{TKE})/dt^2<0$) is chosen as the onset of the saturated growth regime. The 3D evolution of the heavy fluid mole fraction and the velocity during saturated growth is plotted in fig. \ref{Fig:3Devolve_T2} for the low ($A=0.05$) and high ($A=0.75$) \At numbers. Both density and velocity fields start to behave differently for different \At numbers.
    During this period, the stirring within the flow is no longer localized and large scales start to participate in the stirring process. The turbulent intensities become large and density gradients are increased indicating higher mixing rates within the flow. 
As shown in fig. \ref{Fig:Regimes}, the amounts of pure -light and -heavy fluids continue to decrease, but at a rate faster than during explosive growth. In addition, \At number effects become prominent; for larger \At numbers, the amounts of pure light and heavy fluids dramatically deviate from each other, indicating different mixing rates within the different flow regions, similar to results observed by \citep{livescu2008}. This breaks the symmetry of the density PDF, which adds more complexity to the problem. Below, it is shown that for high \At numbers, the asymptotic behavior with respect to the Reynolds number occurs earlier within pure light fluid regions than pure heavy fluid regions, which is taken to signify faster mixing transition within the light fluid regions. 
    
\subsubsection{Energy conversion rates} \label{Sec:Mixing_eff_2}

\begin{figure}
    \centerline{\includegraphics[width=8cm]{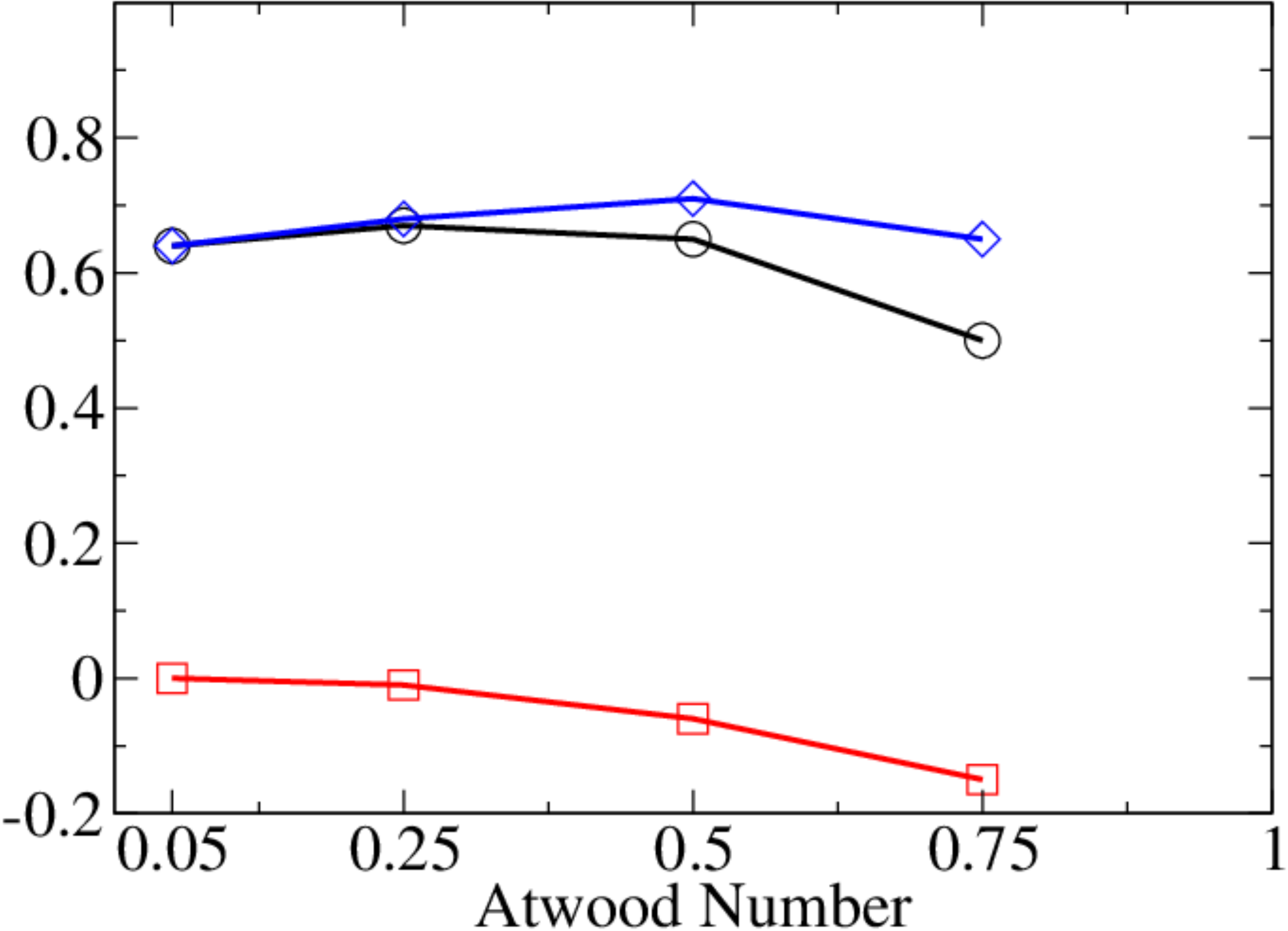}}
\caption{Atwood number effects on energy conversion rates ($\beta_{KE}$ - black line, $\beta_{TKE}$ - blue line and $\beta_{MKE}$ - red line) during saturated growth regime.}
\label{Fig:Eff_At_2}
\end{figure}

\begin{figure}
(\emph{a}) \hspace{6cm}  (\emph{b}) \\  
        \includegraphics[width=6.5cm]{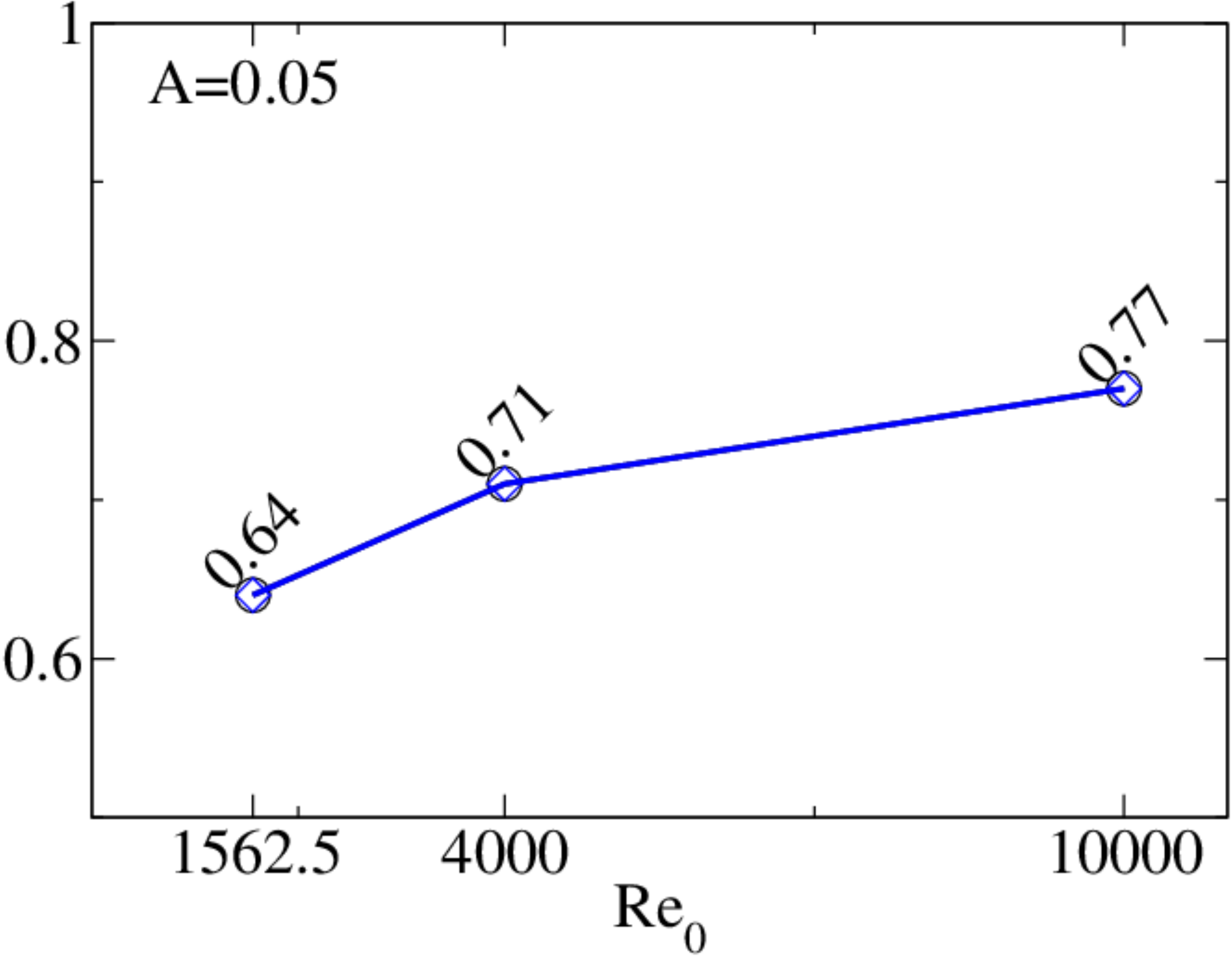}
    \includegraphics[width=6.5cm]{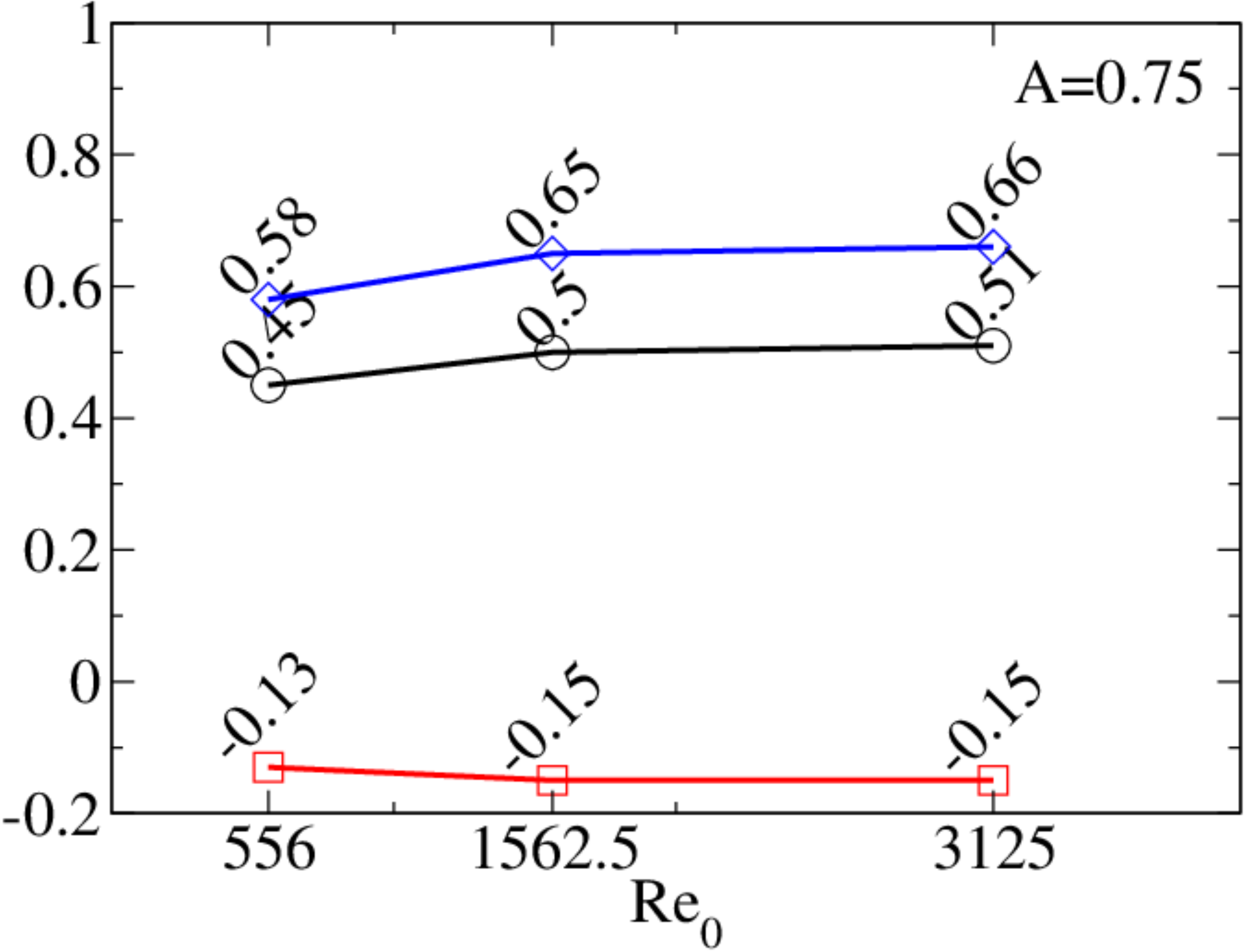}
\caption{Reynolds number effects on energy conversion rates ($\beta_{KE}$ - black line, $\beta_{TKE}$ - blue line and $\beta_{MKE}$ - red line) during saturated growth regime for (a) \At $=0.05$ (blue and black lines overlap and red line is close to zero) and (b) \At $=0.75$.}
\label{Fig:Eff_Re_2}
\end{figure}

During saturated growth, interestingly, for the same \rez values, the relationship between the energy conversion rates and \At number is not monotonic (see Fig. \ref{Fig:Eff_At_2}). Both $\beta_{KE}$ and $\beta_{TKE}$ peak at around \At number values of $0.25$ to $0.5$. The energy conversion rates are highest at moderate \At numbers. In addition, $\beta_{TKE}$ is greater than $\beta_{KE}$ (see Fig. \ref{Fig:Regimes}a). The stored energy in $E_{MKE}$ is released to feed into $E_{TKE}$ which leads to $\beta_{KE}<\beta_{TKE}$ and $\beta_{MKE}<0$. Note that $E_{MKE}$ is decreasing during saturated growth which is also consistent with $\beta_{MKE}<0$ (see Fig. \ref{Fig:Regimes}). It is also found that both low and high \At energy conversion rates tend to converge by increasing \rez, indicating that the flow is beyond the mixing transition during the saturated growth regime in terms of the energy conversion rates (see Fig. \ref{Fig:Eff_Re_2}).

\subsubsection{Atwood number effects on PDF evolutions} \label{Sec:PDF_2}
\vspace{3mm}
\noindent\textit{Density PDF} 

\begin{figure}
   (\emph{a}) \hspace{6.5cm}  (\emph{b}) \\
    \centerline{\includegraphics[width=6.5cm]{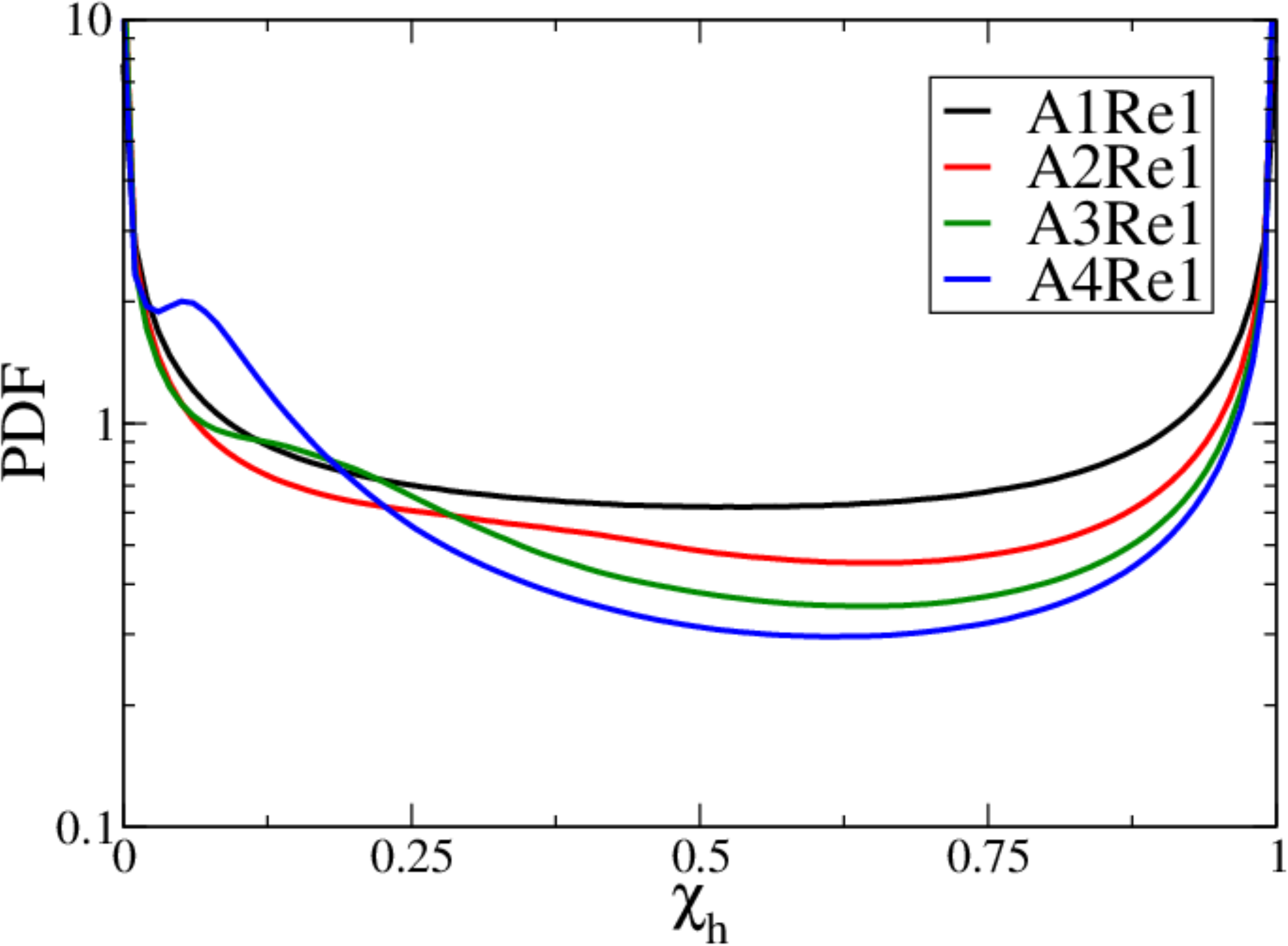}
    \includegraphics[width=6.5cm]{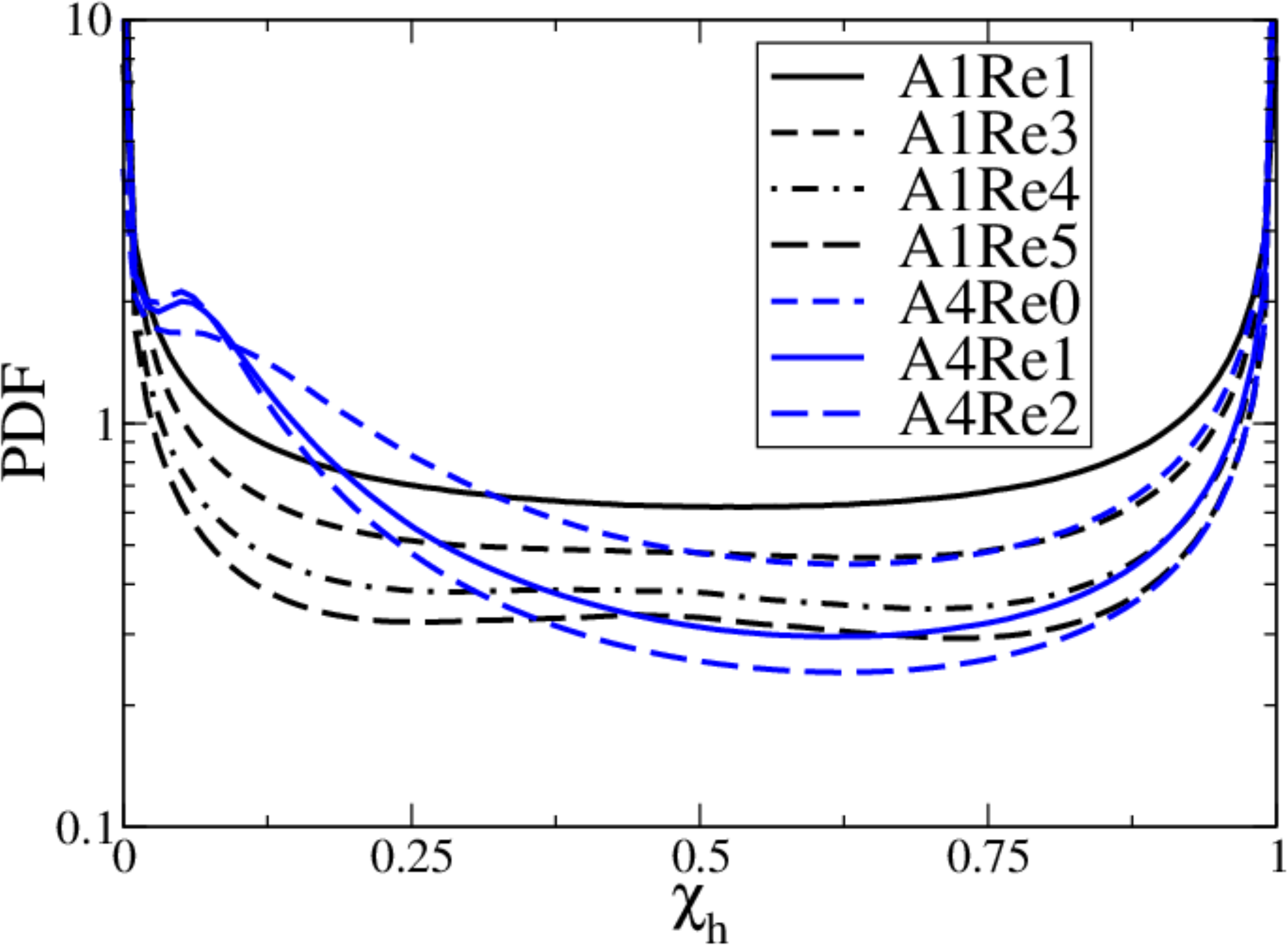}}
 \caption{PDF of density field for the different (a) \At and (b) \rez values at $t/t_r=1.75$.}
\label{Fig:dens_PDF_2}
\end{figure}

    \citet{livescu2008} found that increasing the $A$ number breaks the symmetry of the mixing rates and leads to asymmetric density PDFs. In their study, the reported results are for $A$ number equal to $0.5$, with maximum domain size of $512^3$, which was around the mixing transition threshold. The current work builds on that study to further investigate the asymmetric behavior of the density PDF evolution; density PDFs for both larger $A$ number ($A=0.75$) and larger domain sizes up to $2048^3$ (leading to much larger $Re_t$ values) are presented. Figure \ref{Fig:dens_PDF_2} presents density PDFs; the shape of the PDF is distinctly different for different \At numbers. For the large \At number; the lightest fluid regions mix faster than the heaviest fluid regions. The density PDF exhibits peak values at lighter than average fluid regions; this peak shifts to even lighter fluid regions upon increasing the \At number.
    
As discussed in section 3, in order to determine whether a certain quantity is fully developed (i.e. beyond the mixing transition), we check if the time-evolution of the quantity in question becomes independent of $Re_t$. The authors are of the opinion that for HVDT there is no single threshold that could determine whether the flow is beyond the mixing transition and that different quantities reach this transition differently. In Figure \ref{Fig:dens_PDF_2}.b, for the lowest $A$ number case of $A=0.05$, $Re_0$ has minimal effect on the PDF. However, for large $A$ number cases, it is observed that only the lighter fluid regions tend to become insensitive to an increase in $Re_0$. The variation of the density PDF in the heavier fluid regions is still large for the high $A=0.75$ cases (the cases A075Re3 and A075Re2), even if we compare the cases with the largest two $Re_0$ values. This indicates that mixing transition may occur at earlier times for the lighter fluid regions than the heavier fluid regions when considering the density PDF. This is also consistent with the conditional expectation of $E_{TKE}$ (see Figure 11) which was observed to be much larger in lighter fluid regions compared to heavier fluid regions.

\vspace{3mm}
\noindent\textit{Velocity-density jMDF} 
 
\begin{figure}
\vspace{0.5cm}
\hspace{2.6cm}   Case:A1Re5 \hspace{3.8cm} Case:A4Re2  \\
\centerline{\includegraphics[width=7.0cm]{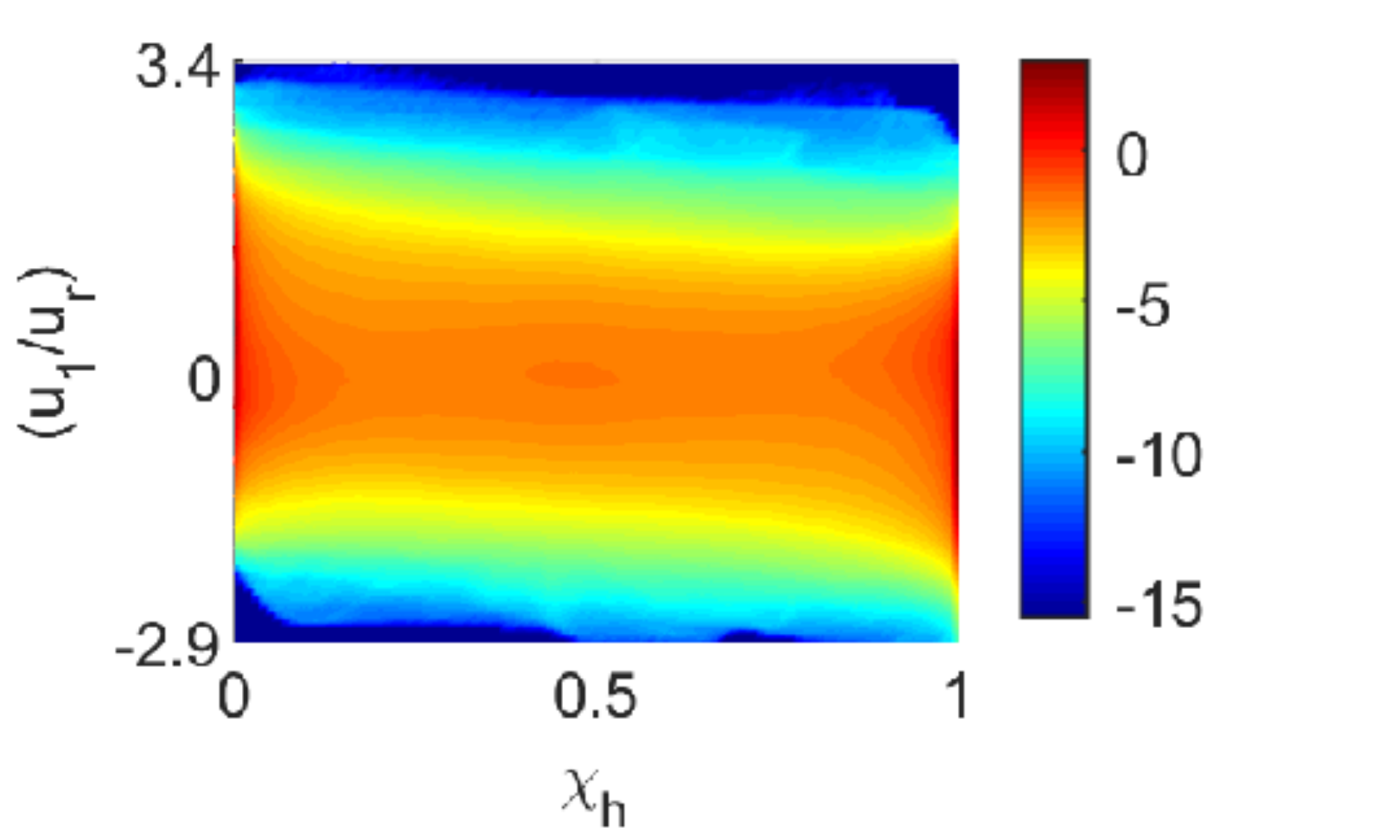}
\includegraphics[width=7.0cm]{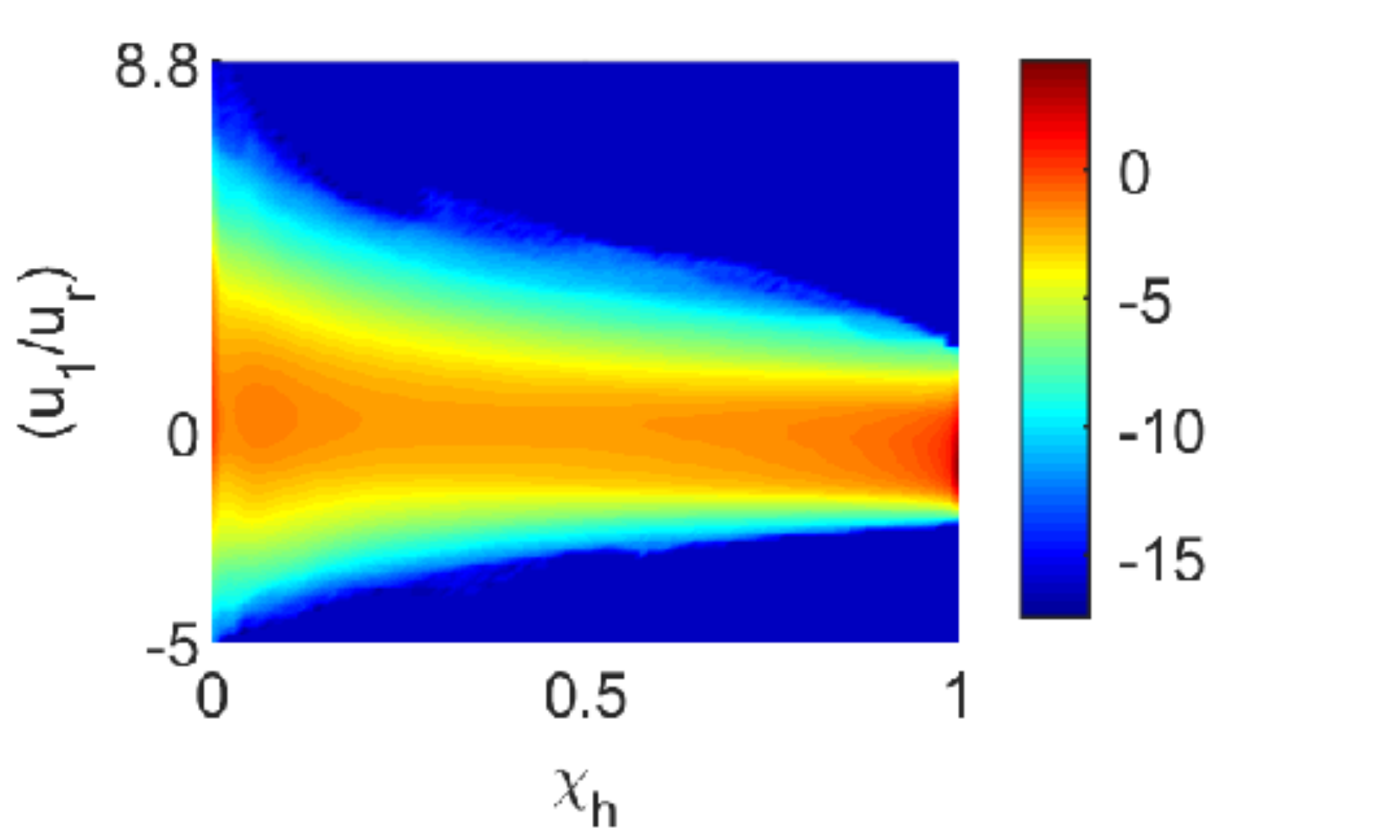}}
 \caption{Normalized jMDFs [$\log(\mathcal{F}/\rhom)$] for (a) $A=0.05$ (A1Re5) and (b) $A=0.75$ (A4Re2) cases displayed at $t/t_r=1.75$.}
\label{Fig:Joint_2}
\end{figure}

The jMDF behavior for low and high \At number cases during the saturated growth regime is shown in fig. \ref{Fig:Joint_2}. For the low \At number case; the jMDF is relatively symmetric with respect to the mole fraction at small velocity magnitudes. There is a slight asymmetry near the largest velocity magnitudes; the jMDF peaks at the light fluid side at large positive velocities and on the heavy fluid side at large negative velocities. In addition, at each heavy fluid mole fraction level, the jMDF remains
quasi-Gaussian, with the peak moving slightly from positive to negative velocities as $\chi_h$ varies from $0$ to $1$.
The jMDF is more complex for the high \At number case. Thus, the jMDF shape is wider at small values of $\chi_h$. At all $\chi_h$, the jMDF is asymmetric and skewed towards positive velocity values, though the asymmetry decreases compared to the explosive growth regime. A longer tail for positive values of the vertical velocity indicates more extreme upward motions in the flow, similar to that observed in the explosive growth.

\subsubsection{Conditional expectations} \label{Sec:cond_2}
\noindent\textit{Conditional expectation of $E_{TKE}$}

\begin{figure}
  (\emph{a}) \hspace{6.5cm}  (\emph{b}) \\
    \centerline{\includegraphics[width=6.4cm]{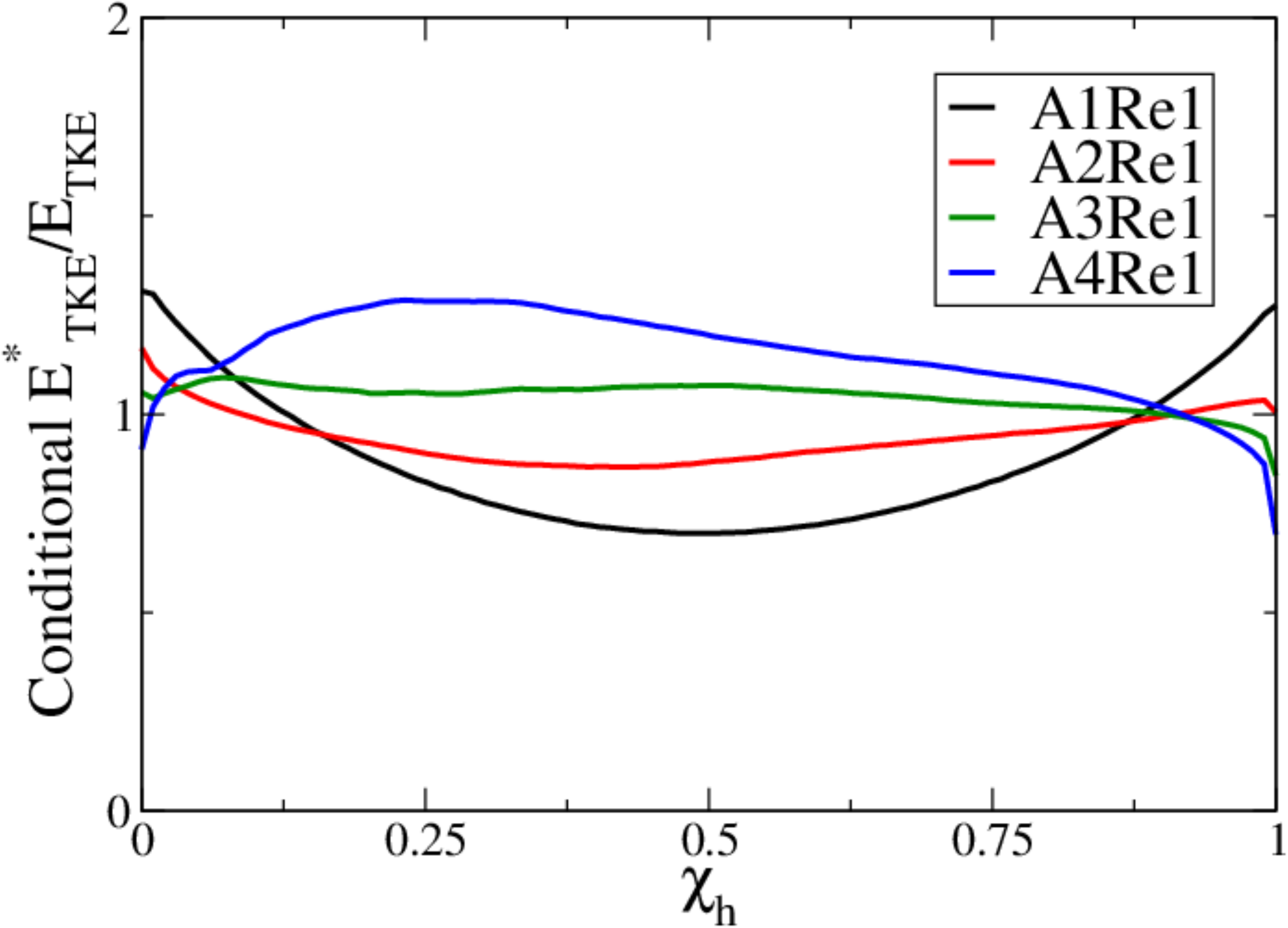}
    \includegraphics[width=6.4cm]{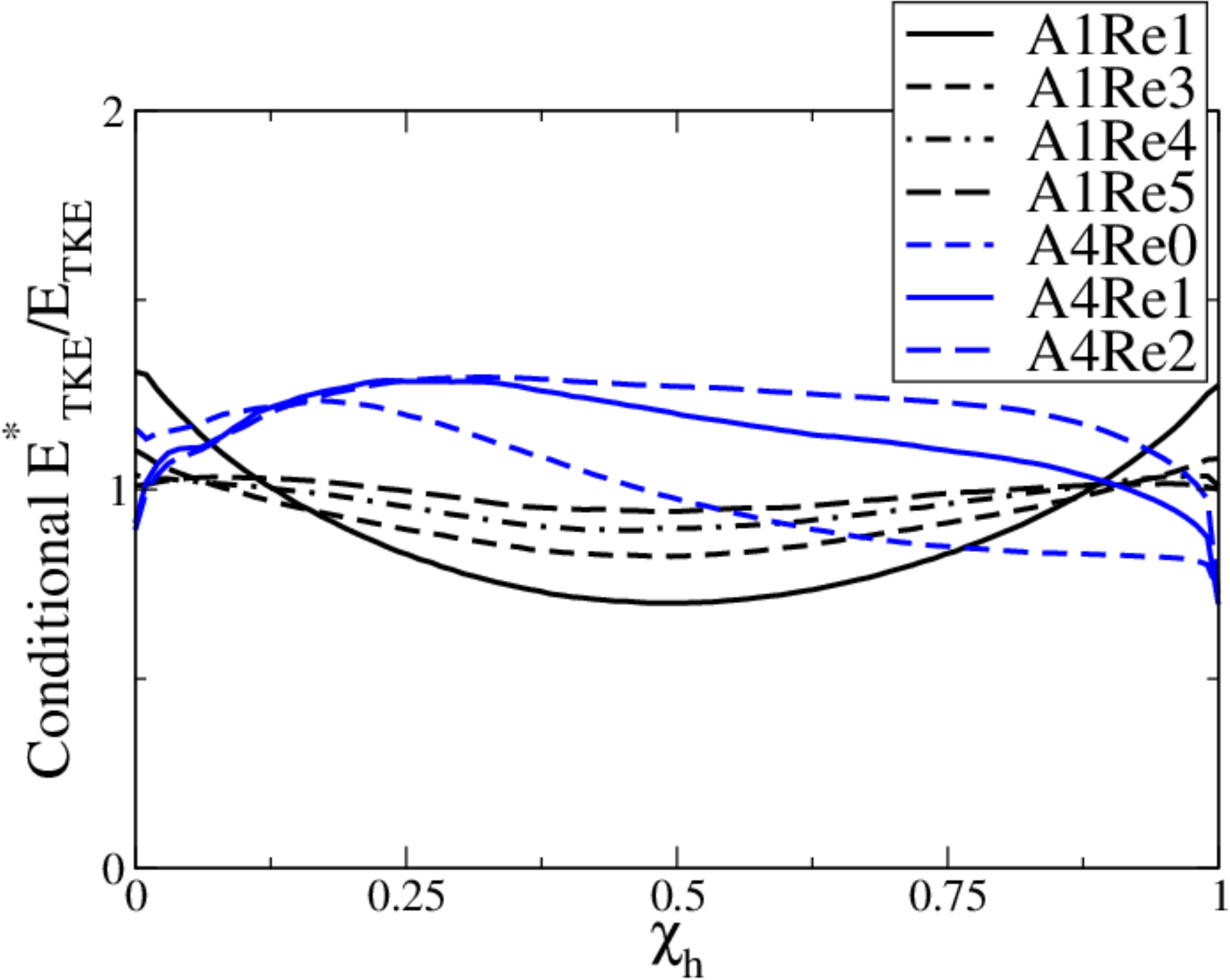}}
\caption{Atwood and Reynolds numbers effects on conditional expectation of $E_{TKE}$ during saturated growth regime (at $t/t_r=1.75$).}
\label{Fig:cond_tke_2}
\end{figure}

The \At and \rez number effects on conditional expectation of $E_{TKE}$ are shown in fig. \ref{Fig:cond_tke_2}. At low \At, $E_{TKE}$ is mostly balanced between lighter and heavier regions of the flow. As \At increases at fixed \rez value, the conditional expectation becomes asymmetric, with larger values in low to moderate density regions (fig. \ref{Fig:cond_tke_2}a). An increase in \rez reduces the asymmetry of conditional expectation. For the higher \At number cases, the conditional PDF converges at the lighter fluid side. This observation is consistent with the notion of mixing transition occurring at earlier times within the lighter fluid regions compared to the heavier fluid regions.

\vspace{3mm}

\noindent\textit{Conditional expectation of dissipation of $E_{TKE}$}

\begin{figure}
  (\emph{a}) \hspace{6.5cm}  (\emph{b}) \\
    \centerline{\includegraphics[width=6.4cm]{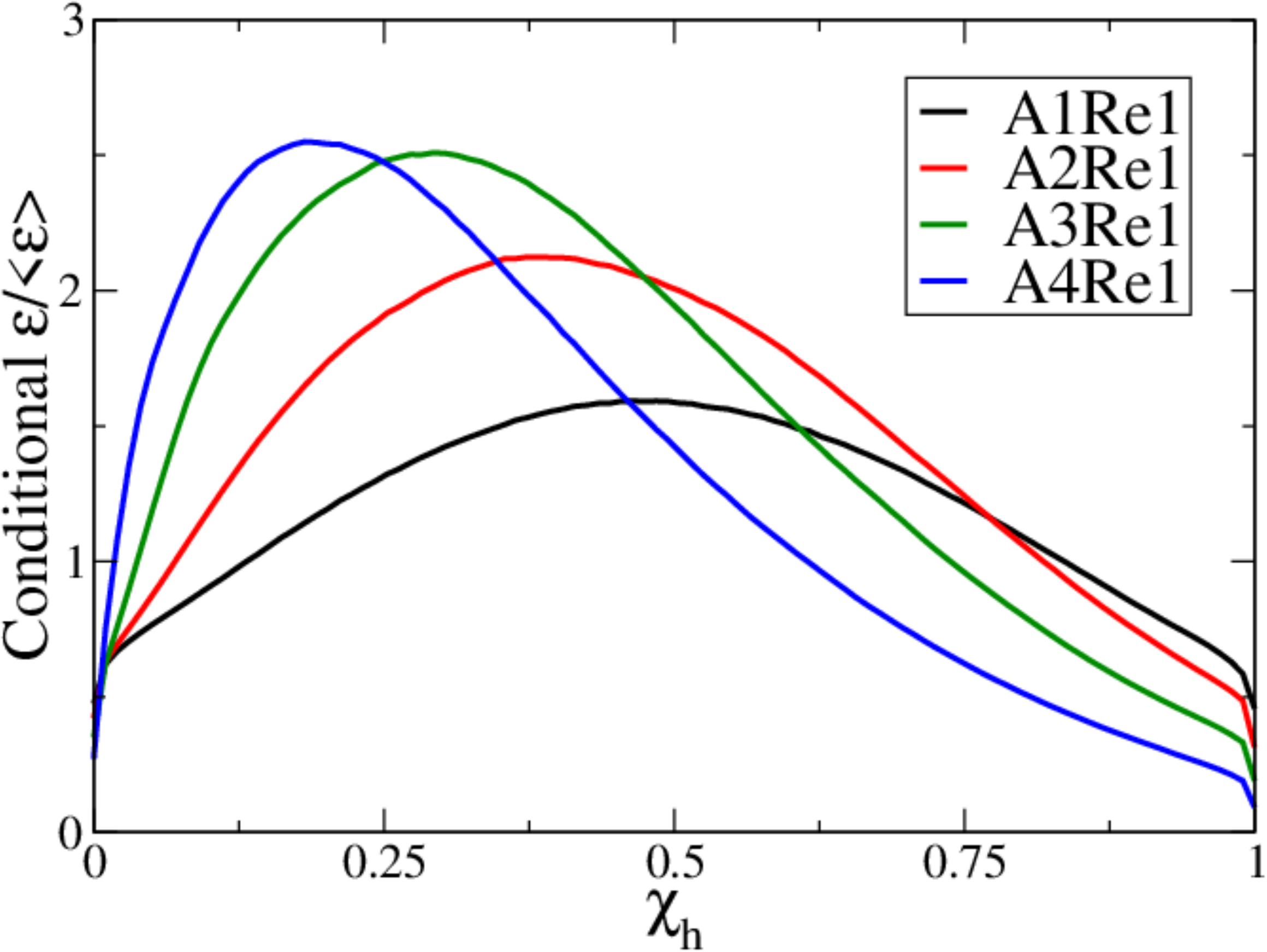}\includegraphics[width=6.4cm]{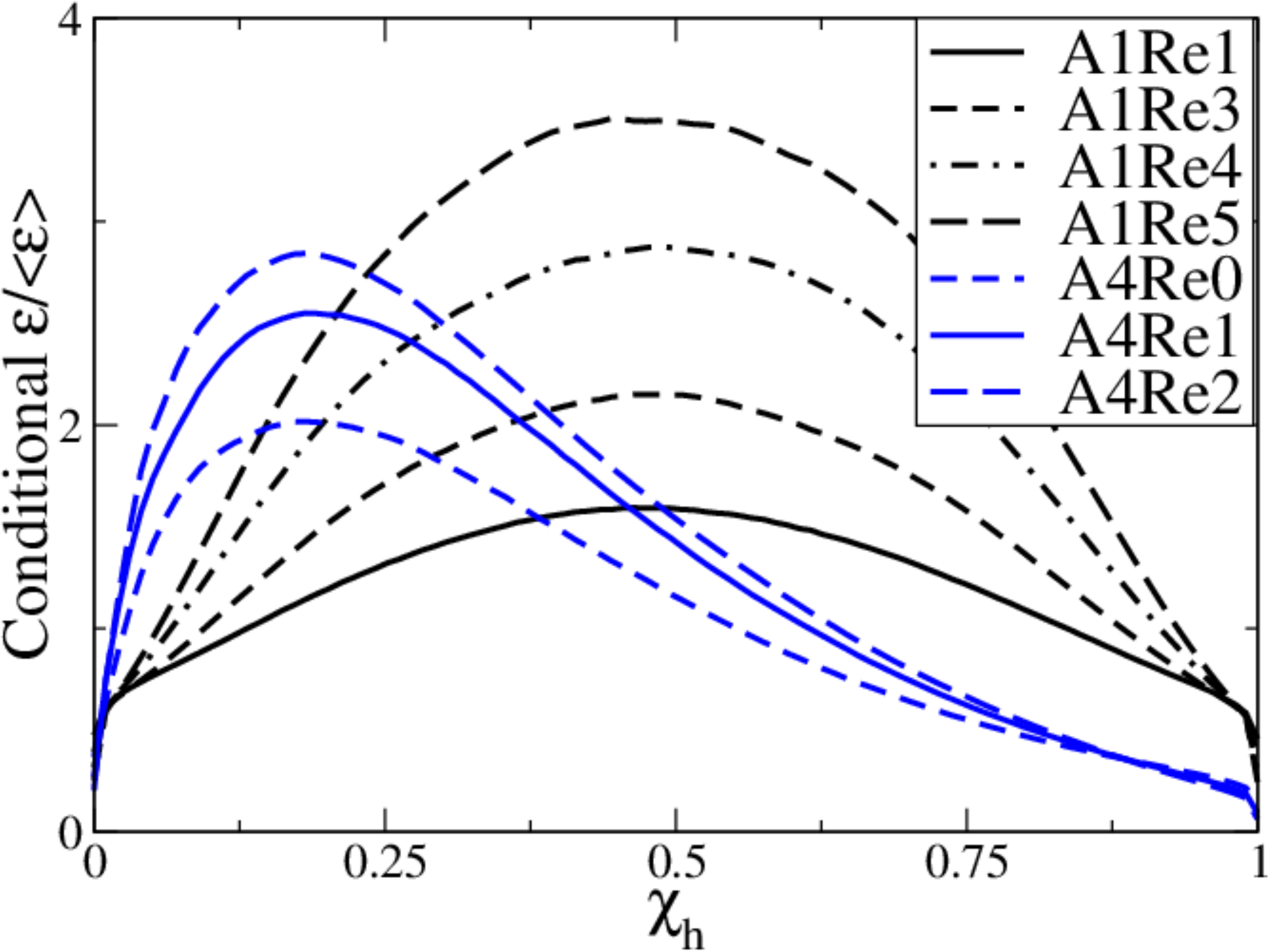}}
\caption{Atwood and Reynolds numbers effects on conditional expectation of $E_{TKE}$ dissipation during saturated growth regime (at $t/t_r=1.75$).}
\label{Fig:cond_eps_2}
\end{figure}

The conditional expectation of the $E_{TKE}$ dissipation remains asymmetric at high \At during this regime, similar to the explosive growth regime (fig. \ref{Fig:cond_eps_2}).  Local dissipation expectations (computed for the specific density values) continue to take much larger values over mean dissipation values for higher \At and \rez numbers. For example, the largest conditional expectation of $E_{TKE}$ dissipation is $\approx 3$ for the A4Re2 case and is $\approx 3.8$ for the A1Re5 case. However, \ret effects decrease during the saturated growth. Furthermore, \rez effects on the peak of the conditional mean is smaller for the higher \At number cases. This might be attributed to the location of the peak, which is within the lighter fluid regions, where the mixing transition occurs relatively earlier than for the fully mixed and heavier fluid regions.

\vspace{3mm}
\noindent\textit{Conditional expectation of enstrophy}

\begin{figure}
(\emph{a}) \hspace{6.5cm}  (\emph{b}) \\
    \centerline{\includegraphics[width=6.4cm]{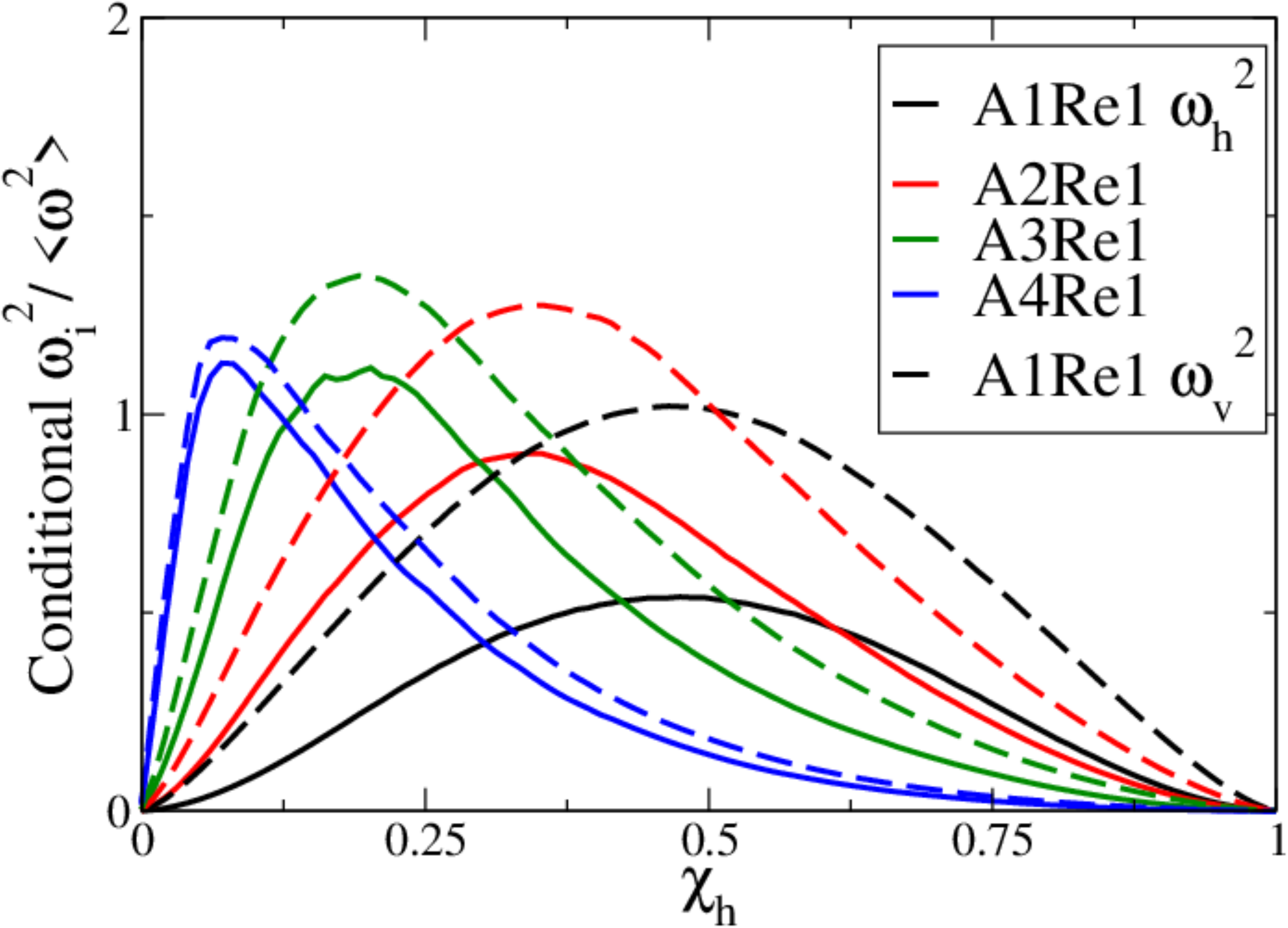}
    \includegraphics[width=6.4cm]{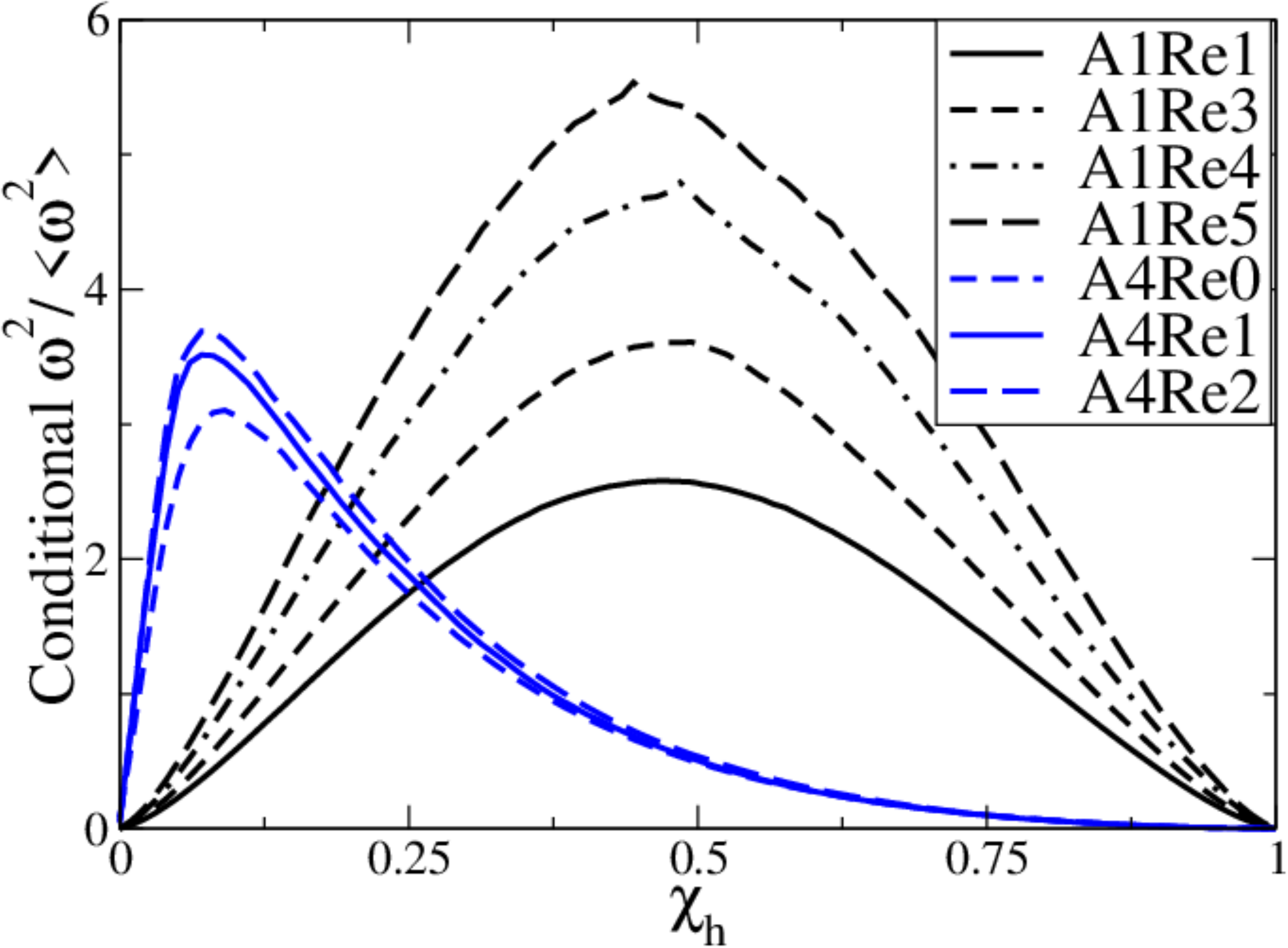}}
\caption{Atwood and Reynolds numbers effects on conditional expectation of (a) vertical ($\omega^2_v=\omega^2_1$) and horizontal ($\omega^2_h=(\omega^2_2+\omega^2_3)/2$) enstrophy components and (b) total enstrophy during saturated growth regime (at $t/t_r=1.75$).}
\label{Fig:cond_ens_2}
\end{figure}

The conditional expectation of enstrophy is shown in Fig. \ref{Fig:cond_ens_2}a-b; the vertical, horizontal and total components are plotted for comparison. The conditional expectation remains strongly skewed towards the light fluid regions for the high \At number cases, similar to observations reported for the explosive growth. The main difference observed in the saturated growth (compared to explosive growth) regime is that the enstrophy becomes almost isotropic at $t/t_r=1.75$ at the highest \At number. Additionally, for the high \At number cases ($A \geq 0.5$), \rez effects become weaker, similar to the behavior of the conditional expectation of the dissipation of $E_{TKE}$. Interestingly, different rates of convergence with \rez are observed for the different metrics discussed in this section. For example, at $t/t_r=1.75$, the velocity PDF is substantially converged for all the cases reported (not shown here). However, the conditional expectation of $E_{TKE}$ is weakly converged for the high \rez cases, while the conditional expectations of dissipation and enstrophy are further away from convergence.

\subsection{Fast decay}\label{Sec:Fast_decay}

The fast decay regime starts when dissipation of $E_{TKE}$ begins to overcome $E_{TKE}$ generation. Since the rate of molecular mixing is high and buoyancy forces decay rapidly, the rate of $E_{TKE}$ decay increases with time such that $d^2(E_{TKE})/dt^2<0$. During this regime, similarities are observed with RTI under reversed acceleration $g \leq 0$. In HVDT, partially mixed fluids become well-mixed within the flow and turbulence starts to decay similar to observations in RTI with acceleration reversal that makes the global flow stable with the presence of local fluid patches that are unstable due to the stirring. These local fluid patches assist the turbulence decay due to local (but limited) turbulence generation \citep{Ramaprabhu_ADA,Denis_PRE}. Furthermore, this regime also has some similarities with shock-driven RMI. In RMI, VD turbulence generated due to the shock decays purely as there is no buoyancy-assistance. However, the VD mixing physics is similar in both flows as turbulent fluid patches with different densities mix efficiently during $E_{TKE}$ decay.

\begin{figure}
\hspace{3.3cm}(\emph{a}) \hspace{6.1cm} (\emph{b})

\vspace{0.5cm}
\hspace{.9cm}   Case:A1Re5 \hspace{1.32cm} Case:A4Re2 \hspace{1.45cm}  Case:A1Re5 \hspace{1.3cm} Case:A4Re2 \\
\rotatebox{90}{\hspace{1cm}$t/t_r=2.95$}\includegraphics[width=3.2cm]{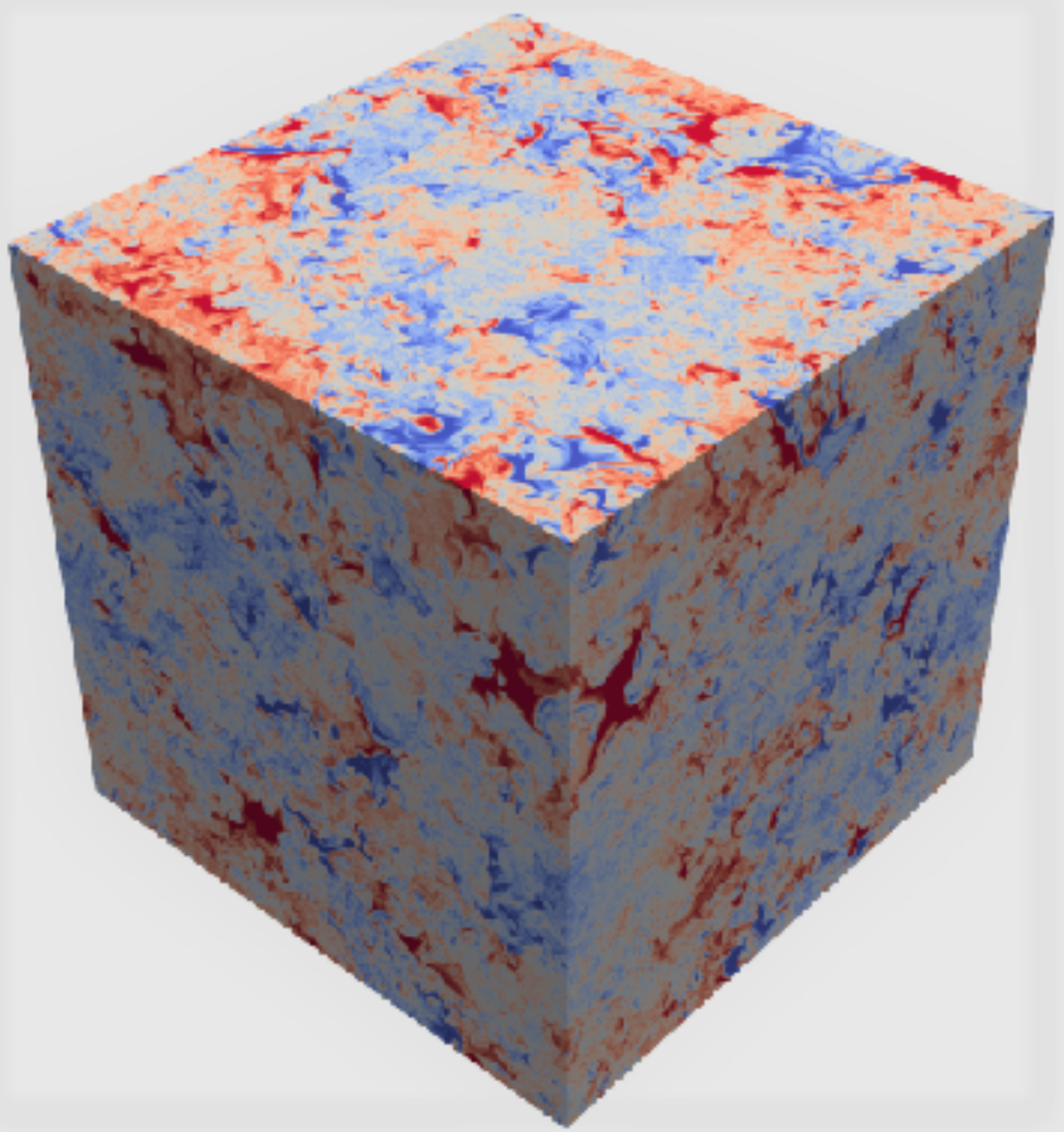}
\includegraphics[width=3.2cm]{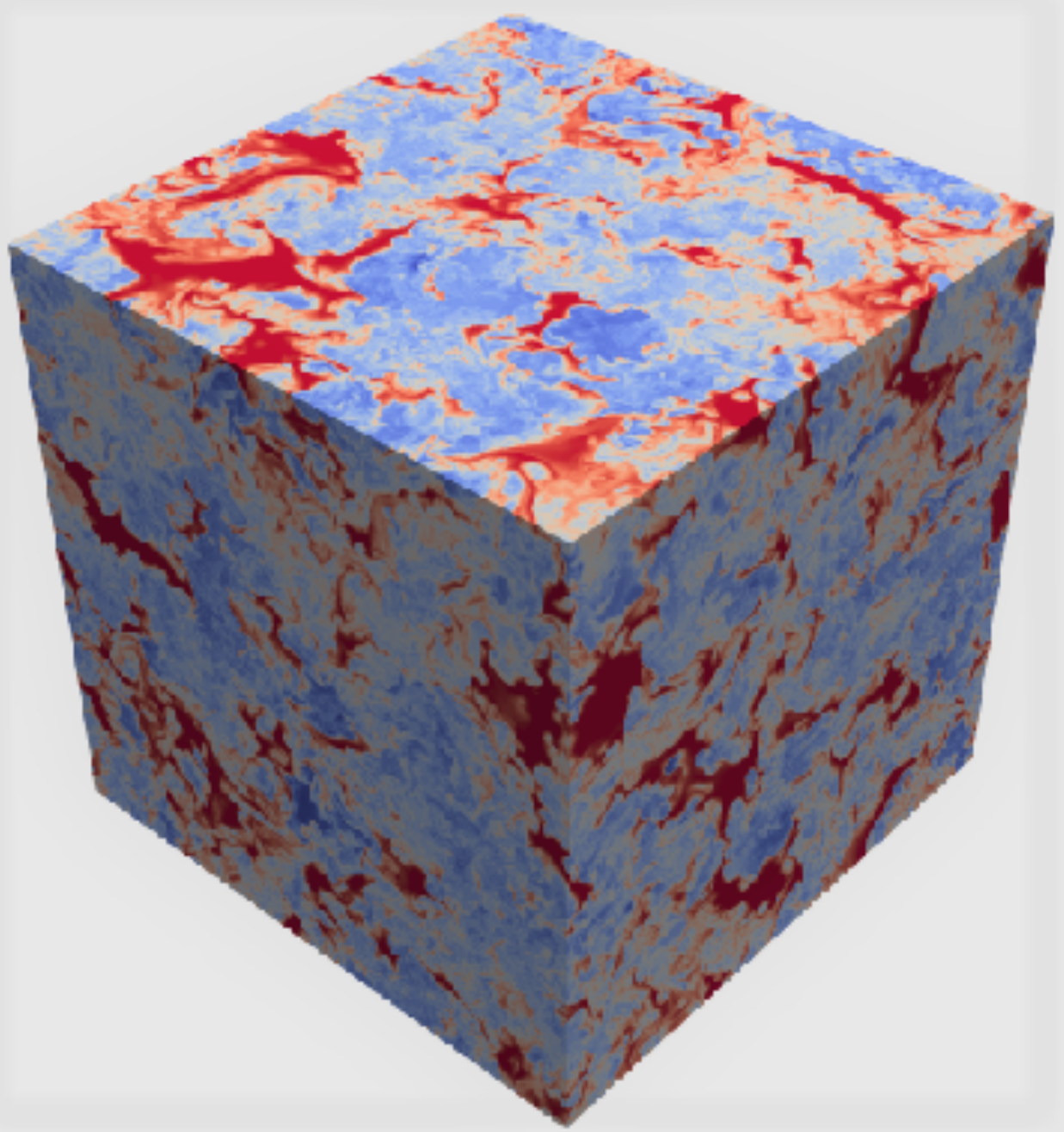}~~
\includegraphics[width=3.2cm]{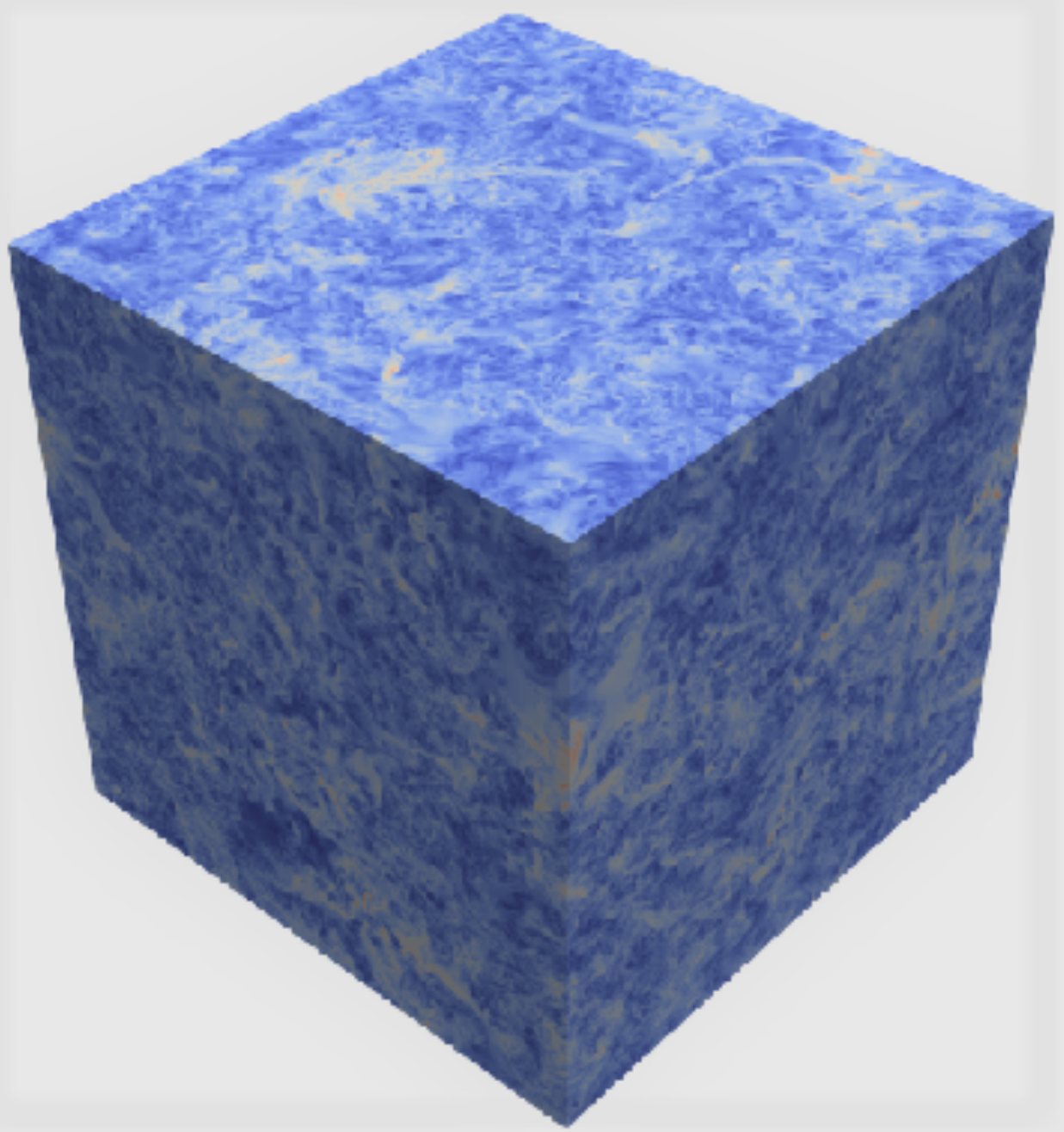}
\includegraphics[width=3.2cm]{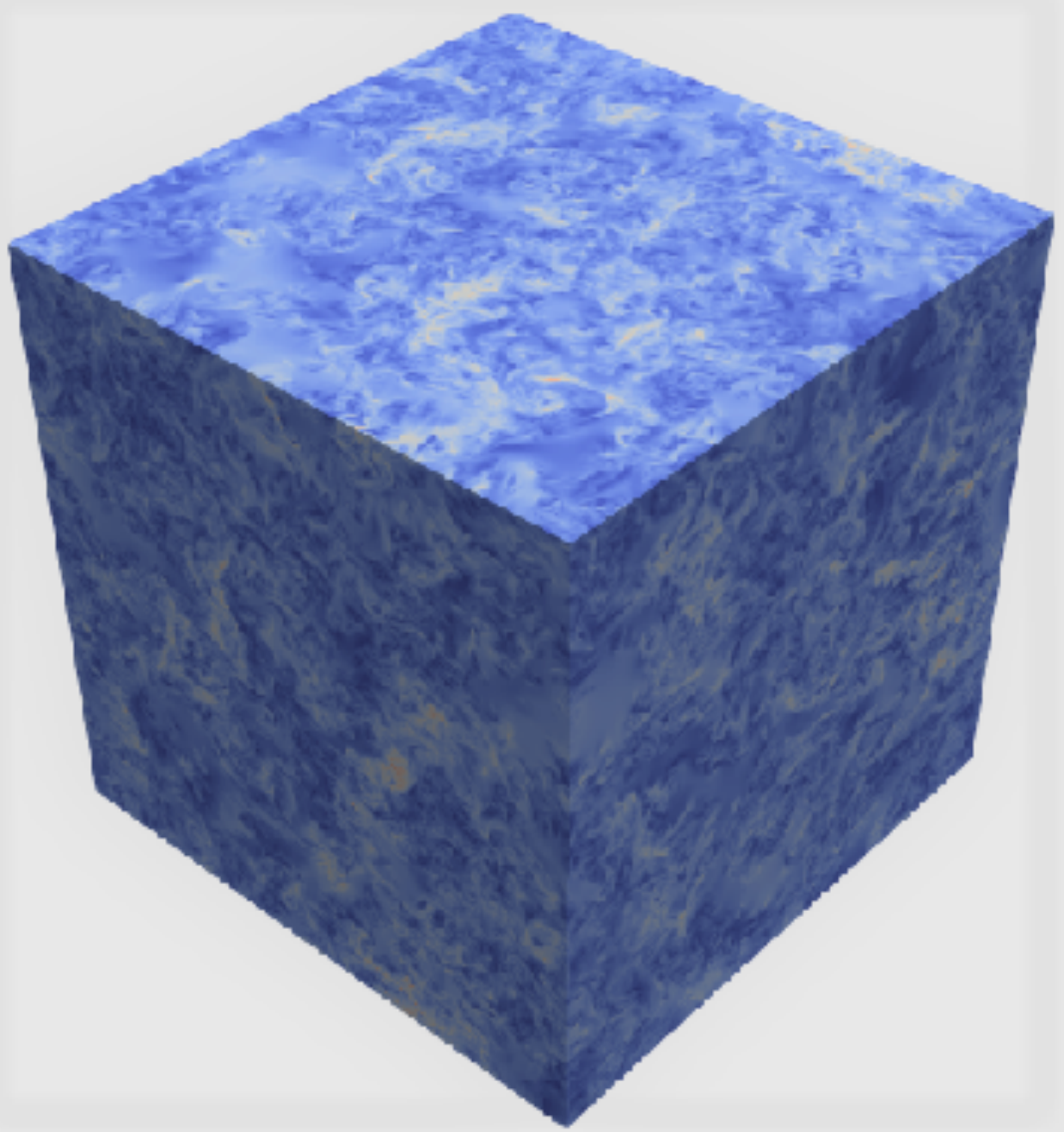}\\
 \caption{3D visualisation of (a) the mole fraction, and (b) the velocity magnitude ($\sqrt{u^2_1+u^2_2+u^2_3}$) for the cases $A=0.05$ (A1Re5) and $A=0.75$ (A4Re2) are displayed at $t/t_r=2.95$.}
\label{Fig:3Devolve_T3}
\end{figure}

\vspace{0.5cm}
\subsubsection{Atwood number effects on PDF evolutions} \label{Sec:PDF_3}
\vspace{3mm}
\noindent\textit{Density PDF}

\begin{figure}
   (\emph{a}) \hspace{6.5cm}  (\emph{b}) \\
    \centerline{\includegraphics[width=6.5cm]{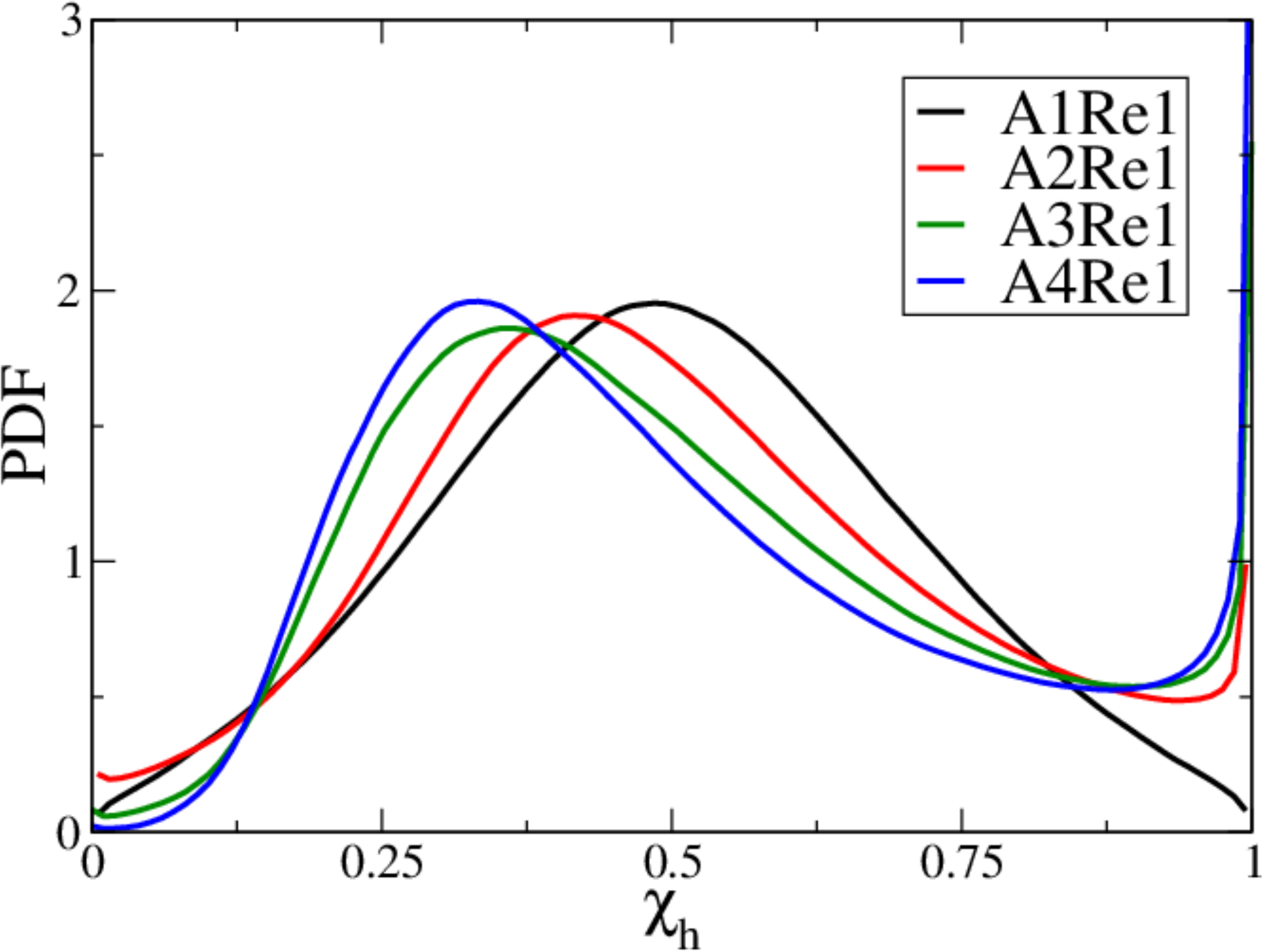}
    \includegraphics[width=6.5cm]{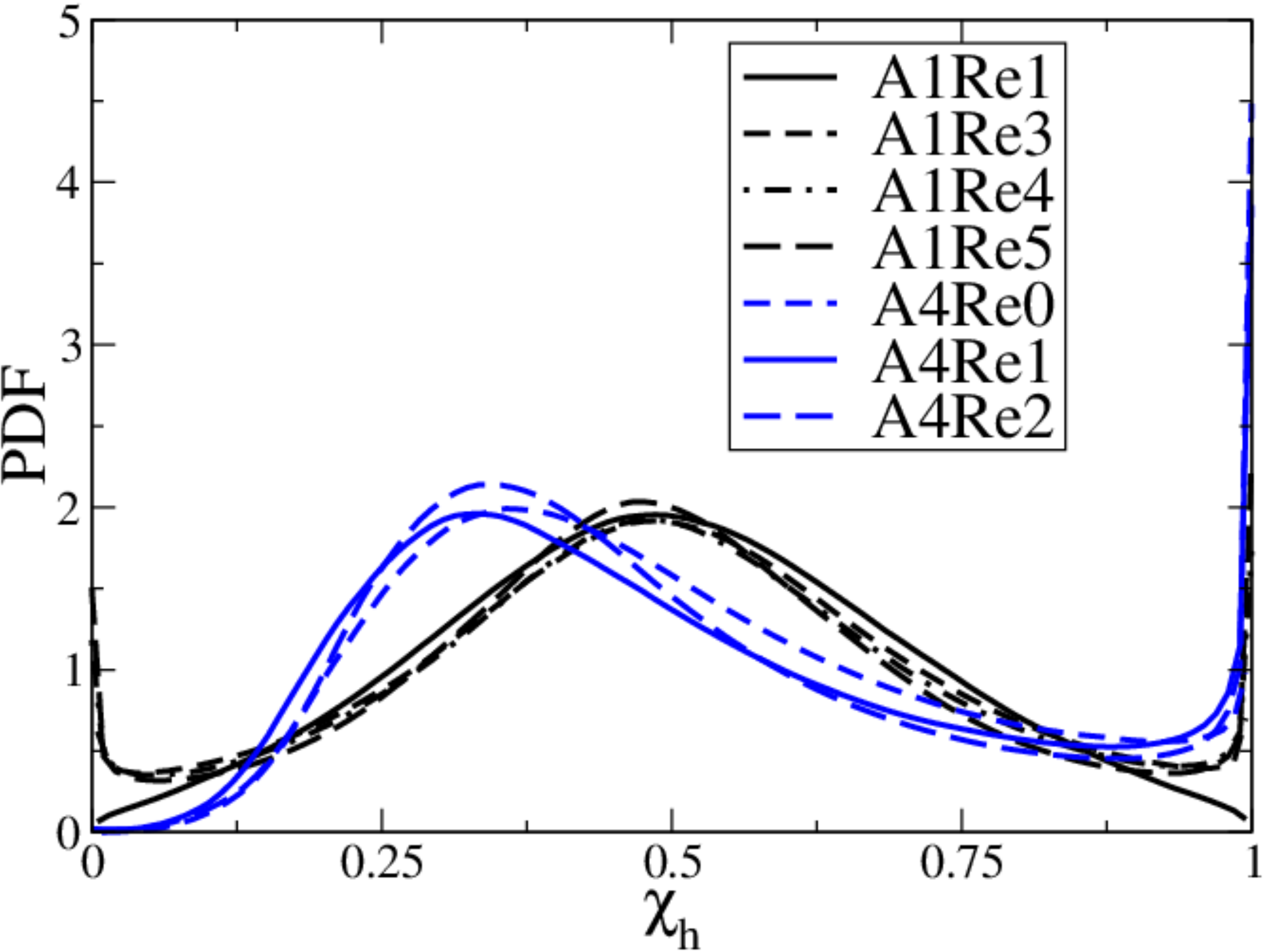}}
 \caption{PDF of the density field for the different (a) \At and (b) \rez values at $t/t_r=2.95$.}
\label{Fig:dens_PDF_3}
\end{figure}

During fast decay, the density PDF is highly asymmetric for large \At number cases. A closer look at the tails of the distribution shows very little amounts of pure light fluid left in the flow. However, there are still substantial amounts of pure heavy fluid within the flow  (see Fig. \ref{Fig:dens_PDF_3}a). The normalized skewness ($S_k$), which identifies the asymmetry level of the PDF distribution and takes positive values for the PDFs with longer right side tails and negative values for the PDFs with longer left side tails, is defined as \citep{ristorcelli_clark_2004}:
\begin{equation}
    \label{Eq:Sk}
    S_k=\frac{\langle\rho^{3}\rangle}{\langle\rho^{2}\rangle^{3/2}}.
\end{equation}
Normalized skewness takes its maximum values during the fast decay regime. $S_k$ tends to slightly increase upon increasing \rez. For all \At number cases (the moderate cases are not shown) \rez number effect is weak during the fast decay indicating this quantity reaches the fully-developed stage. 

\vspace{3mm}
\noindent\textit{Velocity-density jMDF}

\begin{figure}
\vspace{0.5cm}
\hspace{2.6cm}   Case:A1Re5 \hspace{3.8cm} Case:A4Re2  \\
\centerline{\includegraphics[width=7.0cm]{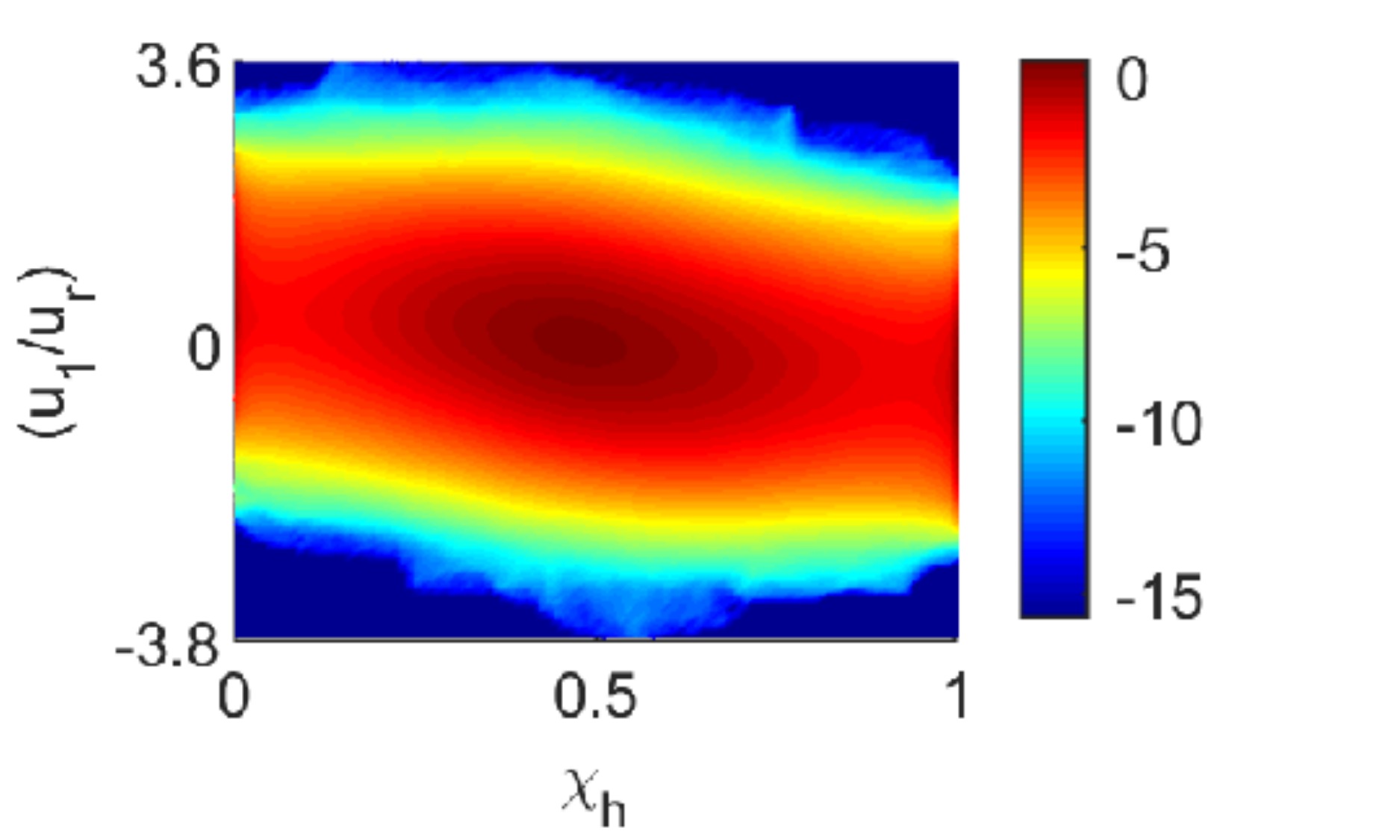}
\includegraphics[width=7.0cm]{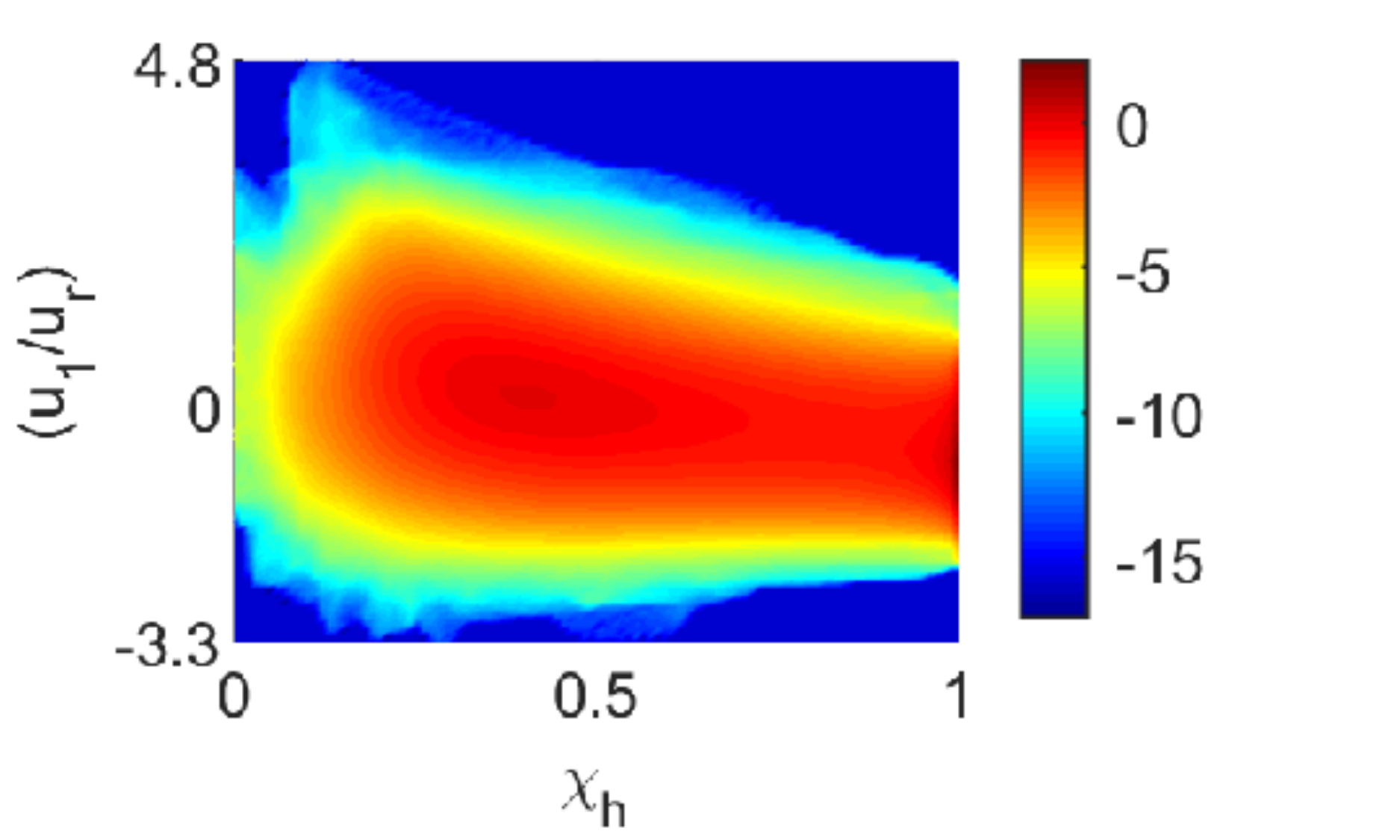}}
 \caption{Normalized jMDFs [$\log(\mathcal{F}/\rhom)$] for (a) $A=0.05$ (A1Re5) and (b) $A=0.75$ (A4Re2) cases displayed at $t/t_r=2.95$.}
\label{Fig:Joint_3}
\end{figure}

The jMDF variations for lowest and highest \At numbers are shown in Fig. \ref{Fig:Joint_3}. As it is seen, in contrast to the density PDF, the asymmetry of the jMDF at each density level tends to decrease, as $P_{,i}$ and variability in $1/\rho^*$ are lower due to mixing. However, the jMDFs are still significantly different for low and high \At number cases. Thus, for the high \At case, there are still significant large positive velocity events. Moreover, due to the existence of large amounts of pure heavy fluid for high \At number cases, the jMDF peaks in two regions; one close to the mean density value and near zero vertical velocity, and the other at the heaviest fluid region and slightly negative vertical velocity.

\subsubsection{Conditional expectations} \label{Sec:cond_3}
\noindent\textit{Conditional expectation of $E_{TKE}$} 

During fast decay, \At number continues to play an important role on the behavior of conditional expectation of $E_{TKE}$. The light and heavy fluid regions have larger $E_{TKE}$ values for the lower \At number cases, as these regions continue to move faster. However, for high \At numbers, the average $E_{TKE}$ values increase slightly within the heavier fluid regions. 

\begin{figure}
(\emph{a}) \hspace{6.5cm}  (\emph{b}) \\
    \centerline{\includegraphics[width=6.4cm]{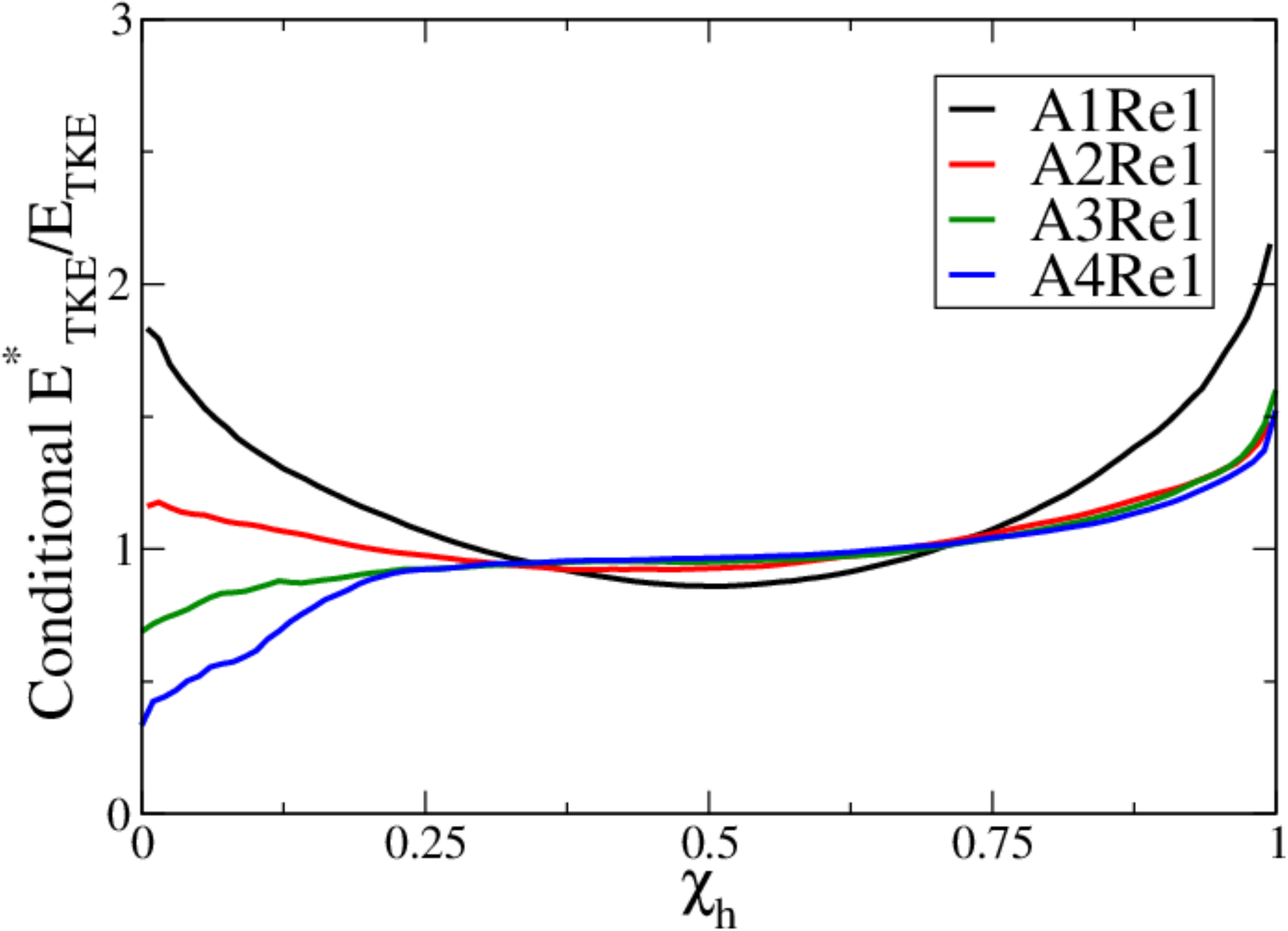}
    \includegraphics[width=6.4cm]{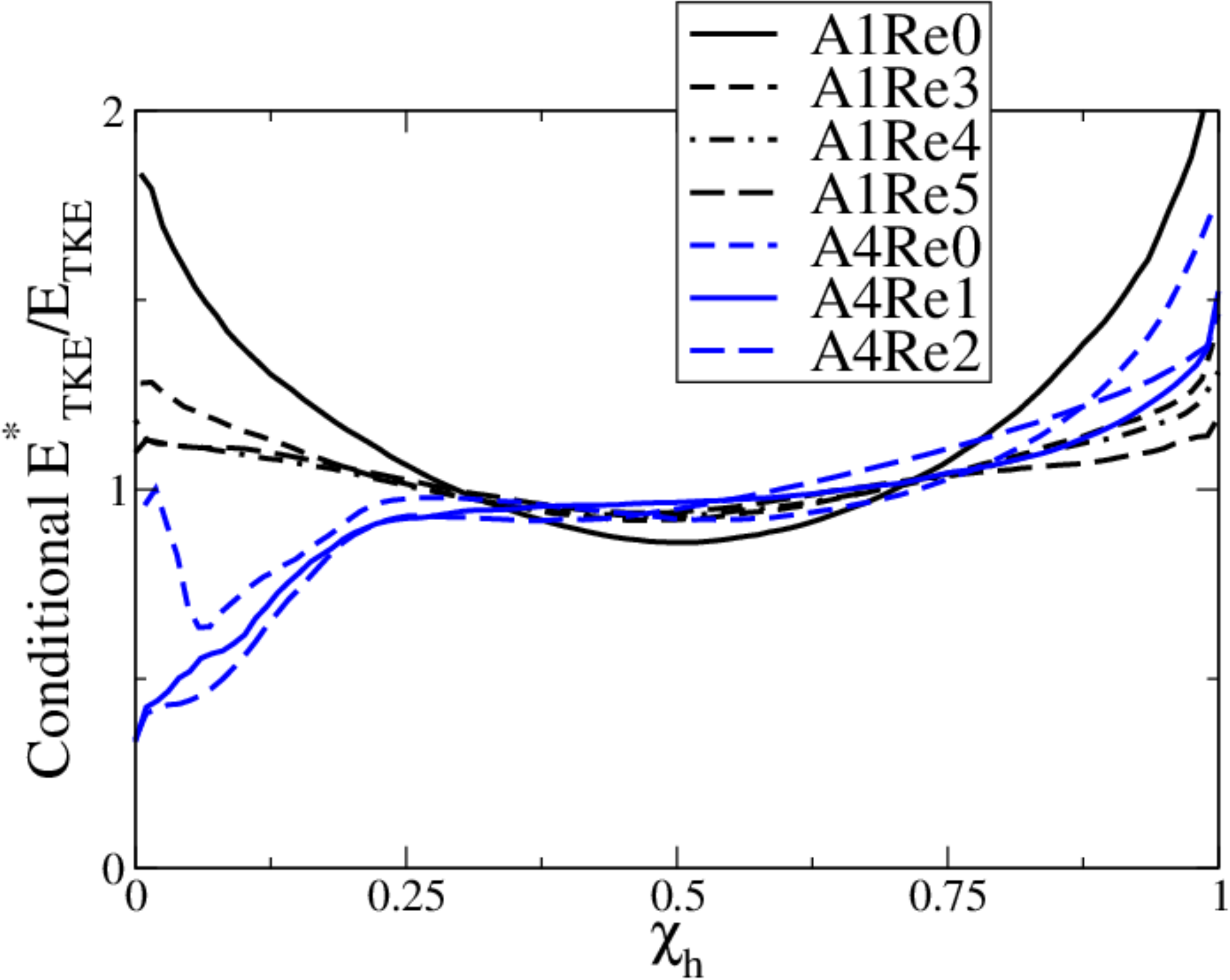}}
\caption{Atwood and Reynolds numbers effects on conditional expectation of $E_{TKE}$ during fast decay regime (at $t/t_r=2.95$).}
\label{Fig:cond_tke_3}
\end{figure}

\vspace{3mm} \noindent\textit{Conditional expectation of $E_{TKE}$ dissipation} 

During fast decay, the conditional mean of dissipation becomes independent of \rez. In addition, the \At number effect becomes weak for the range investigated here. The energy mostly dissipates within flow regions that are fully mixed and the local values are no longer much larger than the mean values.

\begin{figure}
(\emph{a}) \hspace{6.5cm}  (\emph{b}) \\
    \centerline{\includegraphics[width=6.4cm]{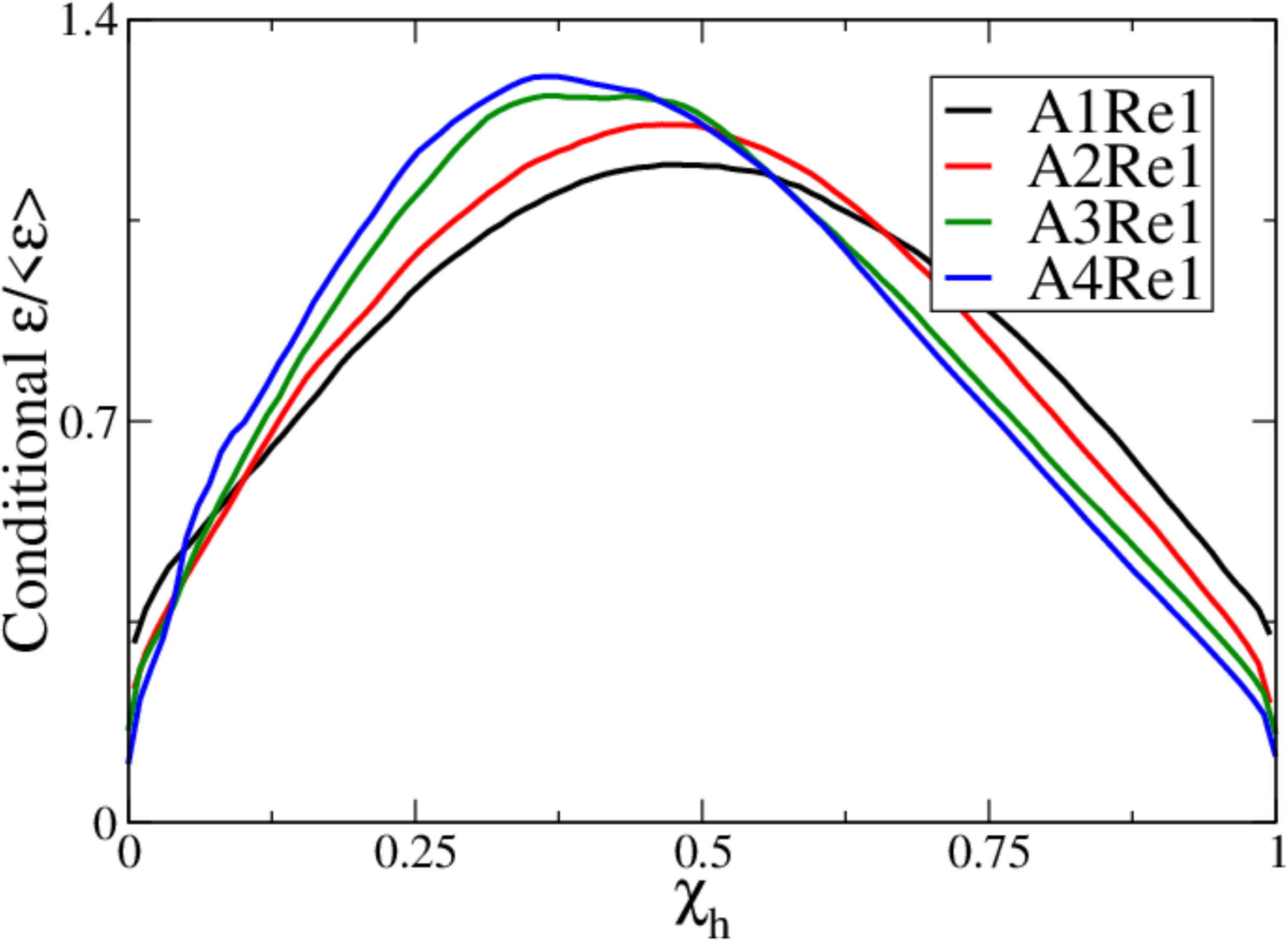}
    \includegraphics[width=6.4cm]{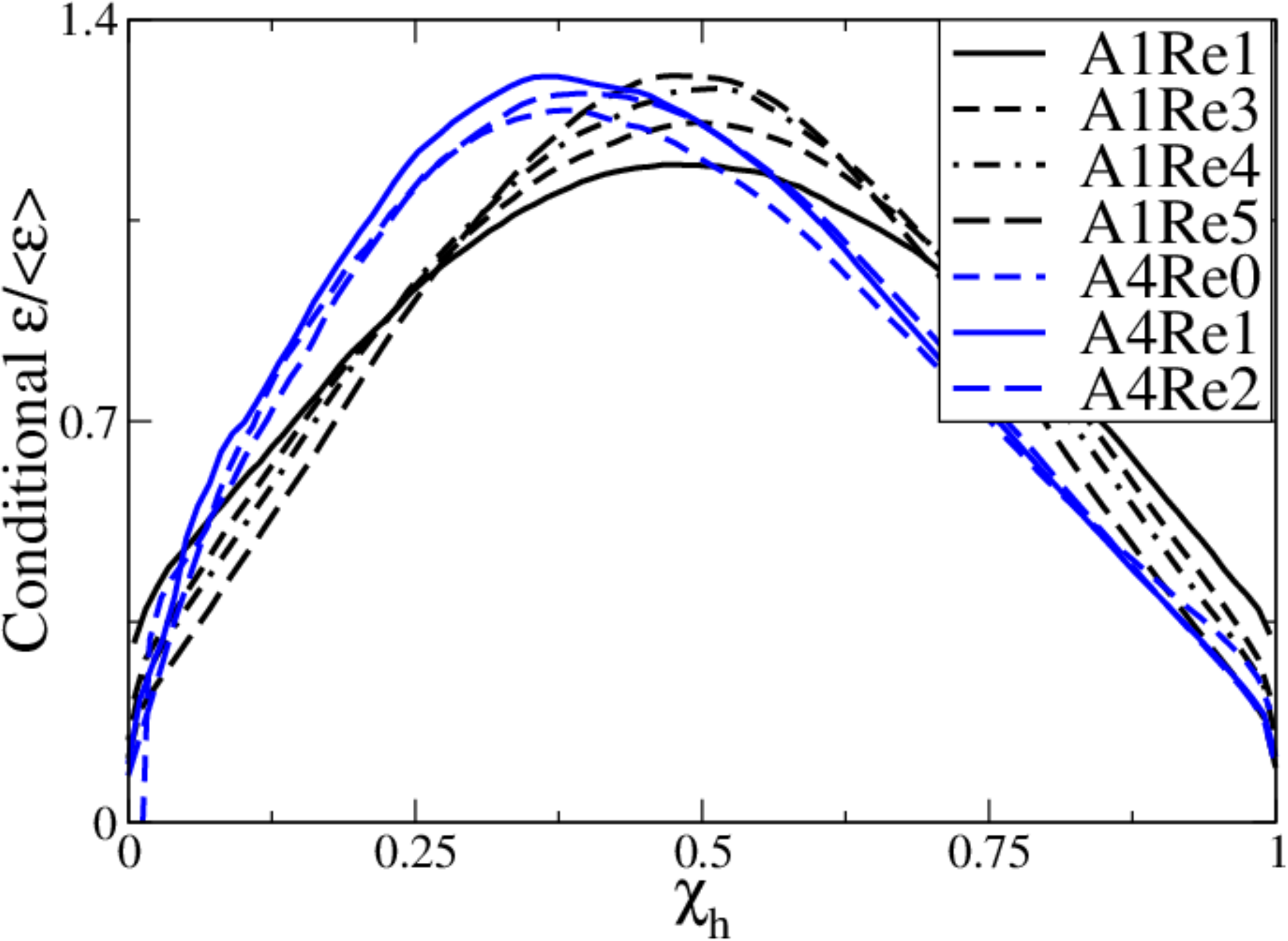}}
\caption{Atwood and Reynolds numbers effects on conditional expectation of $E_{TKE}$ dissipation during fast decay regime (at $t/t_r=2.95$).}
\label{Fig:cond_eps_3}
\end{figure}

 \vspace{3mm} \noindent\textit{Conditional expectation of enstrophy} 

\begin{figure}
(\emph{a}) \hspace{6.5cm}  (\emph{b}) \\
    \centerline{\includegraphics[width=6.5cm]{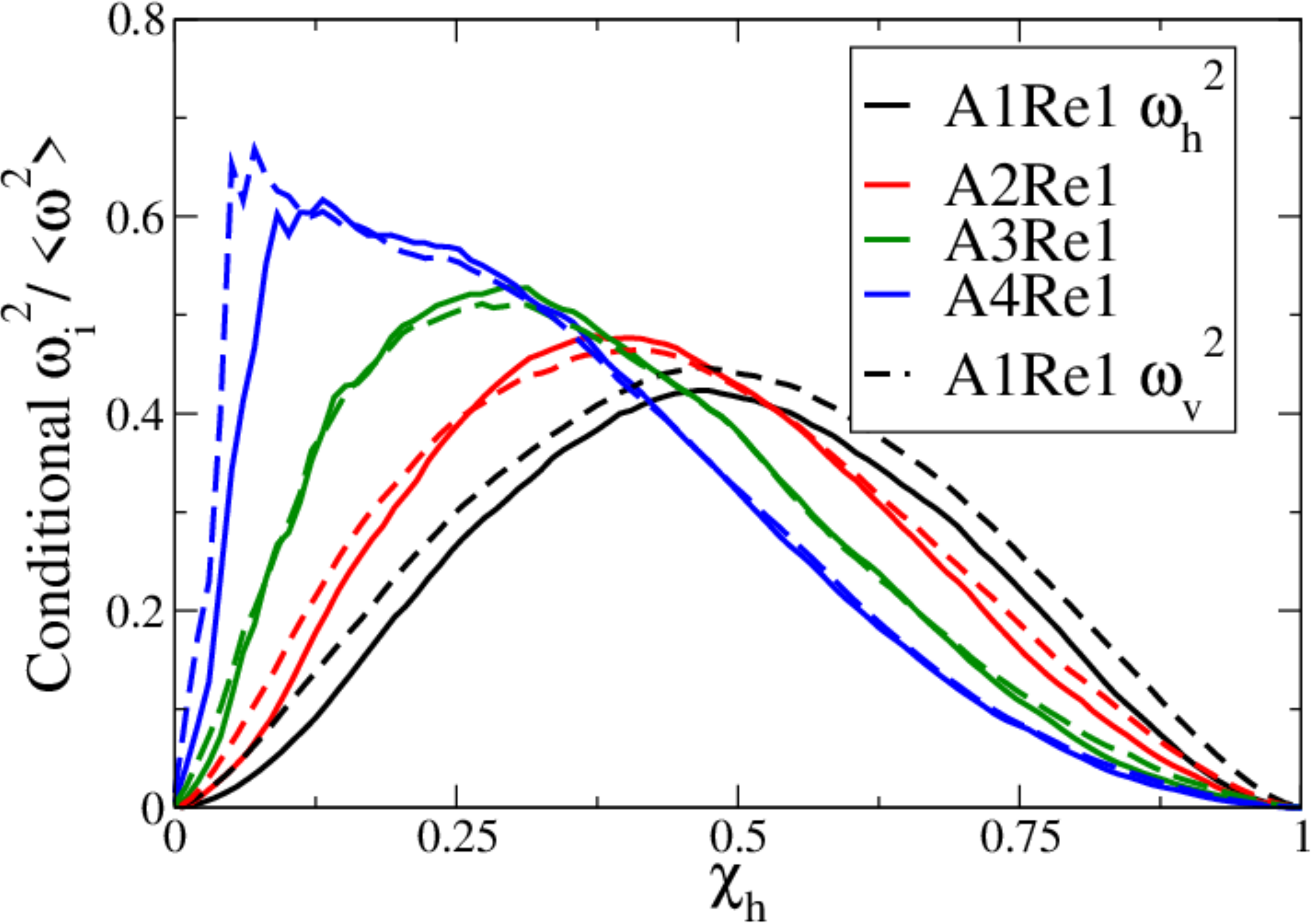}
    \includegraphics[width=6.5cm]{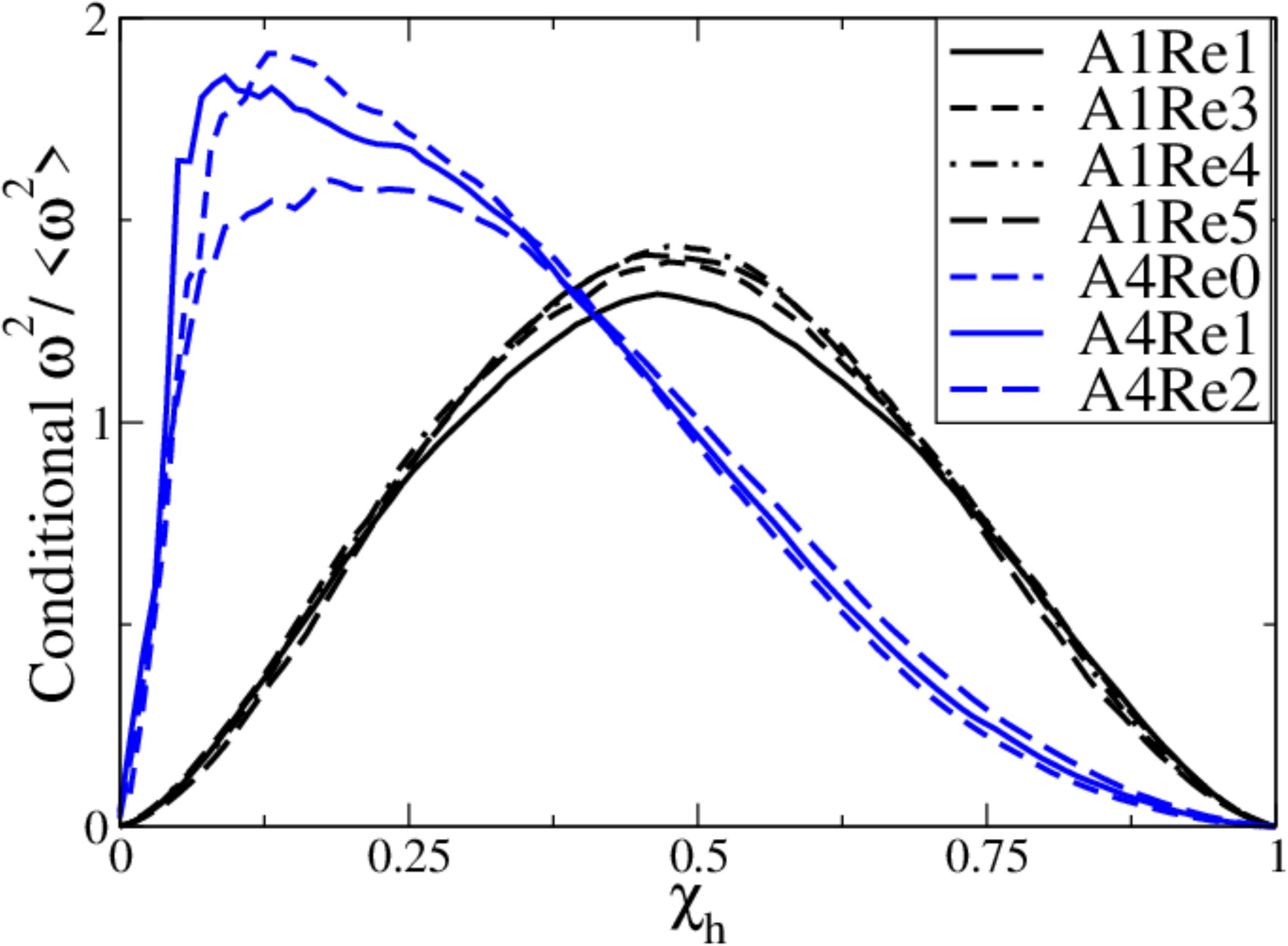}}
\caption{Atwood and Reynolds numbers effects on conditional expectation of (a) vertical ($\omega^2_v=\omega^2_1$) and horizontal ($\omega^2_h=(\omega^2_2+\omega^2_3)/2$) enstrophy components and (b) total enstrophy during fast decay regime at $t/t_r=2.95$.}
\label{Fig:cond_ens_3}
\end{figure}

In contrast to dissipation field, the conditional enstrophy is still dependent on \At as shown in Fig. \ref{Fig:cond_ens_3}. Similar to the dissipation field, the local values are closer to the whole domain average for all cases and \rez effects are not significant. It is also noticeable that components of enstrophy field are similar, which indicates small scale isotropy during this regime.

\subsection{Gradual (slacken) decay}\label{Sec:Gradual_decay}

After the period of fast decay, the flows settles into a late time slow decay regime, as $d^2E_{TKE}/dt^2$ again becomes larger than zero. This regime has some similarities with the core of the mixing region at late time of the RMI and RTI under reversed acceleration ($g \leq 0$). Buoyancy forces continue to weaken and no pure fluids exist within the flow. In addition, as most of the fluids have been molecularly mixed ($\theta>0.95$ for all cases), the VD effects are weak during gradual decay which makes HVDT comparable to the study by \citet{batchelor1992}. Figure \ref{Fig:TKE-Ret} shows that HVDT attains a non-decaying Reynolds number behavior during the late time stage. An interesting comparison can be made between the case of buoyancy-assisted decay and pure decay (where $g_i$ is set  to zero at some time instant). As it can be seen in Fig. \ref{Fig:Re_g}, in the case of pure decay, the \ret number decreases continuously. However, when the decay is assisted by buoyancy forces, the Reynolds number does not decrease. This is consistent with the theoretical analysis of \citet{batchelor1992}, which highlights the importance of density fluctuations in the infrared part of the energy spectrum, even as they asymptotically decrease to zero. 
 In this study, we aim to reach as high of \ret values as possible during the flow evolution, so the initial peak of the density spectrum is chosen to be at low wave-numbers. Unfortunately, this choice of initial conditions leads to insufficient infrared part in the spectra to be able to draw definitive conclusions about the decay laws of $E_\rho$ and $E_{TKE}$ and the behavior of \ret. Shifting the peak of the initial density energy spectrum to higher wave-numbers, while maintaining \ret number above the mixing transition would require much larger resolutions and is outside the scope of the current work.

During the gradual decay regime, as the turbulent kinetic energy behavior no longer changes, the flow also reaches a self-similar behavior. Thus, density and velocity PDFs tend to reach symmetric shape for all \At numbers investigated here.
Gradual decay has strong similarities with sufficiently developed RTI under negative (reversed) acceleration where the flow is stably stratified and is not fed by pure fluids \citep{Denis_PRE}. Similarly, with HVDT during gradual decay, the flow tends to become fully-mixed under the presence of weak buoyancy-forces primarily due to the buoyancy differences between the partially-mixed regions \citep{Ramaprabhu_ADA,Denis_PRE,livescu_variable_accel_2012,livescu_variable_accel_2019}. 
Gradual decay also has similarities with atmospheric and oceanic flows where the flow perturbed by different sources may decay under weak buoyancy-forces.
   
\begin{figure}
\hspace{3.3cm}(\emph{a}) \hspace{6.1cm} (\emph{b})

\vspace{0.5cm}
\hspace{.9cm}   Case:A1Re5 \hspace{1.32cm} Case:A4Re2 \hspace{1.45cm}  Case:A1Re5 \hspace{1.3cm} Case:A4Re2 \\
\rotatebox{90}{\hspace{1cm}$t/t_r=6$}\includegraphics[width=3.2cm]{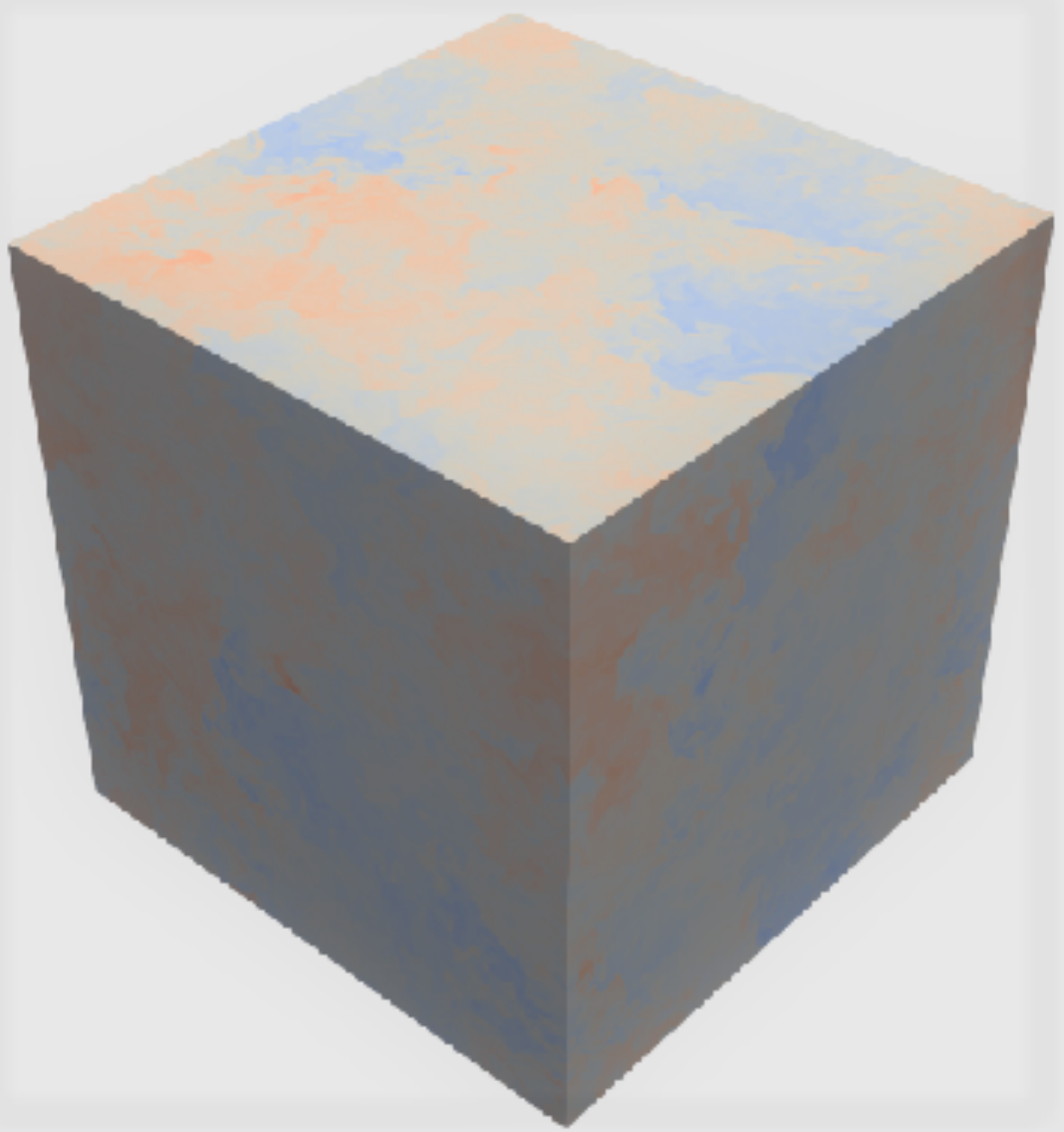}
\includegraphics[width=3.2cm]{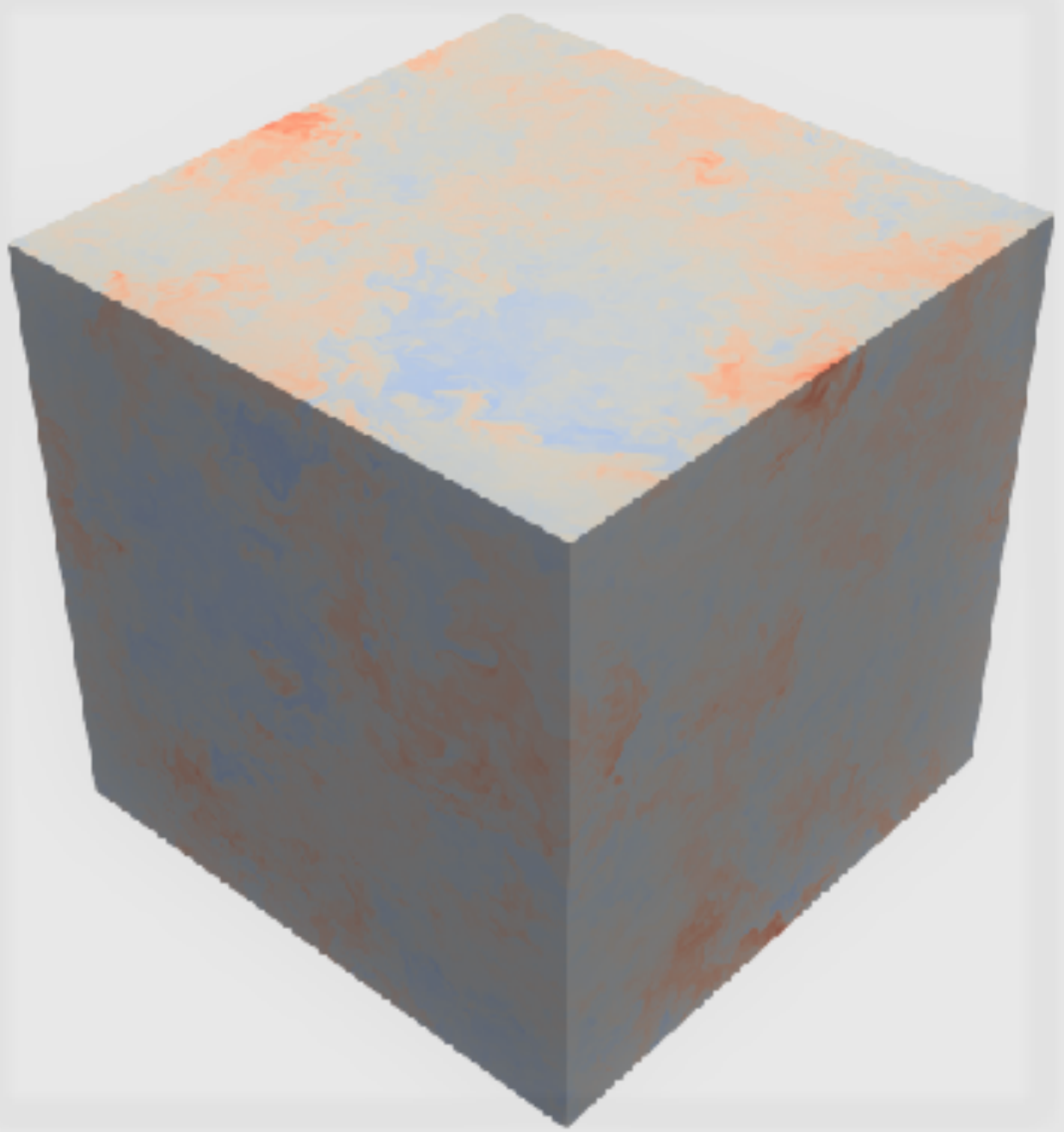}~~
\includegraphics[width=3.2cm]{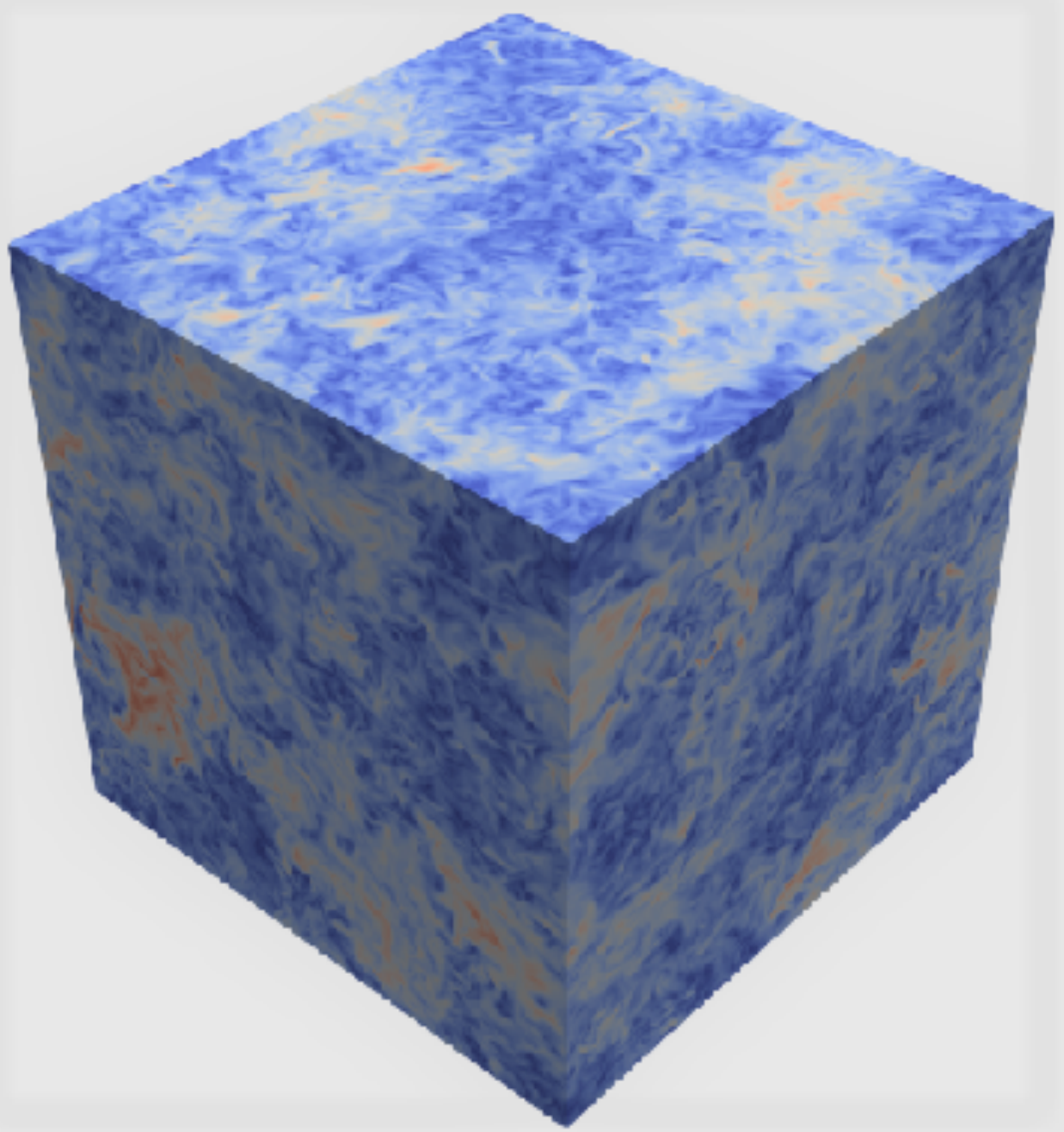}
\includegraphics[width=3.2cm]{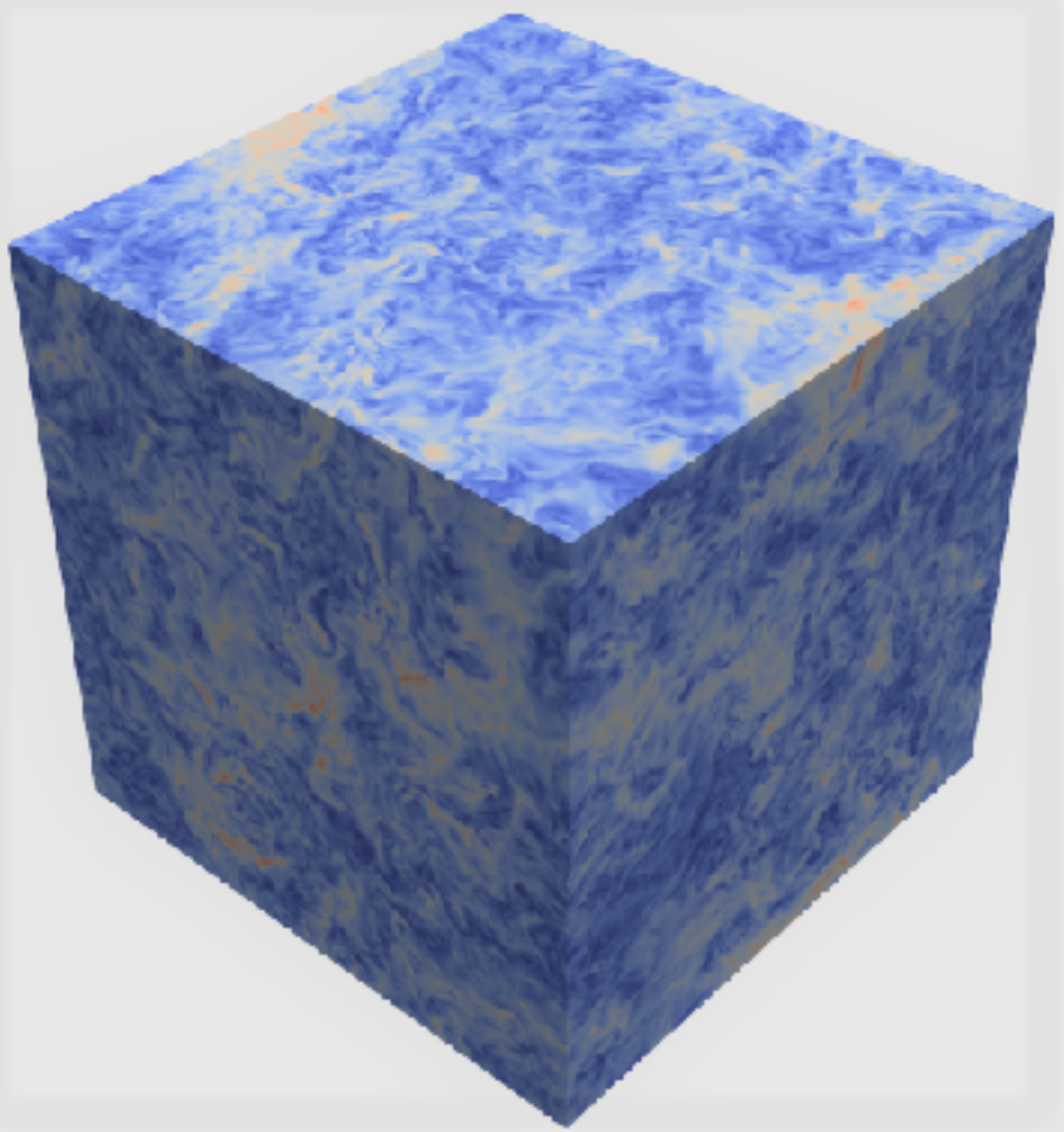}
 \caption{3D visualization of (a) the mole fraction, and (b) the velocity magnitude ($\sqrt{u^2_1+u^2_2+u^2_3}$) for $A=0.05$ case (A1Re5) and $A=0.75$ case (A4Re2) displayed at $t/t_r\approx6$.}
\label{Fig:3Devolve_T4}

\end{figure}

        \begin{figure}
        \centering{\includegraphics[width=8cm]{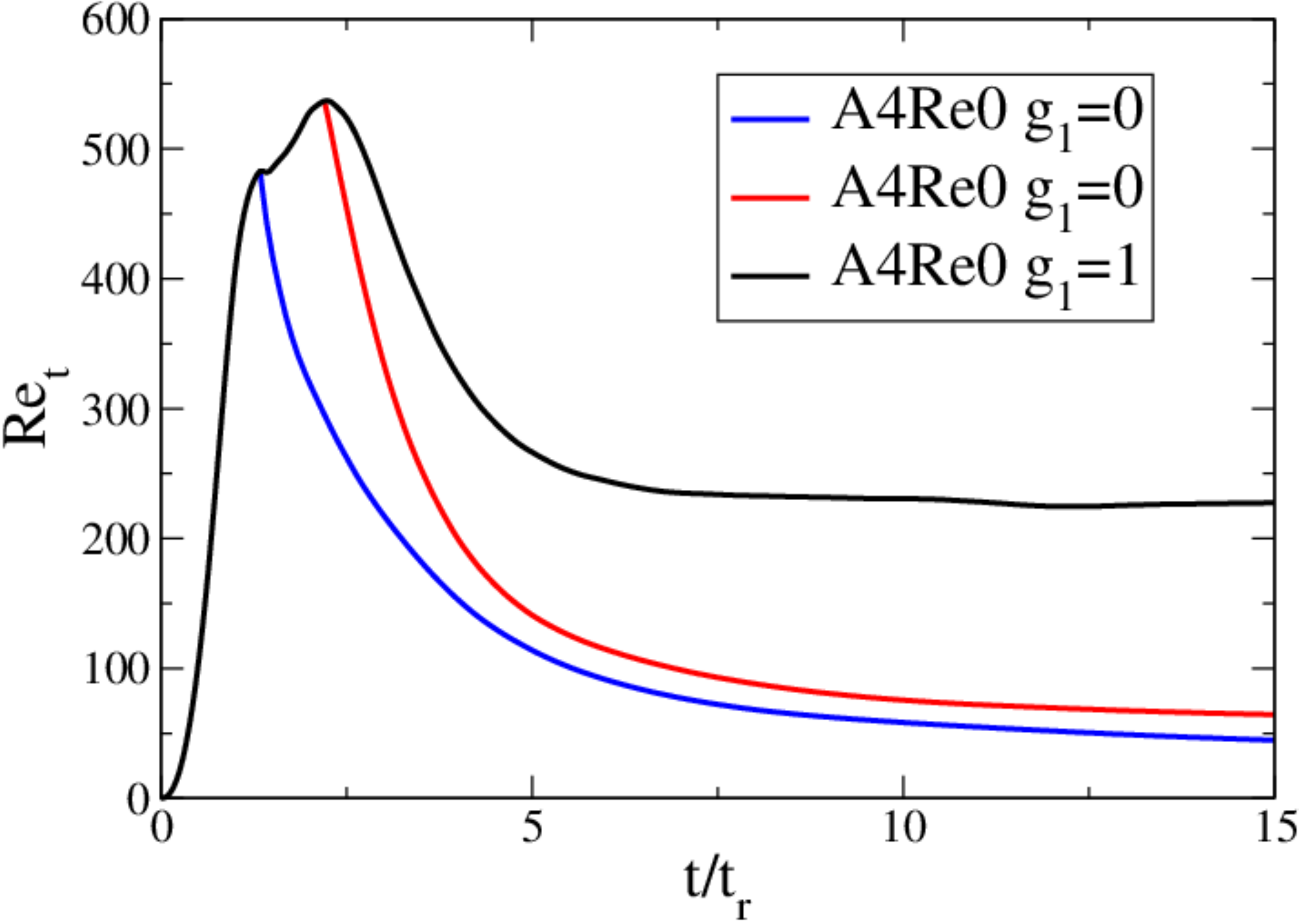}}
        \caption{Reynolds number evolution for pure and buoyancy-assisted decays.}
        \label{Fig:Re_g}
    \end{figure}

\subsubsection{Atwood number effects on PDF evolutions} \label{Sec:PDF_4}
\vspace{3mm}
\noindent\textit{Density PDF} 

\begin{figure}
   (\emph{a}) \hspace{6.5cm}  (\emph{b}) \\
    \centerline{\includegraphics[width=6.5cm]{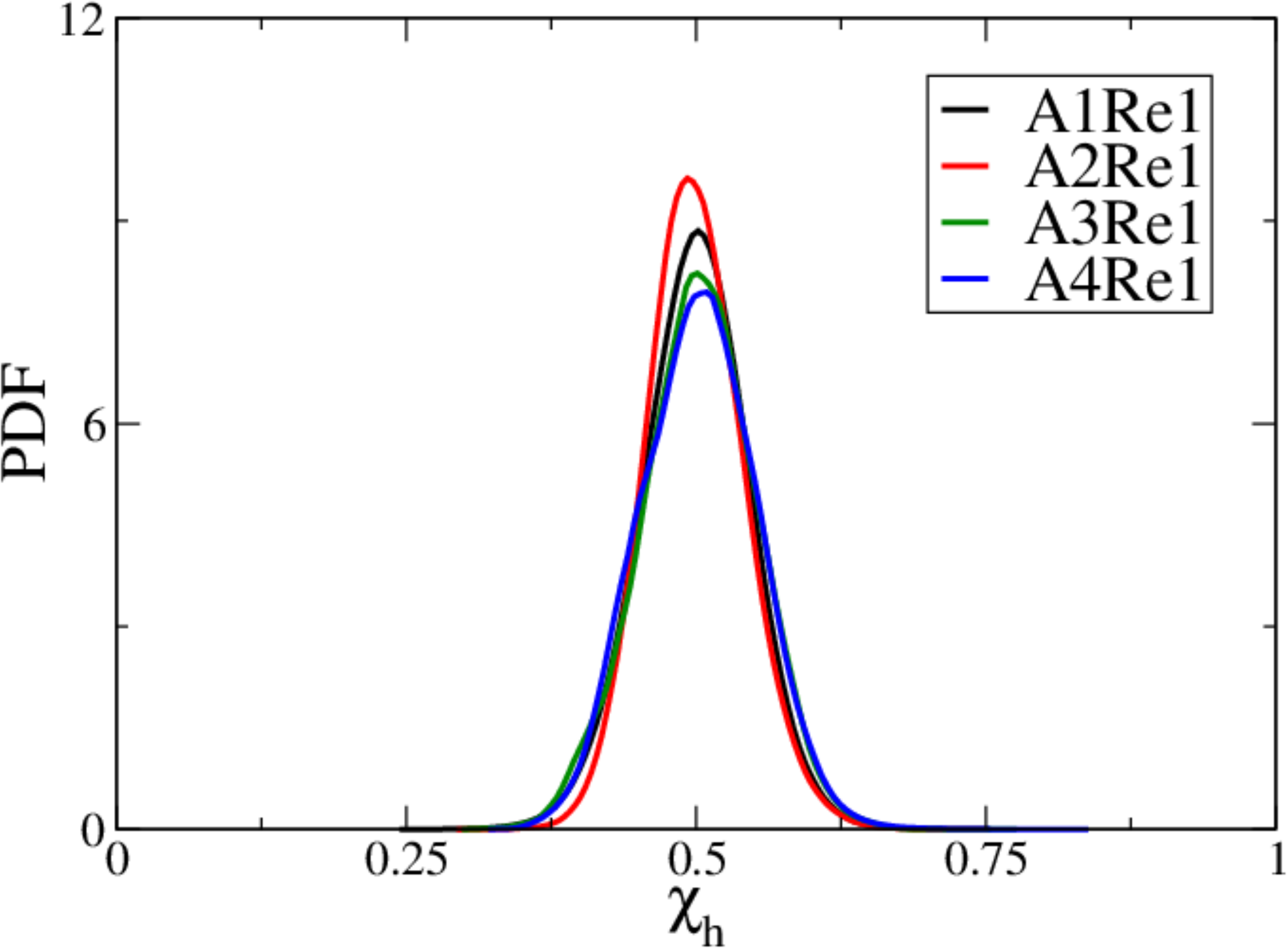}
    \includegraphics[width=6.5cm]{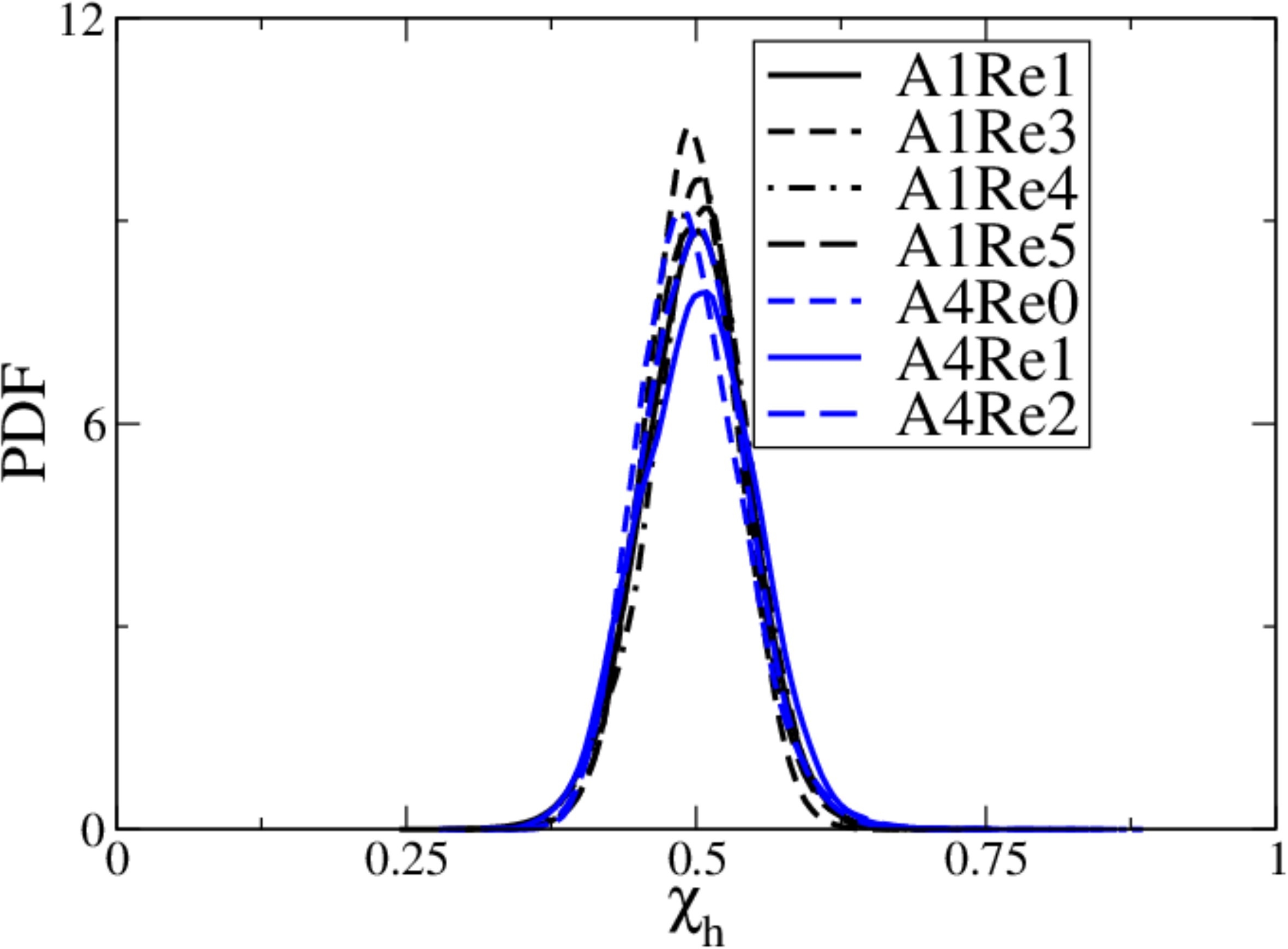}}
 \caption{PDF of the normalized density field for different (a) \At numbers and (b) \ret numbers at $t/t_r\approx6$.}
\label{Fig:dens_PDF_4}
\end{figure}

During gradual decay, as most of the fluids are mixed, the density PDF is accumulated at around mean density (see fig. \ref{Fig:dens_PDF_4}). Both \At and \rez number effects are limited during this regime; however, the density PDF tail is slightly longer at the heavy fluid side for larger \rez and \At numbers.

\vspace{3mm}
\noindent\textit{Velocity-density jMDF} 

\begin{figure}
\vspace{0.5cm}
\hspace{2.6cm}   Case:A1Re5 \hspace{3.8cm} Case:A4Re2  \\
\centerline{\includegraphics[width=7.0cm]{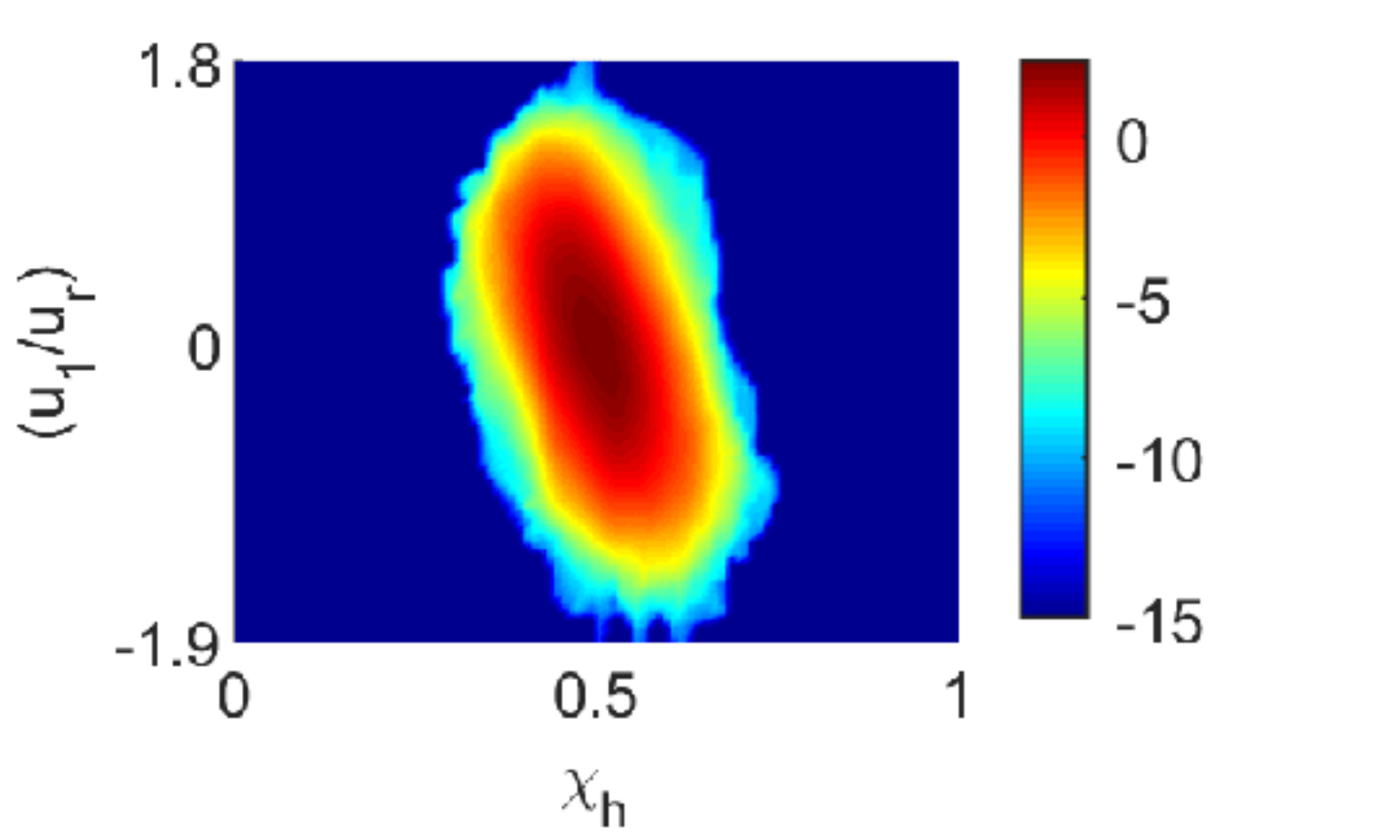}~~
\includegraphics[width=7.0cm]{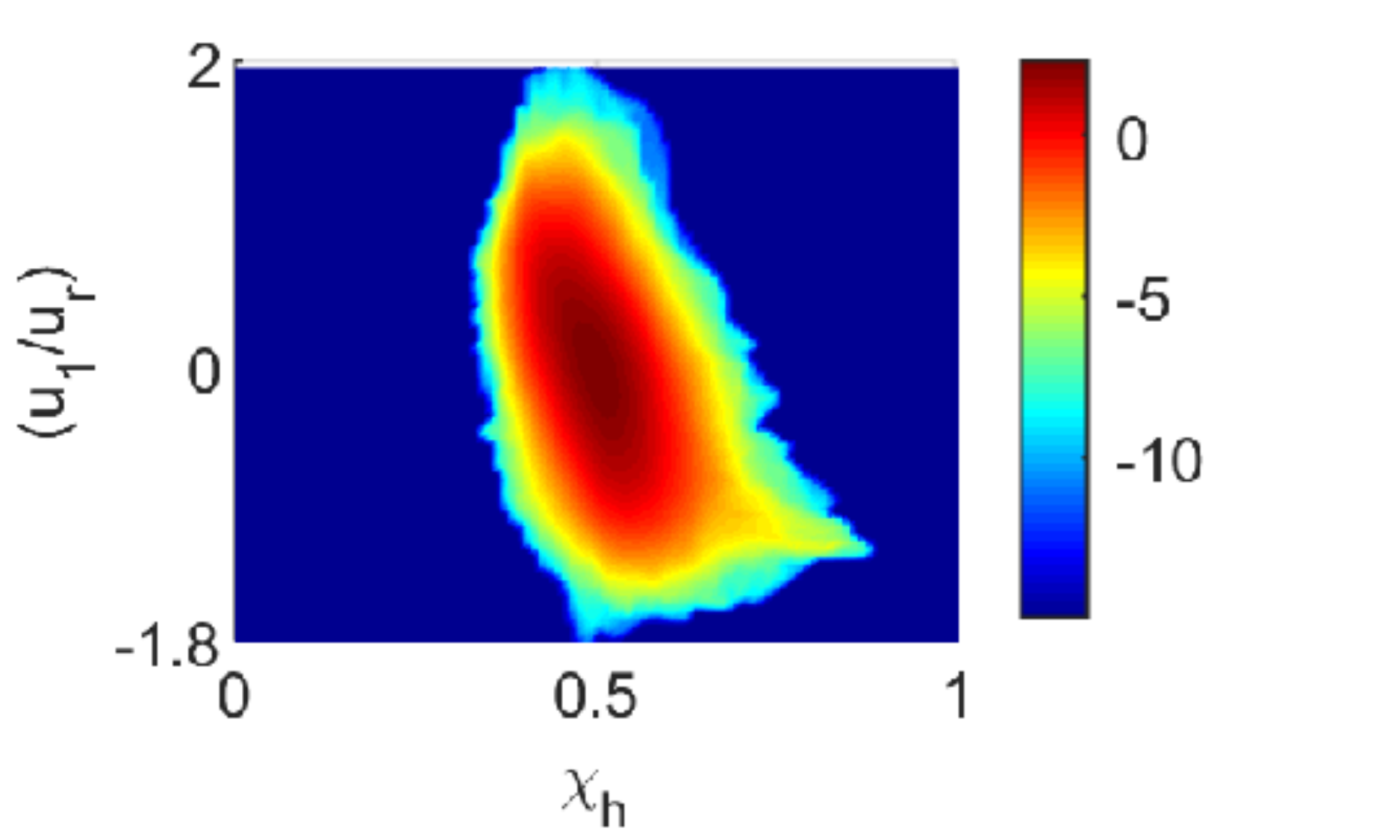}}
 \caption{Normalized jMDFs [$\log(\mathcal{F}/\rhom)$] for (a) $A=0.05$ (A1Re5) and (b) $A=0.75$ (A4Re2) cases displayed at $t/t_r\approx6$.}
\label{Fig:Joint_4}
\end{figure}

The jMDFs of density and velocity fields (see Figure \ref{Fig:Joint_4}) become closer for the low and high Atwood cases. The slightly tilted behavior indicates that during decay, velocity and density fields remain (weakly) negatively correlated. Heavier regions moving close to the maximum downward velocities are still present in the high \At number case, while slightly larger maximum vertical velocities indicates that still the light fluid regions continue to be more stirred by turbulence. 

\subsubsection{Conditional expectations} \label{Sec:cond_4}

\begin{figure}
(\emph{a}) \hspace{5cm}  (\emph{b}) \hspace{5cm}  (\emph{c}) \\  
    \centerline{\includegraphics[height=3cm]{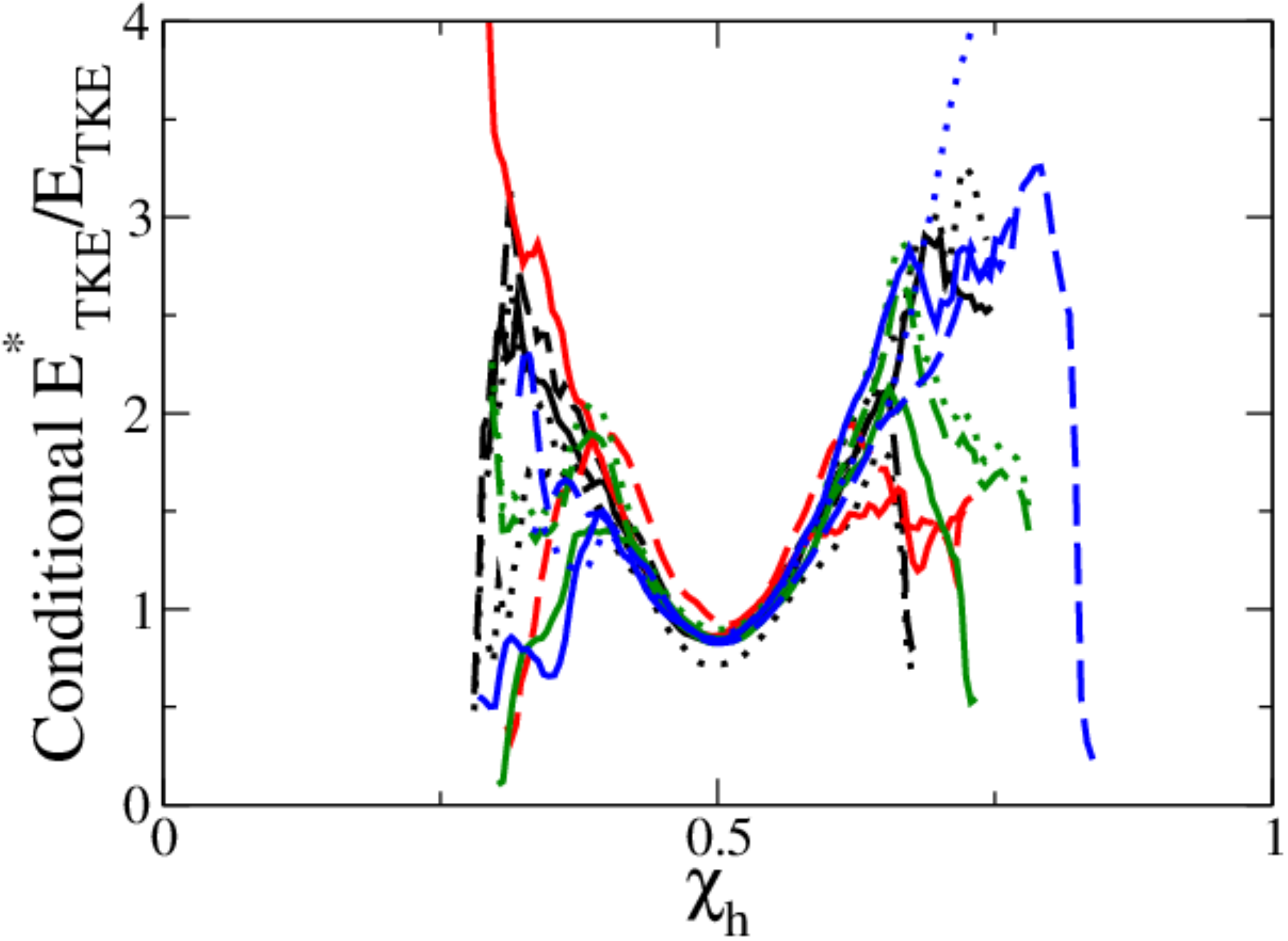}
    \includegraphics[height=3cm]{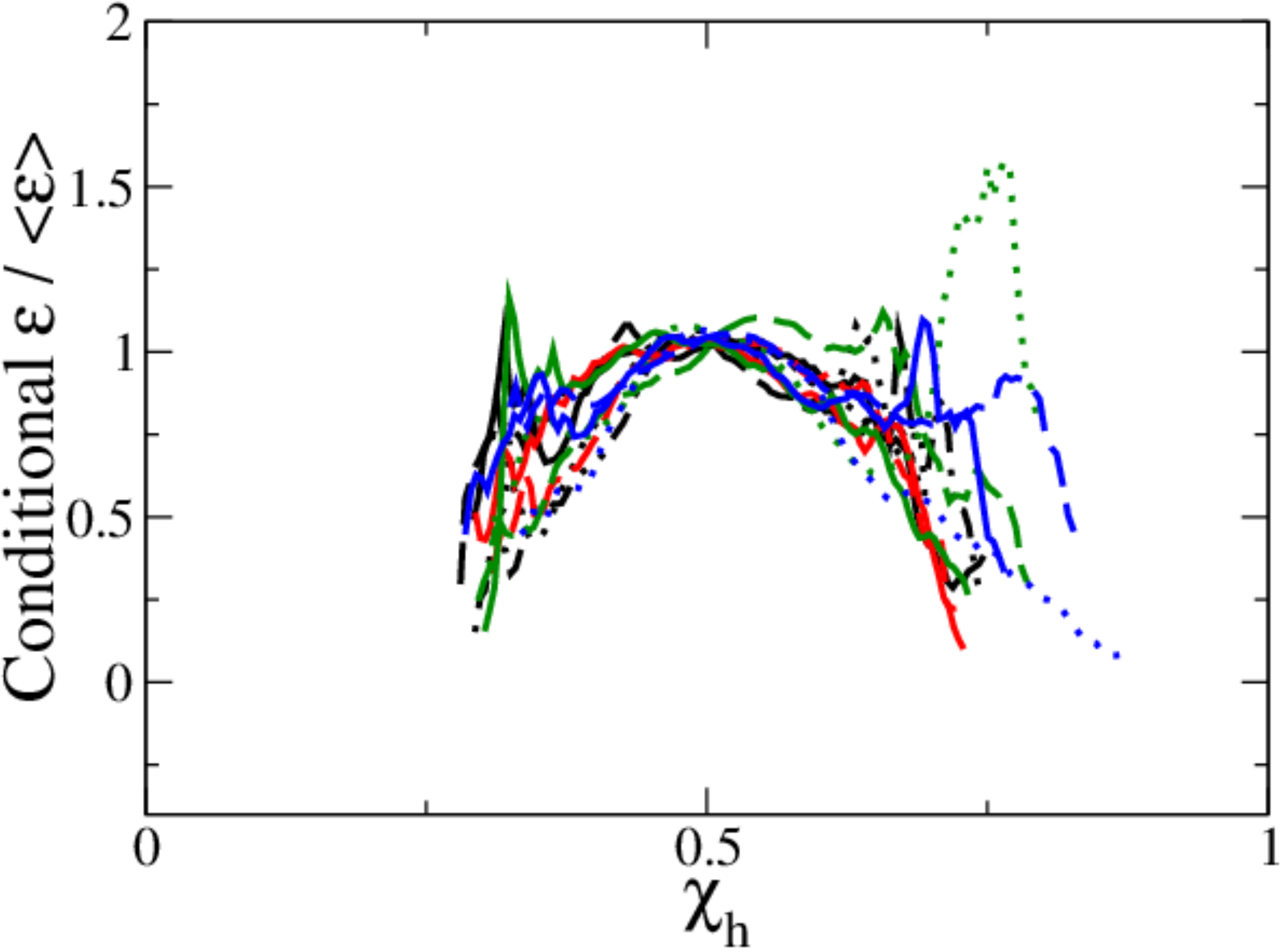}
    \includegraphics[height=3cm]{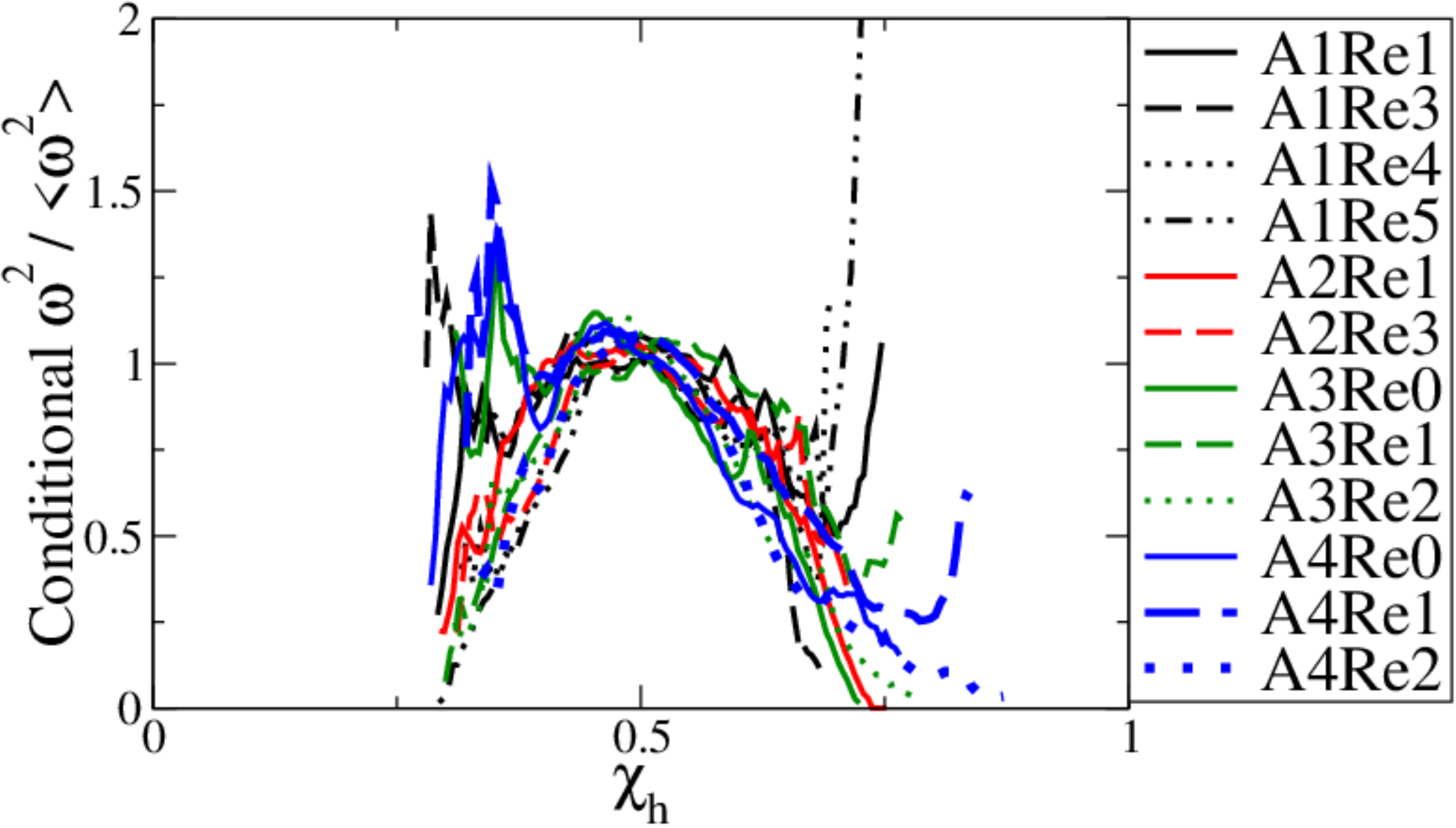}}
\caption{Atwood and Reynolds numbers effects on conditional expectation of normalized (a) $E_{TKE}$, (b) $E_{TKE}$ dissipation, and (c) total enstrophy during gradual decay (at $t/t_r\approx6$).}
\label{Fig:cond_4}
\end{figure}

During gradual decay, the conditional means of $E_{TKE}$, dissipation of $E_{TKE}$ and enstrophy become less dependent on \rez and \At numbers. For all \At numbers studied here, $E_{TKE}$ is slightly larger within the lightest and heaviest flow regions as buoyancy-forces are still present compared to the fully-mixed flow (see Fig. \ref{Fig:cond_4}a). Moreover, all three quantities have values close to their means in well-mixed regions, showing the importance of these regions to the global statistics. However, a lot of variability in still observed in lighter and heavier than average fluid regions. In particular, while dissipation fluctuates and tends to be lower than average in light and heavy fluid regions, enstrophy tends to be lower than average in heavy fluid regions, but higher than average in light fluid regions for larger \At (see Fig. \ref{Fig:cond_4}b, c). Thus, even at late times, lighter fluid pockets maintain stronger vortical motions, whereas in heavier fluid regions vorticity can become small.

\section{Spectral evolution of buoyancy-driven VDT} \label{Sec:Spectral_evolve}

For variable-density flows, there is no unique way to define a kinetic energy spectrum. Quadratic expressions with the same units can be constructed from kinetic energy definitions per unit mass, i.e. $\mbox{TKE}^I\equiv \langle | \sqrt{\rhom} \mathbf{u}^{''} |^2 \rangle$, or per unit volume, i.e. $\mbox{TKE}^{II}\equiv \langle | \sqrt{\rho^*} \mathbf{u}^{''} |^2 \rangle$ \citep{Kida90,Cook02,livescu2008}. Non-quadratic forms can also be constructed, using products between momentum per unit volume, $\rho^* \mathbf{u}^{''}$, or per unit mass, $\mathbf{u}^{''}$ \citep{Clark05,Lai18,Pal18}.
However, in order for the kinetic energy spectrum to develop an inertial range, the viscous effects need to be restricted to the small scales. For VD turbulence, such scale decomposition is not as straightforward as in incompressible flows. In particular, it is possible that certain quantities exhibit viscous effects at all scales, while others do not \citep{Zhao_Hussein_2018}. Using a coarse-grained filtering, \cite{Zhao_Hussein_2018} showed that the filtered Favre turbulent kinetic energy has the property that viscous effects vanish at large scales for sufficiently large Reynolds numbers, which they called the inviscid scale decomposition criterion. The filtered forms of kinetic energies $TKE^I$ and $TKE^{II}$ do not have this property, when the density variations are large. These considerations fully apply only when the viscosity is constant. The variable viscosity case does not satisfy an inviscid scale decomposition criterion for arbitrary variations of viscosity with density. While it is outside the goals of the paper to discuss this general case, here we explore several forms of the velocity power spectra.

A quadratic form of the turbulent kinetic energy that satisfies the inviscid scale decomposition criterion for constant viscosity is proposed here as $\mbox{TKE}^{III}\equiv \langle | \rho^* \mathbf{u}^{''} /\sqrt{\rhom}|^2 \rangle$. Table \ref{table:spectra_opts} lists two commonly used definitions together with the new energy form and the viscous terms in the corresponding spectral transport equations, where \(\widehat{~~}\) denotes a Fourier transform and \(^{\dagger}\) the complex conjugate. As shown by \cite{Zhao_Hussein_2018}, terms of the type $\tau_{ij,j}/\rho^*$ or $\tau_{ij,j}/\sqrt{\rho^*}$ can not be proven to decrease at large scales for sufficiently large density variations. Such terms appear in the spectral transport equations for $\mbox{TKE}^{I}$ and $\mbox{TKE}^{II}$, so it is expected that these kinetic energies will have viscous contributions at all scales in VD turbulence. Similar terms also appear in the spectral transport equations for the non-quadratic forms of the kinetic energy used in the literature. On the other hand, the viscous term in the spectral transport equation for $\mbox{TKE}^{III}$ is expected to vanish at large scales (for constant viscosity), irrespective of the \At number.

\begin{table}
\centering
\def~{\hphantom{0}}
\setlength{\tabcolsep}{18pt}
\begin{tabular}{ccc}
Options      &       formula         & corresponding spectral viscous term \\
$\mbox{TKE}^I$ & $\langle | \sqrt{\rhom} \mathbf{u}^{'} |^2 \rangle$ & $\widehat{\left(\sqrt{\rhom}u_i^{''}\right)}^{\dagger} \widehat{\left(\sqrt{\rhom}\tau_{ij,j}/\rho^*\right)}+\widehat{\left(\sqrt{\rhom}u_i^{''}/\sqrt{\rhom}\right)} \widehat{\left(\sqrt{\rhom}\tau_{ij,j}/\rho^*\right)}^{\dagger}$ \\
$\mbox{TKE}^{II}$ & $\langle | \sqrt{\rho^*} \mathbf{u}^{''} |^2 \rangle$ & $\widehat{\left(\sqrt{\rho^*}u_i^{''}\right)}^{\dagger} \widehat{\left(\tau_{ij,j}/\sqrt{\rho^*}\right)}+(\widehat{\sqrt{\rho^*}u_i^{''}}) \widehat{\left(\tau_{ij,j}/\sqrt{\rho^*}\right)}^{\dagger}$ \\
$\mbox{TKE}^{III}$ & $\langle | \rho^* \mathbf{u}^{''} /\sqrt{\rhom}|^2 \rangle$ & $ \widehat{\left(\rho^*u_i^{''}/\sqrt{\rhom}\right)}^{\dagger} \widehat{\left(\tau_{ij,j}/\sqrt{\rhom}\right)}+\widehat{\left(\rho^*u_i^{''}/\sqrt{\rhom}\right)} \widehat{\left(\tau_{ij,j}/\sqrt{\rhom}\right)}^{\dagger}$
\end{tabular}
\caption{Different approaches to calculate VD kinetic energy spectra. Only the new form proposed here, $\mbox{TKE}^{III}$, has a viscous term in its corresponding spectral transport equation that vanishes at large scale for large density variations.}
\label{table:spectra_opts}
\end{table}

\begin{figure}
    \centerline{\includegraphics[width=7.2cm]{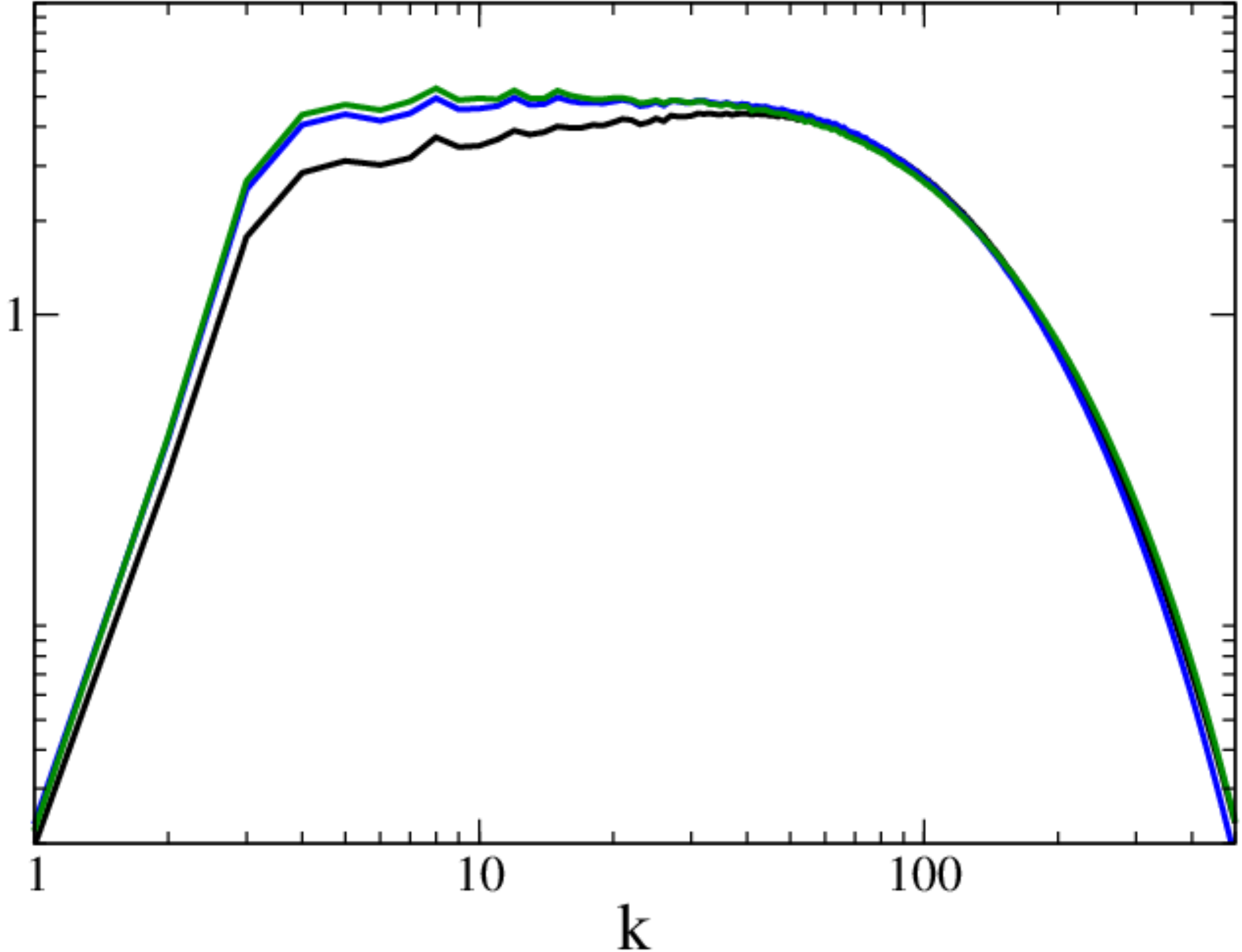}}
    \caption{Comparison of the compensated energy spectrum ($E_kk^{5/3}$) calculated based on $\mbox{TKE}^{I}$ (black line), $\mbox{TKE}^{II}$ (blue line), and $\mbox{TKE}^{III}$ (green line) for the $A4Re2$ case at the start of fast decay at $t/t_r=2.35$.}
    \label{Fig:Spectra_opt}
\end{figure}

\begin{figure}
        \centerline{\includegraphics[width=7.8cm]{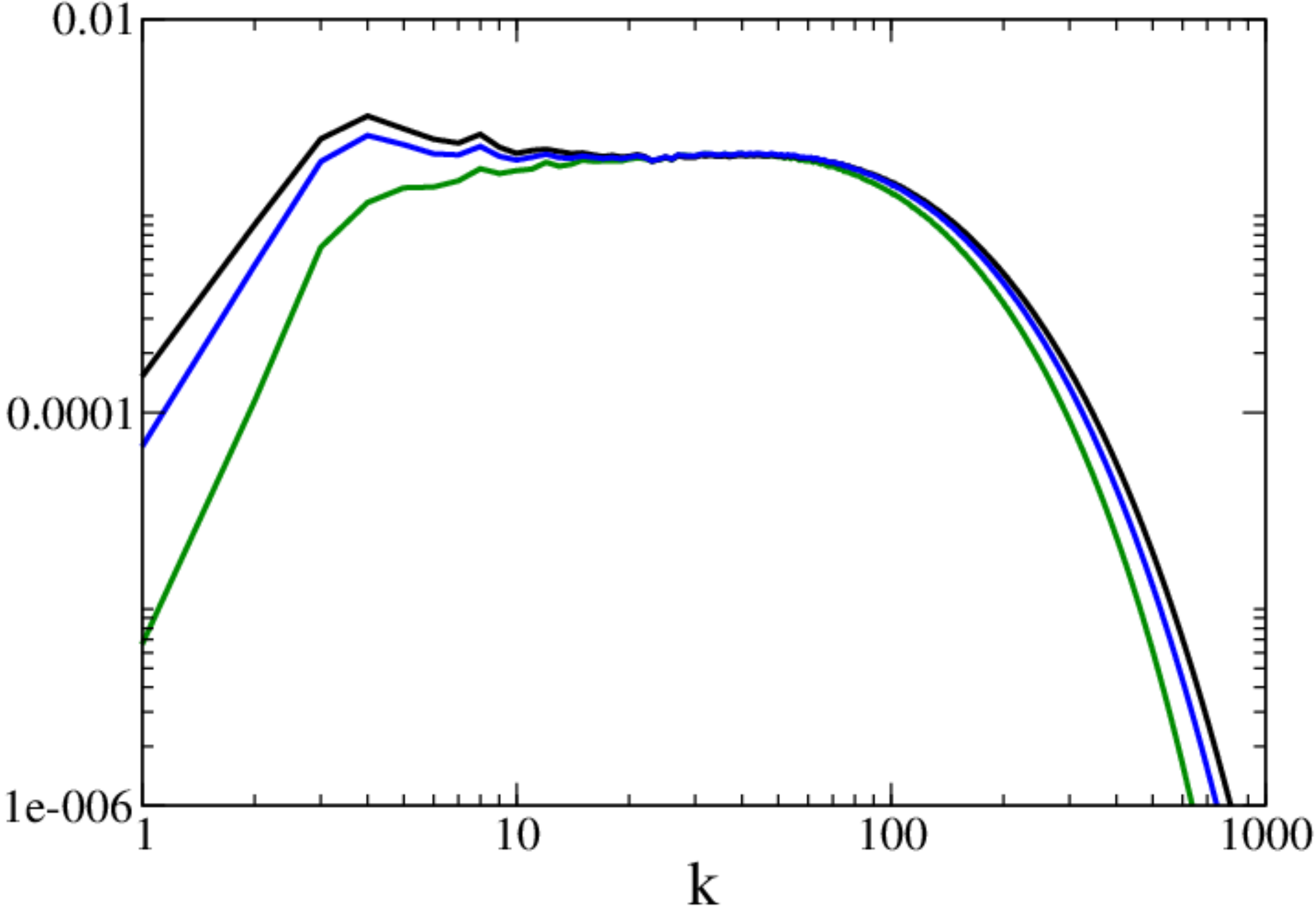}}
    \caption{Comparison of the viscous terms in the spectral transport equations for $TKE^I$ (blue line), $TKE^{II}$ (green line), and $TKE^{III}$ (black line)  for the $A4Re2$ case at $t/t_r=2.35$. The values are multiplied by constants such that the viscous terms have the same magnitudes at $k=50$ for clarity.}
    \label{Fig:Spectra_Vis_opt}
\end{figure}

Figure \ref{Fig:Spectra_opt} presents compensated energy spectra calculated based on the $\mbox{TKE}^{I}$, $\mbox{TKE}^{II}$, and $\mbox{TKE}^{III}$ definitions; Case $A4Re2$ is chosen for illustration. The spectra are plotted at the end of the saturated growth, where production and dissipation are equal. As can be seen in figure \ref{Fig:Spectra_opt}, the large scales are suppressed for definitions I and II and they have a flatter slope than the classical $-5/3$ slope. The differences are small for definition II for the range of density variations considered, as these are attenuated by the square root. In addition, fig. \ref{Fig:Spectra_Vis_opt} plots the corresponding viscous term for each energy spectrum option. The viscous term for the new definition, $\mbox{TKE}^{III}$, is smaller for the entire range; more importantly, the viscous effects decay at large scales. However, the viscous terms estimated using the definitions I and II continue to affect the evolution of large scales. This leads to a relative suppression of the large scales and a flatter slope than -5/3, which is consistent with findings by \citet{Zhao_Hussein_2018}. As the \At number decreases and the flow becomes Boussinesq, the density fluctuations also decrease and all the three definitions give similar spectra. For strongly variable density flows with variable viscosity coefficient, there is no general methodology to construct a quadratic energy form which ensures that viscous effects remain confined at small scales. While $\mbox{TKE}^{III}$ only has the property that the viscous effects vanish at large scales for the constant viscosity case, the results show that it is a better choice to examine the kinetic energy spectrum than $\mbox{TKE}^{I}$, $\mbox{TKE}^{II}$ for the cases considered here.

\subsection{Spectral Energy evolution}\label{Sec:Spectral_evolv_1}

As the flow undergoes rapid changes at early times and transitions from turbulent kinetic energy growth to decay, it is expected that the energy spectrum follows a complex evolution. In particular, the shape of the spectrum at early times should be strongly dependent on the production term, while at late times, during gradual decay, it should relax to fully developed conditions. In the previous section, a new quadratic form of the turbulent kinetic energy was shown to have reduced viscous effects at large scales for the cases considered here, while for the constant viscosity case those effects should vanish at large scales. This form, i.e. $\mbox{TKE}^{III}$, is used throughout this section.

\begin{figure}
(\emph{a}) \hspace{6.5cm}  (\emph{b}) \\
  \includegraphics[width=6.5cm]{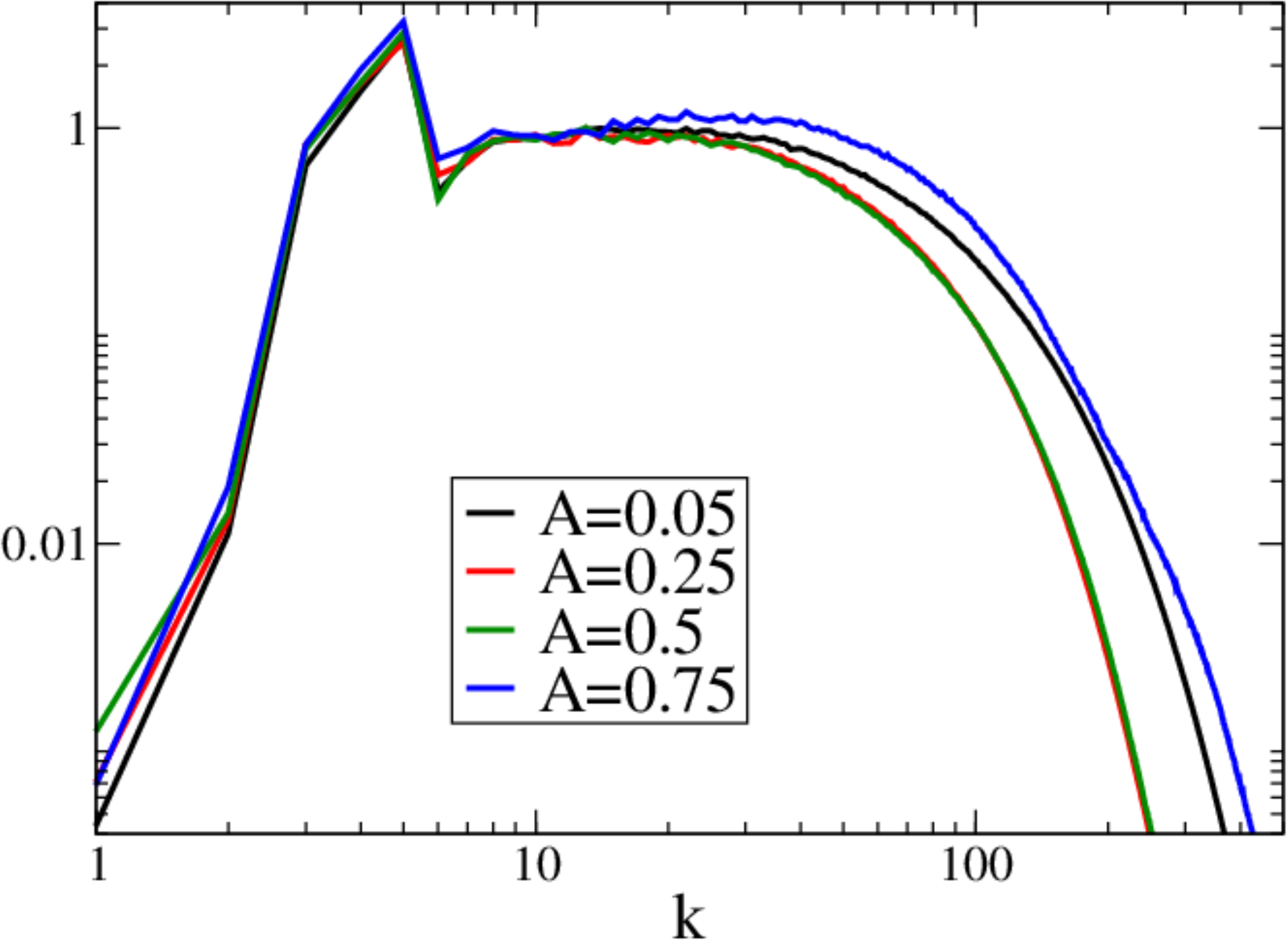}
  \includegraphics[width=6.5cm]{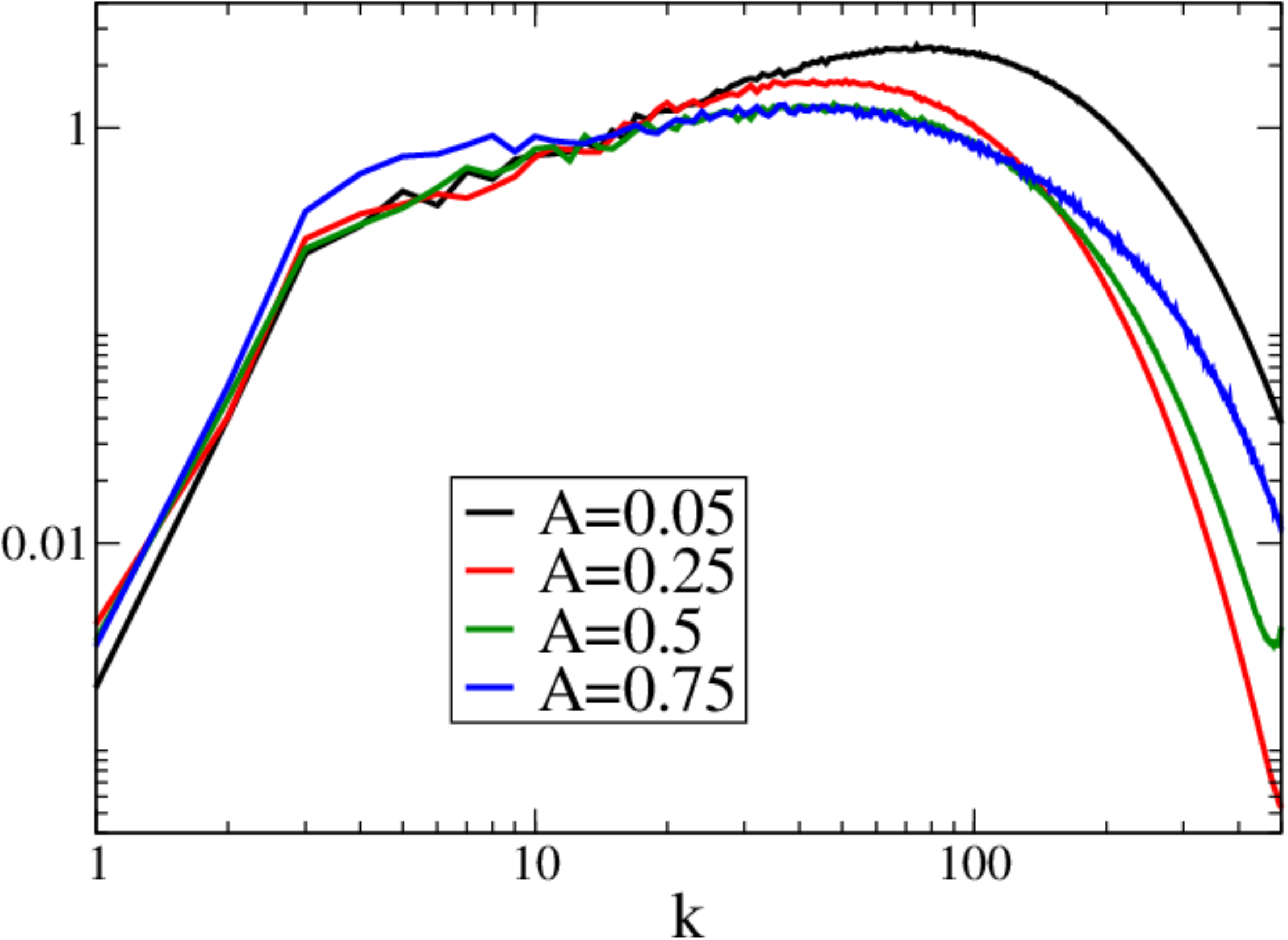}
  \caption{Compensated spectra of the normalized mass flux ($a_1(k)k^{7/3}/(A \rhom U_r)$) for different \At numbers at (a) $t/t_r=1.15$ and (b) $t/t_r=2.35$.}
\label{Fig:mass_1}
\end{figure}

For the Boussinesq limit, the buoyancy-production term, which is proportional to the mass flux ($\rho^*u_1$ from eq. \ref{Eq:KE_evolv}) has been reported to have $(-7/3)$ slope in the inertial range \citep{Lumley_67}; similar observations have also been made in a stationary version of HVDT \citep{chung_pullin_2010}. Figure \ref{Fig:mass_1} presents the mass flux spectra (normalized by $A \rhom U_r$) for different \At numbers at the start of the saturated growth $t/t_r=1.2$ and fast decay $t/t_r=2.4$ regimes. The data are consistent with previous studies as the mass flux has $-7/3$ slope at intermediate scales for all \At numbers.
At early times, the $-7/3$ slope covers a relatively wide range of scales at low \At number. However, upon an increase in \At number, the mass flux spectrum has a sharper decay at small scales and the $-7/3$ slope is restricted to a shorter range of scales. As turbulence starts to decay though, the $-7/3$ slope moves to higher wave numbers and becomes restricted to a shorter range of scales at lower \At number (\ref{Fig:mass_1}b). This behavior underlies the non-monotonic dependence of the flow evolution on the \At number. As the \At number is increased, the light and heavy fluid regions are accelerated faster, which increases the stirring, and ultimately leads to faster molecular mixing. 

\begin{figure}
(\emph{a}) \hspace{6.5cm}  (\emph{b}) \\
  \includegraphics[width=6.5cm]{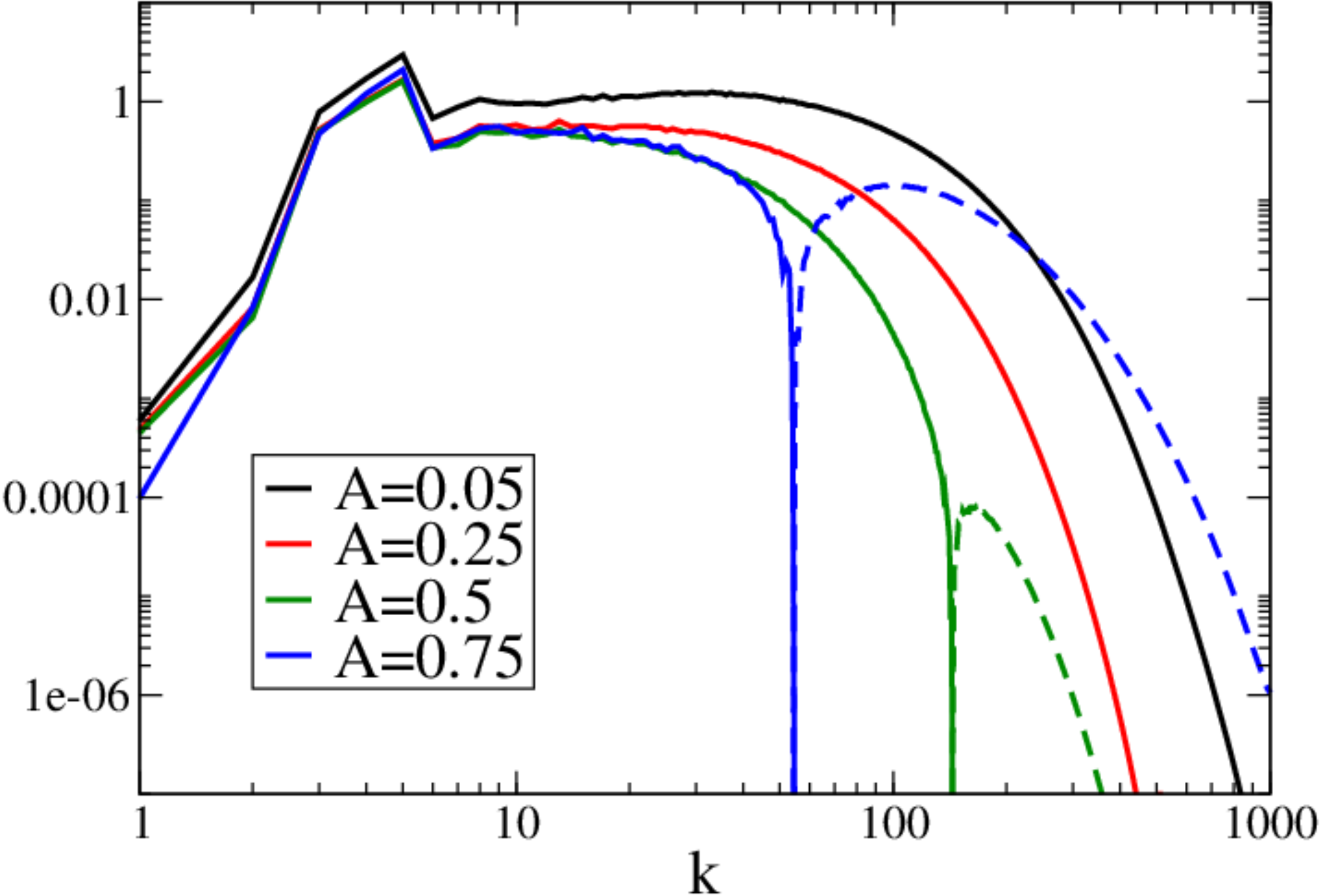}
  \includegraphics[width=6.5cm]{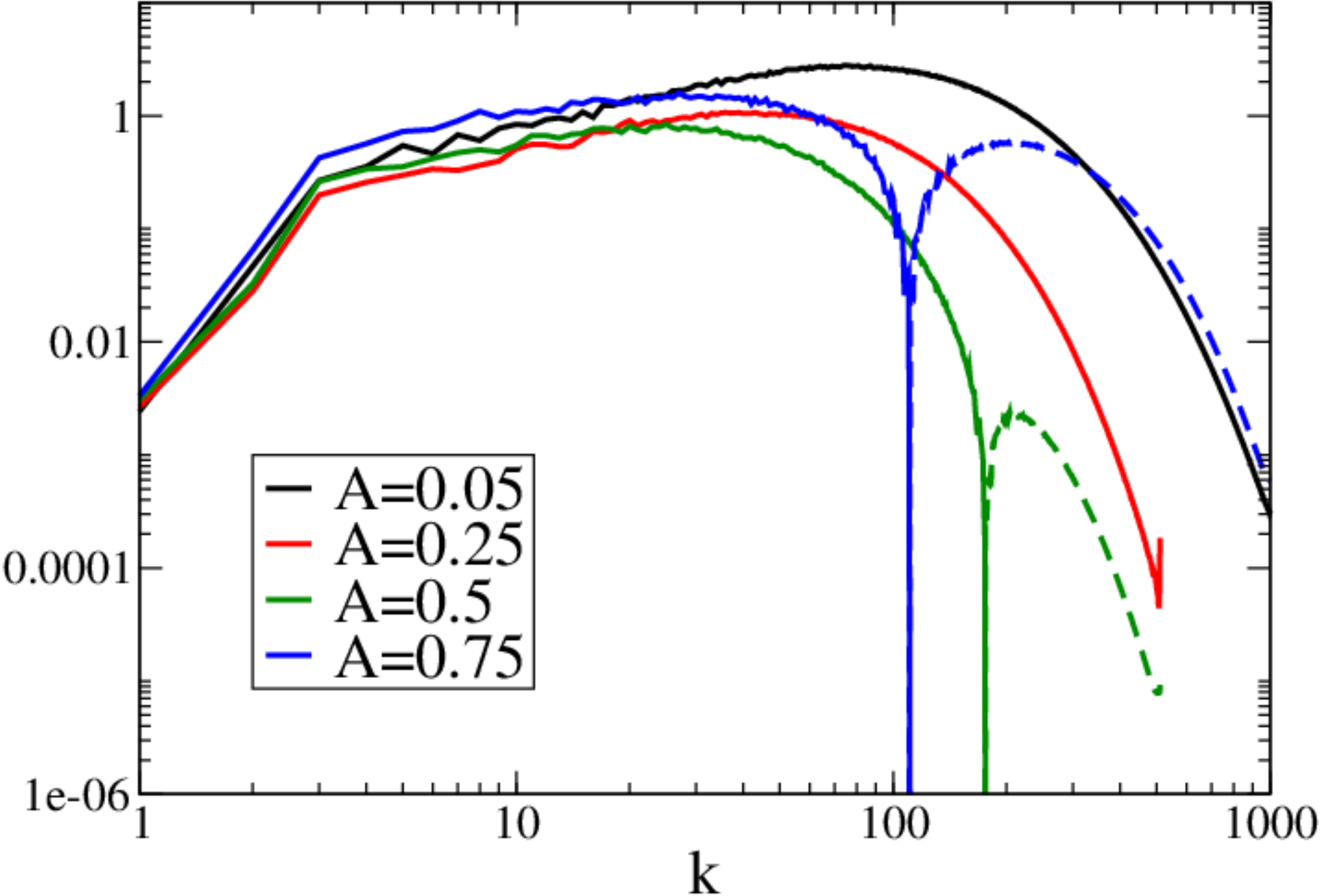}
  \caption{Compensated spectra of the normalized production term for $\mbox{TKE}^{III}$ option ($P_{TKE^{III}}=P_{,1}u_1^{''}\rho^{*}\rho/(4 \rhom A^2)k^{7/3}$) for different \At numbers at (a) $t/t_r=1.15$ and (b) $t/t_r=2.35$. Dashed lines represent negative $P_{TKE^{III}}$.}
\label{Fig:Production_for_opt3}
\end{figure}

For the $\mbox{TKE}^{III}$ definition, the production term, $P_{TKE^{III}}$, is different than the mass flux and is proportional to $P_{,1} \rho u_1^{''}\rho^{*}$. Figure \ref{Fig:Production_for_opt3} presents the spectrum of $P_{TKE^{III}}$, normalized by $4 \rhom A^2$. At small \At numbers, $P_{TKE^{III}}$ spectral shapes are close to those of the mass flux. However, at larger \At numbers, the production term quickly decreases at intermediate scales and even becomes negative at very small scales. This behavior can be related to the weighting with the instantaneous density compared to the mass flux. At large scales, both the heavy and light fluid regions move coherently in the direction of and opposite to gravity, respectively. Thus, $u''$ and $\rho$ have opposite sign and, since the mean pressure gradient is negative, the contributions to the production are positive in both heavy and light fluid regions. However, at early times, the mixing layers between pure fluid regions contain significant inversion regions, where $u''$ and $\rho$ have the same sign, resulting in negative production term at small scales. As the stirring encompasses the whole flow and the pure fluid regions vanish in the later stages of the flow development, the production term becomes positive at all scales (not shown).

\begin{figure}
(\emph{a}) \hspace{6.5cm}  (\emph{b}) \\
  \includegraphics[width=6.5cm]{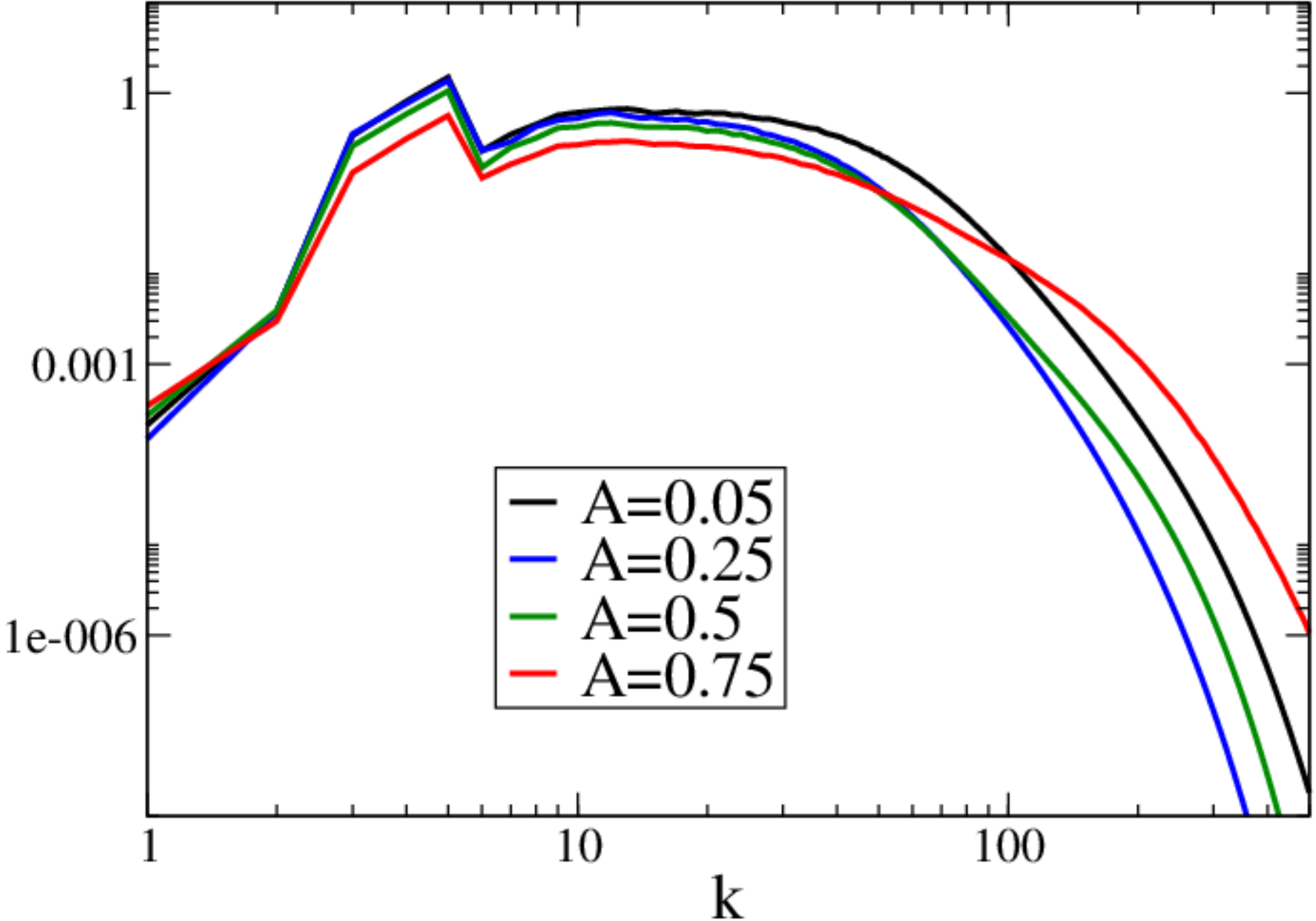}
  \includegraphics[width=6.5cm]{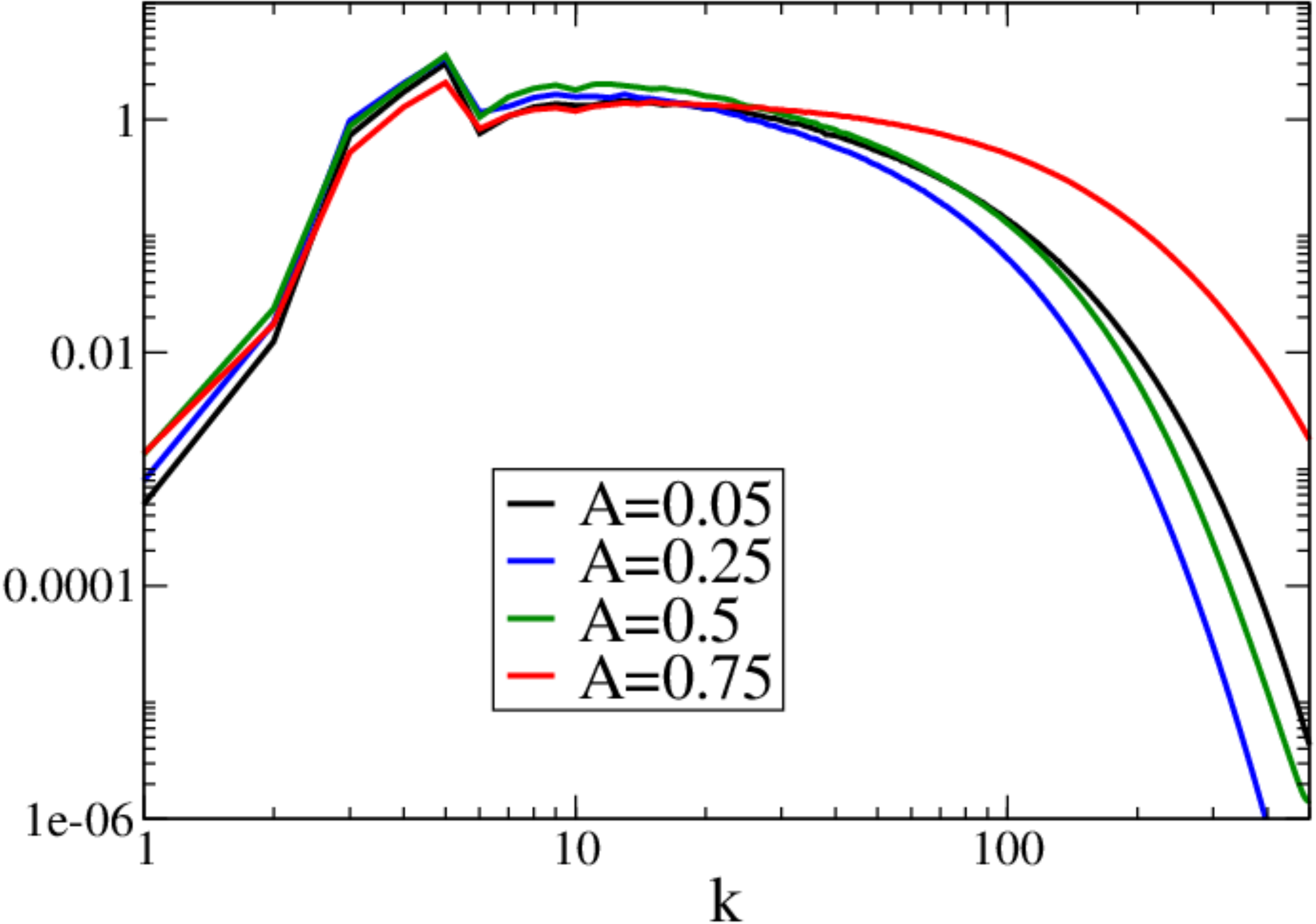}
  \caption{Compensated Energy spectra $E(k)k^{7/3}$ for different \At numbers (a) during explosive growth at $t/t_r=0.65$ and (b) beginning of saturated growth at $t/t_r=1.15$.}
\label{Fig:energy_spectra_1}
\end{figure}

At all \At numbers, during explosive growth, the production term dominates 
the spectral kinetic energy balance, and the turbulent kinetic energy develops a $-7/3$ slope (see figure \ref{Fig:energy_spectra_1}). This slope extends over a larger range for the $A=0.75$ during this regime; however, the lower \At number results again underlie the non-monotonic dependence on \At number.

\begin{figure}
(\emph{a}) \hspace{6.5cm}  (\emph{b}) \\
  \includegraphics[width=6.5cm]{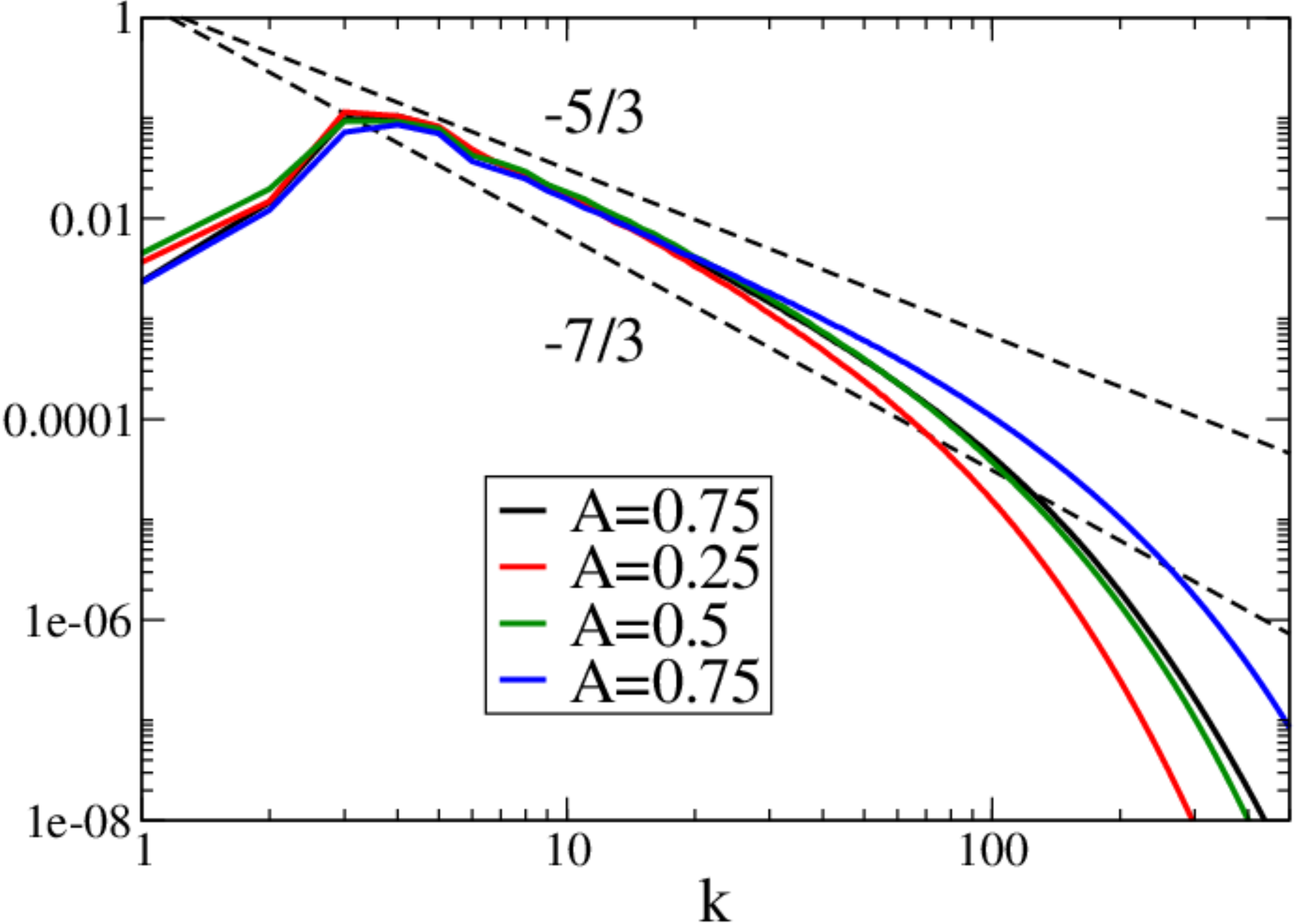}
  \includegraphics[width=6.5cm]{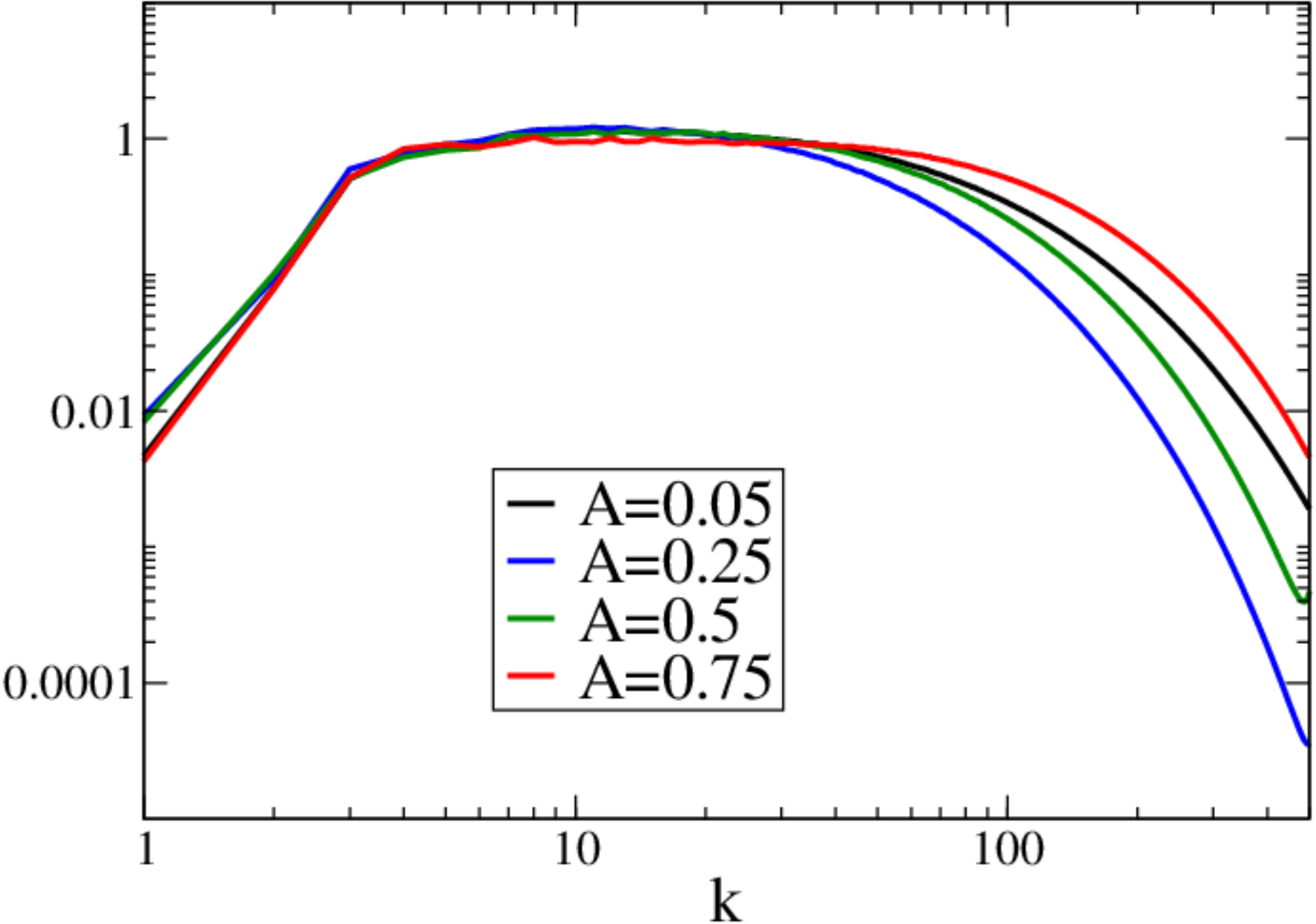}
  \caption{Energy spectra for different \At numbers during (a) saturated growth at $t/t_r=1.75$ and (b) beginning of fast decay at $t/t_r=2.35$ ($E(k)k^{5/3}$).}
\label{Fig:energy_spectra_2}
\end{figure}

\begin{figure}
(\emph{a}) \hspace{6.5cm}  (\emph{b}) \\
  \includegraphics[width=6.5cm]{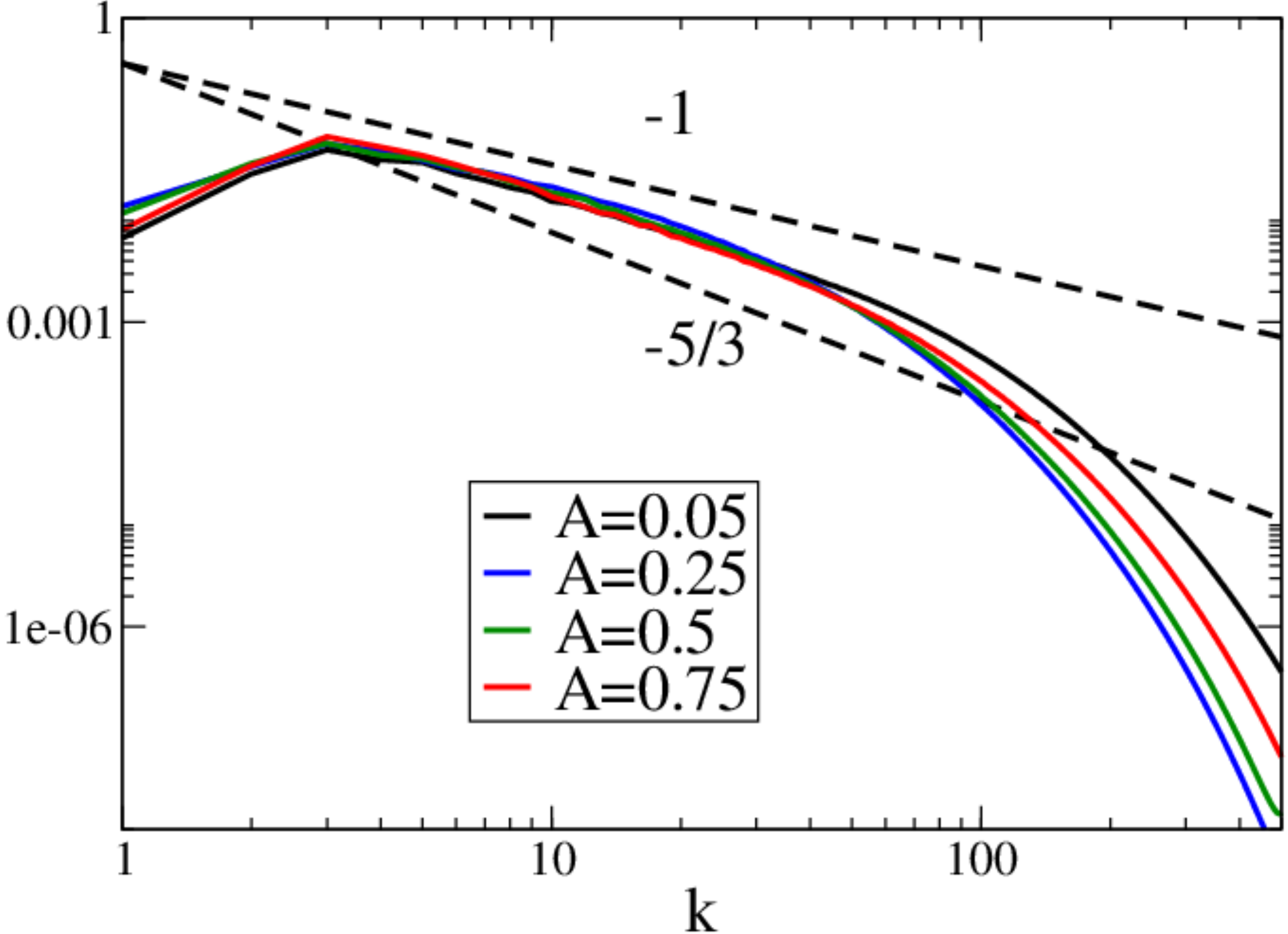}
  \includegraphics[width=6.5cm]{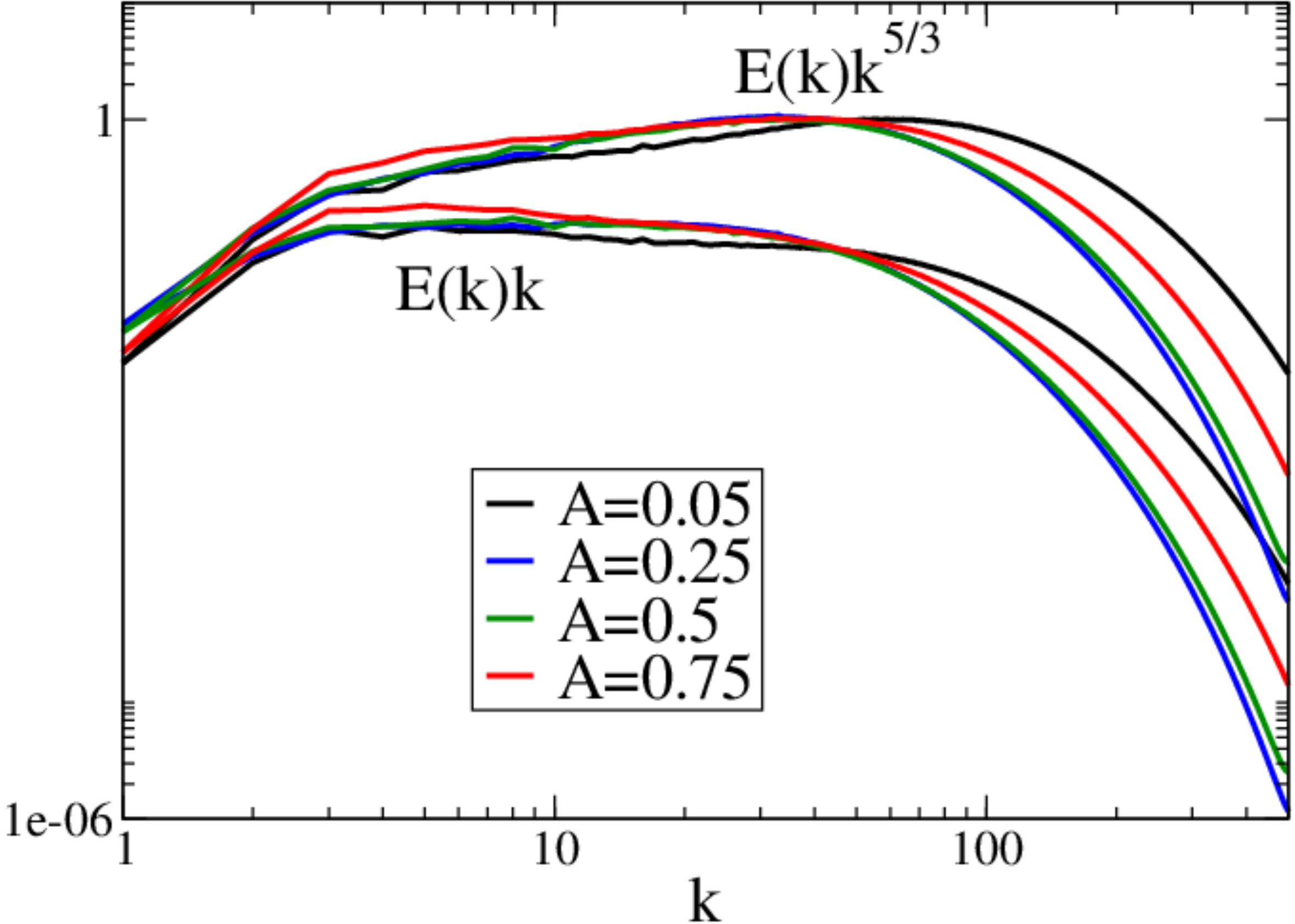}
  \caption{Energy spectra for different \At numbers during fast decay at (a) $t/t_r=2.95$ and (b) $t/t_r=3.35$.}
\label{Fig:energy_spectra_3}
\end{figure}

\begin{figure}
  \centerline{\includegraphics[width=8cm]{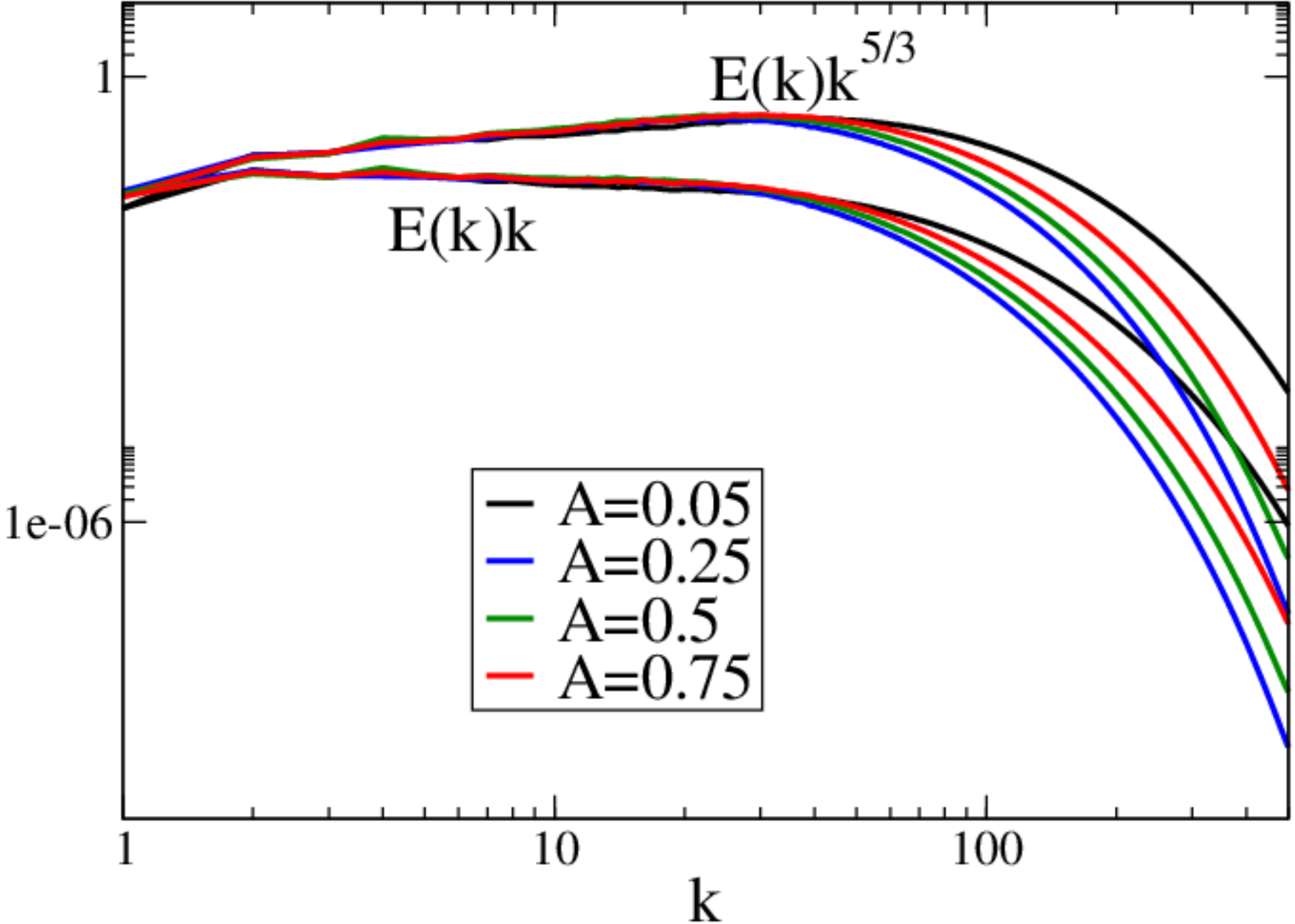}}
  \caption{Energy spectra for different \At numbers during gradual decay at $t/t_r=6$.}
\label{Fig:energy_spectra_4}
\end{figure}

Subsequently, during the saturated growth, due to the enhanced molecular mixing, the production term starts to weaken and the production to dissipation ratio decays to a value of $1$. As the equilibration between turbulent kinetic energy production and dissipation occurs, the kinetic energy spectrum transitions from a $-7/3$ slope at intermediate scales, to a $-5/3$ slope for all \At numbers (see figure \ref{Fig:energy_spectra_2}). 
The fast decay represents a transition period from evolving turbulence to fully-developed turbulence. While at the beginning of this regime the  $-5/3$ slope extends over a large range of scales, this is not the fully developed spectral shape.  Indeed, the upper intermediate scales are still evolving and reach a $-1$ slope at the end of this regime. However, Kolmogorov mechanism dominates the lower intermediate scales of the flow and they retain the $-5/3$ slope (see Figure \ref{Fig:energy_spectra_3}). Furthermore, for the higher-resolution DNS ($1024^3$ and $2048^3$) which reach higher values of \ret and thus a better separation between the energetic and dissipation scales, we observe a longer range of scales that have a slope of $-5/3$ at the end of the fast decay regime.
During gradual decay, the spectral shape remains very similar as seen in Fig. \ref{Fig:energy_spectra_4}, indicating that the flow has reached its fully developed state. Similar results are reported by \citet{batchelor1992} for their energy spectra of Boussinesq case as well.

 Several studies of buoyancy driven turbulence have reported flatter slopes than $-5/3$ at intermediate scales. For example, \citet{gat_2017}  reported a $-4/3$ slope in their study of a buoyancy driven shear flow. They attributed this flatter slope to the turbulence intermittency; however, we believe that the presence of buoyancy causes the emergence of two spectral ranges at intermediate scales, with a $-5/3$ slope at smaller intermediate scales and a flatter slope at larger scales due to the assistance of buoyancy production. 
 
\subsection{Spectral anisotropy} \label{Sec:Spectral_ani}

The Reynolds stress anisotropy tensor ($B_{ij}$) is defined here consistent with the turbulent kinetic energy expression as: 
\begin{equation} \label{Eq:anisotropy}
    B_{ij}=\frac{\langle(\rho^* u^{''}_i)(\rho^* u^{''}_j)\rangle}{\langle(\rho^* u^{''}_i)(\rho^* u^{''}_i)\rangle}.
\end{equation} 
For isotropic turbulence, $B_{11}=B_{22}=B_{33}=1/3$, while the maximum value of each component is 1. To investigate the scale dependence of the normal stresses anisotropy, the definition above is extended using the kinetic energy spectrum in each direction. Due to the presence of energy production, it is expected that the large scales remain anisotropic. This influence extends deep into the intermediate scales \citep{Soulard12}. However, at sufficiently large \ret, the spectrum should become broad enough that the influence of energy production vanishes at high enough wavenumbers, so that a classical inertial range may emerge. However, under strongly non-equilibrium conditions, this picture might not hold. In particular, it is possible that the dissipation range may become anisotropic even when intermediate scales are isotropic.  Thus, \citet{livescu2008} showed a direct connection between the large and dissipation scales during the growth of HVDT at small to moderate \At. This finding was later confirmed in a stationary but inhomogeneous version of the flow with a non-zero mean density gradient \citep{chung_pullin_2010}.  Here, we explore the Atwood and Reynolds numbers effects on this anisotropic behavior of intermediate and viscous ranges during the growth and decay regimes at higher Reynolds numbers and for a larger range of \At numbers than \citet{livescu2008}.

\begin{figure}
(\emph{a}) \hspace{6.5cm}  (\emph{b}) \\
    \includegraphics[width=6.5cm]{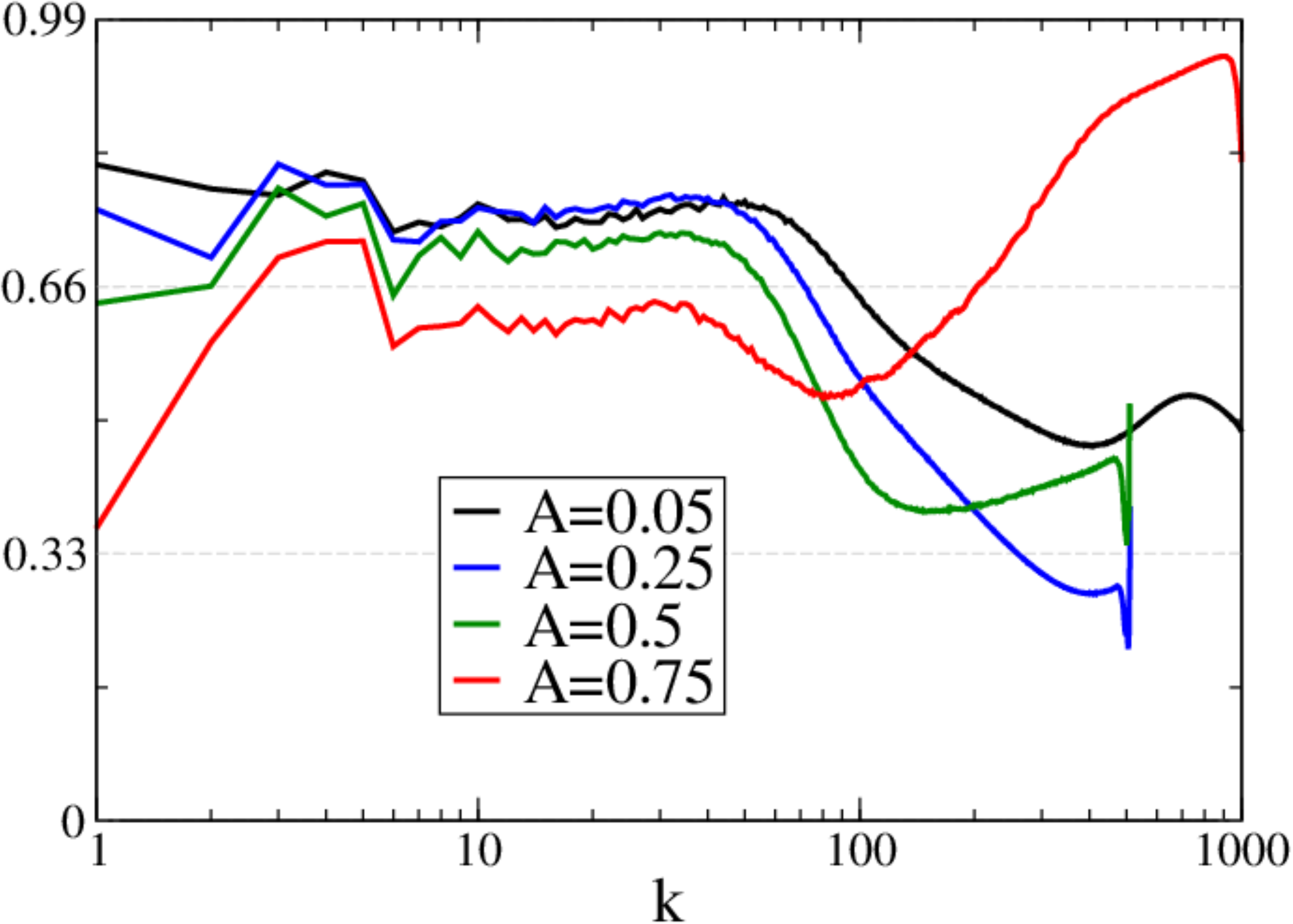}
    \includegraphics[width=6.5cm]{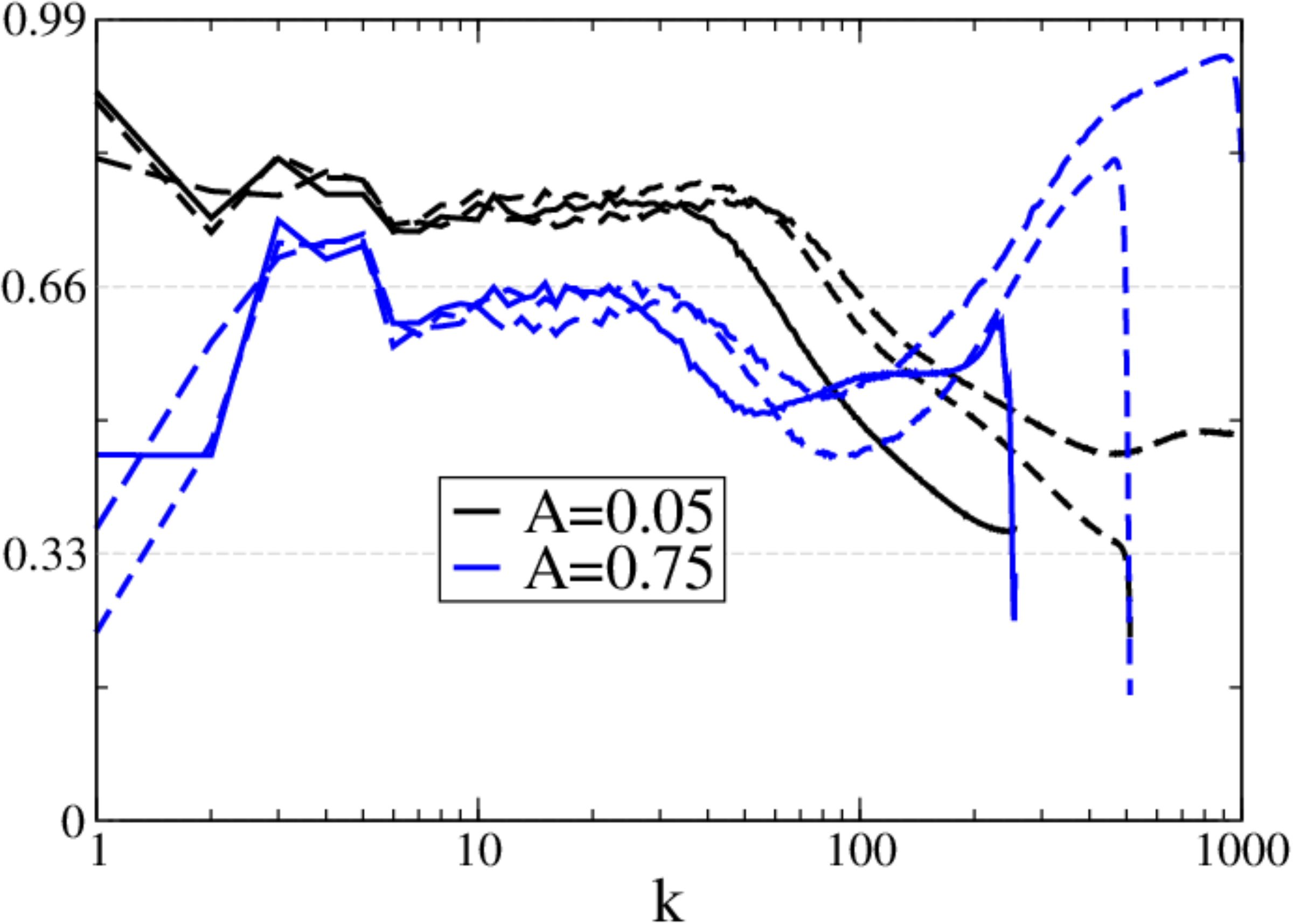}
 \caption{Vertical component of the anisotropy tensor as a function of the wavenumber for different (a) \At and (b) \rez numbers during explosive growth at $t/t_r=0.65$. In (b), longer spectrum ranges correspond to higher resolution and higher Reynolds number cases.}
\label{Fig:ani_spectra_1}
\end{figure}

Figure \ref{Fig:ani_spectra_1} shows the vertical component of the anisotropy tensor for different \At and \rez numbers as a function of the wave number during explosive growth. Due to the 
much larger $E_{TKE}$ production than $E_{TKE}$ dissipation, the kinetic energy spectrum is dominated by the production term and the flow is anisotropic at all scales. Up to about $k\sim60$, the vertical component has as much energy as the horizontal components combined, similar with the overall anisotropy of the flow. While the normal stresses anisotropy decreases for larger wavenumbers, it remains significant in the viscous range, especially at $A=0.75$, where it reaches higher levels than at large scales. Increasing \rez has a relatively minor influence on the normal stresses anisotropy, and the viscous range anisotropy extends to the smallest scales of the flow. 

Furthermore, compared to the cases with lower \At numbers, cases with larger \At numbers have lower anisotropy values at large scales. This observation is consistent with the $E_{TKE}$ conversion rates which decrease upon increase in \At number. $E_{TKE}$ production over dissipation ratio is lower for the higher \At number cases, which reduces the anisotropy levels at intermediate scales. 
        
\begin{figure}
(\emph{a}) \hspace{6.5cm}  (\emph{b}) \\
    \includegraphics[width=6.5cm]{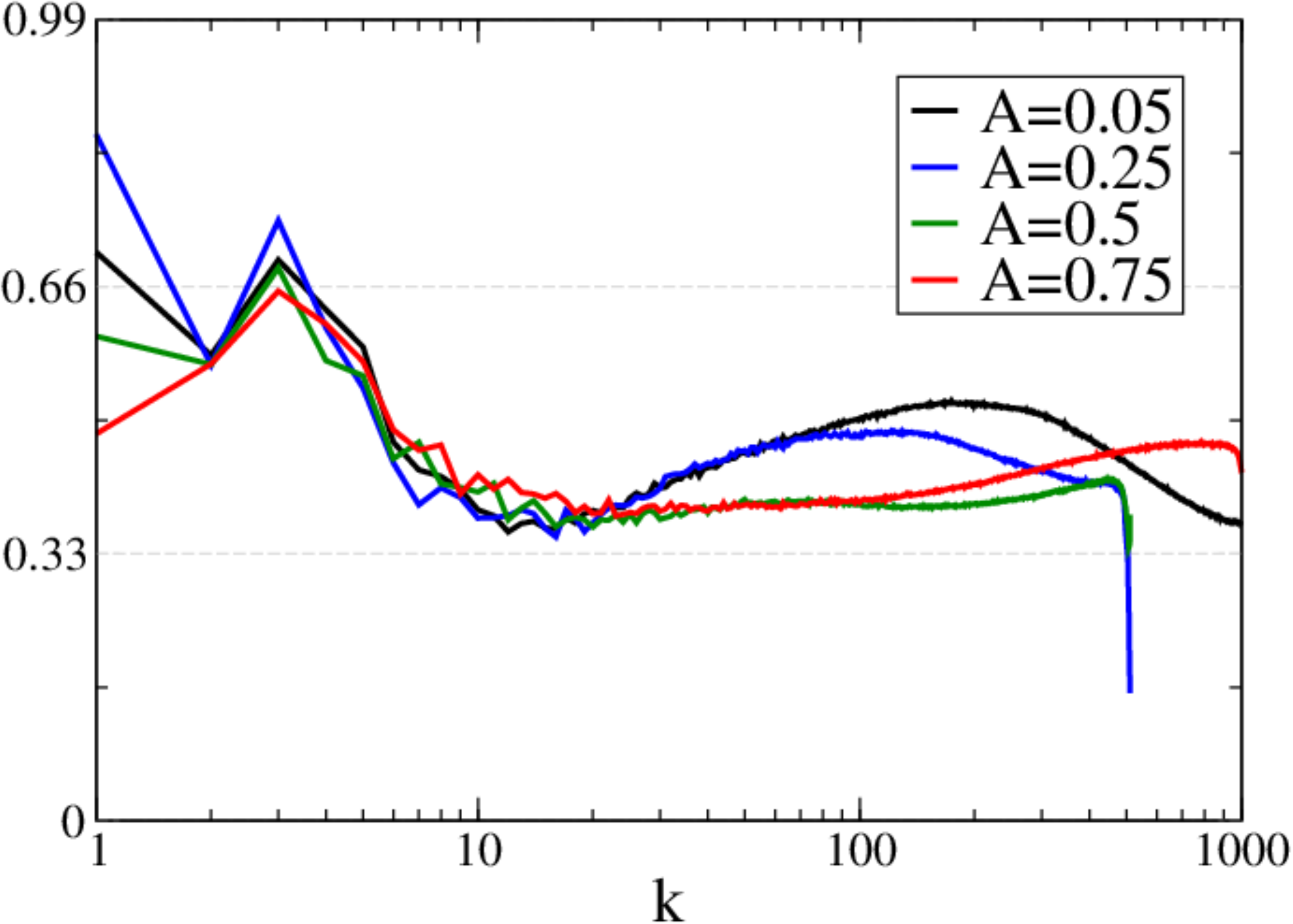}
    \includegraphics[width=6.5cm]{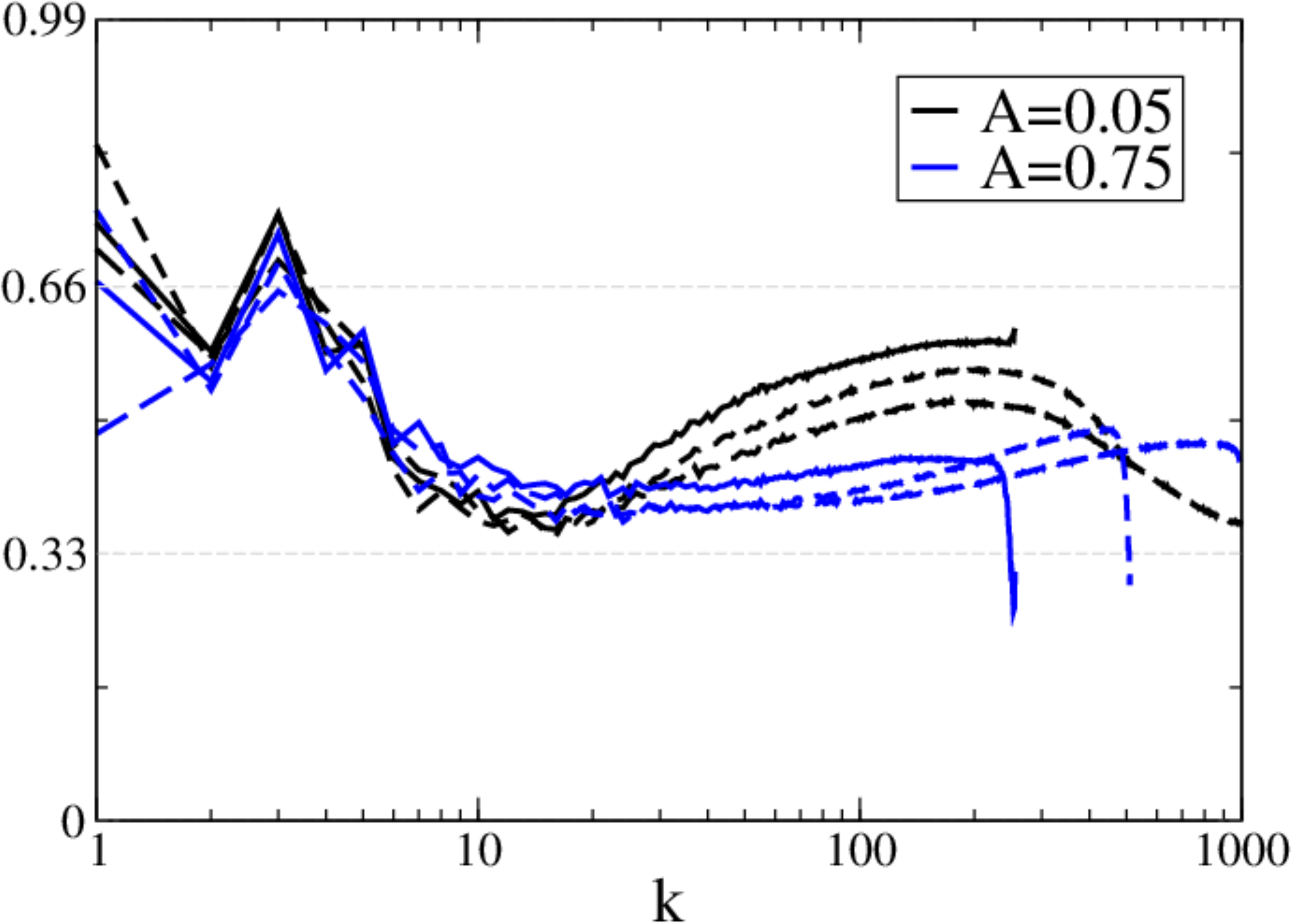}
 \caption{Vertical component of the anisotropy tensor as a function of the wavenumber for different (a) \At and (b) \ret numbers during saturated growth at $t/t_r=1.75$.}
\label{Fig:ani_spectra_2}
\end{figure}

As the flow develops during the saturated growth and $P/D$ ratio decays, the non-linear effects slowly start to dominate at intermediate scales, which become isotropic over all \At and \ret number cases reported in this paper (figure \ref{Fig:ani_spectra_2}). However, there remains a persistent anisotropy in the viscous range, consistent with earlier studies  \citep{livescu2008}. This anisotropy is higher at small \At numbers, underlying the complex relation between stirring and mixing, as described throughout this paper. There seems to be a faster decrease of the anisotropy with the Reynolds number figure \ref{Fig:ani_spectra_2}b), as the early times stronger stirring leads to an accelerated increase in molecular mixing. The classical RTI problem or similar configurations cases where $g>0$ are the most common examples of $P\geq D$ flows that exhibit viscous range anisotropy. Such results are also reported for the classical RTI problem \citep{Livescu_Cabot_Cook} from the DNS data of \citet{Cabot_Cook}.

During decay regimes, production is smaller than the total dissipation and the flow reaches its fully developed structure. consistently, the normal stresses anisotropy extends over the whole small scale range (not shown).

\subsection{Density Spectra}\label{Sec:Spectral_dens_1}
\begin{figure}
(\emph{a}) \hspace{6.5cm}  (\emph{b}) \\
 \centerline{ \includegraphics[width=6.5cm]{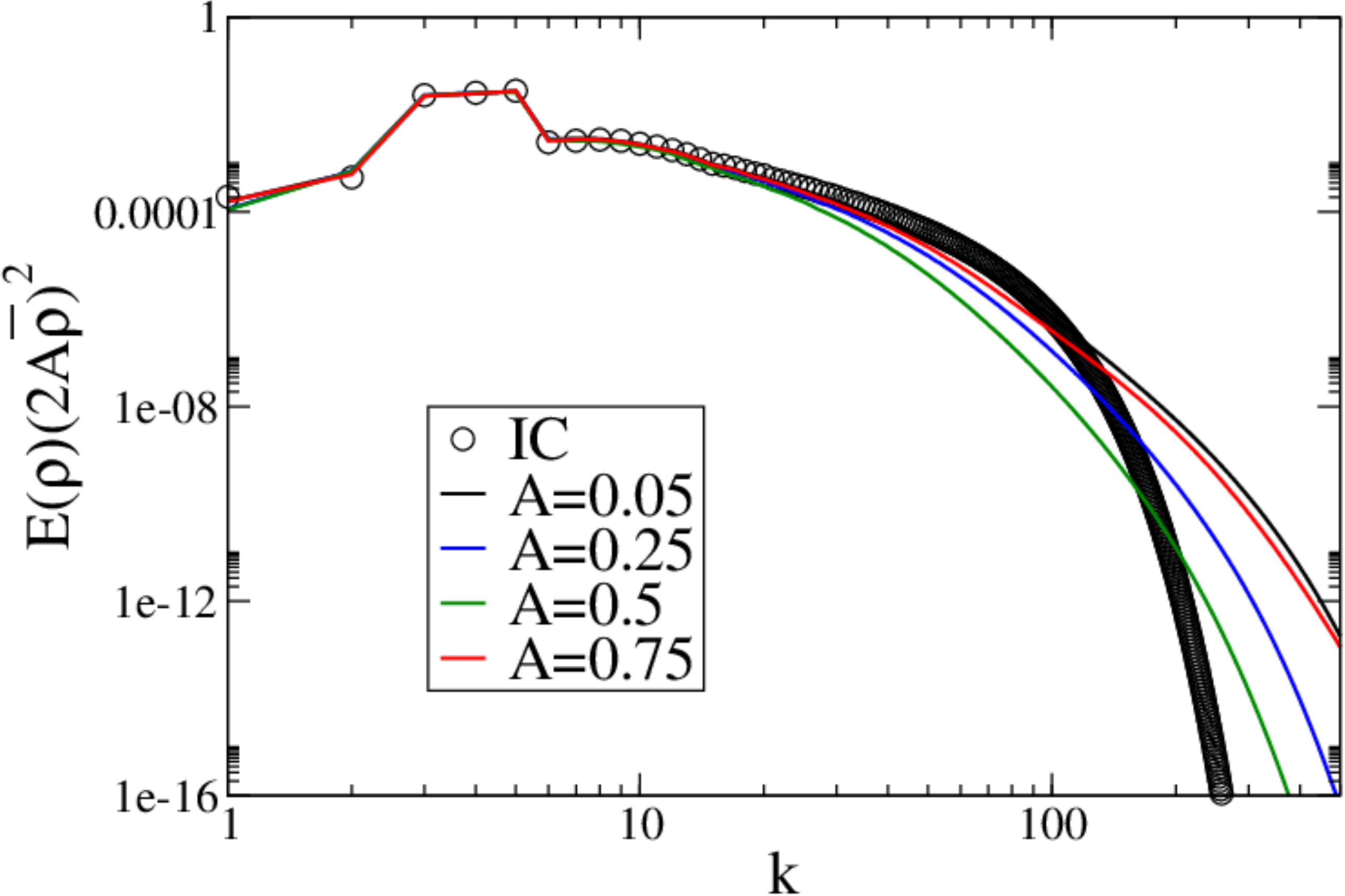}
              \includegraphics[width=6.5cm]{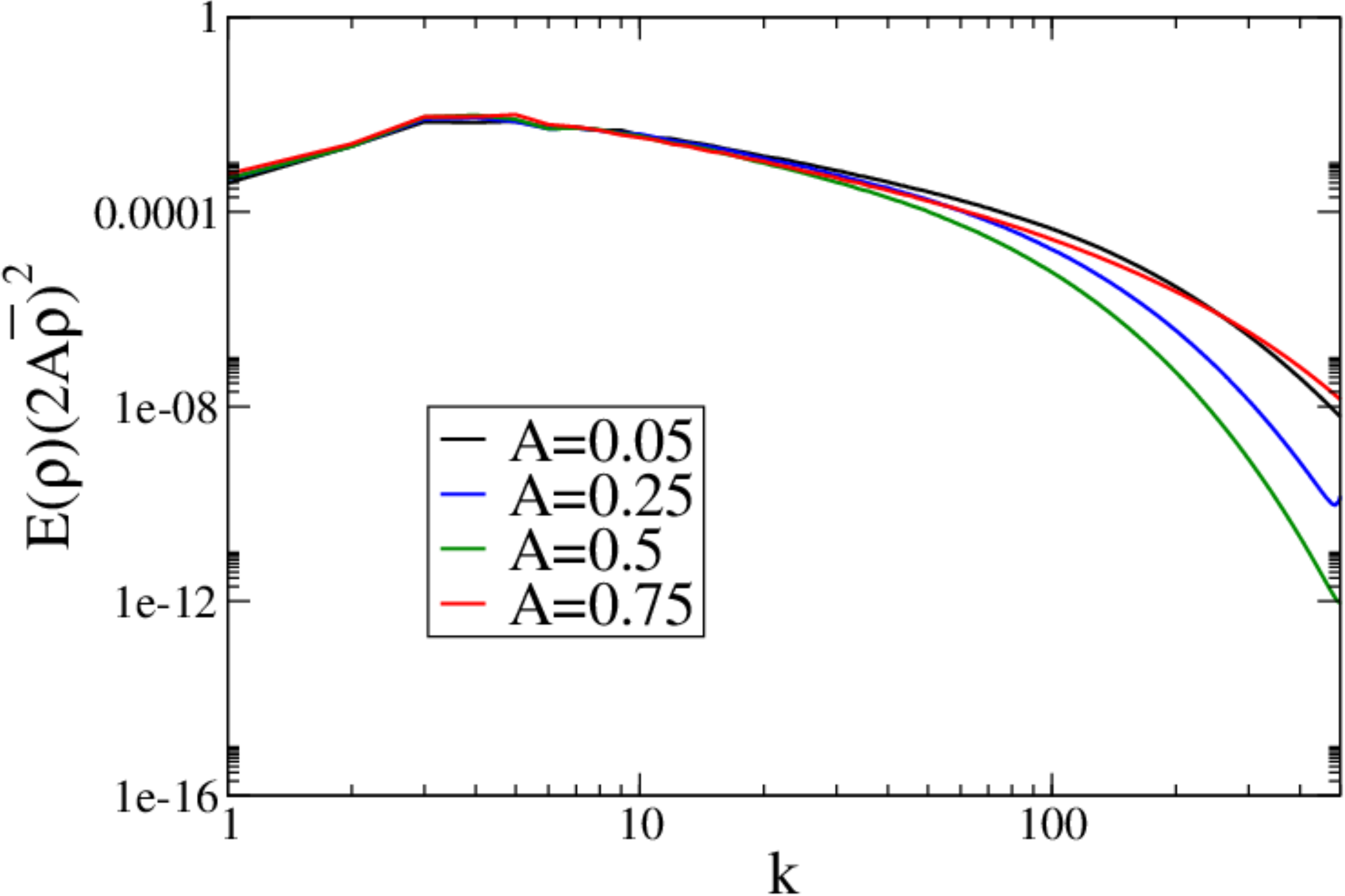}}\\
(\emph{c}) \hspace{6.3cm} (\emph{d})\\
\centerline{\includegraphics[width=6.5cm]{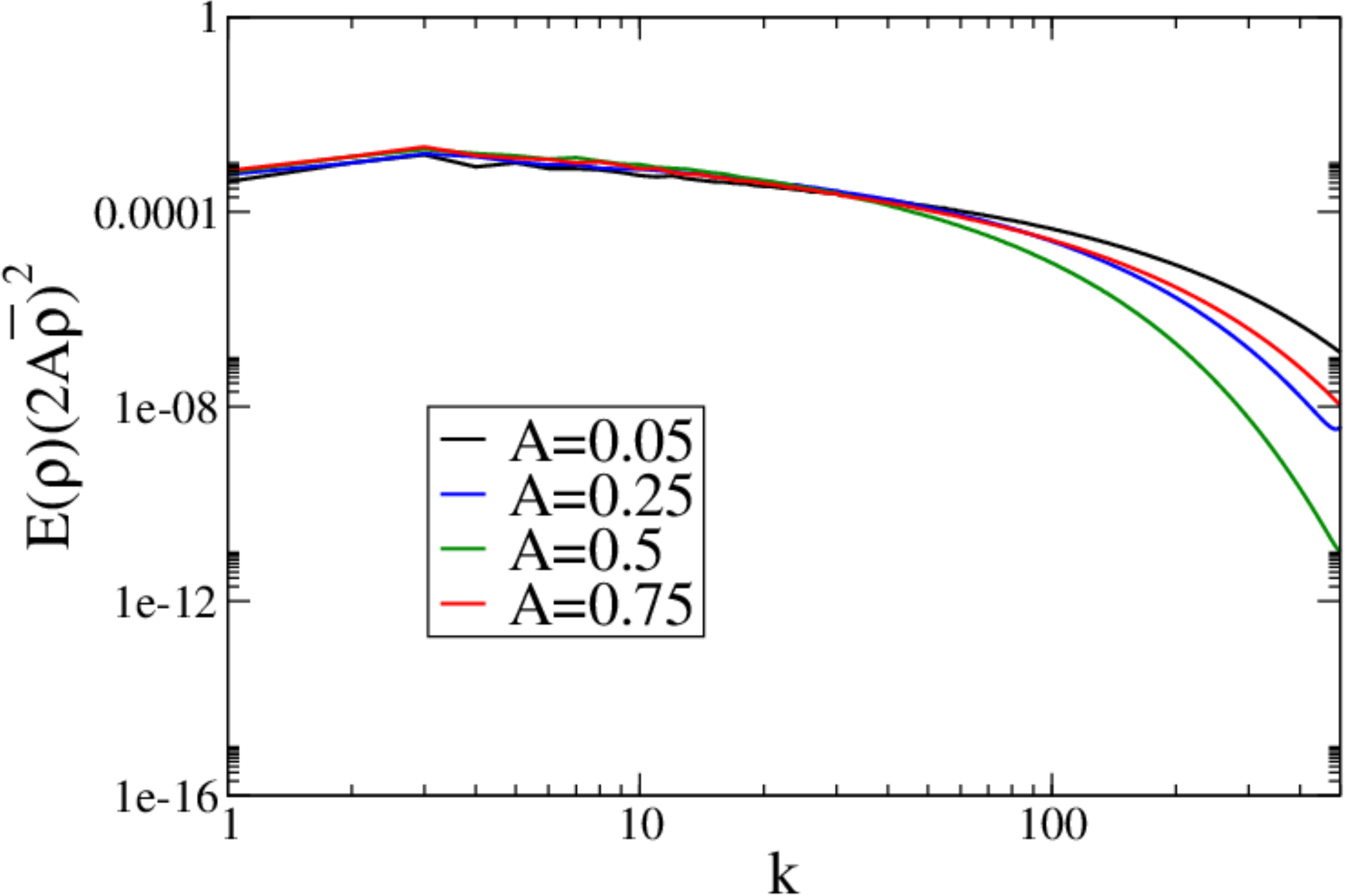}
            \includegraphics[width=6.5cm]{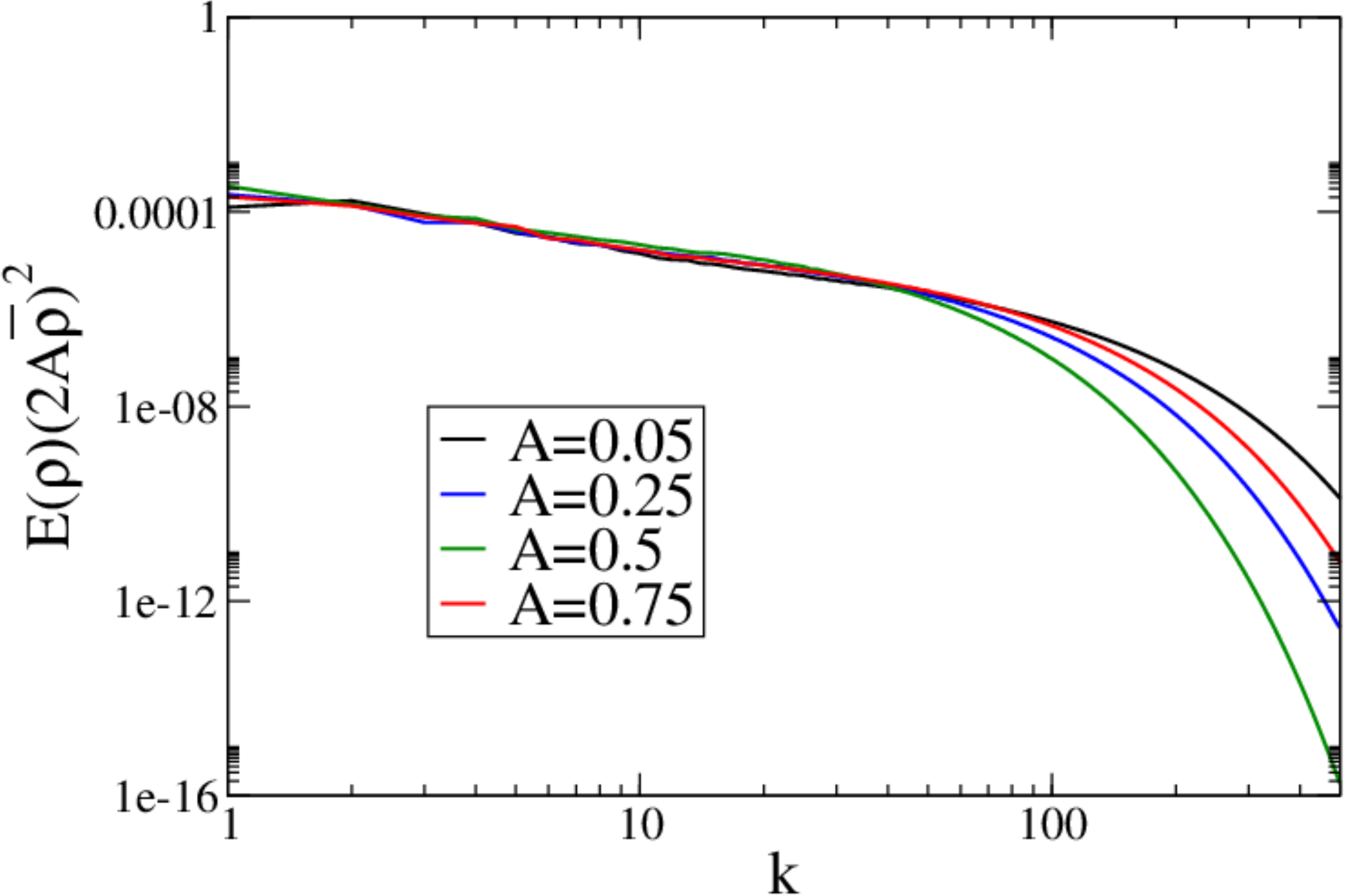}}
\caption{Normalized density energy spectra at (a) $t/t_r=0$ (IC: Initial condition) and $t/t_r=0.65$ (explosive growth), (b) $t/t_r=1.75$ (saturated growth), (c) $t/t_r=2.95$ (fast decay), and (d) $t/t_r=6$ (gradual decay) for different \At numbers.}
\label{Fig:dens_spectra_1}
\end{figure}

    The density energy spectra normalized by by $(A\rhom)^2$, $E_{\rho}(k)/(A \rhom)^2$, are shown in Fig. \ref{Fig:dens_spectra_1} for different \At numbers. In this normalization, the density spectra collapse at large scales for all \At numbers. During the explosive growth, the inertia associated with stirring is not sufficient to disturb the large scales of the flow as observed in Fig. \ref{Fig:dens_spectra_1}a. The initial shapes of the large structures are preserved for all \At numbers, as stirring and molecular mixing are mostly localized at the interface between light and heavy fluid regions. This observation is also consistent with the evolution of the mole fraction shown in Fig. \ref{Fig:3Devolve_T1}. As the flow transitions to saturated growth, the top hat shape of the spectrum flattens out due to enhanced mixing (see Figure \ref{Fig:dens_spectra_1}b). The growth stages are followed by a fast decay where the top hat shape of the density spectrum is removed and the spectra tend to become identical for all cases (see Figure \ref{Fig:dens_spectra_1}c). During late time gradual decay, the slope of the density spectrum at large scales was observed to follow a $k^{-1}$ power law; similar values were first reported by \citet{Passivescalar1doi:10.1063/1.857365} for passive scalar mixing and latter  by \citet{batchelor1992} for HVDT under Boussinesq assumption. 
    
    Even though different mixing rates are observed within different regions of the flow during both saturated growth and fast decay regimes, the density spectra are not able to capture this asymmetric behavior; the evolution of the density spectrum is not affected by the asymmetric behavior of the density PDF.

\section{Conclusions}\label{Sec:Conclusions}

     To better understand the variable-density effects, the highly non-equilibrium nature of HVDT is investigated by dividing the flow evolution into four different regimes based on the time derivatives of $E_{TKE}$. Table \ref{Table:regimes} presents the $E_{TKE}$ behavior during these four regimes. HVDT dynamics is also studied in detail by investigating the \At and \ret numbers effects during these four regimes; the unique characteristics that are observed are summarized below.
     
 Most of the variable density effects can be connected to different inertia of the light and heavy fluid regions, which makes them respond differently to changes in local strain. The inertia, in turn, changes the local stirring and subsequently, the molecular mixing. Overall, the light fluid regions not only develop more intense turbulence faster, but also mix faster. This complex asymmetric behavior leads to non-monotonic flow development with respect to the Atwood number, as highlighted below.

\begin{table}
  \begin{center}
\def~{\hphantom{0}}
\setlength{\tabcolsep}{18pt}
  \begin{tabular}{cc}
      \textbf{Explosive growth}   		& $\frac{\displaystyle dE_{TKE}}{\displaystyle dt}>0$ and $\frac{\displaystyle d^2E_{TKE}}{\displaystyle dt^2}>0$ \\
      \textbf{Saturated growth}          & $\frac{\displaystyle dE_{TKE}}{\displaystyle dt}>0$ and $\frac{\displaystyle d^2E_{TKE}}{\displaystyle dt^2}<0$ \\
      \textbf{Fast decay}                & $\frac{\displaystyle dE_{TKE}}{\displaystyle dt}<0$ and $\frac{\displaystyle d^2E_{TKE}}{\displaystyle dt^2}<0$ \\
      \textbf{Gradual decay}	            & $\frac{\displaystyle dE_{TKE}}{\displaystyle dt}<0$ and $\frac{\displaystyle d^2E_{TKE}}{\displaystyle dt^2}>0$ \\
  \end{tabular}
  \caption{$E_{TKE}$ behaviour during different flow regimes.}
  \label{Table:regimes}
  \end{center}
\end{table}

\subsection{Energy conversion rates}

It is found that energy conversion rates strongly depend on density ratio of the mixing fluids. Cases with lower \At numbers are more efficient to generate $E_{TKE}$; however, all cases investigated are very efficient ($\beta_{KE}>0.9$) to generate $E_{KE}$ during explosive growth. During saturated growth, the \At number has a non-monotonic influence on the energy conversion ratio. Both $E_{KE}$ and $E_{TKE}$ conversion ratios peak at moderate \At number values between 0.25 to 0.5.

\subsection{Asymmetric behavior at high \At number}

Mixing asymmetries that develop in buoyancy-driven HVDT were examined at high \At and \ret numbers. The velocity behavior for high \At number is different than cases with lower \At numbers; different mix rates are observed in flow regions with different density levels.
Due to smaller inertia, lighter fluid regions respond faster to changes in direction associated with stirring than heavier fluid regions. In effect, the turbulent velocity fluctuations are larger in these regions. At $A\geq 0.5$, these asymmetries become noticeable, as observed in the jMDF of density and velocity fields. Thus, the shape of the PDF is significantly wider at lighter fluid regions than heavier fluid regions. The dependency on the Atwood number changes during the flow evolution. The velocity field starts to evolve asymmetrically for high \At number cases, and it tends to lose the \At number dependence and become symmetric before turbulence decay starts. Density PDF also becomes asymmetric during saturated growth; however, it tends to become symmetric again during gradual decay for the high \At number cases.

\subsection{Conditional expectations}

Conditional expectations of $E_{TKE}$, $E_{TKE}$ dissipation, and enstrophy were studied to explore the effects of asymmetric behavior and to study different dynamics of lighter and heavier fluid regions. For higher \At numbers, it is shown that all conditional expectations studied here are skewed to the lighter fluid side, indicating higher $E_{TKE}$, dissipation, and enstrophy levels within lighter regions. However, again, the dependency on the Atwood number evolves differently for each of the quantities examined. In particular, conditional expectation of the dissipation becomes \At number independent noticeably earlier than the conditional expectation of enstrophy.

\subsection{Extreme events in HVDT}

The local flow statistics were found to exhibit much larger values than their whole volume averages during both growth regimes. Increasing \ret and/or \At numbers increases the probability of the presence of such large deviations from the mean within the flow. 

\subsection{Mixing transition in HVDT}

Our results also demonstrate that reaching a high \ret numbers may not be sufficient condition for achieving mixing transition. The flow has to spend a sufficient time at high \ret numbers to become fully-developed and undergo a mixing transition. As a result, different metrics attain the mixing transition threshold at different times; the velocity field becomes fully-developed substantially earlier than the density field. For example, the velocity PDF reaches its asymptotic behavior faster than density PDF and conditional expectations. Moreover, for higher \At number cases, mixing transition occurs at different times for the different regions of the flow. Thus, mixing transition occurs noticeably earlier within lighter fluid regions due to more intense turbulence compared to the heavier fluid regions. In terms of the flow regimes, for the cases with the largest $Re_t$ and low $A$, all investigated quantities reach their fully developed stage during the fast decay everywhere within the flow. However, for the cases with the largest $Re_t$ and high $A$, most of the investigated quantities start to become fully developed in lighter fluid regions during the saturated growth, whereas for the heavier regions it may take until the end of fast decay to become fully developed.

\subsection{HVDT spectral evolution}

\subsubsection{Energy spectra}
The dynamic characteristics of HVDT are also reflected in the evolution of the energy spectrum. It is found that the different HVDT regimes have distinctly different spectral behavior. The discussion follows using the energy spectrum calculated based on the Favre momentum ($\rho^*u^{''}_i / \sqrt{ \rhom}$). While the differences are small for the cases considered here, other alternatives such as ($\sqrt{\rhom}u^{''}_i$) or ($\sqrt{\rho^*}u^{''}_i$) cannot be shown to have vanishing viscous effects at very large scales even for constant viscosity, when the \At number is large enough. During explosive growth, when production dominates the intermediate scales, the slope of the energy spectrum is steeper than the classical Kolmogorov spectrum and was found to be close to $k^{-7/3}$ power law. During saturated growth, as the energy production over dissipation ratio decreases, the slope of the intermediate scales increases from $-7/3$ to $-5/3$. At the end of the growth regimes, production is equal to dissipation and energy spectrum is similar to the classical Kolmogorov spectrum. During fast decay, as the energy production continues to weaken, its effects are restricted to upper-intermediate scales. The scales becomes shallower than a $k^{-5/3}$ power law. Eventually, during gradual decay, energy spectra remain similar to the end of fast decay regime. Upper intermediate scales, which are affected by energy production, show a $k^{-1}$ scaling. At lower intermediate scales, a $k^{-5/3}$ power law spectrum is recovered for the highest \rez examined.

\subsubsection{Spectral anisotropy}
During explosive growth, the normal stresses remain anisotropic at all scales (including inertial and viscous ranges) for all \ret and \At number cases investigated in this work.
During saturated growth, the inertial range isotropy emerge; however, the viscous range continues to be anisotropic.
During fast decay, as energy production is lower than dissipation, both inertial and viscous ranges become isotropic.
\ret number has limited effect on spectral anisotropy during growth regimes. However, higher \ret numbers cause faster isotropy emergence at inertial and small scales during the fast decay regime.

\subsubsection{Density spectra}
The density power spectrum normalized by $(A \rhom)^2$ collapses at large scales for all cases. Since the flow starts from rest, with large heavy and light fluid regions, the imprint of these regions on density spectra is relatively slow to vanish. During the fast decay regime, the density spectrum acquires is asymptotic shape and no longer changes as the flow transitions to gradual decay. During this regime, the density spectrum follows a $k^{-1}$ power law for scales larger than the diffusive scales.
\smallskip

The triply-periodic flow configuration removes edge and wall effects and thus allows to focus on the physics of VD mixing. Compared to classical RTI, HVDT is computationally more efficient and reaches higher Reynolds numbers during flow evolution. HVDT thus allows for exploring VD mixing in regimes further beyond mixing transition. As illustrated, HVDT possesses unique physics that are different from single fluid turbulence due to different inertia of the light and heavy fluid regions. While some of the consequences of this asymmetry, as reflected in joint velocity-density MDFs and conditional expectations, might be obscured in the behavior of global quantities and energy spectra, they are nevertheless important for the evolution of the flow. Modeling variable-density turbulent flows poses unique challenges in any type of modeling strategy. It is our hope that the higher Reynolds and Atwood number runs presented here will help in extending turbulence models to capture turbulent flows with large density variations.
   
 \section{Acknowledgement}\label{Acknowledgement}
 
Arindam Banerjee acknowledge financial support from DOE/NNSA SSAA Program (Grant No. DE-NA0003195) and the U.S. National Science Foundation Early CAREER Program (Grant No. 1453056 from CBET-Fluid Dynamics). This work is co-authored by an employee of Triad National Security, LLC which operates Los Alamos National Laboratory under Contract No. 89233218CNA000001 with the U.S. Department of Energy/National Nuclear Security Administration. Computational resources were provided by the Institutional Computing  Program at Los Alamos National Laboratory and the Argonne Leadership Computing Facility at Argonne National Laboratory through a 2017 ALCC Award.

\textbf{Declaration of interests.} The authors report no conflict of interest.
 
\bibliographystyle{jfm}
\bibliography{jfm_rapid}

\begin{thebibliography}{79}
\expandafter\ifx\csname natexlab\endcsname\relax\def\natexlab#1{#1}\fi
\def\au#1{#1} \def\ed#1{#1} \def\yr#1{#1}\def\at#1{#1}\def\jt#1{\textit{#1}}
  \def\bt#1{#1}\def\bvol#1{\textbf{#1}} \def\vol#1{#1} \def\pg#1{#1}
  \def\publ#1{#1}\def\arxiv#1{#1}\def\org#1{#1}\def\st#1{\textit{#1}}

\bibitem[Adkins {\em et~al.\/}(2002)Adkins, McIntyre \& Schrag]{Adkins1769}
{\sc \au{Adkins, J.~F.}, \au{McIntyre, K.} \& \au{Schrag, D.~P.}} \yr{2002}
  \at{The salinity, temperature, and {$\delta^{18}$O} of the glacial deep
  ocean}.  \jt{Science}  \bvol{298}~(5599),  \pg{1769--1773}.

\bibitem[Ahlers {\em et~al.\/}(2009)Ahlers, Grossmann \&
  Lohse]{RBI_RevModPhys.81.503}
{\sc \au{Ahlers, G.}, \au{Grossmann, S.} \& \au{Lohse, D.}} \yr{2009}  \at{Heat
  transfer and large scale dynamics in turbulent {R}ayleigh-{B}\'enard
  convection}.  \jt{Rev. Mod. Phys.}  \bvol{81},  \pg{503--537}.

\bibitem[Akula \& Ranjan(2016)]{akula_ranjan_2016}
{\sc \au{Akula, B.} \& \au{Ranjan, D.}} \yr{2016}  \at{Dynamics of
  buoyancy-driven flows at moderately high atwood numbers}.  \jt{J. Fluid
  Mech.}  \bvol{795},  \pg{313--355}.

\bibitem[Almagro {\em et~al.\/}(2017)Almagro, Garc\'ia-Villalba \&
  Flores]{almagro_garcia-villalba_flores_2017}
{\sc \au{Almagro, A.}, \au{Garc\'ia-Villalba, M.} \& \au{Flores, O.}} \yr{2017}
   \at{A numerical study of a variable-density low-speed turbulent mixing
  layer}.  \jt{J. Fluid Mech.}  \bvol{830},  \pg{569 -- 601}.

\bibitem[Aslangil {\em et~al.\/}(2016)Aslangil, Banerjee \& Lawrie]{Denis_PRE}
{\sc \au{Aslangil, D.}, \au{Banerjee, A.} \& \au{Lawrie, A. G.~W.}} \yr{2016}
  \at{Numerical investigation of initial condition effects on {Rayleigh-Taylor}
  instability with acceleration reversals}.  \jt{Phys. Rev. E}  \bvol{94},
  \pg{053114}.

\bibitem[Aslangil {\em et~al.\/}(2019)Aslangil, Livescu \&
  Banerjee]{aslangil_book_ch}
{\sc \au{Aslangil, D.}, \au{Livescu, D.} \& \au{Banerjee, A.}} \yr{2019} Flow
  regimes in buoyancy-driven homogeneous variable-density turbulence.  \bt{In
  {\em Progress in Turbulence VIII\/} (ed. \ed{Ramis {\"O}rl{\"u}, Alessandro
  Talamelli, Joachim Peinke \& Martin Oberlack})},  \pg{pp. 235--240}.
  \publ{Cham: Springer International Publishing}.

\bibitem[Aslangil {\em et~al.\/}(2020)Aslangil, Livescu \&
  Banerjee]{aslangilPhysicaD}
{\sc \au{Aslangil, Denis}, \au{Livescu, Daniel} \& \au{Banerjee, Arindam}}
  \yr{2020}  \at{Variable-density buoyancy-driven turbulence with asymmetric
  initial density distribution}.  \jt{Physica D: Nonlinear Phenomena}
  \bvol{406},  \pg{132444}.

\bibitem[Bailie {\em et~al.\/}(2012)Bailie, McFarland, Greenough \&
  Ranjan]{Bailie2012_RMI}
{\sc \au{Bailie, C.}, \au{McFarland, J.~A.}, \au{Greenough, J.~A.} \&
  \au{Ranjan, D.}} \yr{2012}  \at{Effect of incident shock wave strength on the
  decay of {Richtmyer-Meshkov} instability-introduced perturbations in the
  refracted shock wave}.  \jt{Shock Waves}  \bvol{22}~(6),  \pg{511--519}.

\bibitem[Baltzer \& Livescu(2019)]{baltzer_livescu_2018}
{\sc \au{Baltzer, J.~R} \& \au{Livescu, D.}} \yr{2019}  \at{Variable-density
  effects in incompressible non-buoyant shear-driven turbulent mixing layers}.
  \jt{under review J.~Fluid Mech.} .

\bibitem[Banerjee \& Andrews(2009)]{BANERJEE20093906}
{\sc \au{Banerjee, A.} \& \au{Andrews, M.~J.}} \yr{2009}  \at{{3D Simulations
  to investigate initial condition effects on the growth of Rayleigh-Taylor
  mixing}}.  \jt{International Journal of Heat and Mass Transfer}
  \bvol{52}~(17),  \pg{3906 -- 3917}, special Issue Honoring Professor D. Brian
  Spalding.

\bibitem[Banerjee {\em et~al.\/}(2010)Banerjee, Kraft \&
  Andrews]{banerjee_kraft_andrews_2010}
{\sc \au{Banerjee, A.}, \au{Kraft, W.~N.} \& \au{Andrews, M.~J.}} \yr{2010}
  \at{Detailed measurements of a statistically steady {Rayleigh-Taylor} mixing
  layer from small to high atwood numbers}.  \jt{J. Fluid Mech.}  \bvol{659},
  \pg{127--190}.

\bibitem[Batchelor {\em et~al.\/}(1992)Batchelor, Canuto \&
  Chasnov]{batchelor1992}
{\sc \au{Batchelor, G.~K.}, \au{Canuto, V.~M.} \& \au{Chasnov, J.~R.}}
  \yr{1992}  \at{Homogeneous buoyancy-generated turbulence}.  \jt{J. Fluid
  Mech.}  \bvol{235},  \pg{349--378}.

\bibitem[Brouillette(2002)]{Martin_Broillette_RMI_rev_2002}
{\sc \au{Brouillette, M.}} \yr{2002}  \at{{The Richtmyer-Meshkov Instability}}.
   \jt{Annual Review of Fluid Mechanics}  \bvol{34}~(1),  \pg{445--468}.

\bibitem[Cabot \& Cook(2006)]{Cabot_Cook}
{\sc \au{Cabot, W.} \& \au{Cook, A.}} \yr{2006}  \at{Reynolds number effects on
  {Rayleigh-Taylor} instability with possible implications for type ia
  supernovae}.  \jt{Nature Physics}  \bvol{2},  \pg{562--568}.

\bibitem[Charonko \& Prestridge(2017)]{charonko_prestridge_2017}
{\sc \au{Charonko, J.~J.} \& \au{Prestridge, K.}} \yr{2017}
  \at{Variable-density mixing in turbulent jets with coflow}.  \jt{J. Fluid
  Mech.}  \bvol{825},  \pg{887--921}.

\bibitem[Chung \& Pullin(2010)]{chung_pullin_2010}
{\sc \au{Chung, D.} \& \au{Pullin, D.~I.}} \yr{2010}  \at{Direct numerical
  simulation and large-eddy simulation of stationary buoyancy-driven
  turbulence}.  \jt{J. Fluid Mech.}  \bvol{643},  \pg{279--308}.

\bibitem[Clark \& Spitz(2005)]{Clark05}
{\sc \au{Clark, T.~T.} \& \au{Spitz, P.~B.}} \yr{2005}  \bt{Two-point
  correlation equations for variable density turbulence}. {\em Tech. Rep.\/}
  LA-12671-MS.  \org{Los Alamos Technical Report}.

\bibitem[Clemens \& Mungal(1995)]{clemens_mungal_1995}
{\sc \au{Clemens, N.~T.} \& \au{Mungal, M.~G.}} \yr{1995}  \at{Large-scale
  structure and entrainment in the supersonic mixing layer}.  \jt{J. Fluid
  Mech.}  \bvol{284},  \pg{171--216}.

\bibitem[Colgate \& White(1966)]{Colgate_1966_ApJ...143..626C}
{\sc \au{Colgate, S.~A.} \& \au{White, R.~H.}} \yr{1966}  \at{The hydrodynamic
  behavior of supernovae explosions}.  \jt{The Astrophysical Journal}
  \bvol{143},  \pg{626}.

\bibitem[Cook \& Dimotakis(2001)]{cook_dimotakis_2001}
{\sc \au{Cook, A.~W.} \& \au{Dimotakis, P.~E.}} \yr{2001}  \at{Transition
  stages of rayleigh–taylor instability between miscible fluids}.
  \jt{Journal of Fluid Mechanics}  \bvol{443},  \pg{69--99}.

\bibitem[Cook \& Zhou(2002)]{Cook02}
{\sc \au{Cook, A.~W.} \& \au{Zhou, Y.}} \yr{2002}  \at{{Energy transfer in
  Rayleigh-Taylor instability}}.  \jt{Phys. Rev. E.}  \bvol{66},  \pg{026312}.

\bibitem[Daniel {\em et~al.\/}(2018)Daniel, Livescu \&
  Ryu]{DonDaniel_ReactionForcing}
{\sc \au{Daniel, D.}, \au{Livescu, D.} \& \au{Ryu, J.}} \yr{2018}  \at{Reaction
  analogy based forcing for incompressible scalar turbulence}.  \jt{Phys. Rev.
  Fluids}  \bvol{3},  \pg{094602}.

\bibitem[Dimonte \& Schneider(1996)]{Dimontte_Schneider}
{\sc \au{Dimonte, G.} \& \au{Schneider, M.}} \yr{1996}  \at{Turbulent
  {Rayleigh-Taylor} instability experiments with variable acceleration}.
  \jt{Phys. Rev. E}  \bvol{54},  \pg{3740--3743}.

\bibitem[Dimonte {\em et~al.\/}(2004)Dimonte, Youngs, Dimits, Weber, Marinak,
  Wunsch, Garasi, Robinson, Andrews, Ramaprabhu, Calder, Fryxell, Biello,
  Dursi, MacNeice, Olson, Ricker, Rosner, Timmes, Tufo, Young \&
  Zingale]{Alpha_groupRTI2004}
{\sc \au{Dimonte, G.}, \au{Youngs, D.~L.}, \au{Dimits, A.}, \au{Weber, S.},
  \au{Marinak, M.}, \au{Wunsch, S.}, \au{Garasi, C.}, \au{Robinson, A.},
  \au{Andrews, M.~J.}, \au{Ramaprabhu, P.}, \au{Calder, A.~C.}, \au{Fryxell,
  B.}, \au{Biello, J.}, \au{Dursi, L.}, \au{MacNeice, P.}, \au{Olson, K.},
  \au{Ricker, P.}, \au{Rosner, R.}, \au{Timmes, F.}, \au{Tufo, H.}, \au{Young,
  Y.-N.} \& \au{Zingale, M.}} \yr{2004}  \at{A comparative study of the
  turbulent rayleigh–taylor instability using high-resolution
  three-dimensional numerical simulations: The alpha-group collaboration}.
  \jt{Physics of Fluids}  \bvol{16}~(5),  \pg{1668--1693}.

\bibitem[Dimotakis(2000)]{dimotakis_2000}
{\sc \au{Dimotakis, P.~E.}} \yr{2000}  \at{The mixing transition in turbulent
  flows}.  \jt{J. Fluid Mech.}  \bvol{409},  \pg{69--98}.

\bibitem[Gat {\em et~al.\/}(2017)Gat, Matheou, Chung \& Dimotakis]{gat_2017}
{\sc \au{Gat, I.}, \au{Matheou, G.}, \au{Chung, D.} \& \au{Dimotakis, P.~E.}}
  \yr{2017}  \at{Incompressible variable-density turbulence in an external
  acceleration field}.  \jt{J. Fluid Mech.}  \bvol{827},  \pg{506--535}.

\bibitem[Getling(1998)]{RBI_book_doi:10.1142/3097}
{\sc \au{Getling, A.~V.}} \yr{1998} {\em Rayleigh-Be\'nard Convection\/}.
  \publ{World Scientific}.

\bibitem[Givi(1989)]{GIVI1989}
{\sc \au{Givi, P.}} \yr{1989}  \at{Model-free simulations of turbulent reactive
  flows}.  \jt{Progress in Energy and Combustion Science}  \bvol{15}~(1),
  \pg{1 -- 107}.

\bibitem[Gull(1975)]{Gull_1975_doi:10.1093/mnras/171.2.263}
{\sc \au{Gull, S.~F.}} \yr{1975}  \at{The x-ray, optical and radio properties
  of young supernova remnants}.  \jt{Monthly Notices of the Royal Astronomical
  Society}  \bvol{171}~(2),  \pg{263--278}.

\bibitem[Haworth(2010)]{HAWORTH2010168}
{\sc \au{Haworth, D.~C.}} \yr{2010}  \at{Progress in probability density
  function methods for turbulent reacting flows}.  \jt{Progress in Energy and
  Combustion Science}  \bvol{36}~(2),  \pg{168 -- 259}.

\bibitem[Kida \& Orszag(1990)]{Kida90}
{\sc \au{Kida, S.} \& \au{Orszag, S.~A.}} \yr{1990}  \at{Energy and spectral
  dynamics in forced compressible turbulence}.  \jt{J. Sci. Comp.}  \bvol{5},
  \pg{85--125}.

\bibitem[Klimenko \& Pope(2003)]{Klimenko2003}
{\sc \au{Klimenko, A.~Y.} \& \au{Pope, S.~B.}} \yr{2003}  \at{The modeling of
  turbulent reactive flows based on multiple mapping conditioning}.  \jt{Phys.
  Fluids}  \bvol{15}~(7),  \pg{1907--1925}.

\bibitem[Kolla {\em et~al.\/}(2009)Kolla, Rogerson, Chakraborty \&
  Swaminathan]{Kolla_scalar}
{\sc \au{Kolla, H.}, \au{Rogerson, J.~W.}, \au{Chakraborty, N.} \&
  \au{Swaminathan, N.}} \yr{2009}  \at{Scalar dissipation rate modeling and its
  validation}.  \jt{Combustion Science and Technology}  \bvol{181}~(3),
  \pg{518--535}.

\bibitem[Lai {\em et~al.\/}(2018)Lai, Charonko \& Prestridge]{Lai18}
{\sc \au{Lai, C.~C.~K.}, \au{Charonko, J.~J.} \& \au{Prestridge, K.~P.}}
  \yr{2018}  \at{{A Karman-Howarth-Monin equation for variable-density
  turbulence}}.  \jt{J. Fluid Mech.}  \bvol{843},  \pg{382--418}.

\bibitem[Lesieur \& Rogallo(1989)]{Passivescalar1doi:10.1063/1.857365}
{\sc \au{Lesieur, M.} \& \au{Rogallo, R.}} \yr{1989}  \at{Large-eddy simulation
  of passive scalar diffusion in isotropic turbulence}.  \jt{Physics of Fluids
  A: Fluid Dynamics}  \bvol{1}~(4),  \pg{718--722}.

\bibitem[Linden {\em et~al.\/}(1994)Linden, Redondo \&
  Youngs]{linden_redondo_youngs_1994}
{\sc \au{Linden, P.~F.}, \au{Redondo, J.~M.} \& \au{Youngs, D.~L.}} \yr{1994}
  \at{Molecular mixing in {Rayleigh-Taylor} instability}.  \jt{J. Fluid Mech.}
  \bvol{265},  \pg{97--124}.

\bibitem[Lindl(1995)]{Lindl_1995_doi:10.1063/1.871025}
{\sc \au{Lindl, J.}} \yr{1995}  \at{{Development of the indirect-drive approach
  to inertial confinement fusion and the target physics basis for ignition and
  gain}}.  \jt{Physics of Plasmas}  \bvol{2}~(11),  \pg{3933--4024}.

\bibitem[Lindl(1998)]{lindl1998inertial}
{\sc \au{Lindl, J.~D.}} \yr{1998} {\em Inertial Confinement Fusion: The Quest
  for Ignition and Energy Gain Using Indirect Drive\/}.  \publ{AIP Press}.

\bibitem[Livescu(2013)]{livescu2013nst}
{\sc \au{Livescu, D.}} \yr{2013}  \at{Numerical simulations of two-fluid
  turbulent mixing at large density ratios and applications to the
  {Rayleigh-Taylor} instability}.  \jt{Phil. Trans. R. Soc. A}  \bvol{371},
  \pg{20120185}.

\bibitem[Livescu(2020)]{Livescu2020}
{\sc \au{Livescu, D.}} \yr{2020}  \at{Turbulence with large thermal and
  compositional density variations}.  \jt{Annu. Rev. Fluid Mech.}  \bvol{52},
  \pg{309--341}.

\bibitem[Livescu {\em et~al.\/}(2014)Livescu, Canada, Kanov, Burns, {IDIES
  staff} \& Pulido]{JHTD}
{\sc \au{Livescu, D.}, \au{Canada, C.}, \au{Kanov, K.}, \au{Burns, R.},
  \au{{IDIES staff}} \& \au{Pulido, J.}} \yr{2014} Homogeneous buoyancy driven
  turbulence dataset.

\bibitem[Livescu {\em et~al.\/}(2000)Livescu, Jaberi \&
  Madnia]{livescu_jaberi_madnia_2000}
{\sc \au{Livescu, D.}, \au{Jaberi, F.~A.} \& \au{Madnia, C.~K.}} \yr{2000}
  \at{Passive-scalar wake behind a line source in grid turbulence}.
  \jt{Journal of Fluid Mechanics}  \bvol{416},  \pg{117–149}.

\bibitem[Livescu \& Ristorcelli(2007)]{livescu2007}
{\sc \au{Livescu, D.} \& \au{Ristorcelli, J.~R.}} \yr{2007}
  \at{Buoyancy-driven variable-density turbulence}.  \jt{J. Fluid Mech.}
  \bvol{591},  \pg{43--71}.

\bibitem[Livescu \& Ristorcelli(2008)]{livescu2008}
{\sc \au{Livescu, D.} \& \au{Ristorcelli, J.~R.}} \yr{2008}
  \at{Variable-density mixing in buoyancy-driven turbulence}.  \jt{J. Fluid
  Mech.}  \bvol{605},  \pg{145--180}.

\bibitem[Livescu \& Ristorcelli(2009)]{livescu2009mav}
{\sc \au{Livescu, D.} \& \au{Ristorcelli, J.~R.}} \yr{2009}  \at{Mixing
  asymmetry in variable density turbulence}.  \bt{In {\em Advances in
  turbulence XII\/} (ed. \ed{B.~Eckhardt})}, ,  \vol{vol. 132},  \pg{pp.
  545--548}.  \publ{Springer}.

\bibitem[Livescu {\em et~al.\/}(2009)Livescu, Ristorcelli, Gore, Dean, Cabot \&
  Cook]{Livescu_Cabot_Cook}
{\sc \au{Livescu, D.}, \au{Ristorcelli, J.~R.}, \au{Gore, R.~A.}, \au{Dean,
  S.~H.}, \au{Cabot, W.~H.} \& \au{Cook, A.~W.}} \yr{2009}  \at{{High-Reynolds
  number Rayleigh-Taylor turbulence}}.  \jt{Journal of Turbulence}  \bvol{10},
  \pg{N13}.

\bibitem[Livescu {\em et~al.\/}(2010)Livescu, Ristorcelli, Petersen \&
  Gore]{Livescu_HVDT_New_Ph}
{\sc \au{Livescu, D.}, \au{Ristorcelli, J.~R.}, \au{Petersen, M.~R.} \&
  \au{Gore, R.~A.}} \yr{2010}  \at{New phenomena in variable-density
  {Rayleigh-Taylor} turbulence}.  \jt{Physica Scripta}  \bvol{2010}~(T142),
  \pg{014015}.

\bibitem[Livescu \& Wei(2012)]{livescu_variable_accel_2012}
{\sc \au{Livescu, D.} \& \au{Wei, T.}} \yr{2012}  \at{{Direct Numerical
  Simulations of Rayleigh-Taylor instability with gravity reversal}}.  \bt{In
  {\em Proceedings of the Seventh International Conference on Computational
  Fluid Dynamics (ICCFD7)\/}},  \pg{p. paper number 2304}.  \publ{Big Islannd,
  HI, July 9-13, 2012}.

\bibitem[Livescu {\em et~al.\/}(2019)Livescu, Wei \&
  Brady]{livescu_variable_accel_2019}
{\sc \au{Livescu, D.}, \au{Wei, T.} \& \au{Brady, P.T.}} \yr{2019}
  \at{{Rayleigh-Taylor instability with gravity reversal}}.  \jt{Phys. D}
  \pg{p. submitted}.

\bibitem[Livescu {\em et~al.\/}(2011)Livescu, Wei \&
  Petersen]{Livescu_variable_accel_2011}
{\sc \au{Livescu, D.}, \au{Wei, T.} \& \au{Petersen, M.~R.}} \yr{2011}
  \at{Direct numerical simulations of {Rayleigh-Taylor} instability}.
  \jt{Journal of Physics: Conference Series}  \bvol{318}~(8),  \pg{082007}.

\bibitem[Lumley(1967)]{Lumley_67}
{\sc \au{Lumley, J.~L.}} \yr{1967}  \at{Similarity and the turbulent energy
  spectrum}.  \jt{The Physics of Fluids}  \bvol{10}~(4),  \pg{855--858}.

\bibitem[Meshkov(1969)]{Meshkov1969}
{\sc \au{Meshkov, E.~E.}} \yr{1969}  \at{Instability of the interface of two
  gases accelerated by a shock wave}.  \jt{Fluid Dynamics}  \bvol{4}~(5),
  \pg{101--104}.

\bibitem[Molchanov(2004)]{MOLCHANOV2004559}
{\sc \au{Molchanov, O.~A.}} \yr{2004}  \at{On the origin of low- and
  middler-latitude ionospheric turbulence}.  \jt{Physics and Chemistry of the
  Earth, Parts A/B/C}  \bvol{29}~(4),  \pg{559 -- 567}, seismo Electromagnetics
  and Related Phenomena.

\bibitem[Nakai \& Mima(2004)]{Nakai_2004_0034-4885-67-3-R04}
{\sc \au{Nakai, S.} \& \au{Mima, K.}} \yr{2004}  \at{Laser driven inertial
  fusion energy: present and prospective}.  \jt{Reports on Progress in Physics}
   \bvol{67}~(3),  \pg{321}.

\bibitem[Nakai \& Takabe(1996)]{Nakai_1996_0034-4885-59-9-002}
{\sc \au{Nakai, S.} \& \au{Takabe, H.}} \yr{1996}  \at{Principles of inertial
  confinement fusion - physics of implosion and the concept of inertial fusion
  energy}.  \jt{Reports on Progress in Physics}  \bvol{59}~(9),  \pg{1071}.

\bibitem[Nishihara {\em et~al.\/}(2010)Nishihara, Wouchuk, Matsuoka, Ishizaki
  \& Zhakhovsky]{Nishihara_2010_RMI}
{\sc \au{Nishihara, K.}, \au{Wouchuk, J.~G.}, \au{Matsuoka, C.}, \au{Ishizaki,
  R.} \& \au{Zhakhovsky, V.~V.}} \yr{2010}  \at{{Richtmyer-Meshkov instability:
  theory of linear and nonlinear evolution}}.  \jt{Philosophical Transactions
  of the Royal Society of London A: Mathematical, Physical and Engineering
  Sciences}  \bvol{368}~(1916),  \pg{1769--1807},  \arxiv{arXiv:
  http://rsta.royalsocietypublishing.org/content/368/1916/1769.full.pdf}.

\bibitem[Nouri {\em et~al.\/}(2019)Nouri, Givi \& Livescu]{Nouri_etal_PAS19}
{\sc \au{Nouri, A.~G.}, \au{Givi, P.} \& \au{Livescu, D.}} \yr{2019}
  \at{Modeling and simulation of turbulent nuclear flames in type ia
  supernovae}.  \jt{Prog. Aerosp. Sci.}  \bvol{108},  \pg{156--179}.

\bibitem[Overholt \& Pope(1996)]{Overholt_Pope_passive_scalar_mean_gradient}
{\sc \au{Overholt, M.~R.} \& \au{Pope, S.~B.}} \yr{1996}  \at{Direct numerical
  simulation of a passive scalar with imposed mean gradient in isotropic
  turbulence}.  \jt{Physics of Fluids}  \bvol{8}~(11),  \pg{3128--3148}.

\bibitem[Pal {\em et~al.\/}(2018)Pal, Kurien, Clark, Aslangil \&
  Livescu]{Pal18}
{\sc \au{Pal, N.}, \au{Kurien, S.}, \au{Clark, T.~T.}, \au{Aslangil, D.} \&
  \au{Livescu, D.}} \yr{2018}  \at{{Two-point spectral model for
  variable-density homogeneous turbulence}}.  \jt{Phys. Rev. Fluids}  \bvol{3},
   \pg{124608}.

\bibitem[Pope(1985)]{POPE1985119}
{\sc \au{Pope, S.~B.}} \yr{1985}  \at{Pdf methods for turbulent reactive
  flows}.  \jt{Progress in Energy and Combustion Science}  \bvol{11}~(2),
  \pg{119 -- 192}.

\bibitem[Pope(2000)]{pope2000turbulent}
{\sc \au{Pope, S.~B.}} \yr{2000} {\em Turbulent Flows\/}.  \publ{Cambridge
  University Press}.

\bibitem[Ramaprabhu {\em et~al.\/}(2013)Ramaprabhu, Karkhanis \&
  Lawrie]{Ramaprabhu_ADA}
{\sc \au{Ramaprabhu, P.}, \au{Karkhanis, V.} \& \au{Lawrie, A. G.~W.}}
  \yr{2013}  \at{{The Rayleigh-Taylor Instability driven by an
  accel-decel-accel profile}}.  \jt{Physics of Fluids}  \bvol{25}~(11),
  \pg{115104}.

\bibitem[Rao {\em et~al.\/}(2017)Rao, Caulfield \&
  Gibbon]{rao_caulfield_gibbon_2017}
{\sc \au{Rao, P.}, \au{Caulfield, C.~P.} \& \au{Gibbon, J.~D.}} \yr{2017}
  \at{Nonlinear effects in buoyancy-driven variable-density turbulence}.
  \jt{J. Fluid Mech.}  \bvol{810},  \pg{362--377}.

\bibitem[Rayleigh(1884)]{LordRayleigh}
{\sc \au{Rayleigh}} \yr{1884}  \at{Investigation of the character of the
  equilibrium of an incompressible heavy fluid of variable density}.
  \jt{Proceedings of the London Mathematical Society}  \bvol{s1-14}~(1),
  \pg{170--177}.

\bibitem[Richtmyer(1960)]{Richtmey_doi:10.1002/cpa.3160130207}
{\sc \au{Richtmyer, R.~D.}} \yr{1960}  \at{Taylor instability in shock
  acceleration of compressible fluids}.  \jt{Communications on Pure and Applied
  Mathematics}  \bvol{13}~(2),  \pg{297--319}.

\bibitem[Ristorcelli \& Clark(2004)]{ristorcelli_clark_2004}
{\sc \au{Ristorcelli, J.~R.} \& \au{Clark, T.~T.}} \yr{2004}
  \at{{Rayleigh-Taylor} turbulence: self-similar analysis and direct numerical
  simulations}.  \jt{J. Fluid Mech.}  \bvol{507},  \pg{213--253}.

\bibitem[Sandoval(1995)]{SandovalPhd}
{\sc \au{Sandoval, D.~L.}} \yr{1995}  \at{The dynamics of variable density
  turbulence}. PhD thesis, University of Washington.

\bibitem[Sandoval {\em et~al.\/}(1997)Sandoval, Clark \& Riley]{Sandoval1997}
{\sc \au{Sandoval, D.~L.}, \au{Clark, T.~T.} \& \au{Riley, J.~J.}} \yr{1997}
  \at{Buoyancy-generated variable-density turbulence}.  \bt{In {\em {IUTAM}
  Symposium on Variable Density Low-Speed Turbulent Flows: Proceedings of the
  {IUTAM} Symposium held in Marseille, France, 8-10 July 1996\/} (ed.
  \ed{L.~Fulachier, J.~L. Lumley \& F.~Anselmet})},  \pg{pp. 173--180}.
  \publ{Dordrecht: Springer Netherlands}.

\bibitem[Schilling \& Latini(2010)]{SCHILLING2010595}
{\sc \au{Schilling, O.} \& \au{Latini, M.}} \yr{2010}  \at{{High-order WENO
  simulations of three-dimensional reshocked Richtmyer-Meshkov instability to
  late times: dynamics, dependence on initial conditions, and comparisons to
  experimental data}}.  \jt{Acta Mathematica Scientia}  \bvol{30}~(2),  \pg{595
  -- 620}, {Mathematics dedicated to professor James Glimm on the occasion of
  his $75^{th}$ birthday}.

\bibitem[Schilling {\em et~al.\/}(2007)Schilling, Latini \& Don]{O_S_RMI}
{\sc \au{Schilling, O.}, \au{Latini, M.} \& \au{Don, W.~S.}} \yr{2007}
  \at{{Physics of reshock and mixing in single-mode Richtmyer-Meshkov
  instability}}.  \jt{Phys. Rev. E}  \bvol{76},  \pg{026319}.

\bibitem[Schwarzkopf {\em et~al.\/}(2016)Schwarzkopf, Livescu, Baltzer, Gore \&
  Ristorcelli]{schwarzkopf2016tls}
{\sc \au{Schwarzkopf, J.~D.}, \au{Livescu, D.}, \au{Baltzer, J.~R.}, \au{Gore,
  R.~A.} \& \au{Ristorcelli, J.~R.}} \yr{2016}  \at{A two-length scale
  turbulence model for single-phase multi-fluid mixing}.  \jt{Flow Turbul.
  Combust.}  \bvol{96},  \pg{1--43}.

\bibitem[Sellers \&
  Chandra(1997)]{Sellers_Chandra_1997_doi:10.1108/02644409710157596}
{\sc \au{Sellers, C.~L.} \& \au{Chandra, S.}} \yr{1997}  \at{Compressibility
  effects in modelling turbulent high speed mixing layers}.  \jt{Engineering
  Computations}  \bvol{14}~(1),  \pg{5--13}.

\bibitem[Soulard \& Griffond(2012)]{Soulard12}
{\sc \au{Soulard, O.} \& \au{Griffond, J.}} \yr{2012}  \at{{Inertial-range
  anisotropy in Rayleigh-Taylor turbulence}}.  \jt{Phys. Fluids}  \bvol{24},
  \pg{025101}.

\bibitem[Taylor(1950)]{LordTaylor}
{\sc \au{Taylor}} \yr{1950}  \at{The instability of liquid surfaces when
  accelerated in a direction perpendicular to their planes. i}.
  \jt{Proceedings of the Royal Society of London A: Mathematical, Physical and
  Engineering Sciences}  \bvol{201}~(1065),  \pg{192--196}.

\bibitem[Wunsch \&
  Ferrari(2004)]{Wunsch_2004_doi:10.1146/annurev.fluid.36.050802.122121}
{\sc \au{Wunsch, C.} \& \au{Ferrari, R.}} \yr{2004}  \at{Vertical mixing,
  energy, and the general circulation of the oceans}.  \jt{Annual Review of
  Fluid Mechanics}  \bvol{36}~(1),  \pg{281--314}.

\bibitem[Youngs(1991)]{Youngs91}
{\sc \au{Youngs, D.~L.}} \yr{1991}  \at{{Three-dimensional numerical simulation
  of turbulent mixing by Rayleigh-Taylor instability}}.  \jt{Physics of Fluids
  A: Fluid Dynamics}  \bvol{3}~(5),  \pg{1312--1320}.

\bibitem[Zhao \& Aluie(2018)]{Zhao_Hussein_2018}
{\sc \au{Zhao, D.} \& \au{Aluie, H.}} \yr{2018}  \at{Inviscid criterion for
  decomposing scales}.  \jt{Phys. Rev. Fluids}  \bvol{3},  \pg{054603}.

\bibitem[Zhou(2017{\natexlab{{\em a\/}}})]{ZHOU2017_1}
{\sc \au{Zhou, Y.}} \yr{2017{\natexlab{{\em a\/}}}}  \at{{Rayleigh-Taylor} and
  {Richtmyer-Meshkov} instability induced flow, turbulence, and mixing. i}.
  \jt{Physics Reports}  \bvol{720-722},  \pg{1 -- 136}, rayleigh-Taylor and
  {Richtmyer-Meshkov} instability induced flow, turbulence, and mixing. I.

\bibitem[Zhou(2017{\natexlab{{\em b\/}}})]{ZHOU2017_2}
{\sc \au{Zhou, Y.}} \yr{2017{\natexlab{{\em b\/}}}}  \at{{Rayleigh-Taylor} and
  {Richtmyer-Meshkov} instability induced flow, turbulence, and mixing. ii}.
  \jt{Physics Reports}  \bvol{723-725},  \pg{1 -- 160}, {Rayleigh-Taylor} and
  {Richtmyer-Meshkov} instability induced flow, turbulence, and mixing. II.

\end{thebibliography}

\section{Appendix-A}

To derive a transport equation for the jMDF, we start from the fine-grained, one-point, one-time Eulerian velocity–
density joint-PDF ($f^*_{\rho^*,u^*_i}$) which is defined by:
\begin{equation}
    f^*_{\rho^*,u^*_i}=\delta(u^*_i-V_i)\delta(\rho^*-R),
\end{equation}
and use the formula $f_{\rho^*,u^*_i}=\langle f^*_{\rho^*,u^*_i}\rangle$, where $f_{\rho^*,u^*_i}$ is the one-point, one-time Eulerian joint-PDF of the velocity and density fields. Also, $\langle Q(\rho^*,u^*_i)f^*_{\rho^*,u^*_i}\rangle=\langle Q(\rho^*,u^*_i)\vert_{V_i,R}\rangle$ is the conditional expectation of any function $Q(\rho^*,{u^*_i};{x_i}, t)$ at $u^*_i=V_i$ and $\rho^*=R$. For a homogeneous flow, averages are constant in space.

Then, the transport equation for $f^*_{\rho^*,u^*_i}$ can be derived by explicitly calculating the material derivative:
\begin{equation}
    \rho^* \frac{Df^*_{\rho^*,u^*_i}}{Dt}=\rho^*\frac{\partial f^*_{\rho^*,u^*_i}}{\partial t}+\rho^*u^*_{j}\frac{\partial f^*_{\rho^*,u^*_i}}{\partial x_j},
\end{equation}
Using the chain rule:
\begin{equation}\label{Eq:f_t}
    \frac{\partial f^*_{\rho^*,u^*_i}}{\partial t}=-\frac{\partial f^*_{\rho^*,u^*_i}}{\partial V_j}\frac{\partial u^*_j}{\partial t}-\frac{\partial f^*_{\rho^*,u^*_i}}{\partial R}\frac{\partial \rho^*}{\partial t},
\end{equation}
and
\begin{equation}\label{Eq:f_x}
\frac{\partial f^*_{\rho^*,u^*_i}}{\partial x_i}=-\frac{\partial f^*_{\rho^*,u^*_i}}{\partial V_j}\frac{\partial u^*_j}{\partial x_i}-\frac{\partial f^*_{\rho^*,u^*_i}}{\partial R}\frac{\partial \rho^*}{\partial x_i}.
\end{equation}

By using equations \ref{Eq:f_t} and \ref{Eq:f_x}, the transport equation can be written as:
\begin{equation}\label{Eq:transport_f_star}
\begin{split}
    \rho^*\frac{\partial f^*_{\rho^*,u^*_i}}{\partial t}+\rho^*u^*_{j}\frac{\partial f^*_{\rho^*,u^*_i}}{\partial x_j} &=-\rho^*\frac{\partial f^*_{\rho^*,u^*_i}}{\partial V_j}\frac{\partial u^*_j}{\partial t}-\rho^*u^*_i\frac{\partial f^*_{\rho^*,u^*_i}}{\partial V_j}\frac{\partial u^*_j}{\partial x_i}\\
    &-\rho^*\frac{\partial f^*_{\rho^*,u^*_i}}{\partial R}\frac{\partial \rho^*}{\partial t}-\rho^*u^*_i\frac{\partial f^*_{\rho^*,u^*_i}}{\partial R}\frac{\partial \rho*}{\partial x_i}\\
&=-\frac{\partial}{\partial V_j}\Bigg[\rho^*\Bigg(\frac{\partial u^*_j}{\partial t}+u^*_i\frac{\partial u^*_j}{\partial x_i}\Bigg)f^*_{\rho^*,u^*_i}\Bigg]\\
    &-\frac{\partial}{\partial R}\Bigg[\rho^*\Bigg(\frac{\partial \rho^*}{\partial t}+u^*_i\frac{\partial \rho^*}{\partial x_i}\Bigg)f^*_{\rho^*,u^*_i}\Bigg]\\
    &=-\frac{\partial}{\partial V_j}\Bigg[\rho^*\frac{D u^*_j}{Dt}f^*_{\rho^*,u^*_i}\Bigg]-\frac{\partial}{\partial R}\Bigg[\rho^*\frac{D\rho^*}{Dt}f^*_{\rho^*,u^*_i}\Bigg],
    \end{split}
\end{equation}

By using the continuity equation, the left hand side of (\ref{Eq:transport_f_star}) becomes $\frac{\partial (\rho^* f^*_{\rho^*,u^*_i})}{\partial t}+\frac{\partial (\rho^*u^*_{j} f^*_{\rho^*,u^*_i})}{\partial x_j}$. After taking the average and using the sifting property of the delta function, such that $\rho^*$ can be replaced by $R$ and $u^*_i$ by $V_i$, equation (\ref{Eq:transport_f_star}) turns into:
\begin{equation}
    \frac{\partial (R f_{\rho^*,u^*_i})}{\partial t}+ \frac{\partial (RV_j f_{\rho^*,u^*_i})}{\partial x_j}=-\frac{\partial}{\partial V_i}\Bigg[ Rf_{\rho^*,u^*_i}\Big\langle\frac{D u^*_i}{Dt}\Big\rangle\Big\vert_{V_i,R}\Bigg]-\frac{\partial}{\partial R}\Bigg[Rf_{\rho^*,u^*_i}\Big\langle \frac{D\rho^*}{Dt}\Big\rangle\Big\vert_{V_i,R}\Bigg],
\end{equation}

Using the relation between the joint-PDF and MDF, which for homogeneous flows becomes  $\mathcal{F}_{u_i\rho^*}({V_i,R; t})=\mathcal{F}=Rf_{u_i\rho^*}({V_i,R; t})$, the equation turns into:
\begin{equation}\label{Eq:F_t}
    \frac{\partial \mathcal{F}}{\partial t}= -\frac{\partial}{\partial V_i}\Bigg[ \mathcal{F}\Big\langle\frac{D u^*_i}{Dt}\Big\rangle\Big\vert_{V_i,R} \Bigg]-\frac{\partial}{\partial R}\Bigg[\mathcal{F}\Big\langle \frac{D\rho^*}{Dt}\Big\rangle\Big\vert_{V_i,R}\Bigg].
\end{equation}
Finally, from equations \ref{Eq:continuity} and \ref{Eq:moment} the material derivatives of the density and velocity fields can be written as:
\begin{equation}\label{Eq:continuity_ap}
\frac{D\rho^*}{Dt}=\rho^*_{,t}+\rho^*_{,j}u^*_j=-\rho^*u^*_{j,j}   
\end{equation}

\begin{equation} \label{Eq:D_ui_ap}
\frac{Du^*_i}{Dt}=(u^*_i)_{,t}+u^*_ju^*_{i,j}=-\frac{1}{\rho ^*}(p_{,i}+P_{,i}-\tau^*_{ij,j})+\frac{1}{Fr^2}g_i.
\end{equation}
Upon combining equations \ref{Eq:F_t}, \ref{Eq:continuity_ap} and \ref{Eq:D_ui_ap}, the final version of the transport equation of $\mathcal{F}$ can be written as:

\begin{equation}\label{Eq:joint_ap}
\begin{split}
      \frac{d\mathcal{F}}{dt}+ \frac{dV_j\mathcal{F}}{d x_j}&=-\frac{\partial}{\partial V_i}\bigg[\mathcal{F} \bigg\langle-\frac{p_{,i}}{\rho^*}\bigg\vert_{V_i,R} -\frac{P_{,i}}{\rho^*}\bigg\vert V_i, R+\frac{\tau^*_{ij,j}}{\rho^*}\bigg\vert_{V_i,R} +\frac{1}{Fr^2}g_i \bigg\vert_{V_i,R} \bigg\rangle\bigg] -\\
    &\frac{\partial}{\partial R} \bigg[\mathcal{F} \bigg\langle -\rho^*u^*_{j,j} \bigg\vert_{V_i,R} \bigg\rangle\bigg].
\end{split}
\end{equation}
\end{document}